\pdfoutput=1
\documentclass[11pt]{article}

\usepackage[english]{babel}
\usepackage{xcolor}
\usepackage{graphicx}

\usepackage[colorlinks,allcolors=blue,linktocpage=true]{hyperref}

\usepackage{cite}
\usepackage{subfigure}
\usepackage{enumitem}
\usepackage{amsmath}

\usepackage{placeins}
\usepackage{booktabs}

\usepackage[top=15mm,bottom=12mm,left=30mm,right=30mm,foot=12mm,includefoot]{geo
metry}

\allowdisplaybreaks[3]

\numberwithin{equation}{section}

\newcommand{\sect}[1]{section~#1}
\newcommand{\sects}[1]{sections~#1}
\newcommand{\app}[1]{appendix~#1}

\newcommand{\fig}[1]{figure~#1}
\newcommand{\figs}[1]{figures~#1}

\newcommand{\tab}[1]{table~#1}

\newcommand{\eqn}[1]{equation~#1}
\newcommand{\eqs}[1]{equations~#1}

\newcommand{\as}{a_s}

\newcommand{\tev}{\operatorname{TeV}}
\newcommand{\gev}{\operatorname{GeV}}

\newcommand{\ms}{\mskip 1.5mu}
\newcommand{\bs}{\mskip -1.5mu}

\newcommand{\tvec}[1]{\boldsymbol{#1}}

\newcommand{\msbar}{\overline{\text{MS}}}

\newcommand{\conv}[1]{\underset{#1}{\otimes}}

\newcommand{\nf}{n_F}

\newcommand{\mQ}{m_Q}

\newcommand{\qbar}{\overline{q}}
\newcommand{\Qbar}{\overline{Q}}

\newcommand{\sbar}{\overline{s}}
\newcommand{\cbar}{\overline{c}}
\newcommand{\bbar}{\overline{b}}
\newcommand{\tbar}{\overline{t}}
\newcommand{\abar}{\overline{a}}

\newcommand{\rev}[1]{#1}

\begin{document}

\begin{flushright}
DESY-22-197 \\
\href{https://arxiv.org/abs/2212.07736}{arXiv:2212.07736 [hep-ph]}
\end{flushright}

\begin{center}
\vspace{4\baselineskip}
\textbf{\Large Quark mass effects in double parton distributions} \\
\vspace{3\baselineskip}
Markus~Diehl$^{\ms 1}$, Riccardo Nagar$^{\ms 2}$ and Peter Pl{\"o}{\ss}l$^{\ms 1}$
\end{center}

\vspace{2\baselineskip}

\noindent
${}^{1}$ Deutsches Elektronen-Synchrotron DESY, Notkestr.~85, 22607 Hamburg, Germany\\
${}^{2}$ Universit\`a degli Studi di Milano-Bicocca \& INFN Sezione di Milano-Bicocca,\\
\phantom{${}^{2}$} Piazza della Scienza~3, Milano 20126, Italy

\vspace{3\baselineskip}

\parbox{0.9\textwidth}{
}

Double parton distributions can be computed from the perturbative splitting of one parton into two if the distance between the two observed partons is small.  We develop schemes to take into account quark mass effects in this computation, and we study these schemes numerically at leading order in the strong coupling.  Furthermore, we investigate in detail the structure of the next-to-leading order corrections to the splitting kernels that include quark mass effects.

\vfill

\newpage

\tableofcontents

\begin{center}
\rule{0.6\textwidth}{0.3pt}
\end{center}


%
\section{Introduction}
\label{sec:intro}

To make the best physics use of measurements at the Large Hadron Collider, it is important to understand the strong-interaction dynamics of hadron collisions to the best degree possible.  Among many other things, this motivates the study of double parton scattering (DPS), where in a single proton-proton collision two partons in each proton undergo a hard scattering.  Generically, the importance of this mechanism increases with collision energy, so that it is relevant for future hadron colliders at least as much as for the LHC.

Significant progress in the theory description of DPS has been made in the last decade \cite{Gaunt:2009re, Blok:2010ge, Gaunt:2011xd, Ryskin:2011kk, Blok:2011bu, Diehl:2011yj, Manohar:2012jr, Manohar:2012pe, Ryskin:2012qx, Gaunt:2012dd, Blok:2013bpa, Diehl:2015bca, Diehl:2017kgu, Diehl:2018wfy, Cabouat:2019gtm, Cabouat:2020ssr}, and there is an increasing set of DPS measurements from the Tevatron and the LHC, see \cite{Abe:1997xk, Abazov:2015nnn, Aaij:2016bqq, Aaboud:2018tiq, CMS:2022pio} and references therein.  As an example for the huge body of phenomenological work, let us mention the recent study \cite{Fedkevych:2020cmd} of four-jet production.  A detailed account of theoretical and experimental aspects of DPS is given in the monograph~\cite{Bartalini:2017jkk}.

Double parton distributions (DPDs) are the nonperturbative ingredients in the calculation of DPS cross sections.  A corresponding factorisation formalism has been developed in \cite{Diehl:2017kgu} and underlies the present work.  In this formalism, a DPD depends on the momentum fractions $x_1$ and $x_2$ of the two partons extracted from a proton, and on the transverse distance $y$ between these two partons.  It also depends on the renormalisation scales $\mu_1$ and $\mu_2$ associated with each parton.  We assume them to be equal throughout this work, since taking $\mu_1 \neq \mu_2$ would not add much to our discussion.  We will limit ourselves to DPDs without colour correlations between the two partons; an extension of our results to the colour correlated case should be possible but would require additional work.

Consider a DPS process with the two hard scatterings taking place at a scale $\mu$.  The DPDs of the two colliding protons enter the factorisation formula in the form of a double parton luminosity
\begin{align}
   \label{eq:DPD-lumi-0}
\int \mathrm{d}^2 y \; \theta(y - y_{\min}) \,
      F_{a_1 a_2}(x_{1a}, x_{2a}, y; \mu) \,
      F_{b_1 b_2}(x_{1b}, x_{2b}, y; \mu) \,,
\end{align}
where $a_i$ and $b_i$ label the species and polarisation of the scattering partons.  The lower cutoff $y_{\min}$ in the integration over $y$ is to be taken of order $1/\mu$.  In the formalism of \cite{Diehl:2017kgu}, a subtraction term depending on the same cutoff appears in the overall cross section and removes double counting between the cross sections for single and for double parton scattering.  The  dependence on $y_{\text{min}}$ cancels in the sum of terms, up to higher-order corrections beyond the accuracy of the calculation.

It important to understand which distances $y$ are most relevant in the integral \eqref{eq:DPD-lumi-0}.  In the limit of perturbatively small $y$, one may decompose the DPD as $F = F^{\text{spl}} + F^{\text{intr}}$.  The first term represents the case where the two extracted partons originate from a single parton, which can be computed in terms of a perturbative splitting kernel and an ordinary parton distribution (PDF).  The second term represents the ``intrinsic'' two-parton content of the proton and can be expressed in terms of a twist-four distribution.  In fixed-order perturbation theory, one obtains a power behaviour $F^{\text{spl}} \sim y^{-2}$ and $F^{\text{intr}} \sim y^{0}$ at small $y$.  This suggests that the double parton luminosity \eqref{eq:DPD-lumi-0} should be strongly dominated by the region where $y$ is close to $y_{\text{min}}$.  In the overall cross section, however, the contribution from this region is largely cancelled by the subtraction term just mentioned.  \rev{Furthermore,} it was found in \cite{Diehl:2017kgu} that the stated power behaviour can be substantially flattened by evolution from the scale $\sim 1/y$ at which a fixed-order calculation of $F^{\text{spl}}$ is reliable to the scale $\mu$ of the hard scatterings.  Finally, for some parton combinations, $F^{\text{spl}}$ is suppressed compared with $F^{\text{intr}}$ by $\alpha_s^2$ rather than by $\alpha_s$.  In summary, the relevant $y$ region in a cross section with hard scale $\mu$ extends to values substantially larger than $1/\mu$, with details depending on kinematics and the parton channels.  This region may include both perturbative distances, where one has a natural decomposition $F = F^{\text{spl}} + F^{\text{intr}}$, and nonperturbatively large $y$, where one must resort to modelling $F$.  Guidance for such modelling may for instance be obtained from quark model studies \cite{Chang:2012nw, Rinaldi:2013vpa, Broniowski:2013xba,  Rinaldi:2014ddl, Broniowski:2016trx, Kasemets:2016nio, Rinaldi:2016jvu, Rinaldi:2016mlk, Rinaldi:2018zng, Courtoy:2019cxq, Broniowski:2019rmu}, from calculations in lattice QCD \cite{Bali:2020mij, Bali:2021gel}, or from the sum rules that DPDs must obey \cite{Gaunt:2009re, Golec-Biernat:2014bva, Golec-Biernat:2015aza, Diehl:2020xyg, Fedkevych:2022myf, Golec-Biernat:2022wkx}.

In the present work, we focus on the perturbative splitting contribution $F^{\text{spl}}$.  Its relevance for DPS phenomenology has been highlighted in different theory formalisms \cite{Ryskin:2012qx, Snigirev:2014eua, Blok:2013bpa, Blok:2015rka, Blok:2015afa, Golec-Biernat:2014nsa, Gaunt:2014rua}, including the one we are using here \cite{Diehl:2017kgu}.

Heavy-quark masses play a quite nontrivial role in the computation of DPS.
The number~$\nf$ of active quark flavours in a double parton luminosity depends on $\mu$ and on the particular scheme used to compute the hard-scattering cross sections.  If one has for instance $m_b \ll \mu \ll m_t$, then it is appropriate to use DPDs with $\nf = 5$ active flavours, and $m_b$ and $m_c$ can be neglected in the hard scattering.  However, as explained above, the relevant $y$ range in the double parton luminosity may include regions where $y$ is comparable to $1/m_b$ or $1/m_c$.  In these regions, computing $F^{\text{spl}}$ with massless charm and bottom quarks is a poor approximation.  This calls for a more realistic scheme for evaluating $F^{\text{spl}}$ across the relevant range of $y$ in the perturbative region.  To develop and assess such schemes is the purpose of the present work.

There is a certain similarity between the problem just stated and the role of $\nf$ in the computation of transverse-momentum dependent single-parton distributions (TMDs).  Taken in impact parameter space, a TMD $f(x,b;\mu)$ depends on a transverse distance just like $F(x_1,x_2,y;\mu)$, and at small impact parameter $b$ the TMD can be computed in terms of a perturbative matching kernel and a PDF.  The treatment of heavy-quark masses in this case was investigated in detail in \cite{Pietrulewicz:2017gxc}.  An important difference with the present case is that the relevant distances of $b$ in a TMD cross section are of order $1/q_T$, where $q_T$ is a measured transverse momentum, whereas there is no such simple relation between the process kinematics and the relevant distances $y$ in DPS.

This paper is structured as follows.  In \sect{\ref{sec:basics}}, we recall some theory results that will be needed in our work.  In \sect{\ref{sec:schemes}}, we present general schemes for treating heavy flavours in splitting DPDs.  In \sect{\ref{sec:LO-numerics}} these schemes are studied numerically at leading order (LO) in  $\alpha_s$.  In \sects{\ref{sec:NLO-kernels-constraints}} and \ref{sec:NLO-kernels-constraints-two-flavours} we analyse the structure of massive splitting kernels at next-to-leading order (NLO) in $\alpha_s$, first for the case of a single heavy quark and then for the case of charm and bottom.  In \sect{\ref{sec:sum-rules}}, we show how the number and momentum sum rules for DPDs imply corresponding sum rules for the massive splitting kernels.  This is used in \sect{\ref{sec:NLO-kernels-model}} to construct a model ansatz for these kernels at NLO.  Our main results are summarised in \sect{\ref{sec:conclusions}}.  Additional numerical examples are given in \app{\ref{app:add-numerics}}, and technical material relevant to the model in \sect{\ref{sec:NLO-kernels-model}} can be found in \app{\ref{app:basis-functions}}.

%
\section{Theory background}
\label{sec:basics}

In this section, we set up our notation and recall basic results about scale dependence, flavour matching, and the splitting mechanism for DPDs.

To denote PDFs and DPDs for $\nf$ active flavours, we respectively write $f_{a_1}^{\nf}(x; \mu)$ and $F^{\nf}_{a_1 a_2}(x_1,x_2,y; \mu)$, where $a_1$ and $a_2$ specify the flavour and polarisation of a parton.  A generic massive quark flavour is denoted by $Q$, and light quark flavours by $q$ or $q'$.

We consider only colour-singlet DPDs in this work.  Unless stated otherwise, our arguments apply to both unpolarised and polarised partons.  For ease of notation, we will in general give explicit relations for the unpolarised case and note how they generalise to the polarised one.
Notice that for transverse quark or linear gluon polarisation, DPDs carry one or more Lorentz indices and depend on the transverse vector $\tvec{y}$ rather than its length $y$.  For ease of writing, we will not indicate this explicitly.

We will explicitly indicate sums over flavour labels, where it is understood that the flavour sums run over the active flavours of the quantities being summed.  We use the convention that
\begin{align}
   \label{eq:def-nf-1}
      f_{a_1}^{\nf} = 0 \,, \quad
    & F_{a_1 a_2}^{\nf} = 0 \,, \quad
      P_{a_1 a_0}^{\nf} = 0 \,, \quad
      V_{a_1 a_2, a_0}^{\nf} = 0 \,, \quad
      U_{a_1 a_2, a_0}^{\nf} = 0
\nonumber \\[0.2em]
& \quad
   \text{if $a_0$, $a_1$, or $a_2$ is heavier than the $\nf$ active flavours,}
\end{align}
where $P$ denotes the DGLAP evolution kernels and the kernels $V$ and $U$ are introduced in \sects{\ref{sec:splitting-basics}} and \ref{sec:sum-rules}, respectively.  For kernels involving $\nf$ light flavours and a heavy flavour $Q$, we set
\begin{align}
   \label{eq:def-nf-2}
      A_{a_1 Q}^{Q, \nf}
   &= 0 \,,
   &&
      V_{a_1 a_2, Q}^{Q, \nf} = 0 \,,
\end{align}
where $A^{Q}$ is the flavour matching kernel for parton distributions (section~\ref{sec:matching-basics}) and $V^{Q}$ the massive DPD splitting kernel defined in section~\ref{sec:massive-kernels}.

In general, kernels like $P^{\nf}$ of $A^{Q, \nf}$ have an explicit $\nf$ dependence beyond the conditions \eqref{eq:def-nf-1} and \eqref{eq:def-nf-2} on their flavour indices.  This dependence is absent in some channels at low perturbative orders.  When this is the case, the superscript $\nf$ is omitted, i.e.~its absence signals that the corresponding quantity is $\nf$ independent.

Due to charge conjugation invariance, all kernels in \eqref{eq:def-nf-1} and \eqref{eq:def-nf-2} remain the same if one changes quarks to antiquarks and vice versa.  Furthermore, permutation symmetry implies $F_{a_1 a_2}(x_1, x_2, y) = F_{a_2 \ms a_1}(x_2, x_1, y)$ and corresponding relations for the kernels $V$ and $U$.  We will give explicit results only for channels that are independent w.r.t.\ these symmetry operations.
%
%
\subsection{Scale dependence}
\label{sec:dglap}

Throughout this work, we assume that PDFs and DPDs are renormalised by the $\msbar$ prescription for all of the $\nf$ active flavours, regardless of whether a flavour is considered as massive or massless.  The corresponding DGLAP evolution kernels are denoted by $P_{a_1 a_0}^{\nf}$.  For PDFs, we then have
\begin{align}
   \label{eq:RGE-0}
    \frac{\text{d}}{\text{d} \log \mu^2} \,
    f^{\nf}_{a_1}(x; \mu)
  & = \sum_{a_0} P_{a_1 a_0}^{\nf}(\mu) \conv{} f^{\nf}_{a_0}(\mu)
  \,,
\end{align}
where $\otimes$ denotes the usual Mellin convolution.  In the factors of a convolution product we omit the momentum fractions that are integrated over, and it is understood that the product depends on one momentum fraction (which can be deduced from the context).

As mentioned in the introduction, we consider DPDs with the same factorisation scale $\mu$ for both partons.  The evolution equation then reads
\begin{align}
  \label{eq:RGE-1}
    \frac{\text{d}}{\text{d} \log \mu^2} \,
    F^{\nf}_{a_1 a_2}(x_1,x_2,y; \mu)
  & = \sum_{b_1} P_{a_1 b_1}^{\nf}(\mu) \conv{1} F^{\nf}_{b_1 a_2}(y; \mu)
    + \sum_{b_2} P_{a_2 b_2}^{\nf}(\mu) \conv{2} F^{\nf}_{a_1 b_2}(y; \mu)
\end{align}
with separate Mellin convolutions
\begin{align}
   \label{conv-1-def}
   P \conv{1} F
   &= \int_{x_1}^{1} \frac{dz}{z}\; P(z)\,
      F\biggl( \frac{x_1}{z}, x_2 \biggr) \,,
   &
   P \conv{2} F
   &= \int_{x_2}^{1} \frac{dz}{z}\; P(z)\,
      F\biggl( x_1, \frac{x_2}{z} \biggr)
\end{align}
for each parton momentum fraction.  It is understood that convolution products with a subscript $1$ or $2$ depend on two momentum fractions ($x_1$ and $x_2$ in the present case).

We write the perturbative expansion of the evolution kernels as
\begin{align}
  \label{eq:P-expanded}
    P_{a_1 a_0}^{\nf}(z;\mu)
  & = \sum_{k = 0}^\infty \bigl[ a_s^{\nf}(\mu) \bigr]^{k + 1} \,
      P_{a_1 a_0}^{\nf (k)}(z) \,,
\end{align}
where
\begin{align}
   \label{eq:as}
      a^{\nf}_{s}(\mu)
   &= \frac{\alpha^{\nf}_{s}(\mu)}{2 \pi}
\end{align}
and $\alpha^{\nf}_{s}$ is the strong coupling constant for $\nf$ flavours renormalised in the $\msbar$ scheme.
The scale dependence of the coupling reads
\begin{align}
   \label{eq:beta-def}
      \frac{\text{d}}{\text{d}\log \mu^2} \, a_s^{\nf}(\mu)
   &= \frac{\beta^{\nf}\bigl( a_s^{\nf}(\mu) \bigr)}{2 \pi}
\end{align}
with
\begin{align}
   \label{eq:beta-LO}
      \frac{\beta^{\nf}(a_s^{\nf})}{2 \pi}
   &= {}- \frac{\beta_0^{\nf}}{2}\, (a_s^{\nf})^2
      + \mathcal{O}\bigl( (a_s^{\nf})^3 \bigr)
\end{align}
and
\begin{align}
   \label{eq:beta-0}
      \beta_0^{\nf}
   &= \frac{11}{3} \ms C_A - \frac{4}{3} \ms T_F \ms \nf
   \,,
\end{align}
where $C_A = N_c$ for $N_c$ colours, and $T_F = 1/2$.

At LO, the $\nf$ dependence of the DGLAP kernel for gluon splitting resides only in $\beta_0$:
\begin{align}
   \label{eq:Pgg-LO}
P_{g g}^{\nf (0)}(z)
   &= \widetilde{P}_{g g}^{(0)}(z)
      + \frac{\beta_0^{\nf}}{2}\, \delta(1-z) \,.
\end{align}
Corresponding relations hold for polarised gluons, with different functions $\widetilde{P}_{\Delta g \Delta g}^{(0)}$ and $\widetilde{P}_{\delta g \delta g}^{(0)}$ for longitudinal and linear gluon polarisation, but the same term ${\beta_0^{\nf}} \delta(1-z) \big/ {2}$ in all cases.
For all other parton channels, the LO coefficients $P_{a_1 a_0}^{(0)}$ are $\nf$ independent.
%

%
\subsection{Flavour matching}
\label{sec:matching-basics}

We refer to changing the number of active flavours as ``flavour matching''.  When matching from $\nf$ to $\nf + 1$ active flavours, we call the first $\nf$ flavours ``light'' and the $(\nf + 1)$st flavour $Q$ ``heavy''.

The matching relation for the strong coupling is
\begin{align}
   \label{eq:a-matching-expanded}
      a_s^{\nf}(\mu)
   &= \sum_{k = 0}^{\infty} \bigl[ a_s^{\nf + 1}(\mu) \bigr]^{k + 1} \,
      A^{\nf (k)}_{\alpha}(\mQ; \mu) \,,
\end{align}
where $\mQ$ is the mass of the heavy quark.  The LO and NLO coefficients are $\nf$ independent and read
\begin{align}
   \label{eq:Aalpha-1}
A^{(0)}_{\alpha} &= 1 \,,
&
A^{(1)}_{\alpha}(\mQ; \mu)
   &= \frac{\Delta \beta_0}{2} \, \log \frac{\mu^2}{\mQ^2}
\end{align}
with
\begin{align}
   \label{eq:delta-beta0}
      \Delta\beta_0
   &= \beta_0^{\nf + 1} - \beta_0^{\nf}
    = -\frac{4}{3} \ms T_F \,.
\end{align}
One readily verifies that \eqref{eq:Aalpha-1} is consistent with the scale dependence of $a_s(\mu)$ given by \eqs{\eqref{eq:beta-def}} to \eqref{eq:beta-0}.  The higher-order coefficients $A^{\nf (2)}_\alpha$ and $A^{\nf (3)}_\alpha$ can be found in \cite{Chetyrkin:1997sg}.

The matching relation for PDFs reads
\begin{align}
   \label{eq:PDF-matching}
   f_{a_1}^{\nf + 1}(x; \mu)
      &= \sum_{a_0} A_{a_1 a_0}^{Q, \nf}(\mQ; \mu) \otimes f_{a_0}^{\nf}(\mu)
\end{align}
with
\begin{align}
   \label{eq:A-general-form}
   A^{Q, \nf}_{a_1 a_0}(z, \mQ; \mu)
      &= \sum_{k = 0}^\infty \bigl[ a_{s}^{\nf + 1}(\mu) \bigr]^k \,
      A_{a_1 a_0}^{Q, \nf (k)}(z, \mQ; \mu) \,.
\end{align}
As is customary, we expand the matching kernels $A^{Q, \nf}$ in $a_s^{\nf + 1}$ rather than in $a_s^{\nf}$.
The matching coefficients in \eqref{eq:A-general-form} contain explicit logarithms:
\begin{align}
   \label{eq:A-general-logs}
   A_{a_1 a_0}^{Q ,\nf (k)}(z, \mQ; \mu)
      &= \sum_{\ell = 0}^{k - 1} \log^{\ms \ell \bs} \frac{\mu^2}{\mQ^2} \;
         A_{a_1 a_0}^{Q, \nf [k, \ell\ms]}(z) \,.
\end{align}

At LO, the matching kernels have the simple form
\begin{align}
   \label{eq:A-LO}
   A_{a_1 a_0}^{Q (0)}(z)
      &= \delta^{\nf}_{a_1 \ms l} \; \delta^{}_{a_1 a_0}\, \delta(1-z) \,,
\end{align}
where
\begin{align}
   \label{eq:delta-light-def}
   \delta^{\nf}_{a \ms l}
   &= \begin{cases}
      1  & \text{if $a$ is one of the $\nf$ light flavours} \\
      0  & \text{otherwise}
      \end{cases}
\end{align}
enforces the condition \eqref{eq:def-nf-2}.
The NLO matching kernels are $\nf$ independent and can be written as
\begin{align}
   \label{eq:A-NLO}
   A_{a_1 a_0}^{Q (1)}(z, \mQ; \mu)
      &= \Bigl[ P^{\nf + 1 (0)}_{a_1 a_0}(z)
         - P^{\nf (0)}_{a_1 a_0}(z) \Bigr] \, \log \frac{\mu^2}{\mQ^2} \,,
   \end{align}
where we recall the convention specified in \eqn{\eqref{eq:def-nf-1}}.
This implies $A^{Q (1)}_{a_1 a_0} = 0$ if $a_1$ or $a_0$ is a light quark or antiquark.  In the other channels, we have
\begin{align}
   \label{eq:AQg1}
   A^{Q (1)}_{Q g}(z, \mQ; \mu)
      &= P^{(0)}_{q g}(z) \, \log \frac{\mu^2}{\mQ^2}
      = T_F \ms \Bigl[ z^2 + (1-z)^2 \ms\Bigr] \, \log \frac{\mu^2}{\mQ^2} \,,
   \\
   \label{eq:Agg1}
   A^{Q (1)}_{g g}(z, \mQ; \mu)
      &= \frac{\Delta\beta_0}{2} \, \delta(1-z) \, \log \frac{\mu^2}{\mQ^2} \,,
\end{align}
for unpolarised partons, as well as
\begin{align}
   \label{eq:AQg1-pol}
   A^{Q (1)}_{\Delta Q \ms \Delta g}(z, \mQ; \mu)
      &= P^{(0)}_{\Delta q \ms \Delta g}(z) \, \log \frac{\mu^2}{\mQ^2}
      = T_F \ms \Bigl[ z^2 - (1-z)^2 \ms\Bigr] \, \log \frac{\mu^2}{\mQ^2} \,,
\end{align}
and
\begin{align}
   \label{eq:Agg1-pol}
   A^{Q (1)}_{\Delta g \ms \Delta g}
      &= A^{Q (1)}_{\delta g \ms \delta g}
       = A^{Q (1)}_{g g \phantom{\delta}}
\end{align}
for polarised ones.  Note that there is no transition from linearly polarised gluons to transversely polarised quarks, i.e.\ $A^{Q (1)}_{\delta Q \ms \delta g} = 0$.
The two-loop matching coefficients $A^{Q (2)}$ for unpolarised partons can be found in references \cite{Buza:1996wv, Ablinger:2014lka, Behring:2014eya, Ablinger:2014vwa}.  They are all independent of $\nf$.

Flavour matching for DPDs proceeds in full analogy to the PDF case and reads
\begin{align}
   \label{eq:DPD-matching}
   F_{a_1 a_2}^{\nf + 1}(x_1,x_2,y; \mu)
      &= \sum_{b_1} A_{a_1 b_1}^{Q, \nf}(\mQ; \mu)
         \conv{1} F_{b_1 a_2}^{\nf}(y;\mu)
       + \sum_{b_2} A_{a_2 b_2}^{Q, \nf}(\mQ; \mu)
         \conv{2} F_{a_1 b_2}^{\nf}(y;\mu) \,.
\end{align}
One way to see this is to rewrite the matching equation \eqref{eq:PDF-matching} for PDFs as a matching relation between the twist-two operators that define PDFs with $\nf + 1$ or $\nf$ active flavours.  DPDs are defined in terms of the same twist-two operators, containing the product of an operator for parton $a_1$ and an operator for parton $a_2$ at relative transverse distance~$y$.  This readily implies the matching relation \eqref{eq:DPD-matching} as a generalisation of \eqref{eq:PDF-matching}, just as the renormalisation group equation for the twist-two operators implies the evolution equation \eqref{eq:RGE-1} for DPDs as a generalisation of the DGLAP equation \eqref{eq:RGE-0} for PDFs.

%
%
\subsection{DPDs at short distance}
\label{sec:splitting-basics}
As we just mentioned, DPDs are defined as the matrix elements of the product between two twist-two operators that are separated by a distance $y$ in the transverse plane.  In the limit where this distance $y$ is much smaller than a typical hadronic scale, one can perform an operator product expansion and thus express DPDs in terms of short-distance coefficients and matrix elements of single operators with definite twist.   The leading operators have twist two, and their matrix elements are PDFs.  This  corresponds to the first term in the decomposition $F = F^{\text{spl}} + F^{\text{intr}}$ we already mentioned in the introduction.  From now on, we focus on this term and omit the superscript ``spl'' for brevity.  A detailed account of its properties can be found in \cite{Diehl:2019rdh, Diehl:2021wpp}, where the associated short-distance coefficients (called ``DPD splitting kernels'') are computed up to NLO for unpolarised massless partons.

The general form of the splitting formula for DPDs with $\nf$ massless flavours reads
\begin{align}
   \label{eq:small-y-DPD}
      F^{\nf}_{a_1 a_2}(x_1,x_2,y; \mu)
   &= \frac{1}{\pi y^2} \sum_{a_0}
      V^{\nf}_{a_1 a_2, a_0}(y; \mu) \conv{12} f^{\nf}_{a_0}(\mu) \,,
\end{align}
where
\begin{align}
   \label{eq:Mellin-12}
V \conv{12} f
   &= \int\limits_{x_1 + x_2}^{1} \!\!\! \frac{dz}{z^2}\;
      V\biggl( \frac{x_1}{z}, \frac{x_2}{z} \biggr) \, f(z)
\end{align}
is a generalised Mellin convolution depending on two momentum fractions.  A useful relation is
\begin{align}
   \label{eq:conv-12-relation}
   D \conv{12} [C \conv{} f] = [ D \conv{12} C ] \conv{12} f \,,
\end{align}
where $D$ depends on two momentum fractions and $C$ and $f$ depend on one momentum fraction.

The DPD splitting kernels can be expanded in the strong coupling as
\begin{align}
   \label{eq:VnF-expanded}
      V^{\nf}_{a_1 a_2, a_0}(z_1, z_2, y; \mu)
   &= \sum_{k = 1}^\infty \bigl[ a^{\nf}_{s}(\mu) \bigr]^k \;
      V^{\nf (k)}_{a_1 a_2, a_0}(z_1, z_2, y; \mu)
\intertext{with coefficients of the form}
   \label{eq:VnF-logs}
      V^{\nf (k)}_{a_1 a_2, a_0}(z_1, z_2, y; \mu)
   &= \sum_{\ell = 0}^{k-1} \log^{\ms \ell \bs} \frac{\mu^2}{\mu_y^2} \;
      V^{\nf [k,\ell\ms]}_{a_1 a_2, a_0}(z_1, z_2)\,.
\end{align}
Here we have introduced the mass scale corresponding to the distance $y$,
\begin{align}
   \label{eq:mu-y-def}
   \mu_y &=  \frac{b_0}{y} \,,
\end{align}
where $b_0 = 2 \text{e}^{-\gamma_E} \approx 1.12$ with the Euler-Mascheroni constant $\gamma_E$.

The LO splitting kernels are $\nf$ independent and have the kinematic constraint
\begin{align}
   \label{eq:DPD-split-LO}
V^{(1)}_{a_1 a_2, a_0}(z_1, z_2)
   &= \delta(1 - z_1 - z_2)\,
      V^{(1)}_{a_1 a_2, a_0}(z_1)
   \,,
\end{align}
where the function $V$ on the right-hand side depends on only one momentum fraction.  With this constraint, the general form \eqref{eq:small-y-DPD} turns into
\begin{align}
   \label{eq:small-y-DPD-LO}
      F^{\nf}_{a_1 a_2}(x_1,x_2,y; \mu)
   &= \frac{a^{\nf}_s(\mu)}{\pi y^2} \;
      V^{(1)}_{a_1 a_2, a_0}\Bigl( \frac{x_1}{x_1 + x_2} \Bigr) \,
      \frac{f^{\nf}_{a_0}(x_1 + x_2; \mu)}{x_1 + x_2}
      + \mathcal{O}\bigl( (a_s^{\nf})^2 \bigr)
   \,.
\end{align}
The functions $V^{(1)}_{a_1 a_2, a_0}(z)$ are equal to the LO DGLAP kernels $P^{(0)}_{a_1 a_0}(z)$ with all plus-distri\-bu\-tions and $\delta(1-z)$ terms removed.
The structure of the DPD splitting kernels at NLO (i.e.\ at order $a_s^2$) will be discussed at the beginning of section~\ref{sec:NLO-kernels-constraints}.

The relations from \eqref{eq:small-y-DPD} to \eqref{eq:small-y-DPD-LO} hold if the partons $a_1$ and $a_2$ are unpolarised or polarised.

%
\graphicspath{{Figures/Scheme/}}
%
%
\section{Schemes for treating heavy quarks in splitting DPDs}
\label{sec:schemes}
As discussed in the introduction, the computation of a DPS cross section requires the double parton luminosities \eqref{eq:DPD-lumi-0}, which involve DPDs for all distances $y$ above a value of order $1/\mu$, where $\mu$ is the typical scale of the hard-scattering processes.  The DPDs need to be evolved to this scale, starting from a scale at which they can either be computed (for small $y$) or modelled (for large $y$).  In the present section, we discuss how to treat massive quarks in different regions of small $y$.

Before doing so, let us briefly discuss the transition from small to large $y$.  For small $y$ and massless partons, one should compute the DPDs at a scale proportional to $\mu_y$ given in \eqref{eq:mu-y-def}, since this avoids large logarithms in the coefficients \eqref{eq:VnF-logs} of the splitting formula \eqref{eq:small-y-DPD}.  For large $y$, it is natural to model the DPDs at a scale in the GeV region, which must be large enough to use the perturbative expansion of the DGLAP kernels when evolving to higher scales.  To interpolate between the two regimes, one may take the starting scale for DPD evolution proportional to
\begin{align}
   \label{eq:mu-y-star}
   \mu_{y^*}
   &= \frac{b_0}{y^*(y)}
   &\text{~with~}
   y^*(y)
      &\to
      \begin{cases}
         y & \text{ for $y \to 0$} \\
         y_{\text{max}} & \text{ for $y\to \infty$}
      \end{cases}
\end{align}
which tends to $\mu_y$ for small $y$ and to $\mu_{\text{min}} = b_0 / y_{\text{max}}$ for large $y$.  In the numerical study of \sect{\ref{sec:LO-numerics}}, we will take $\mu_{\text{min}} = 1 \gev$ and the functional form
\begin{align}
   \label{eq:y-star}
y^*(y)
   &= \frac{y}{\bigl( 1 + y^p_{\phantom{|}} \big/ y_{\text{max}}^p \bigr)^{1/p}} \,.
\end{align}
With $p=2$, this function is widely used in the phenomenology of transverse-momentum dependent parton distributions.  We will instead take the form with $p=4$, which tends more rapidly to $\mu_y$ for small $y$.   A rather similar function was used in \cite{Bacchetta:2015ora}.
%
%
\subsection{Splitting kernels including mass effects}
\label{sec:massive-kernels}
The DPD splitting formula \eqref{eq:small-y-DPD} is applicable if all quark masses in the perturbative splitting process can be neglected.  The corresponding LO kernels are given in reference \cite{Diehl:2011yj} for all polarisations, and the unpolarised NLO kernels can be found in \cite{Diehl:2019rdh, Diehl:2021wpp}.  This version of the splitting formula is appropriate if the characteristic mass scale $\mu_y = b_0 / \mu$ of the splitting is much bigger than the masses of the active quark flavours in the DPD.

If $\mu_y$ is much larger than the masses of the first $\nf$ quark flavours but similar in size to the mass $\mQ$ of the $\nf + 1$st flavour, then the latter should be treated as massive in the computation of the $1 \to 2$ splitting kernels $V$.  In this case, we can use the splitting formula
\begin{align}
   \label{eq:small-y-DPD-massive}
      F_{a_1 a_2}^{\nf + 1}(x_1, x_2, y; \mu)
   &= \frac{1}{\pi y^2} \; \sum_{a_0} V_{a_1 a_2, a_0}^{Q, \nf}(y, \mQ; \mu)
      \conv{12} f_{a_0}^{\nf}(\mu)
\end{align}
with a perturbative expansion
\begin{align}
   \label{eq:VmQ-expanded}
      V^{Q, \nf}_{a_1 a_2, a_0}(z_1, z_2, y, \mQ; \mu)
   &= \sum_{k = 1}^\infty \bigl[ a^{\nf + 1}_{s}(\mu) \bigr]^k \,
      V^{Q, \nf (k)}_{a_1 a_2, a_0}(z_1, z_2, y, \mQ; \mu) \,.
\end{align}
The label $\nf$ on $V^{Q}$ indicates the number of flavours that are treated as massless.  Notice that only the $\nf$ light quark flavours are taken as active in the PDF on the r.h.s.\ of \eqref{eq:small-y-DPD-massive}, i.e.\ the heavy flavour $Q$ only appears in the splitting kernel.  In analogy to the flavour matching kernels \eqref{eq:A-general-form}, we expand the massive splitting kernels in $a^{\nf + 1}_{s}$.

At LO one finds only one channel where heavy quarks can be produced by the splitting, namely $g \to Q \Qbar$.  The corresponding kernels can readily be obtained by extending the calculation in section 5.2.2 of reference \cite{Diehl:2011yj}.  We find
\begin{align}
   \label{eq:VQQbarg-LO}
      V_{Q \Qbar, g}^{Q (1)}(z_1, z_2, y, \mQ)
   &= T_F \, (y \ms \mQ)^2 \,
      \Bigl[
      (z_1^2 + z_2^2) \, K_1^2(y \ms \mQ) + K_0^2(y \ms \mQ)
      \Bigl] \delta(1 - z_1 - z_2) \,,
   \nonumber\\
      V_{\Delta Q \Delta \Qbar, g}^{Q (1)}(z_1, z_2, y, \mQ)
   &= T_F \, (y \ms \mQ)^2 \,
      \Bigl[
      - (z_1^2 + z_2^2) \, K_1^2(y \ms \mQ) + K_0^2(y \ms \mQ)
      \Bigr] \, \delta(1 - z_1 - z_2) \,,
   \nonumber\\
      V_{\delta Q \delta \Qbar, g}^{Q (1)}(z_1, z_2, y, \mQ)
   &= T_F \, (y \ms \mQ)^2 \,
      \Bigl[ - 2 z_1 z_2\, K_1^2(y \ms \mQ) \Bigr] \, \delta(1 - z_1 - z_2)
   \,,
\end{align}
where $K_i(w)$ denotes the modified Bessel functions of the second kind.
Kernels with exactly one observed heavy flavour are zero at this order, whereas kernels with only light flavours are equal to their massless counterparts,
\begin{align}
   \label{eq:VQ-light-LO}
   V_{a_1 a_2, a_0}^{Q (1)}(z_1, z_2, y, \mQ)
      &= V_{a_1 a_2, a_0}^{(1)}(z_1, z_2)
      & \text{ if $a_1$ and $a_2$ are light.}
\end{align}
The situation at NLO is significantly more involved and will be analysed in section \ref{sec:NLO-kernels-constraints}.

Consider now the unpolarised $g \to Q \Qbar$ kernel from \eqref{eq:VQQbarg-LO} in the limits $\mu_y \ll \mQ$ and $\mu_y \gg \mQ$. In the first case --- which corresponds to $y \gg 1 / \mQ$ --- the massive kernel vanishes exponentially,
\begin{align}
   \label{eq:VQQbarg-LO-large-y}
   V_{Q \Qbar, g}^{Q (1)}(z_1, z_2, y, \mQ)
      &\overset{\mu_y \ll \mQ}{\longrightarrow}
         y \ms \mQ \, \exp(-2 y \ms \mQ)
      \overset{y \to \infty}{\longrightarrow} 0 \,,
\end{align}
so that the production of heavy quarks is strongly suppressed.  In the second case --- which corresponds to $y \ll 1 / \mQ$ --- one obtains
\begin{align}
   \label{eq:VQQbarg-LO-small-y}
   V_{Q \Qbar, g}^{Q (1)}(z_1, z_2, y, \mQ)
      &\overset{\mu_y \gg \mQ}{\longrightarrow}
         V_{Q \Qbar, g}^{(1)}(z_1, z_2)
         + \mathcal{O}\bigl( (y \ms \mQ)^2 \log^2(y \ms \mQ) \bigr)
   \,.
\end{align}
Analogous limiting expressions hold for polarised quarks, except that for transverse polarisation the corrections to the small $y$ limit are of order $(y \ms \mQ)^2 \log(y \ms \mQ)$.
As one expects, the massive kernels approach the massless ones for $\mu_y \gg \mQ$, with power corrections in the small parameter $y \ms \mQ$.  Notice that the massless LO kernel on the right-hand side of \eqref{eq:VQQbarg-LO-small-y} does not depend on the number of active flavours, but the coupling $a_s$ it is multiplied with in the perturbative expansion \eqref{eq:VnF-expanded} does.

\paragraph{Two heavy flavours}
If $\mu_y$ is comparable in size to both $m_c$ and $m_b$, one may want to treat both charm and bottom quarks as massive when computing the $1\to 2$ splitting.  The DPD with five active flavours is then given by
\begin{align}
      F_{a_1 a_2}^{5}(x_1, x_2, y; \mu)
   &= \frac{1}{\pi y^2} \; \sum_{a_0} V_{a_1 a_2, a_0}^{c b}(y, m_c, m_b; \mu)
      \conv{12} f_{a_0}^{3}(\mu)
\end{align}
with the perturbative expansion
\begin{align}
   \label{eq:Vcs-expanded}
      V^{c b}_{a_1 a_2, a_0}(z_1, z_2, y, m_c, m_b; \mu)
   &= \sum_{k = 1}^\infty \bigl[ a^{5}_{s}(\mu) \bigr]^k \;
      V^{c b}_{a_1 a_2, a_0}(z_1, z_2, y, m_c, m_b; \mu) \,.
\end{align}
For brevity, we do not indicate that $V^{c b}$ assumes the presence of 3 light quark flavours.

At LO, the only nonzero kernels for observed heavy flavours are
\begin{align}
   V^{c b (1)}_{c \cbar, g}(z_1, z_2, y, m_c, m_b)
      &= V^{Q (1)}_{Q \Qbar, g}(z_1, z_2, y, m_c) \,,
   \nonumber \\
   V^{c b (1)}_{b \bbar, g}(z_1, z_2, y, m_c, m_b)
      &= V^{Q (1)}_{Q \Qbar, g}(z_1, z_2, y, m_b) \,,
\end{align}
and their polarised counterparts, whereas for observed light flavours one has $V^{c b (1)}_{a_1 a_2, a_0} = V^{Q (1)}_{a_1 a_2, a_0} = V^{(1)}_{a_1 a_2, a_0}$.

%

\subsection{Schemes for one massive flavour}
\label{sec:schemes-one-flavour}

In this subsection, we consider a setting with $\nf$ light flavours and one heavy flavour $Q$.  The DPD for $\nf + 1$ active flavours is characterised by three mass scales: the scale $\Lambda$ of nonperturbative interactions, the scale $\mu_y$ associated with the distance between the two observed partons, and the mass of the heavy quark, which satisfies $\mQ \gg \Lambda$.  The DPD factorises in different ways depending on the size of $\mu_y$.

For $\mu_y \sim \Lambda$, the dynamics for the production of light flavours is purely nonperturbative and encoded in the $\nf$ flavour DPD $F^{\nf}$.  The DPD for $\nf + 1$ active flavours, including the production of heavy quarks, is then given by flavour matching for each of the two partons and given by
equation \eqref{eq:DPD-matching}.

\begin{figure}[t]
   \subfigure[\label{fig:splitting-large-y}$\mu_y \ll \mQ$]{
      \includegraphics[width=0.3\linewidth]{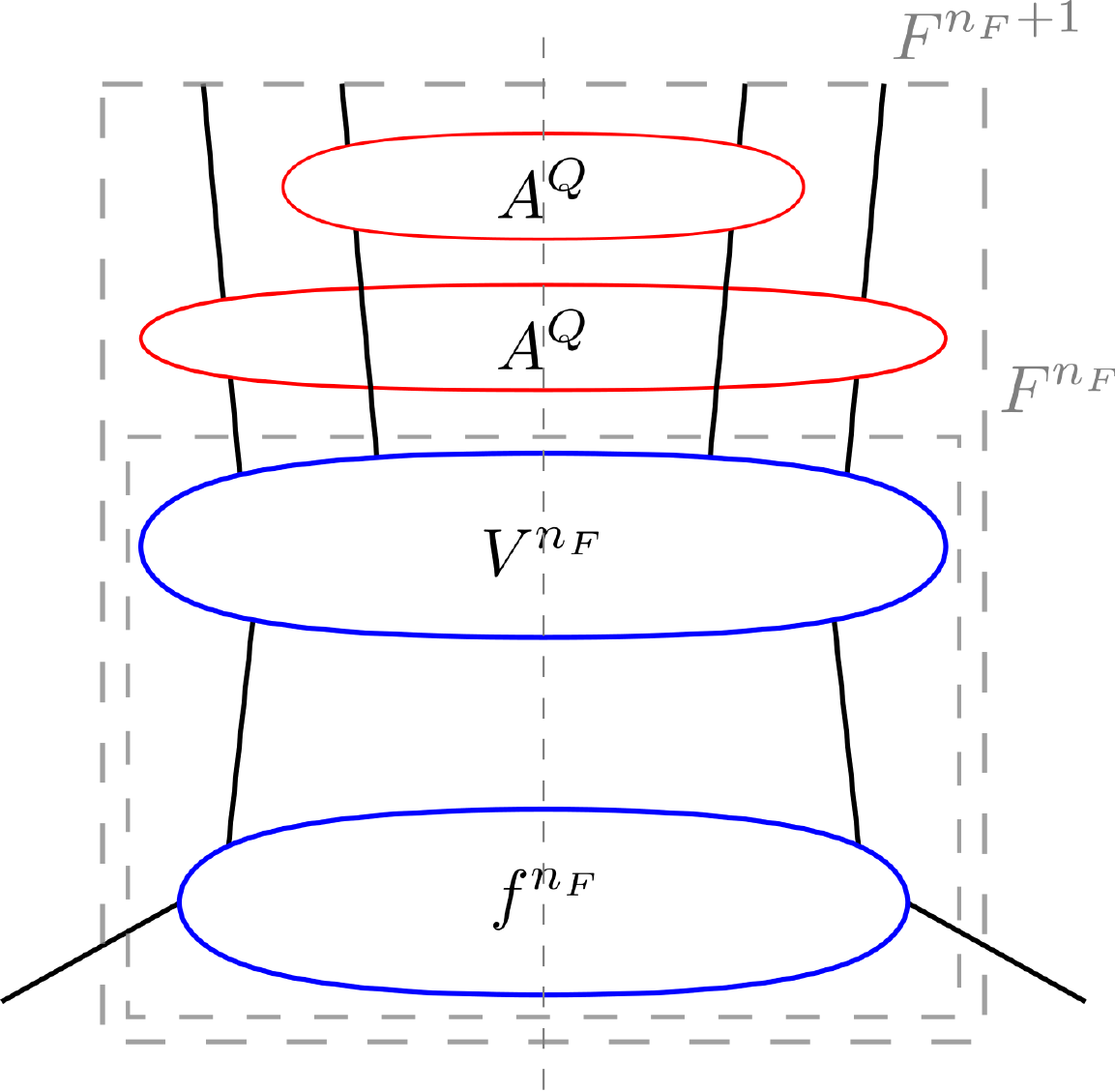}
   }
   \subfigure[\label{fig:splitting-intermediate-y}$\mu_y \sim \mQ$]{
      \includegraphics[width=0.3\linewidth]{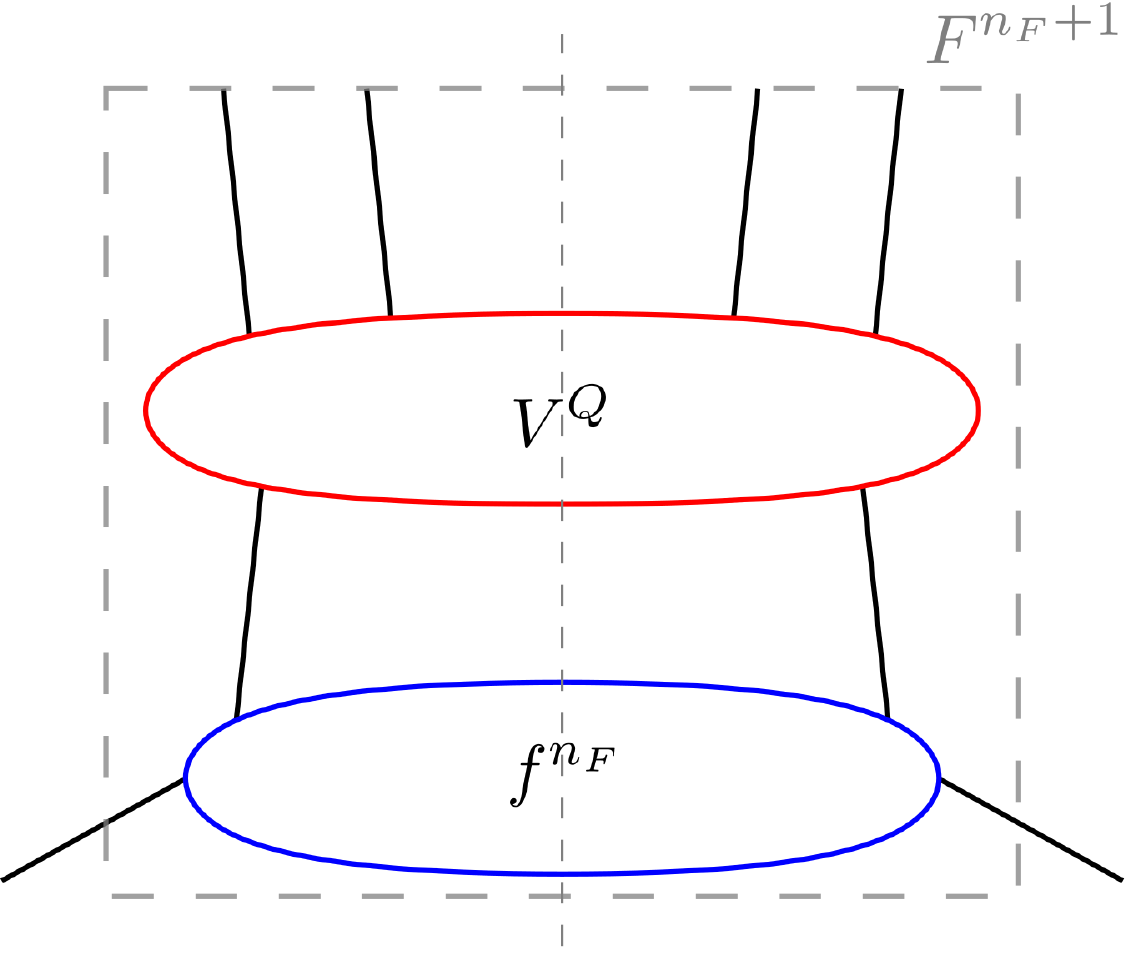}
   }
   \subfigure[\label{fig:splitting-small-y}$\mu_y \gg \mQ$]{
      \includegraphics[width=0.3\linewidth]{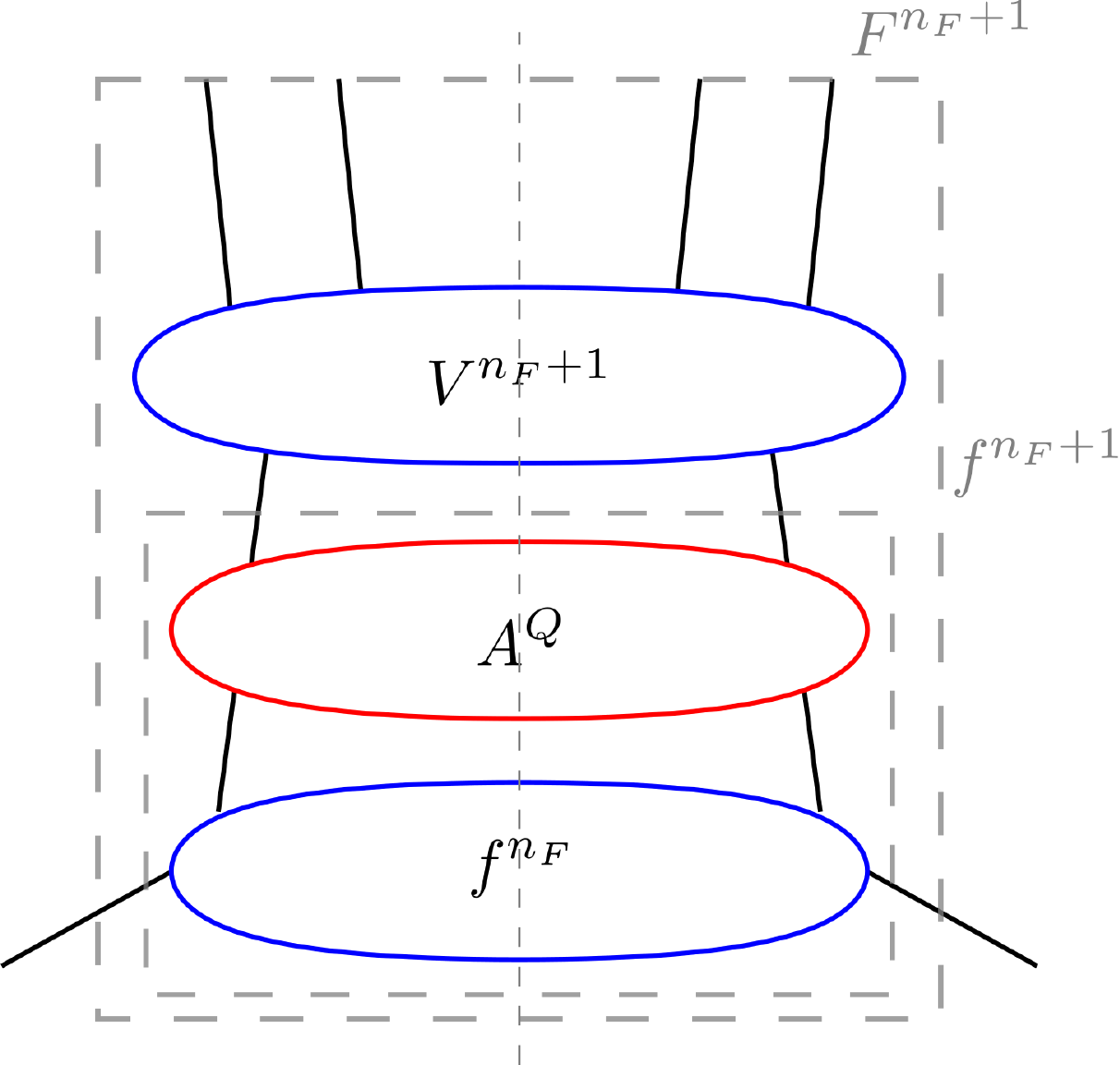}
   }
   \caption{\label{fig:splitting-regions} The factorisation regimes for an $\nf + 1$ flavour DPD in different regions of $\mu_y \gg \Lambda$. Kernels with superscript $Q$ are computed with the heavy flavour $Q$ treated as massive, whereas kernels with superscripts $\nf$ and $\nf + 1$ are computed for $\nf$ or $\nf + 1$ massless quark flavours, respectively.  For brevity, we omit the superscript $\nf$ in $A^Q$ and $V^Q$.}
\end{figure}

For $\mu_y \gg \Lambda$, the dynamics at nonperturbative scales is contained in PDFs for the $\nf$ light flavours, and we can distinguish three different factorisation regimes.
\begin{enumerate}
\item For $\mu_y \ll \mQ$, the dynamics at scales $\sim \mu_y$ is contained in the DPD splitting kernel $V^{\nf}$ that describes the transition from the parton in the PDF to two light partons in an $\nf$ flavour DPD.  The dynamics at scales $\sim \mQ$ is contained in the flavour matching kernels for the transition from $F^{\nf}$ to $F^{\nf + 1}$.  This corresponds to factorised graphs as shown in \fig{\ref{fig:splitting-large-y}}.
\item For $\mu_y \sim \mQ$, splitting and heavy-quark excitation take place at the same scale, and the transition from the light parton in the PDF to the observed partons of the DPD is described by the massive splitting kernel $V^{Q, \nf}$ introduced in the previous subsection.  The structure of the corresponding graphs is given in \fig{\ref{fig:splitting-intermediate-y}}.
\item For $\mu_y \gg \mQ$, the quark mass effects are contained in the matching kernel for the transition from PDFs $f^{\nf}$ to PDFs $f^{\nf + 1}$.  The $1\to 2$ splitting happens at much higher scales $~ \mu_y$ and is described by the massless splitting kernel $V^{\nf + 1}$.  The factorised graphs have the form of \fig{\ref{fig:splitting-small-y}}.
\end{enumerate}
In \sect{\ref{sec:NLO-kernels-limits}} we will use the preceding analysis to derive the limiting forms of $V^{Q, \nf}$ for small and for large $y$.  In the following, we discuss two schemes to compute the DPD in the full range of perturbative $y$.  We will briefly comment on the transition to nonperturbative $y$ in \sect{\ref{sec:two-massive}}.
%
%
\subsubsection{Scheme with massless splitting kernels}
\label{sec:massless}
We start with a simplified scheme in which only the splitting formula \eqref{eq:small-y-DPD} for massless quarks is used.  Not surprisingly, this requires rather coarse approximations.  One may nevertheless want to use such a scheme, e.g.\ because at NLO accuracy only massless splitting kernels are presently available.

In this purely massless scheme, one directly switches from the description of \fig{\ref{fig:splitting-large-y}} to the one of \fig{\ref{fig:splitting-small-y}} at a scale $\mu_y = b_0 / y = \gamma \ms \mQ$.  The choice of the scheme parameter $\gamma$ will be discussed in \sect{\ref{sec:DPD-lumis-params}}; we anticipate that values $\gamma \sim 1$ are most appropriate.

\begin{figure}[t]
   \begin{center}
      \includegraphics[height = 0.65\linewidth]{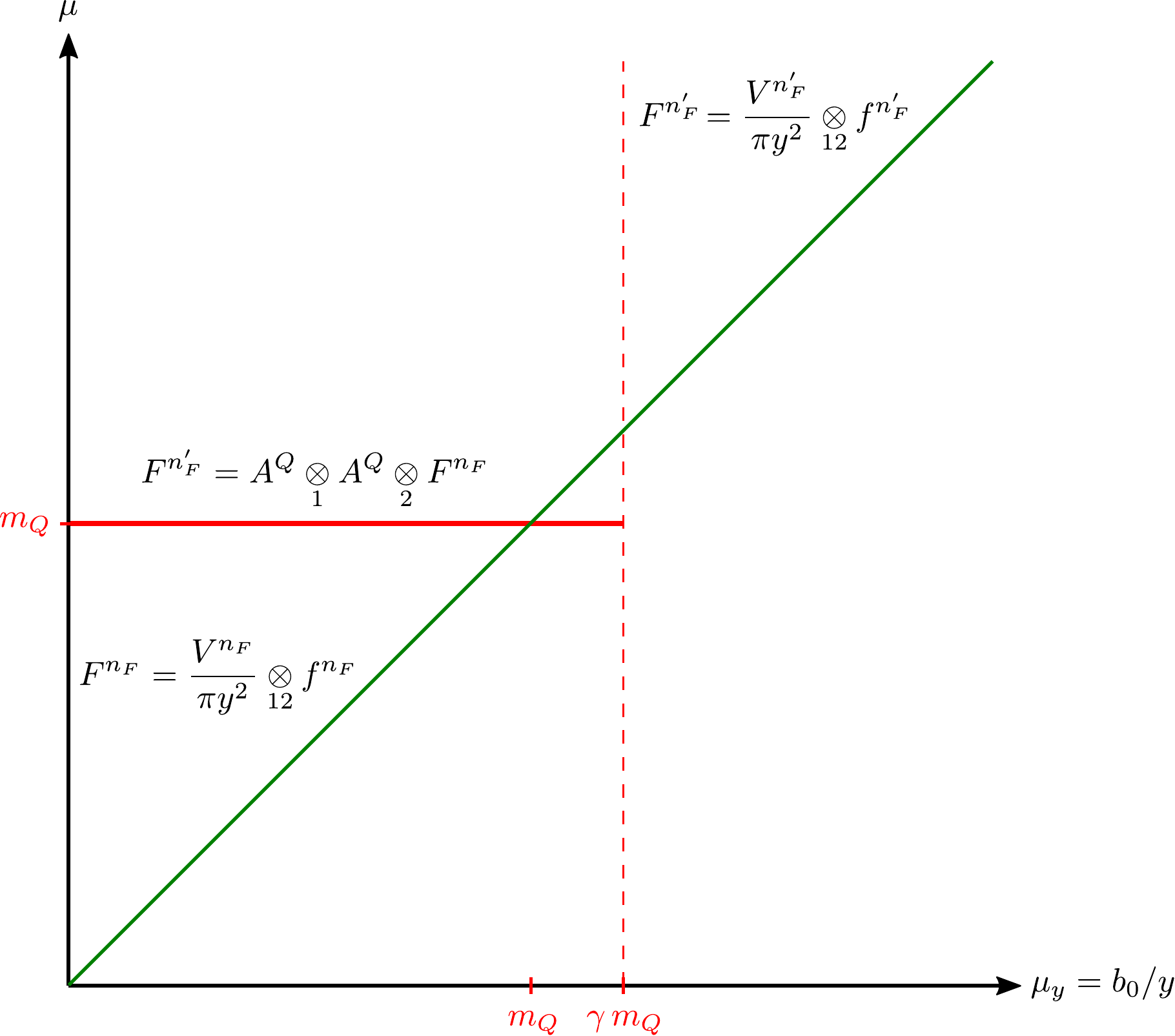}
   \end{center}
   \caption{\label{fig:scheme-massless} Illustration of the scheme with only massless splitting kernels. The green line indicates the scale $\mu_{\text{split}} = \mu_y$ at which the DPD splitting formulae are evaluated, and the solid red line indicates the scale $\mu_Q = \mQ$ used for DPD flavour matching.  For brevity, we write $\nf' = \nf + 1$ and omit the superscript $\nf$ in $A^Q$.  The region of nonperturbatively small $\mu_y$ is not included in the plot (i.e.\ the zero point is suppressed on the axes for $\mu_y$ and $\mu$).}
\end{figure}

For $\mu_y < \gamma \ms \mQ$, the splitting DPD is computed for $\nf$ massless flavours from an $\nf$ flavour PDF.  This is done at a renormalisation scale
\begin{align}
   \label{eq:mu-split-def}
   \mu_{\text{split}} & \sim \mu_y
\end{align}
to avoid large logarithms spoiling the perturbative expansion of the splitting kernel.  The DPD is then evolved to the scale
\begin{align}
   \mu_Q & \sim \mQ
   \,,
\end{align}
where flavour matching of the DPD from $\nf$ to $\nf + 1$ active flavours is performed.  The corresponding formulae read
\begin{align}
   \label{eq:massless-1}
      F_{a_1 a_2}^{\nf}(y; \mu_{\text{split}})
   &= \frac{1}{\pi y^2} \;
      \sum_{a_0} V_{a_1 a_2, a_0}^{\nf}(y; \mu_{\text{split}}) \conv{12}
      f_{a_0}^{\nf}(\mu_{\text{split}})\,,
   \nonumber \\
      F_{a_1 a_2}^{\nf + 1}(y; \mu_{Q})
   &= \sum_{b_1, b_2} A_{a_1 b_1}^{Q, \nf}(\mQ; \mu_Q)
         \conv{1} A_{a_2 b_2}^{Q, \nf}(\mQ; \mu_Q) \conv{2}
         F_{b_1 b_2}^{\nf}(y; \mu_{Q})
   \nonumber \\[-0.6em]
   & \hspace{22em} \text{ for $\mu_y < \gamma \ms \mQ$.}
\end{align}
For $\mu_y > \gamma \ms \mQ$, the splitting DPD is computed for $\nf + 1$ massless flavours from an $\nf + 1$ flavour PDF:
\begin{align}
   \label{eq:massless-3}
      F_{a_1 a_2}^{\nf + 1}(y; \mu_{\text{split}})
   &= \frac{1}{\pi y^2} \; \sum_{a_0}
      V_{a_1 a_2, a_0}^{\nf + 1}(y; \mu_{\text{split}}) \conv{12}
      f_{a_0}^{\nf + 1}(\mu_{\text{split}})
   & \text{ for $\mu_y > \gamma \ms \mQ$}
\end{align}
with $\mu_{\text{split}}$ as in \eqref{eq:mu-split-def}.
The $\nf + 1$ flavour DPD at any other scale is then obtained by evolution from the starting conditions in \eqref{eq:massless-1} or \eqref{eq:massless-3}.
The transition from $\nf$ to $\nf + 1$ flavours in the PDF is obtained by flavour matching:
\begin{align}
   \label{eq:PDF-matching-in-schemes}
      f_{a_0}^{\nf + 1}(\mu_{Q})
   &= \sum_{b_0} A_{a_0 b_0}^{Q, \nf}(\mQ; \mu_Q) \conv{} f_{b_0}(\mu_{Q}) \,.
\end{align}
Here and in the following it is understood that flavour matching for the strong coupling is also performed at the scale $\mu_{Q}$.
A graphical representation of this scheme is given in \fig{\ref{fig:scheme-massless}}.

The approach just described has an obvious shortcoming.  It lacks the heavy-quark contributions that can be produced by splitting for $\mu_y < \gamma \mQ$, and it neglects the effects of finite $\mQ$ in the splitting kernels for $\mu_y > \gamma \mQ$, even though these effects can only be neglected for $\mu_y \gg \mQ$.  An indication of these shortcomings is that the DPDs for some flavour combinations have large unphysical discontinuities at $\mu_y = \gamma \ms \mQ$, as we will see in \sect{\ref{sec:massless-DPDs}}.
Nevertheless, we will show in \sect{\ref{sec:DPD-lumis-params}} that, after one integrates over $y$ to obtain a double parton luminosity \eqref{eq:DPD-lumi-0}, the massless scheme can actually give a fair approximation of the more realistic scheme presented next.
%
%
\subsubsection{Scheme with massive splitting kernels}
\label{sec:one-massive}
We now consider the case where the massive splitting kernels are used.
For the transition between the three regimes shown in \fig{\ref{fig:splitting-regions}}, we introduce two scales $\alpha \mQ$ and $\beta \mQ$ with $\alpha \ll 1$ and $\beta \gg 1$.  For $\mu_y < \alpha \mQ$ or $\mu_y > \beta \mQ$, it is appropriate to use the two-step factorisation of \fig{\ref{fig:splitting-large-y}} or \fig{\ref{fig:splitting-small-y}}, respectively.  In the intermediate region $\alpha \mQ < \mu_y < \beta \mQ$, one uses the massive splitting formula \eqref{eq:small-y-DPD-massive}.

The choice of the scheme parameters $\alpha$ and $\beta$ is a matter of compromise.  To minimise the errors inherent in the two-step factorisation schemes, one should take $\alpha$ small and $\beta$ large enough.  On the other hand, the splitting kernels $V^{Q, \nf}$ in the intermediate regime contain logarithms $\log(\mu / \mQ)$ \emph{and} $\log(\mu / \mu_y)$ at higher orders in $a_s$, which cannot be kept small  for any choice of renormalisation scale $\mu$ if the interval from $\alpha$ to $\beta$ becomes too large.

\begin{figure}[t]
   \begin{center}
      \includegraphics[height = 0.65\linewidth]{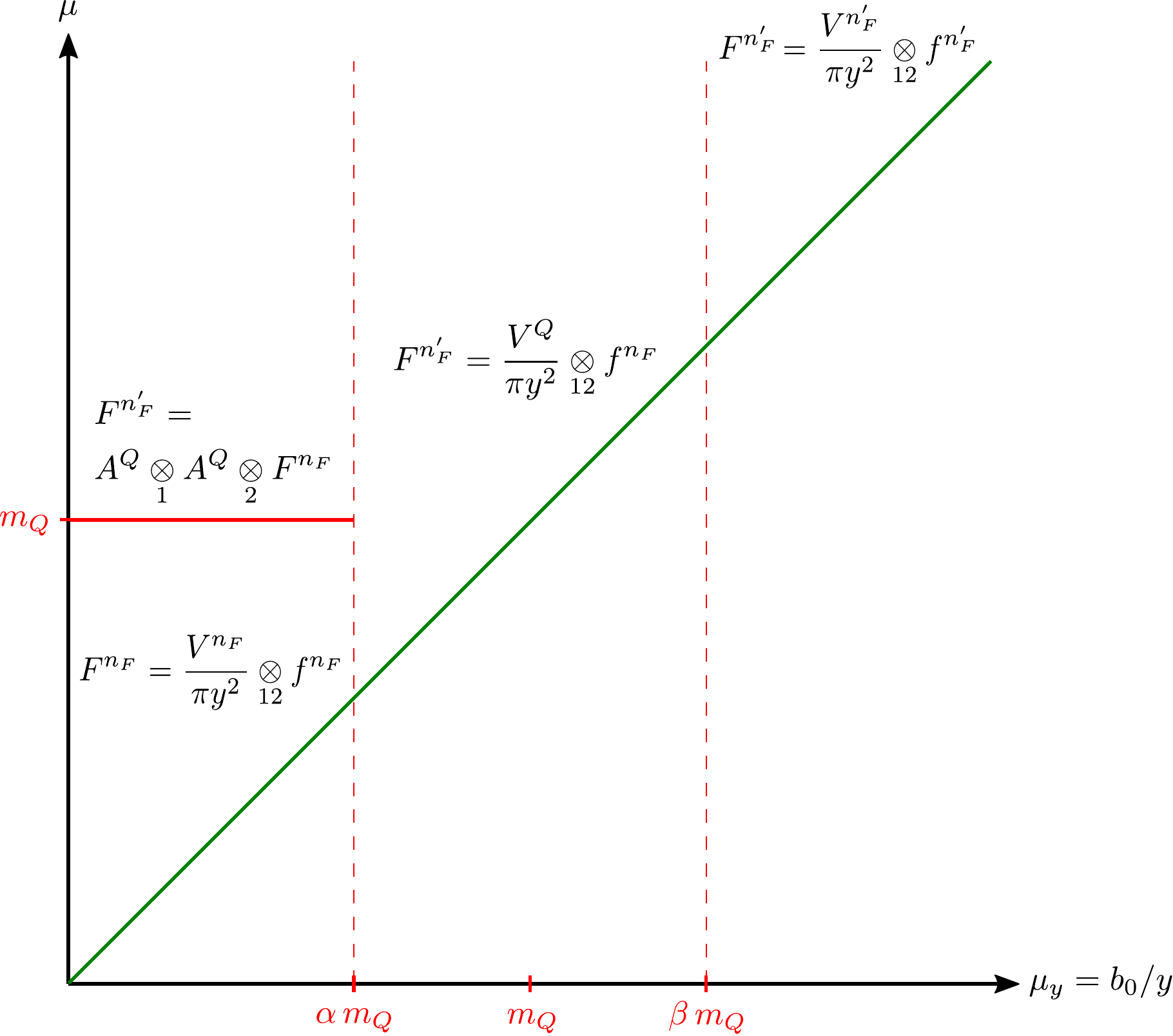}
   \end{center}
   \caption{\label{fig:scheme-one-massive} As \fig{\protect{\ref{fig:scheme-massless}}}, but for the scheme with massless and massive splitting kernels.  For brevity, we write $\nf' = \nf + 1$ and omit the superscript $\nf$ in $A^Q$ and $V^Q$.}
\end{figure}

The scheme just described is represented in \fig{\ref{fig:scheme-one-massive}}.
The splitting DPD is computed from
\begin{align}
   \label{eq:one-massive-1}
   F_{a_1 a_2}^{\nf}(y; \mu_{\text{split}})
      &= \frac{1}{\pi y^2} \;
         \sum_{a_0} V_{a_1 a_2, a_0}^{\nf}(y; \mu_{\text{split}}) \conv{12}
         f_{a_0}^{\nf}(\mu_{\text{split}}) \,,
   \nonumber \\
   F_{a_1 a_2}^{\nf + 1}(y; \mu_{Q})
      &= \sum_{b_1, b_2} A_{a_1 b_1}^{Q, \nf}(\mQ; \mu_Q)
            \conv{1} A_{a_2 b_2}^{Q, \nf}(\mQ; \mu_Q) \conv{2}
            F_{b_1 b_2}^{\nf}(y; \mu_{Q})
   \nonumber \\[-0.6em]
   & \hspace{22em} \text{ for $\mu_y < \alpha \ms \mQ$,}
\end{align}
from
\begin{align}
   \label{eq:one-massive-3}
      F_{a_1 a_2}^{\nf + 1}(y; \mu_{\text{split}})
      &= \frac{1}{\pi y^2} \;
         \sum_{a_0} V_{a_1 a_2, a_0}^{Q, \nf}(y, \mQ; \mu_{\text{split}})
         \conv{12} f_{a_0}^{\nf}(\mu_{\text{split}})
   & \text{ for $\alpha \ms \mQ \! < \! \mu_y \! < \! \beta \ms \mQ$,}
\end{align}
and from
\begin{align}
   \label{eq:one-massive-4}
      F_{a_1 a_2}^{\nf + 1}(y; \mu_{\text{split}})
   &= \frac{1}{\pi y^2} \;
      \sum_{a_0} V_{a_1 a_2, a_0}^{\nf + 1}(y; \mu_{\text{split}}) \conv{12}
      f_{a_0}^{\nf + 1}(\mu_{\text{split}})
   & \text{ for $\mu_y > \beta \ms \mQ$}
\end{align}
with $f^{\nf + 1}$ given by \eqref{eq:PDF-matching-in-schemes}.
In all cases, we take renormalisation scales $\mu_{\text{split}} \sim \mu_y$ and $\mu_{Q} \sim \mQ$ as before.  We will explain in \sect{\ref{sec:NLO-kernels-large-logs}} why we prefer the choice $\mu_{\text{split}} \sim \mu_y$ in the massive kernels $V^{Q, \nf}$, where one may alternatively consider scales that depend on $\mQ$ or on both $y$ and $\mQ$.
%
%
\subsection{Two heavy flavours (charm and bottom)}
\label{sec:two-massive}
For two quark masses that are far apart, such as $m_b$ and $m_t$, the schemes described above can easily be combined sequentially.  However, the masses $m_c$ and $m_b$ are not well separated, and there is a region of $y$ where it is appropriate to keep both of them in the $1\to 2$ splitting kernels.

\begin{figure}[t]
   \begin{center}
      \includegraphics[height = 0.65\linewidth]{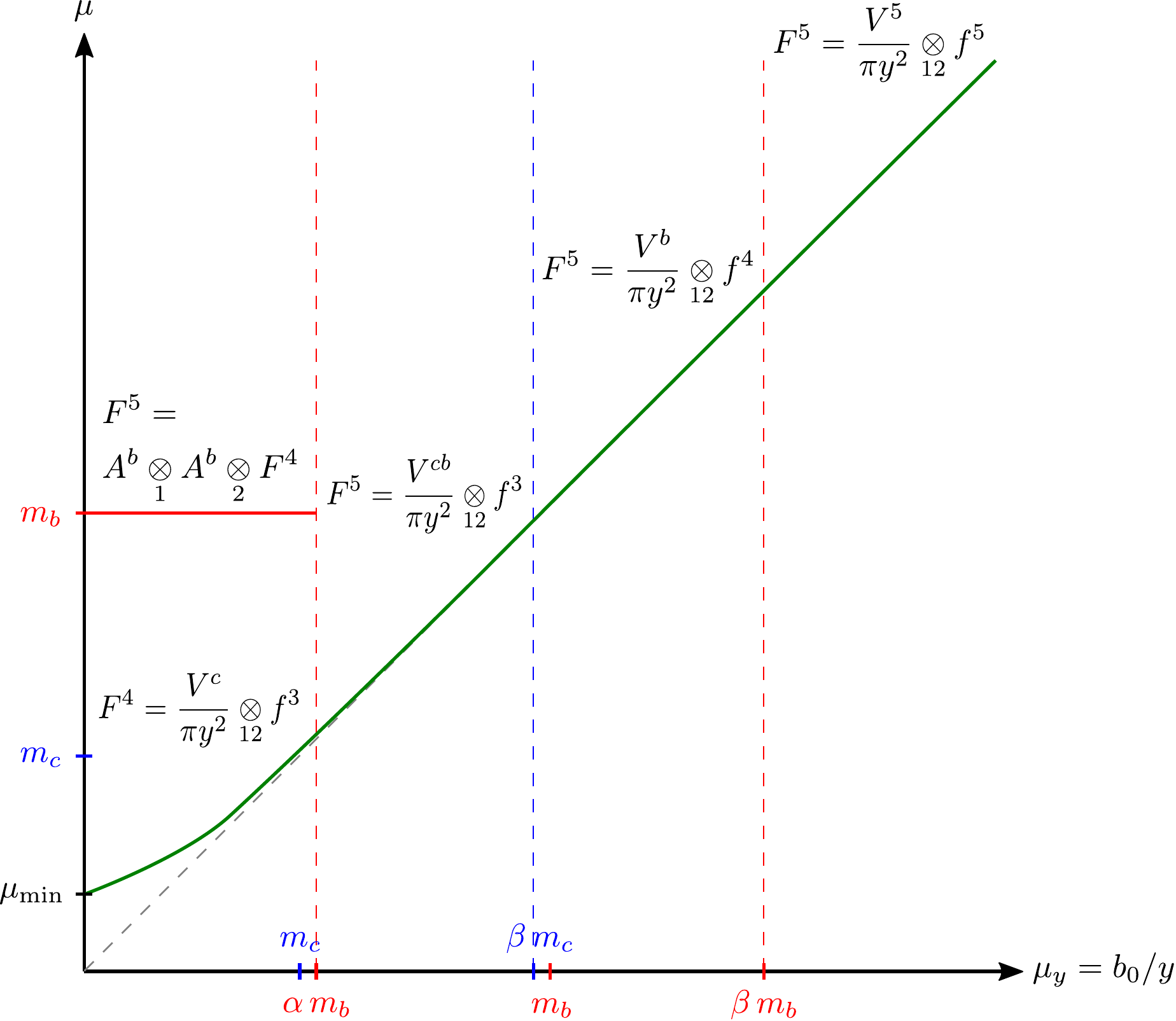}
   \end{center}
   \caption{\label{fig:scheme-two-massive} Illustration of the massive scheme for charm and bottom quarks.  The green line indicates the splitting scale
   $\mu_{\text{split}} = \mu_{y^*}$ specified in \eqn{\protect\eqref{eq:mu-y-star}} and below, and the solid red line indicates the flavour matching scale $\mu_b = m_b$.}
\end{figure}

A scheme that takes this into account is shown in \fig{\ref{fig:scheme-two-massive}}.
The general idea remains the same as in the case of a single heavy flavour, with the main difference being that now there is a regime where both the $c$ and $b$ quarks are treated as massive.  Notice that this scheme does \emph{not} include a separate region $\mu_y < \alpha \ms m_c$ where the charm quark would be absent in the $1\to 2$ splitting process and generated by flavour matching from $F^3$ to $F^4$.  This is because for $\alpha \ll 1$, the value of $\alpha \ms m_c$ is not large compared with nonperturbative scales.  The $\nf = 4$ splitting DPD is thus computed with the massive kernel $V^{c, 3} = V^{Q, 3}$.

In addition, the figure explicitly shows that for low $\mu_y$ the splitting is to be computed at the modified scale $\mu_{y^*}$ of \eqn{\eqref{eq:mu-y-star}} rather than at $\mu_y$, in order to keep scales in the perturbative region.  Moreover, for values of $y$ where $\mu_{y*}$ differs appreciably from $\mu_y$, the perturbative splitting formula for DPDs loses its validity.  One way to deal with this issue is described in \sect{\ref{sec:LO-DPDs}}.

The computation of the DPD at perturbative values of $y$ thus proceeds as follows:
\begin{align}
   \label{eq:two-massive-1}
      F_{a_1 a_2}^{4}(y; \mu_{\text{split}})
   &= \frac{1}{\pi y^2} \;
      \sum_{a_0} V_{a_1 a_2, a_0}^{c, 3}(y, m_c; \mu_{\text{split}}) \conv{12}
      f_{a_0}^{3}(\mu_{\text{split}}) \,,
\nonumber \\
      F_{a_1 a_2}^{5}(y; \mu_{b})
   &= \sum_{b_1, b_2} A_{a_1 b_1}^{b, 4}(m_b; \mu_b) \conv{1}
      A_{a_2 b_2}^{b, 4}(m_b; \mu_b) \conv{2}
      F_{b_1 b_2}^{4 \phantom{b,}}(y; \mu_{b})
& \text{ for $\mu_y < \alpha \ms m_b$}
\end{align}
with $\mu_{\text{split}} \sim \mu_{y^*}$ and $\mu_b \sim m_b$,
\begin{align}
   \label{eq:two-massive-3}
      F_{a_1 a_2}^{5}(y; \mu_{\text{split}})
   &= \frac{1}{\pi y^2} \;
      \sum_{a_0} V_{a_1 a_2, a_0}^{c b}(y, m_c, m_b; \mu_{\text{split}})
      \conv{12} f_{a_0}^{3}(\mu_{\text{split}})
   & \text{\! for $\alpha \ms m_b \! < \! \mu_y \! < \! \beta \ms m_c$,}
\end{align}
\begin{align}
   \label{eq:two-massive-4}
      F_{a_1 a_2}^{5}(y; \mu_{\text{split}})
   &= \frac{1}{\pi y^2} \;
      \sum_{a_0} V_{a_1 a_2, a_0}^{b, 4}(y, m_b; \mu_{\text{split}}) \conv{12}
      f_{a_0}^{4}(\mu_{\text{split}})
   & \text{ for $\beta \ms m_c < \mu_y < \beta \ms m_b$,}
\end{align}
and
\begin{align}
   \label{eq:two-massive-5}
      F_{a_1 a_2}^{5}(y; \mu_{\text{split}})
   &= \frac{1}{\pi y^2} \;
      \sum_{a_0} V_{a_1 a_2, a_0}^{5}(y; \mu_{\text{split}}) \conv{12}
      f_{a_0}^{5}(\mu_{\text{split}})
   & \text{ for $\mu_y > \beta \ms m_b$.}
\end{align}
In \eqs{\eqref{eq:two-massive-3}} to \eqref{eq:two-massive-5} one may use either $\mu_{\text{split}} \sim \mu_{y}$ or $\mu_{\text{split}} \sim \mu_{y*}$.  The PDFs in \eqref{eq:two-massive-4} and \eqref{eq:two-massive-5} are obtained from $f^{3}$ by flavour matching at the appropriate scales.

%
\graphicspath{{Figures/LO_numerics/}}
%
%
\section{Numerical studies with LO splitting kernels}
\label{sec:LO-numerics}
We now present numerical studies of the schemes introduced in the previous section.  Since the massive $1 \to 2$ splitting kernels are at present only known at leading order, we limit ourselves to that order.  The initialisation and scale evolution of the splitting DPDs are performed with the \textsc{ChiliPDF} library \cite{Diehl:2021gvs, Diehl:2023cth}.   Unless stated otherwise, scale evolution and flavour matching of DPDs, PDFs, and $\alpha_s$ are all performed at LO, with flavor matching done at $\mu_Q = \mQ$.  We restrict ourselves to unpolarised partons.

To cover effects from all three heavy quarks, we consider DPDs evolved to the scale $\mu = Q$ at momentum fractions associated with one of the settings
\begin{align}
   \label{eq:dijet-setting}
\nf &= 5 \,, & Q &= 25 \gev \,, & \sqrt{s} &= 14 \tev \,,
\\
   \label{eq:ttbar-setting}
\nf &= 6 \,, & Q &= 1 \tev \,, & \sqrt{s} &= 100 \tev \,,
\end{align}
where $Q$ is the invariant mass of the system produced in each of the hard-scattering processes, and $\sqrt{s}$ is the c.m.\ energy of the proton-proton collision.
Examples of corresponding DPS processes are \rev{the production of two dijets with heavy flavors} at the LHC for the first setting and double $t \tbar$ production at a future hadron collider for the second one.  In both cases, $Q$ is significantly larger than the mass of the heaviest active parton in the DPD, so that in the hard-scattering cross section all active partons may be treated as massless.

Given the special interest of like-sign $W$ production for DPS studies \cite{Kulesza:1999zh, Gaunt:2010pi, Ceccopieri:2017oqe, Cotogno:2018mfv, Cotogno:2020iio, Cabouat:2019gtm, Sirunyan:2019zox, CMS:2022pio}, we also considered the setting
\begin{align}
   \label{eq:WW-setting}
\nf &= 5 \,, & Q &= m_W \,, & \sqrt{s} &= 14 \tev \,.
\end{align}
We did not find any features in this setting that are not also present in the one of \eqn{\eqref{eq:dijet-setting}} and therefore refer to \app{\ref{app:add-numerics}} for its discussion.
%
%
\subsection{Splitting DPDs}
\label{sec:LO-DPDs}
We start by looking at the splitting DPDs.  For each of the three settings just discussed, the parton momentum fractions $x_1$ and $x_2$ are chosen such that the final-state system corresponding to each hard scatter is produced at central rapidity, i.e.
\begin{align}
   \label{eq:mom-fracs-central-rap}
      x_1 = x_2 = Q / \sqrt{s} \,.
\end{align}
As discussed in \sect{\ref{sec:two-massive}}, we initialise the splitting DPDs at the scale $\mu_{\text{split}} = \mu_{y^*}$, where $\mu_{y^*}$ is specified by \eqref{eq:mu-y-star} with minimal value $\mu_{\text{min}} = 1 \gev$ and by \eqref{eq:y-star} with power $p=4$.  In addition, we modify the perturbative splitting formula by multiplying with an exponential damping factor,

\begin{align}
   \label{eq:small-y-DPD-LO-mod}
      F_{a_1 a_2}(x_1,x_2,y; \mu)
   &= \exp \left[\frac{-y^2}{4 h_{a_1 a_2}}\right] \;
      \frac{a_s(\mu)}{\pi y^2} \,
      V_{a_1 a_2, a_0}\left(\frac{x_1}{x_1 + x_2}\right)
      \frac{f_{a_0}(x_1 + x_2; \mu)}{x_1 + x_2}
   \,,
\end{align}
where $V$ is either $V^{(1)}$ or $V^{Q (1)}$.  For the damping constants, we take
\begin{align}
   \label{eq:damping-parameters}
      h_{g g} = 4.66 \gev^{-2} \,,
   && h_{g q} = h_{q g} = 5.86 \gev^{-2} \,,
   && h_{q q} = 7.06 \gev^{-2} \,.
\end{align}
The exponential gives a more realistic behaviour at nonperturbative $y$ while not significantly affecting the perturbative region.  For a motivation of the values in \eqref{eq:damping-parameters} we refer to \sect{9.2.1} of reference \cite{Diehl:2017kgu}.

For the PDFs in the splitting formula, we take the default LO set of the MSHT20 parame\-trisation \cite{Bailey:2020ooq}, with the associated values $\alpha_s(m_Z) = 0.13$ for the strong coupling and
\begin{align}
\label{eq:MSHT-masses}
m_c &= 1.4 \gev \,,
&
m_b &= 4.75 \gev \,,
&
m_t &= 172.5 \gev
\end{align}
for the quark masses.  We also produced plots for LO sets from HERAPDF2.0 and NNPDF4.0 \cite{H1:2015ubc, NNPDF:2021njg} and find them to be quite similar to the ones for MSHT20.\footnote{%
Specifically, we used the sets named \texttt{MSHT20lo\_as130}, \texttt{HERAPDF20\_LO\_EIG}, and \texttt{NNPDF40\_lo\_pch\_as\_01180} in the LHAPDF interface \cite{Buckley:2014ana}.
}
In particular, they lead us to the same conclusions regarding the values of our scheme parameters $\alpha$, $\beta$, and $\gamma$.

The focus of our attention is the $y$ dependence of the DPDs evolved to scale $\mu$ for different parton combinations.  We will plot against $\mu_y = b_0 / y$ rather than $y$, in order to facilitate comparison with the graphs in \figs{\ref{fig:scheme-massless}} to \ref{fig:scheme-two-massive}.  Let us recall that we need the DPDs for $\mu_y$ up to order $\mu$ when computing DPS observables.

For a given number $\nf$ of active flavours, we expect that the DPDs are smooth functions of $y$ (and hence of $\mu_y$), since $y$ corresponds to a space-like distance between different fields, and there are no physical production thresholds associated with such a distance.  In both the massless and the massive schemes we introduced earlier, we will however find discontinuities at the transition points between regimes where different approximations are made in computing the DPDs.  While this is unavoidable, we regard it as desirable to minimise such discontinuities.  We will use this as our default criterion to select the parameters $\alpha$ and $\beta$ in the massive scheme.
%
%
\subsubsection{Massless scheme}
\label{sec:massless-DPDs}
To begin with, we discuss the DPDs in the massless scheme of \sect{\ref{sec:massless}}, where the splitting DPDs are initialised for $\nf = 3$, $4$, $5$ or $6$ active flavours as $\mu_y$ increases, with transition points at $\gamma m_c$, $\gamma m_b$, and $\gamma m_t$.  The qualitative features we wish to discuss depend weakly on $\gamma$, and we find it sufficient to show plots with $\gamma = 1$ in the following.

\begin{figure}[t]
   \begin{center}
      \subfigure[\label{subfig:Fccbar-massless}${c \cbar}$]{
         \includegraphics[width=0.475\linewidth, trim=0 0 0 50, clip]{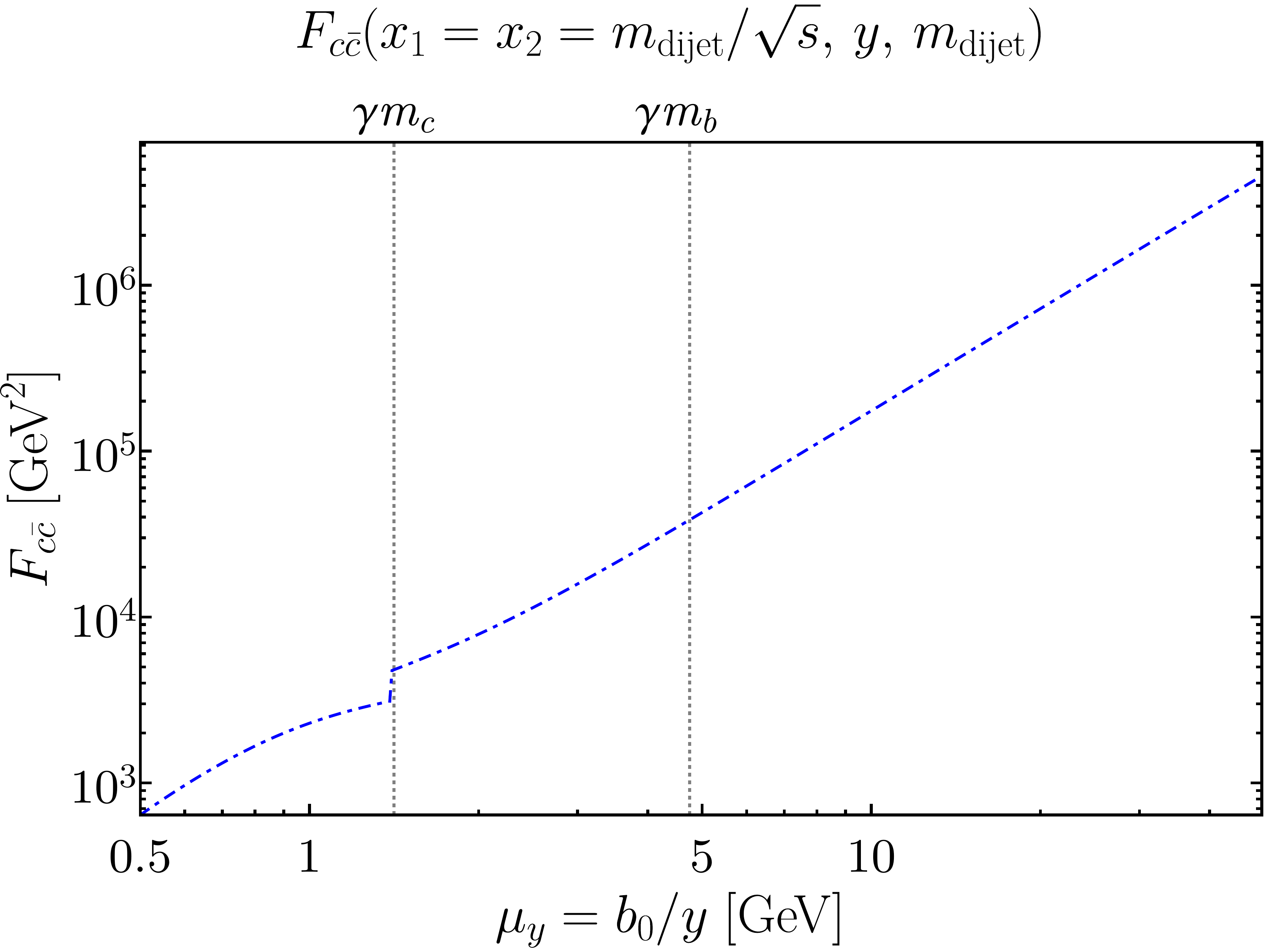}
      }
      \subfigure[\label{subfig:Fbbbar-massless}${b \bbar}$]{
         \includegraphics[width=0.475\linewidth, trim=0 0 0 50, clip]{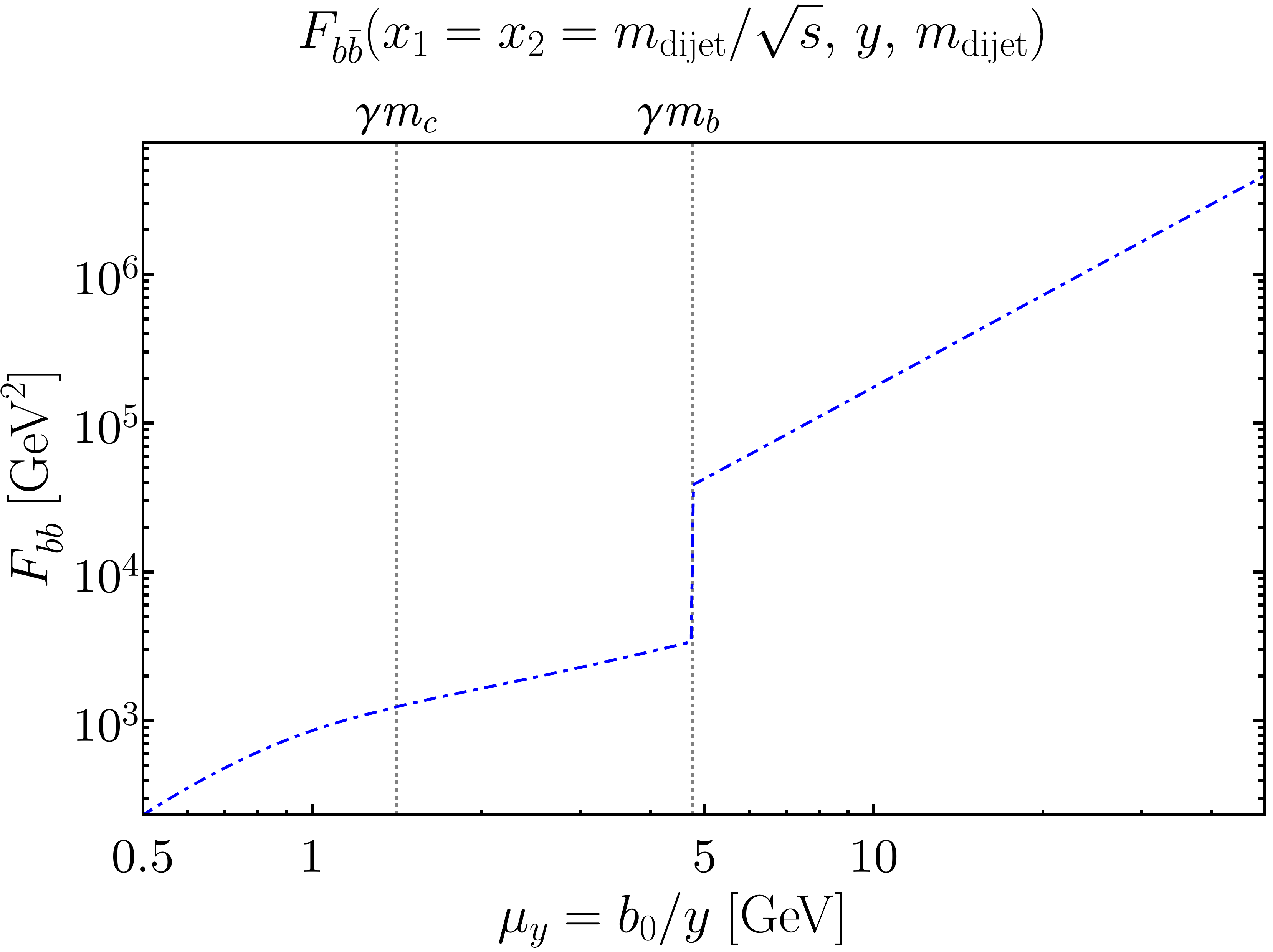}
      }
      \\
      \subfigure[\label{subfig:Fcb-massless}${c b}$]{
         \includegraphics[width=0.475\linewidth, trim=0 0 0 50, clip]{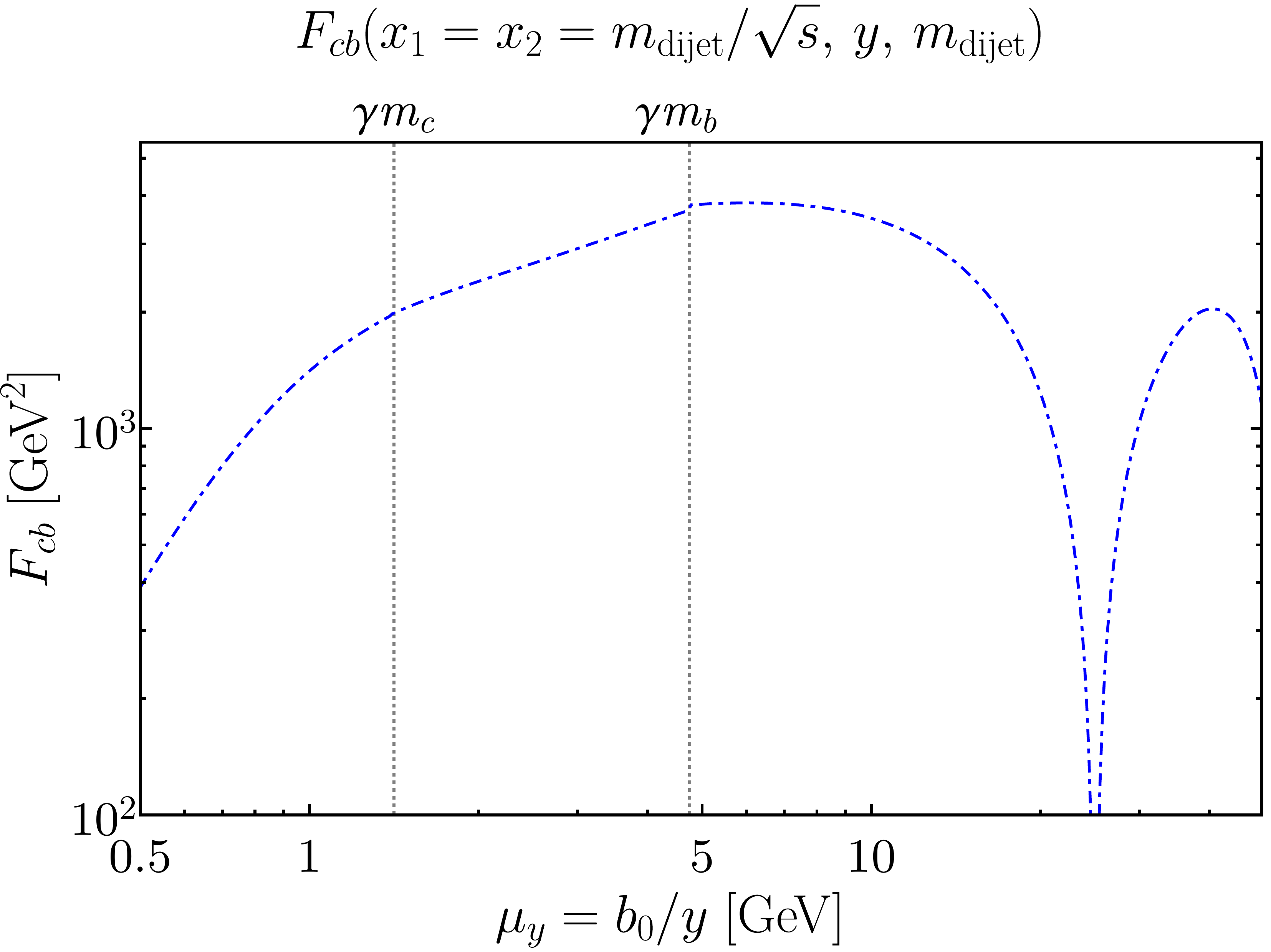}
      }
      \subfigure[\label{subfig:Fgb-massless}${g b}$]{
         \includegraphics[width=0.475\linewidth, trim=0 0 0 50, clip]{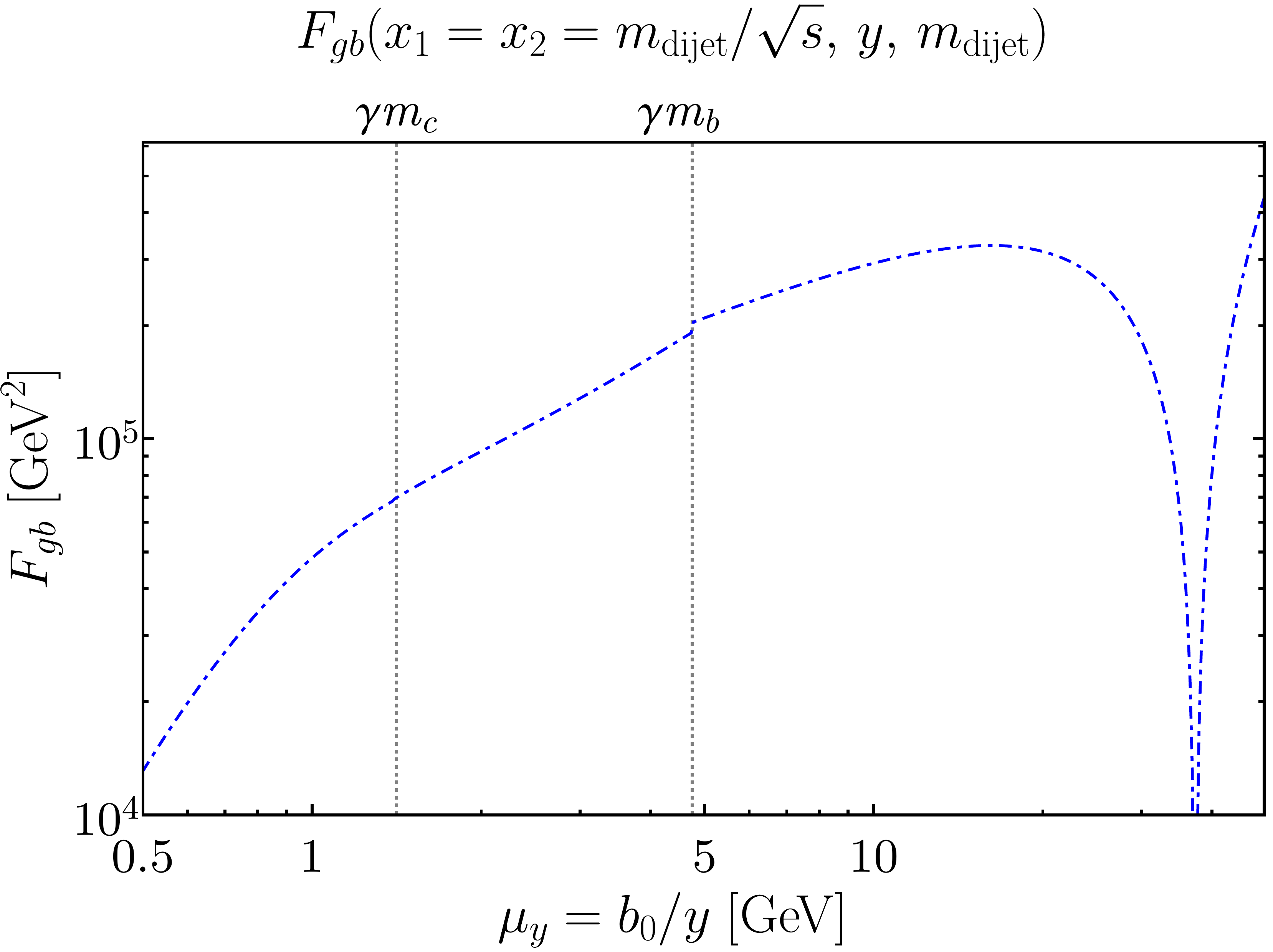}
      }
   \end{center}
   \caption{\label{fig:F-massless} $\nf = 5$ splitting DPDs at $\mu = 25 \gev$ for the setting \protect\eqref{eq:dijet-setting}, computed in the purely massless scheme of \sect{\protect\ref{sec:massless}} with $\gamma = 1$. The momentum fractions are $x_1 = x_2 \approx 1.8 \times 10^{-3}$ according to \protect\eqref{eq:mom-fracs-central-rap}.  Here and in similar figures $|F_{a_1 a_2}|$ is shown.}
\end{figure}

In \fig{\ref{fig:F-massless}} we show the DPDs for the jet production setting at $\mu = 25 \gev$ introduced above.
We see in figure \ref{subfig:Fccbar-massless} that the $c \cbar$ distribution has a  discontinuity at $\mu_y = \gamma \ms m_c$. This is readily understood: above this value, charm is treated as massless and $F_{c \cbar}$ can be directly produced by $g \to  c \cbar$ splitting.  Below this value, charm is only produced by evolution above the flavour matching scale $\mu_c = m_c$.  For $F_{c \cbar}$ this requires two $g \to  c \cbar$ splittings in the evolution chain. The discussion of $F_{b \smash{\bbar}}$, shown in \ref{subfig:Fbbbar-massless}, proceeds in full analogy.  The discontinuity is even stronger in this case.

A very different behaviour is seen for the $c b$ distribution in figure \ref{subfig:Fcb-massless}.  This parton combination is not produced by direct splitting at LO, but only by scale evolution, which explains the zero crossing at $\mu_y = \mu = 25 \gev$, where no evolution takes place.  Above this value, the DPD is actually negative, because one has to evolve backwards from $\mu_y$ to $\mu$.
To understand the tiny discontinuity of the distribution at $\gamma \ms m_b$, we note that above this value $F_{\, \smash{\bbar} b}$ is directly produced by splitting (see  \fig{\ref{subfig:Fbbbar-massless}}), so that $F_{c b}$ is produced more abundantly by evolution. The corresponding effect at $\gamma \ms m_c$ is too small to be visible in the plot.  We remark in passing that when the $1\to 2$ splitting is evaluated at NLO, the $c b$ channel can directly be produced by $c \to c b$ and $b \to c b$ for $\mu_y > \gamma \ms m_b$, so that one can expect a more pronounced discontinuity in that case.

As an example for a channel with one observed heavy flavour, we show the $g b$ distribution in \fig{\ref{subfig:Fgb-massless}}. We see a small discontinuity at $\mu_y = \gamma \ms m_b$, above which $F_{g b}$ can be produced directly by $b \to g b$ splitting, or by one evolution step from $F_{\, \smash{\bbar} b}$. The fact that the discontinuity is not very pronounced suggests that the dominant contribution in this region comes from the evolution of $F_{g g}$ via a $g \to b$ splitting.  Indeed, one finds that $F_{g g}$ is larger than $F_{\, \smash{\bbar} b}$ by more than a factor $100$ around $\mu_y = \gamma \ms m_b$.

For DPDs with only light flavours, we find a smooth behaviour at the transition points in $\mu_y$, as one may expect.
The behaviour of the $\nf = 6$ DPDs in the $t \tbar$ setting \eqref{eq:ttbar-setting} is analogous to the one just discussed and therefore not shown here.
%
%
\subsubsection{Massive scheme}
\label{sec:massive-DPDs}
We now investigate the scheme where massive splitting kernels are used along with massless ones.  DPDs with $\nf = 5$ are computed as laid out in \sect{\ref{sec:two-massive}}.  DPDs with $\nf = 6$ are obtained from these by flavour matching if $\mu_y < \alpha m_t$, whereas for $\mu_y > \alpha m_t$ the prescription of \sect{\ref{sec:one-massive}} is used.  We computed distributions with different scheme parameters, namely $\alpha = 1/2, 1/3, 1/4$ and $\beta = 2, 3, 4$, and will compare a subset of these in the following.  We do not consider smaller $\alpha$ or larger $\beta$ in order to avoid large logarithms in the intermediate mass region, as explained in \sect{\ref{sec:one-massive}}.

\begin{figure}[p]
   \begin{center}
      \subfigure[\label{subfig:Fbbbar-massive-2}${b \bbar}, \; 1/\alpha = \beta = 2$]{
         \includegraphics[width=0.475\linewidth, trim=0 0 0 50, clip]{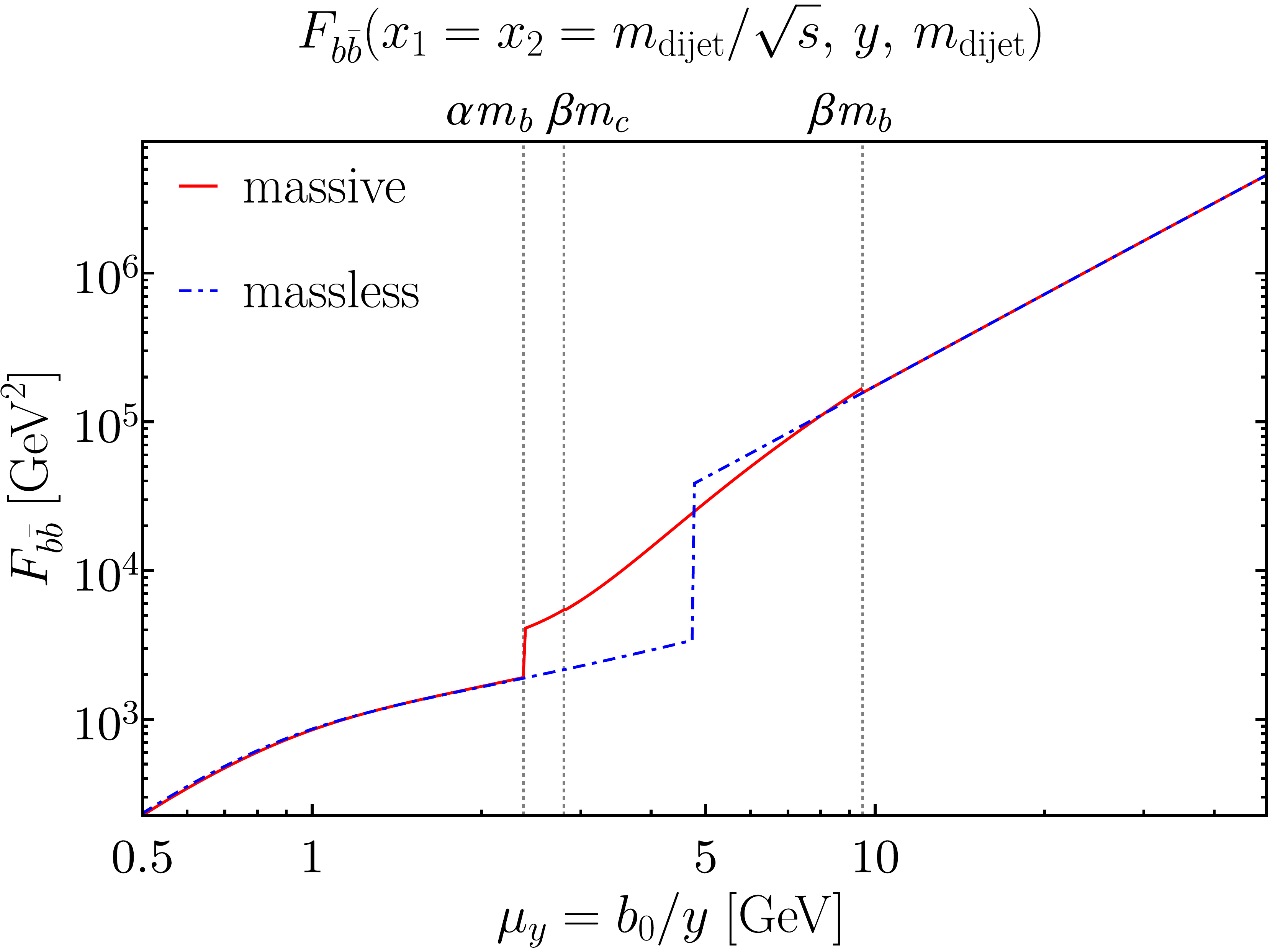}
      }
      \subfigure[\label{subfig:Fbbbar-massive-4}${b \bbar}, \; 1/\alpha = \beta = 4$]{
         \includegraphics[width=0.475\linewidth, trim=0 0 0 50, clip]{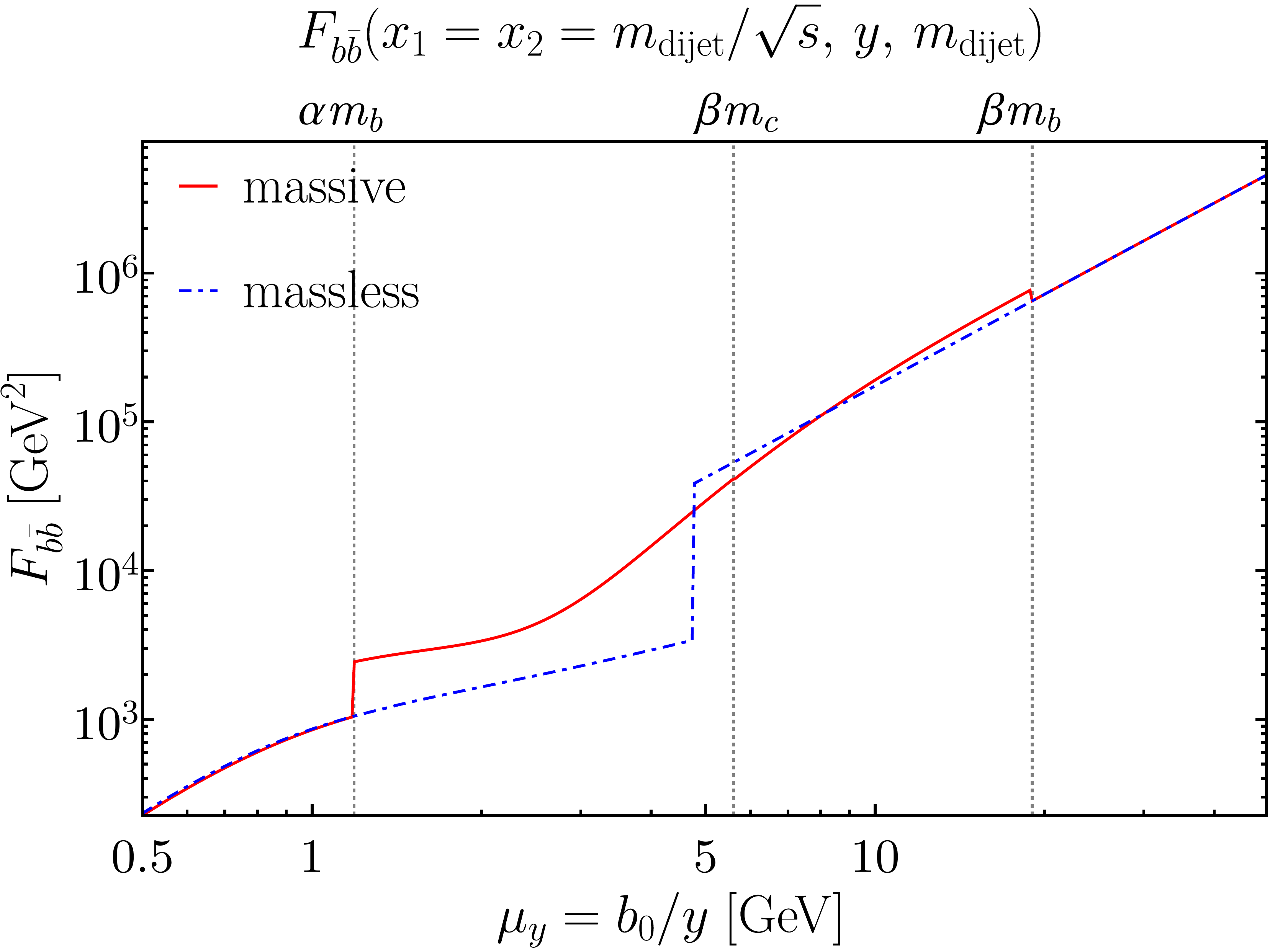}
      }
      \\
      \subfigure[\label{subfig:Fgb-massive-2}${g b}, \; 1/\alpha = \beta = 2$]{
         \includegraphics[width=0.475\linewidth, trim=0 0 0 50, clip]{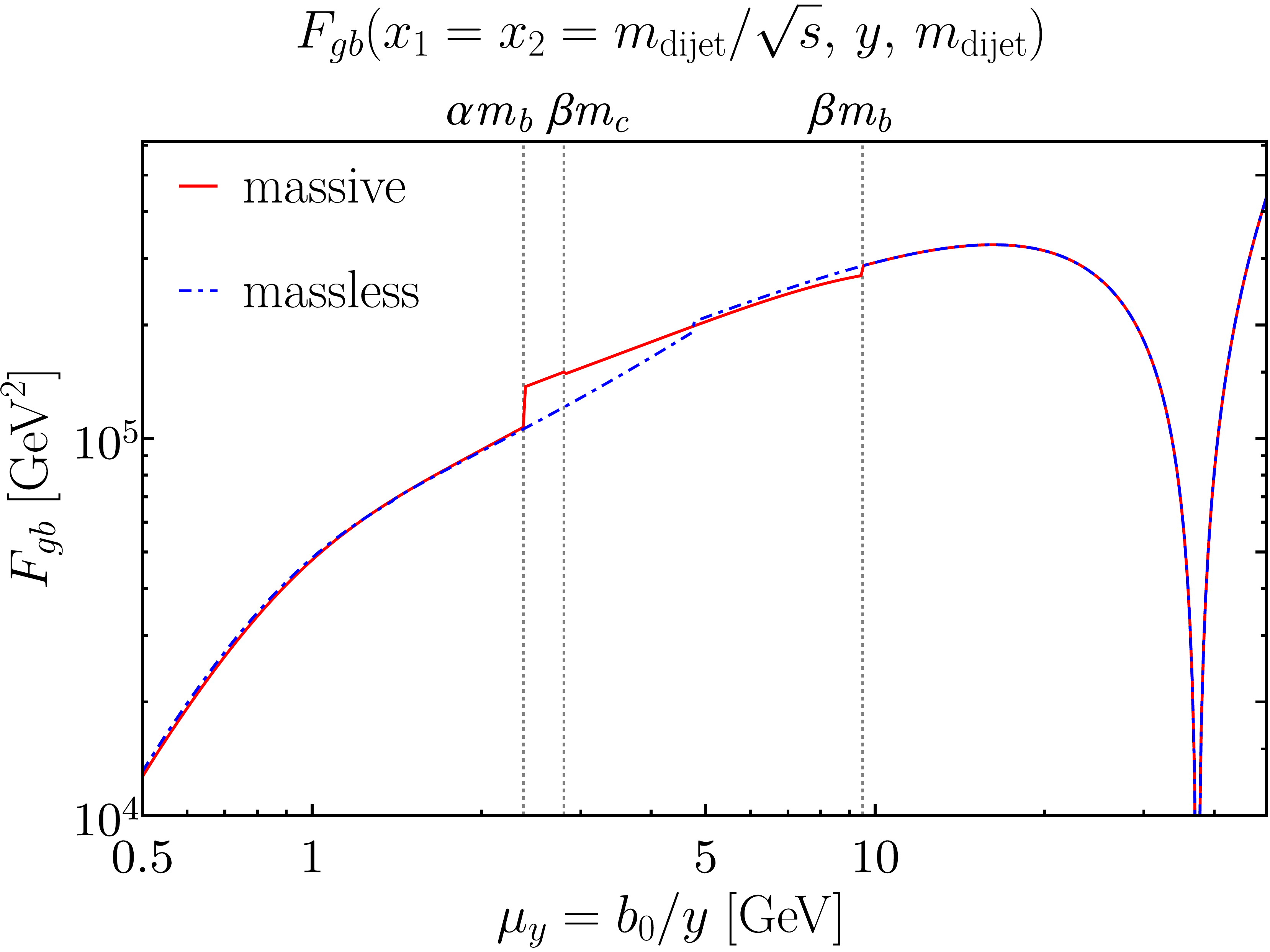}
      }
      \subfigure[\label{subfig:Fgb-massive-4}${g b}, \; 1/\alpha = \beta = 4$]{
         \includegraphics[width=0.475\linewidth, trim=0 0 0 50, clip]{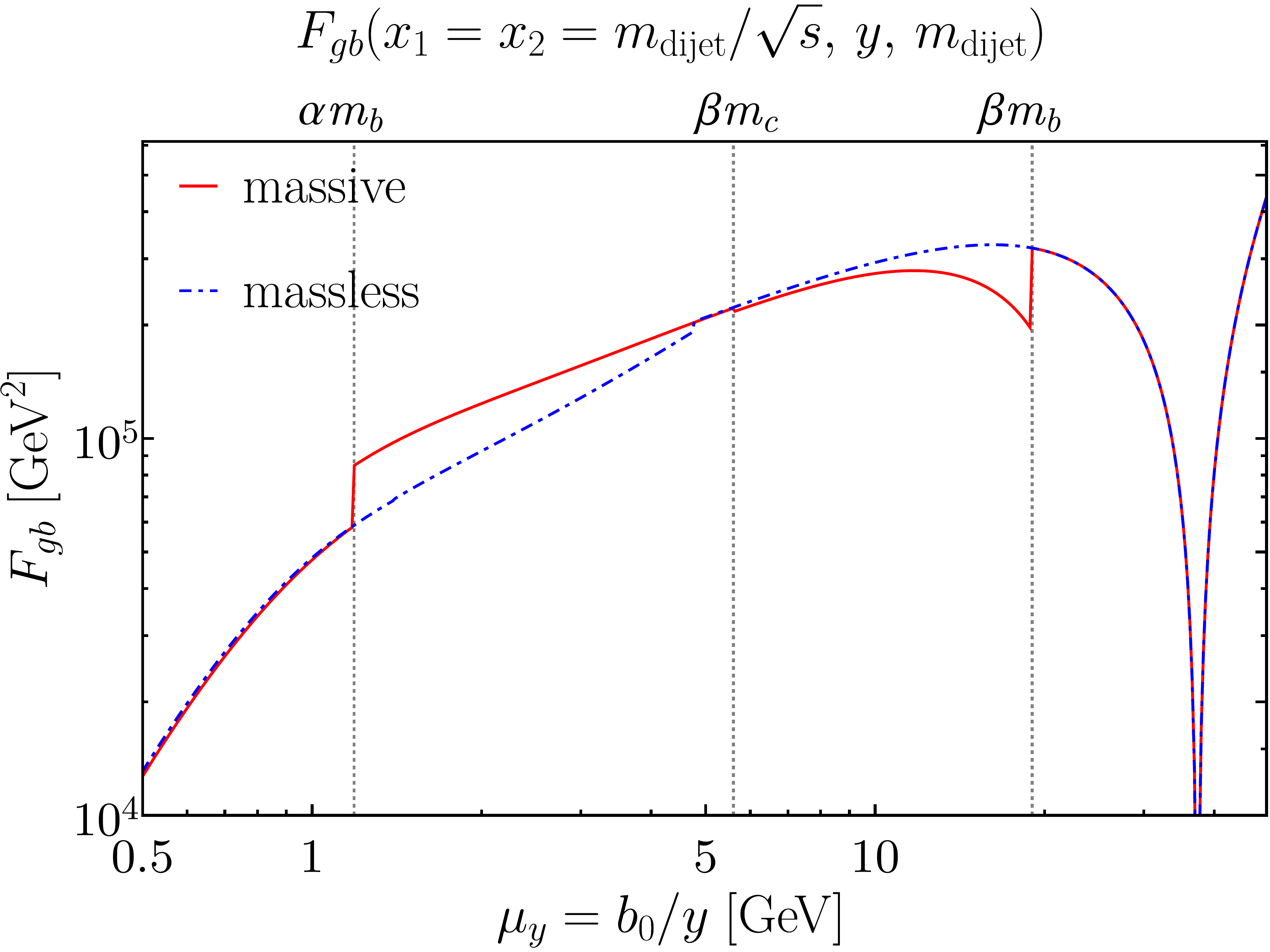}
      }
      \\
      \subfigure[\label{subfig:Fgg-massive-2}${g g}, \; 1/\alpha = \beta = 2$]{
         \includegraphics[width=0.475\linewidth, trim=0 0 0 50, clip]{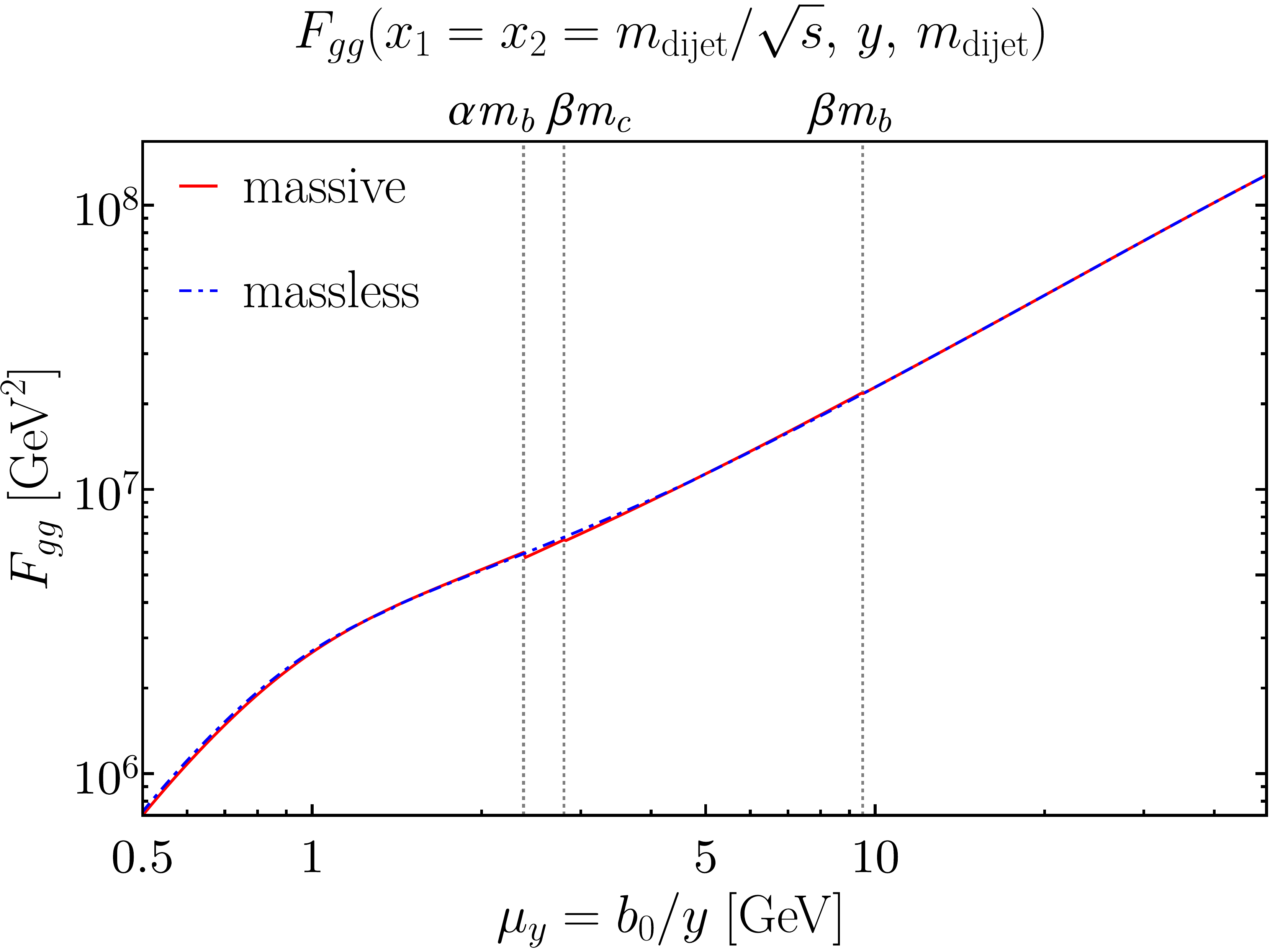}
      }
      \subfigure[\label{subfig:Fgg-massive-4}${g g}, \; 1/\alpha = \beta = 4$]{
         \includegraphics[width=0.475\linewidth, trim=0 0 0 50, clip]{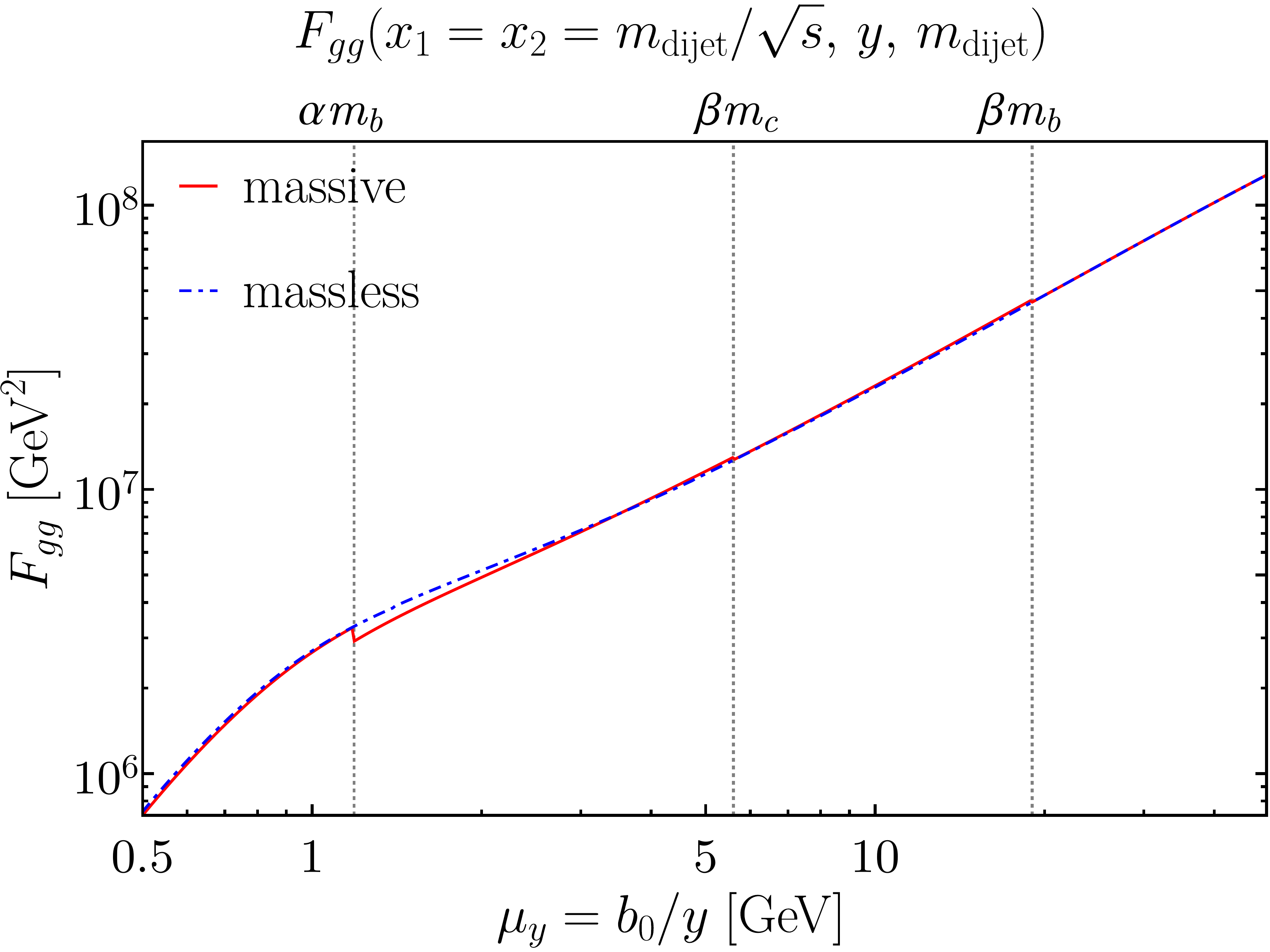}
      }
   \end{center}
   \caption{\label{fig:F-alpha-beta-comp} $\nf = 5$ splitting DPDs at $\mu = 25 \gev$ for the setting \protect\eqref{eq:dijet-setting}.  Solid lines are for the massive scheme of section \ref{sec:two-massive}, and dashed lines are for the massless scheme with $\gamma = 1$ (already shown in \fig{\protect\ref{fig:F-massless}}).}
 \end{figure}

Let us first look at distributions in setting \eqref{eq:dijet-setting}, beginning with $F_{b \smash{\bbar}}$ shown in \figs{\ref{subfig:Fbbbar-massive-2}} and \ref{subfig:Fbbbar-massive-4}.  Instead of the large discontinuity at $\gamma \ms m_b$ in the massless scheme, the massive scheme has a much smaller discontinuity at $\alpha \ms m_b$ and a tiny one at $\beta \ms m_b$.  The tiny jump at $\beta \ms m_b$ is due to switching from massive to massless DPD splitting kernels \rev{---} which according to \eqref{eq:VQQbarg-LO-small-y} is a small effect \rev{---} and to changing from a $4$ to a $5$ flavour gluon density in the splitting formula.  Quantitatively, we find that this jump is somewhat smaller for $\beta = 2$ than for $\beta = 4$.
The discontinuity at $\alpha \ms m_b$ can also be ascribed to two effects.  For $\mu_y$ above this value, $b \bar{b}$ can be produced by direct splitting, although the massive splitting kernels are exponentially suppressed at $\mu_y \ll m_b$ according to \eqref{eq:VQQbarg-LO-large-y}.  Furthermore, the $b \bar{b}$ distribution is initialised at scale $\mu_y$ to the right of $\alpha \ms m_b$, whereas to the left of this point it can be produced by evolution only above the flavour matching scale $\mu_b = m_b$, i.e.\ with a shorter evolution path towards the final scale $\mu$.  Quantitatively, we find that the absolute size of the discontinuity is smaller for $\alpha = 1/4$ than for $\alpha = 1/2$.  Among the scheme parameters we considered, the mildest discontinuities for the $b \bbar$ distribution are thus obtained for the combination $\alpha = 1/4$ and $\beta = 2$.

\begin{figure}[p]
   \begin{center}
      \subfigure[\label{subfig:Fgb-massless-scheme}$F_{g b}$ from $f_b$]{
         \includegraphics[width=0.25\linewidth, trim=0 0 0 0, clip]{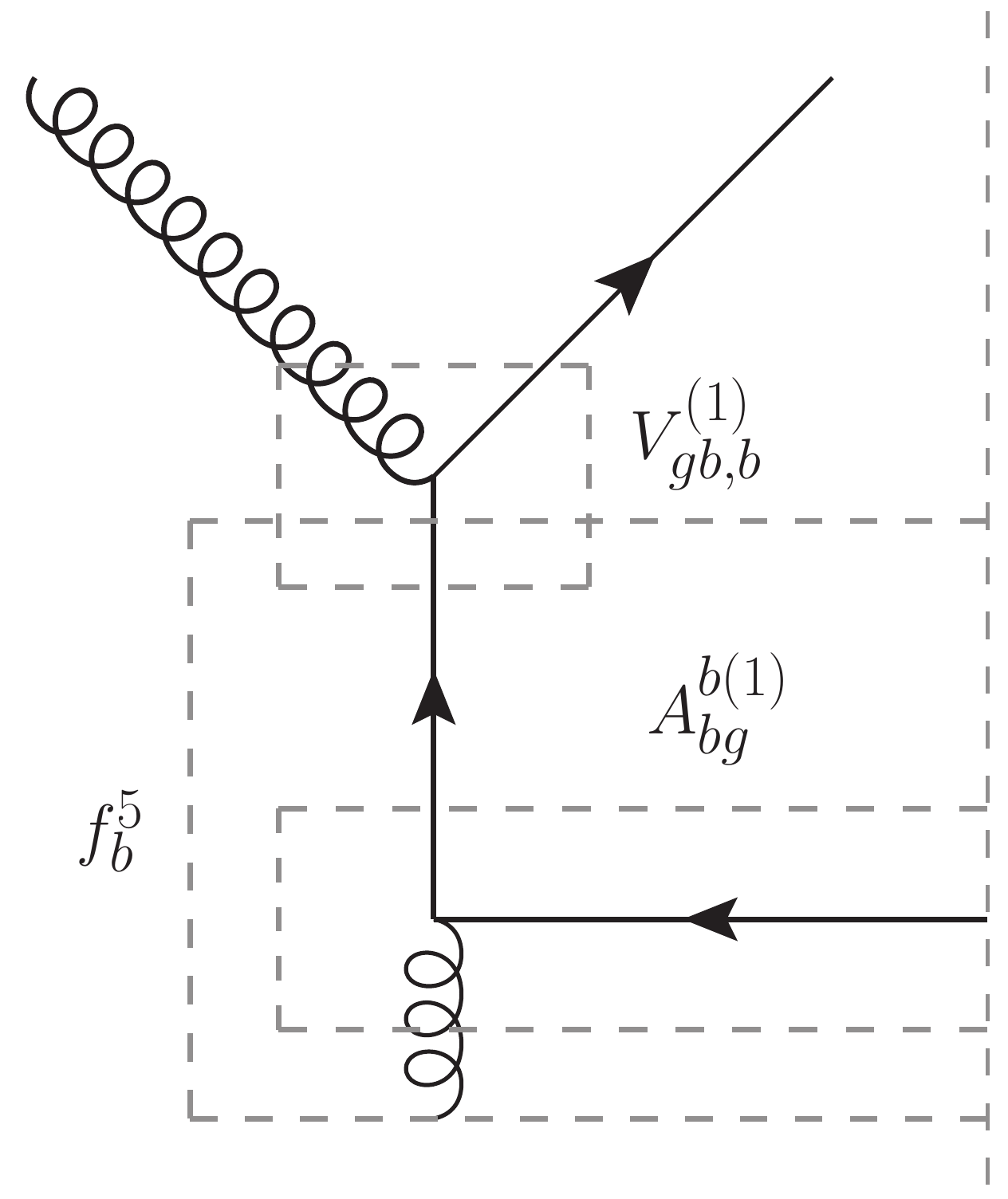}
      }
      \hfill
      \subfigure[\label{subfig:Fbg-bbbar}$F_{b g}$ from $F_{\smash{b \bbar}}$]{
         \includegraphics[width=0.25\linewidth, trim=0 0 0 0, clip]{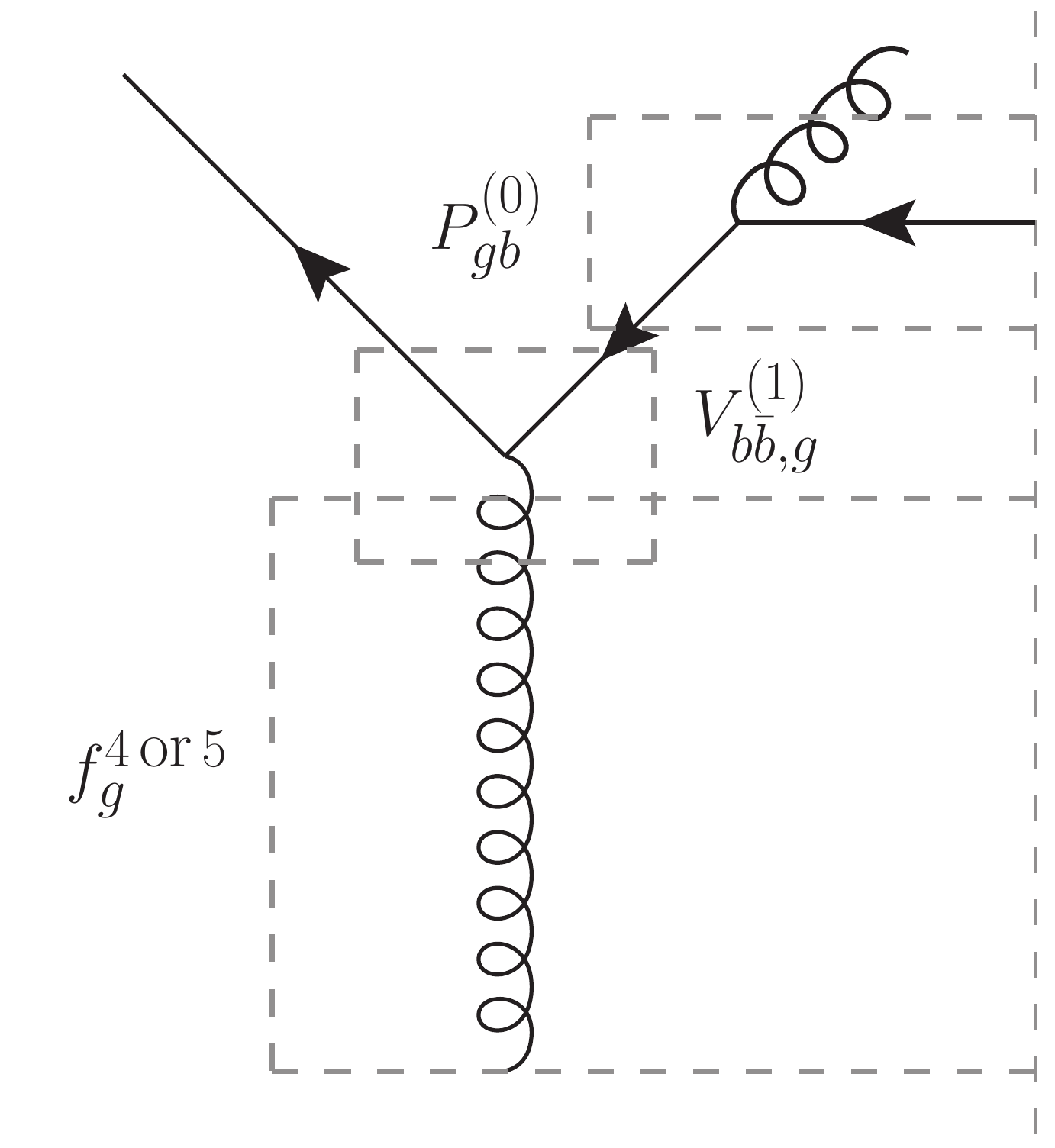}
      }
      \hfill
      \subfigure[\label{subfig:Fgb-gg}$F_{g b}$ from $F_{g g}$]{
         \includegraphics[width=0.25\linewidth, trim=0 0 0 0, clip]{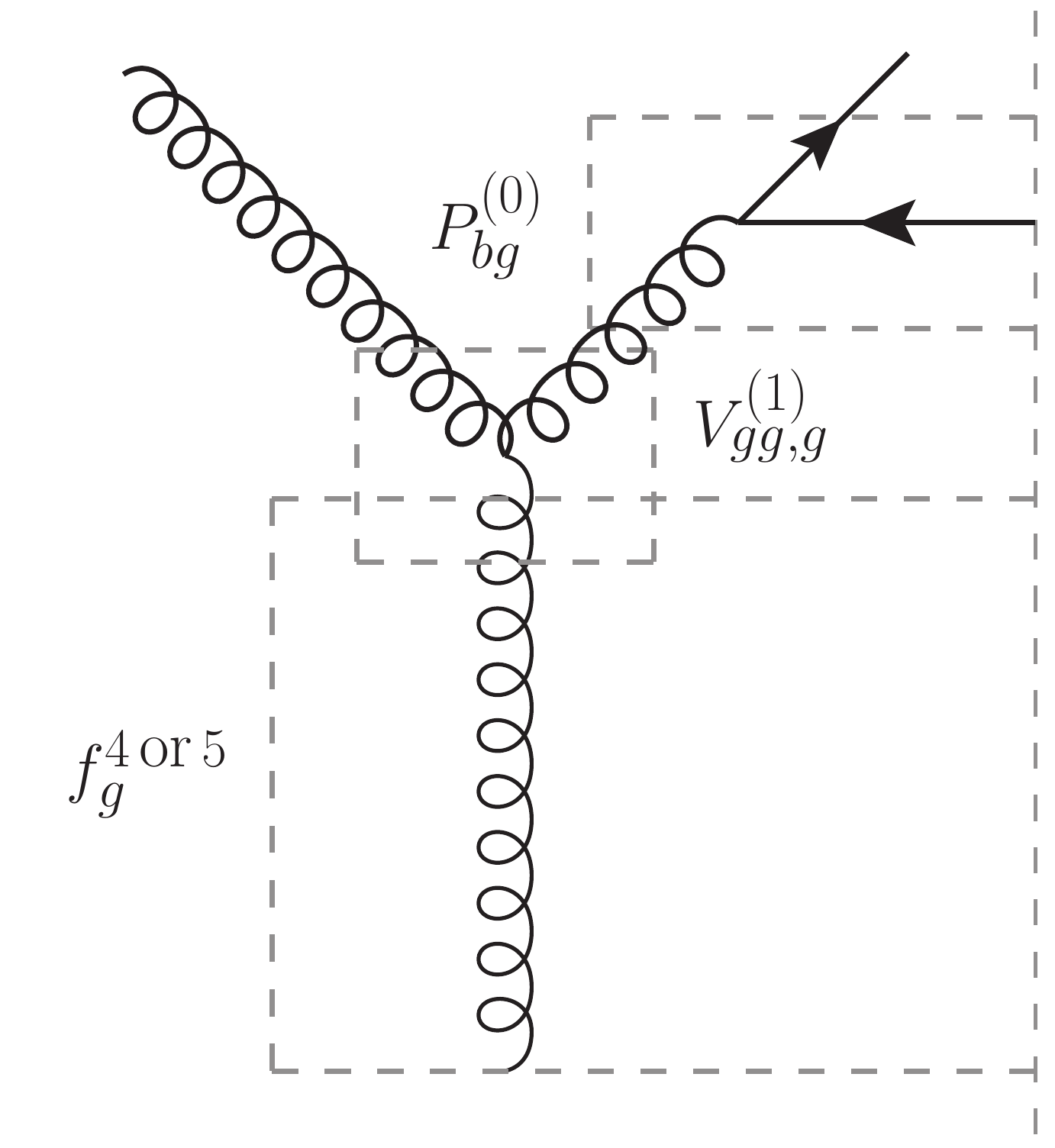}
      }
   \end{center}
   \caption{\label{fig:Fgb-diagrams} Production modes of $F_{g b}$ for $\mu_y > \gamma \ms m_b$ in the massless scheme and for $\beta \ms m_c < \mu_y < \beta \ms m_b$ in the massive scheme. For $\mu_y > \beta \ms m_b$ the two schemes coincide.  Horizontal parton lines correspond to unobserved spectators. In panel \hyperref[subfig:Fbg-bbbar]{(b)} $F_{b g}$ is shown for clarity of presentation --- from this $F_{g b}$ is readily obtained by swapping the two partons.  In the same panel, $V^{(1)}$ must be replaced by $V^{b (1)}$ for $\beta \ms m_c < \mu_y < \beta \ms m_b$ in the massive scheme.}
   \begin{center}
      \subfigure[\label{subfig:Fccbar-massive-24}${c \cbar}$]{
         \includegraphics[width=0.475\linewidth, trim=0 0 0 50, clip]{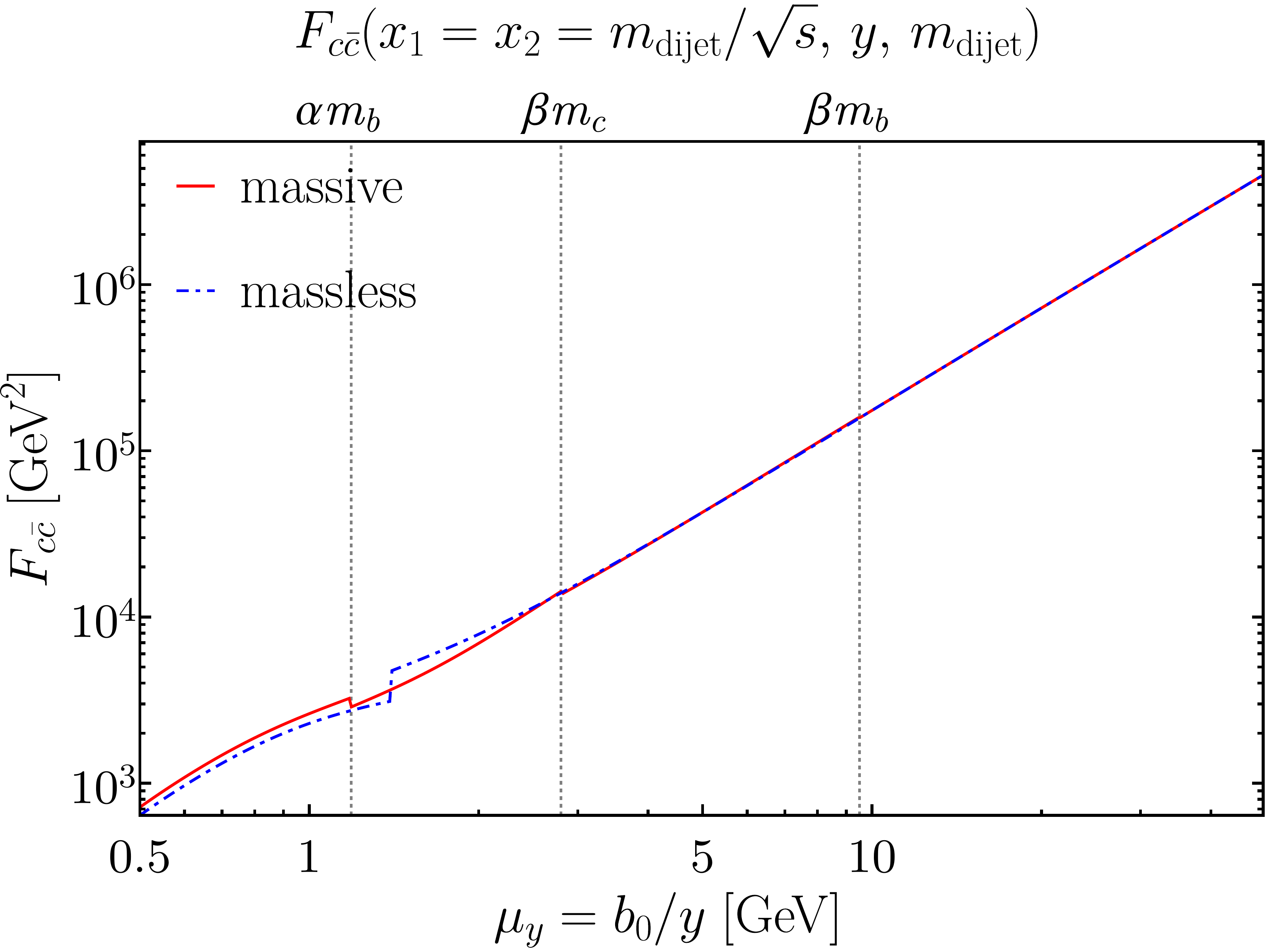}
      }
      \subfigure[\label{subfig:Fbbbar-massive-24}${b \bbar}$]{
         \includegraphics[width=0.475\linewidth, trim=0 0 0 50, clip]{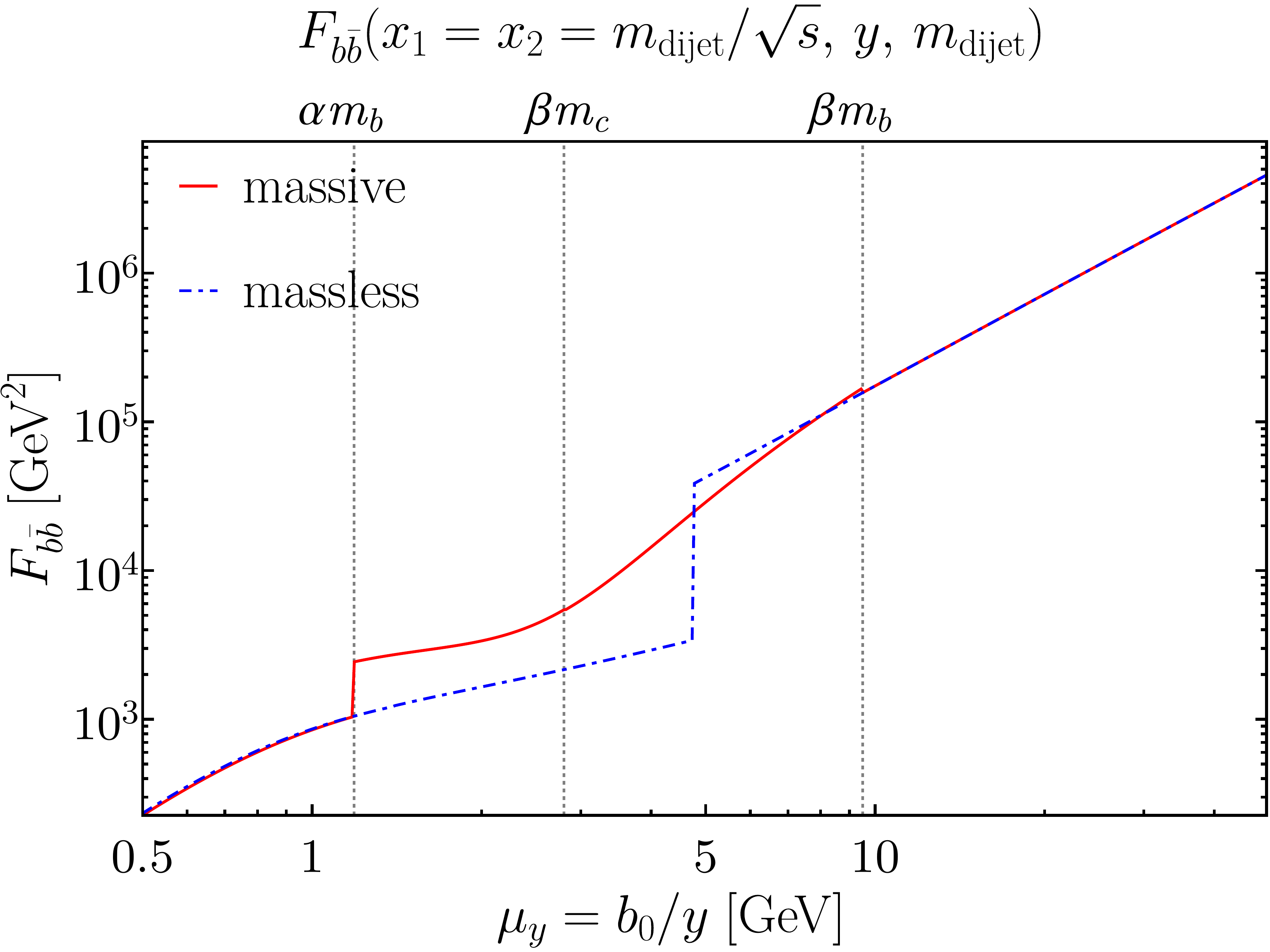}
      }
      \\
      \subfigure[\label{subfig:Fcb-massive-24}${c b}$]{
         \includegraphics[width=0.475\linewidth, trim=0 0 0 50, clip]{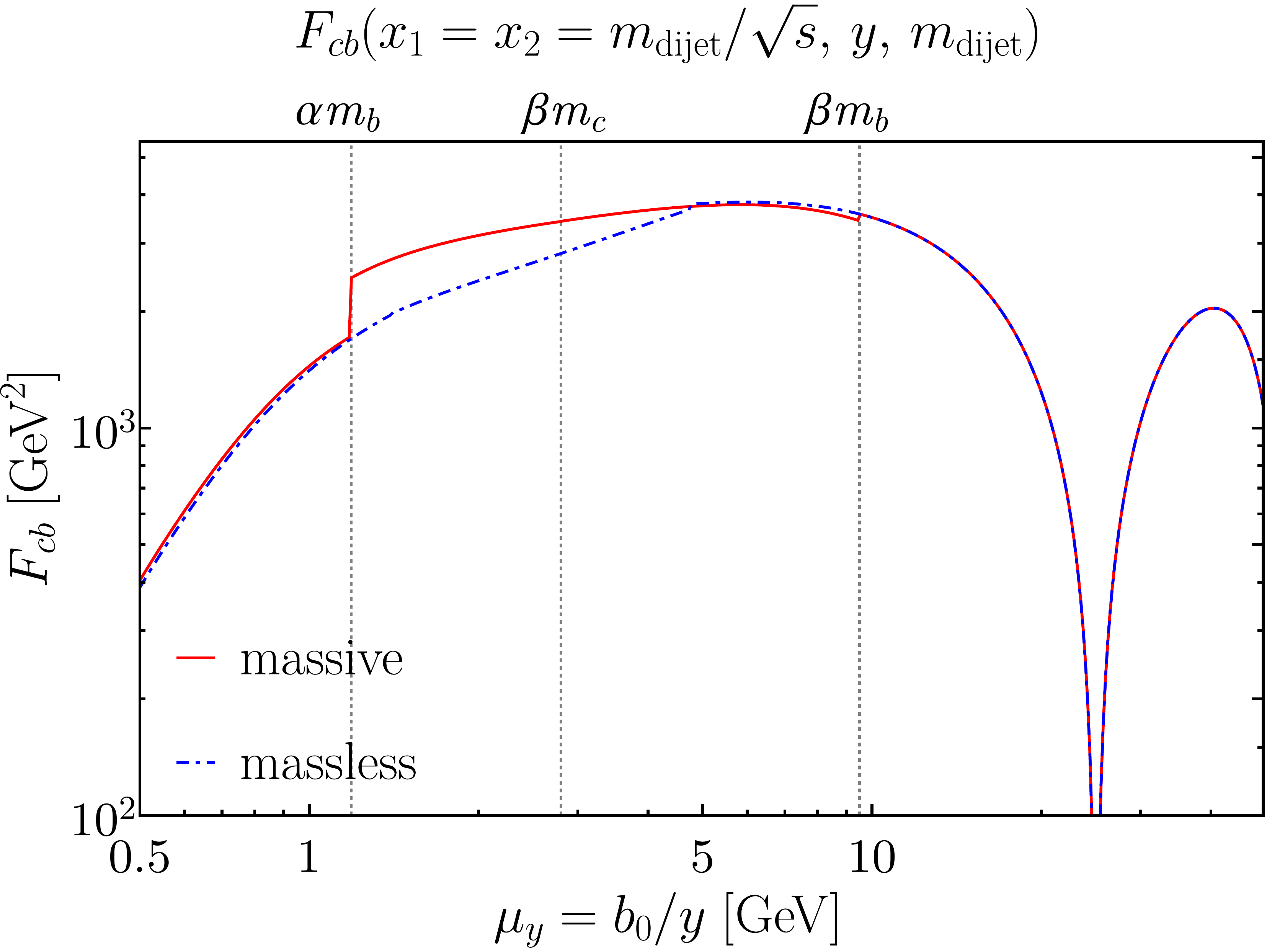}
      }
      \subfigure[\label{subfig:Fgb-massive-24}${g b}$]{
         \includegraphics[width=0.475\linewidth, trim=0 0 0 50, clip]{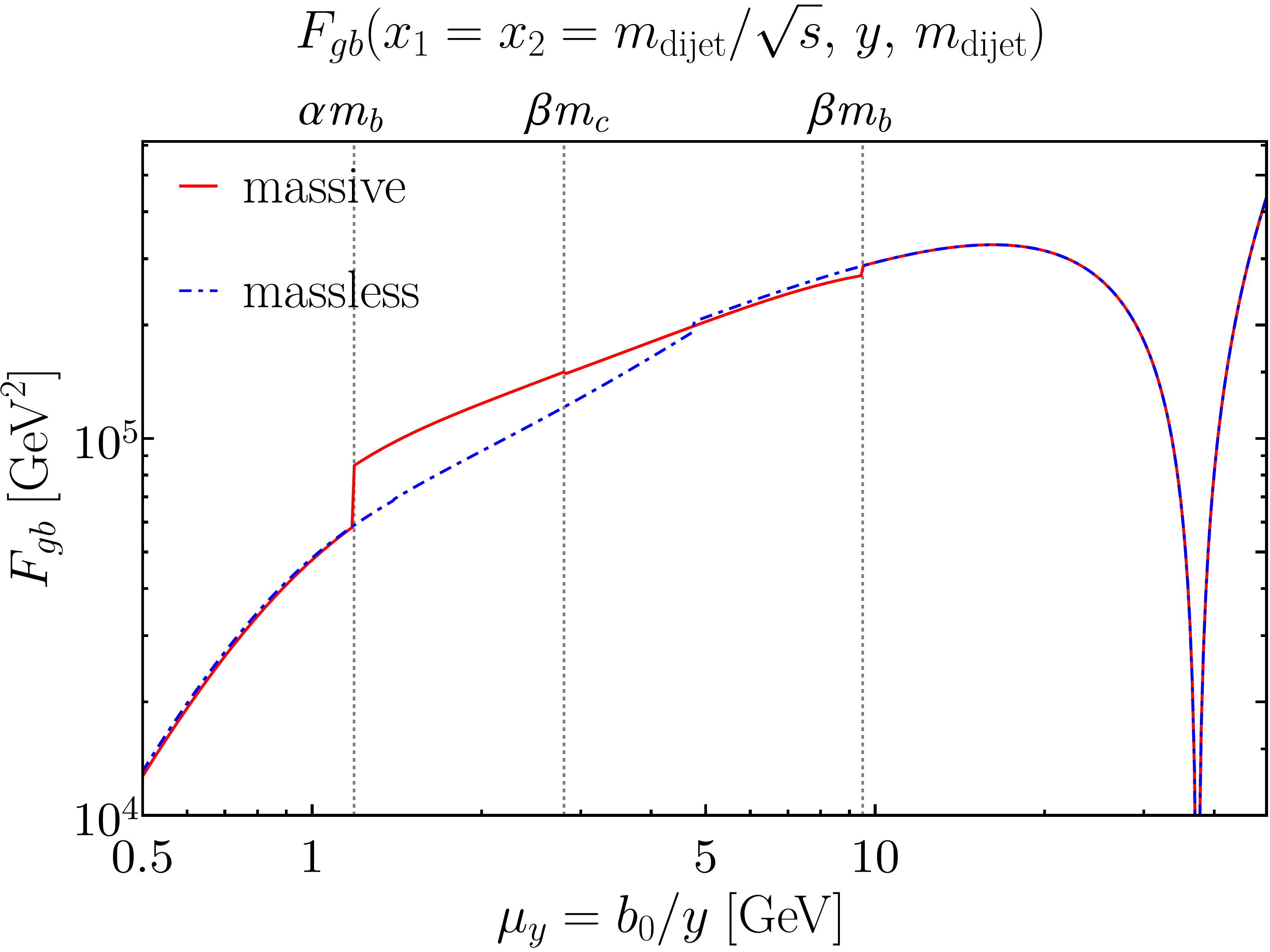}
      }
   \end{center}
   \caption{\label{fig:F-massive-24} $\nf = 5$ splitting DPDs at $\mu = 25 \gev$ for the setting \protect\eqref{eq:dijet-setting}.  Solid lines are for the massive scheme of section \ref{sec:two-massive} with our preferred values $\alpha = 1/4$ and $\beta = 2$, and dashed lines are for the massless scheme with $\gamma = 1$.}
\end{figure}

We now turn to the $g b$ distribution shown in figures \ref{subfig:Fgb-massive-2} and \ref{subfig:Fgb-massive-4}, which has only small discontinuities in the massless scheme.
In the massive scheme, a discontinuity appears at $\mu_y = \alpha \, m_b$, and its absolute size varies only weakly with $\alpha$.  To explain this discontinuity, we recall that $g b$ can be produced by one evolution step from the very large $g g$ distribution.  Just to the right of the discontinuity, this evolution starts at scale $\mu_y \approx \alpha \ms m_b$, whereas just to the left it starts only at $\mu_b = m_b$.  This large difference in evolution length is absent in the massless scheme, where the transition happens at $\mu_y = \gamma \ms m_b$ with $\gamma \sim 1$.

A second (and perhaps more intriguing) feature of the $g b$ distribution in the massive scheme is the significant discontinuity at $\mu_y = \beta \ms m_b$, which strongly increases with $\beta$.  To understand this behaviour, let us compare how $g b$ can be produced in the massive scheme for the region $\beta \ms m_c < \mu_y < \beta \ms m_b$ and in the massless scheme for $\mu_y > \gamma \ms m_b$ (which coincides with the massive scheme for $\mu_y > \beta \ms m_b$). In the massless scheme, all production modes shown in \fig{\ref{fig:Fgb-diagrams}} contribute for $\mu_y > \gamma \ms m_b$ and are evaluated with massless $1 \to 2$ splitting kernels. The PDFs in the splitting formula are evaluated for $5$ flavours, and the $b$ quark PDF has been obtained from a four-flavour gluon PDF by flavour matching and subsequent evolution as illustrated in the lower box of \fig{\ref{subfig:Fgb-massless-scheme}}. For the contributions in \figs{\ref{subfig:Fbg-bbbar}} and \ref{subfig:Fgb-gg}, evolution is necessary to produce the final $g b$ distribution, as indicated by the DGLAP splitting in the upper boxes. This implies that these contributions vanish if the $1 \to 2$ splitting is evaluated at the target scale, i.e.\ for \rev{$\mu_y$ equal to $\mu = 25 \gev$. As $\mu_y$ approaches this value}, the direct splitting contribution in \fig{\ref{subfig:Fgb-massless-scheme}} therefore becomes dominant. In the massive scheme, the PDFs in the splitting formula are evaluated for four flavours for $\beta \ms m_c < \mu_y < \beta \ms m_b$, so that only the modes in \figs{\ref{subfig:Fbg-bbbar}} and \ref{subfig:Fgb-gg} --- now evaluated with four-flavour PDFs and massive $1 \to 2$ splitting kernels --- contribute to the $g b$ distribution. This explains why the discrepancies between the massless and massive scheme become larger \rev{as $\mu_y$ comes closer to $\mu = 25 \gev$} (for any $\beta \le 25 \gev / m_b$).

At which value of $\beta$ should one switch between the two regions within the massive scheme?  For the contribution in \fig{\ref{subfig:Fgb-gg}}, the two regions differ only by the number of active flavours in the gluon distribution, which is a higher-order effect.  The contribution in \fig{\ref{subfig:Fbg-bbbar}} is evaluated more precisely for $\mu_y < \beta \ms m_b$, because mass effects are included in the kernel for $g\to b \bbar$.  For equal momentum fractions of $b$ and $\bbar$, the massive kernel differs from the massless one by more than $100\%$ if $\mu_y = m_b$, but for $\mu_y \ge 2 m_b$ the difference is less than $15\%$.  Most importantly, the contribution of \fig{\ref{subfig:Fgb-massless-scheme}} is missing for $\mu_y < \beta \ms m_b$ if one works with LO kernels, which becomes a serious deficiency if $\beta$ is too large.  Our preferred choice is therefore $\beta = 2$, which we already favoured in the description of the $b \bbar$ distribution.  Notice that the shortcoming of the massive scheme for larger values of $\beta$ should be mitigated if one could work with splitting kernels at NLO, which include the process $g \to g b$ of \fig{\ref{subfig:Fgb-massless-scheme}}.  This provides a particular motivation to study the massive NLO kernels in \sects{\ref{sec:NLO-kernels-constraints}} and \ref{sec:NLO-kernels-model}.  \rev{We return to this particular issues of the $g b$ distribution in \sect{\ref{sec:nlo_numerics}}.}

The features of the $c b$ distribution in the massive scheme are similar to the $g b$ case, including a discontinuity at $\beta \ms m_b$ that grows with $\beta$ (but is less pronounced than for $g b$).  The explanation of this finding is similar to the one just presented.
For the $c \cbar$ and $g c$ distributions, we find only tiny discontinuities in the massive scheme.  We recall that there is no transition point $\alpha \ms m_c$ in this case.

The two-gluon distribution, shown in the bottom row of \fig{\ref{fig:F-alpha-beta-comp}},  differs only a little between the massive and the massless schemes.  Nevertheless, we can see a small discontinuity at $\alpha \ms m_b$, which is more pronounced for smaller $\alpha$.  Visibly, it matters that to the left of this point, gluons evolve with $\nf = 4$ flavours up to the scale $\mu_b = m_b$ and with $\nf = 5$ flavours above, whereas to the right of this point, they evolve with $\nf = 5$ starting at the initialisation scale $\mu_y$.  Similar small discontinuities are seen in DPDs for light quark flavours, showing their sensitivity to the gluon distribution during evolution.
Of course, the presence of these small discontinuities in the massive scheme does \emph{not} lead us to conclude that it is less realistic than the massless one (which makes more drastic approximations).  Clearly, the absence of discontinuities in the $y$ dependence does not guarantee that an approximation for the DPDs is very precise.

To avoid the largest discontinuities across different parton channels for the dijet production setting, we choose $\alpha = 1/4$ and $\beta = 2$ as our preferred scheme parameters.  In \fig{\ref{fig:F-massive-24}} we show the corresponding DPDs for the same parton combinations that we already presented in \fig{\ref{fig:F-massless}} for the massless scheme.

\begin{figure}[t!]
   \begin{center}
      \subfigure[\label{subfig:Fttbar-massive-2}${t \tbar}, \; 1/\alpha = \beta = 2$]{
         \includegraphics[width=0.475\linewidth, trim=0 0 0 50, clip]{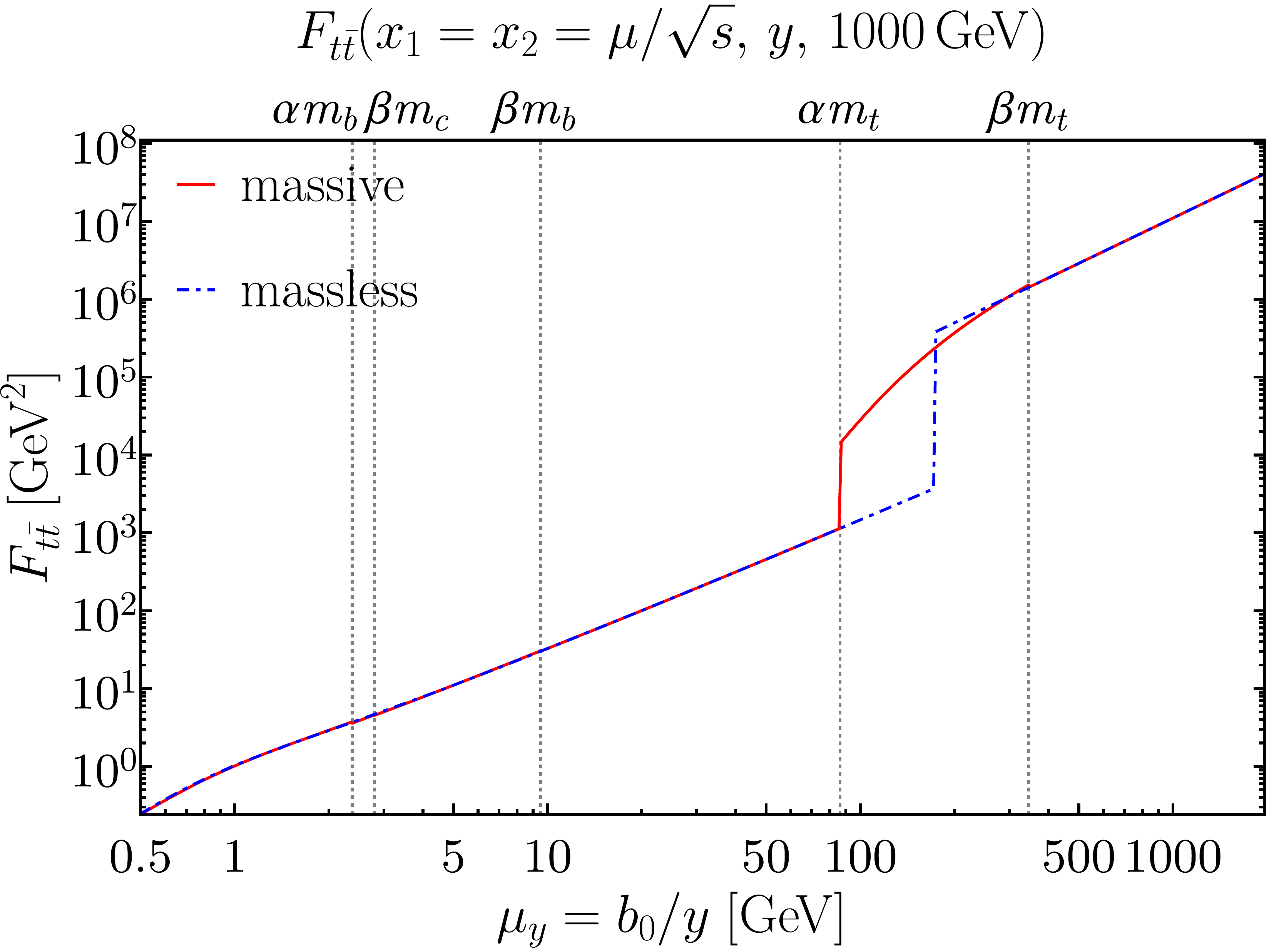}
      }
      \subfigure[\label{subfig:Fttbar-massive-4}${t \tbar}, \; 1/\alpha = \beta = 4$]{
         \includegraphics[width=0.475\linewidth, trim=0 0 0 50, clip]{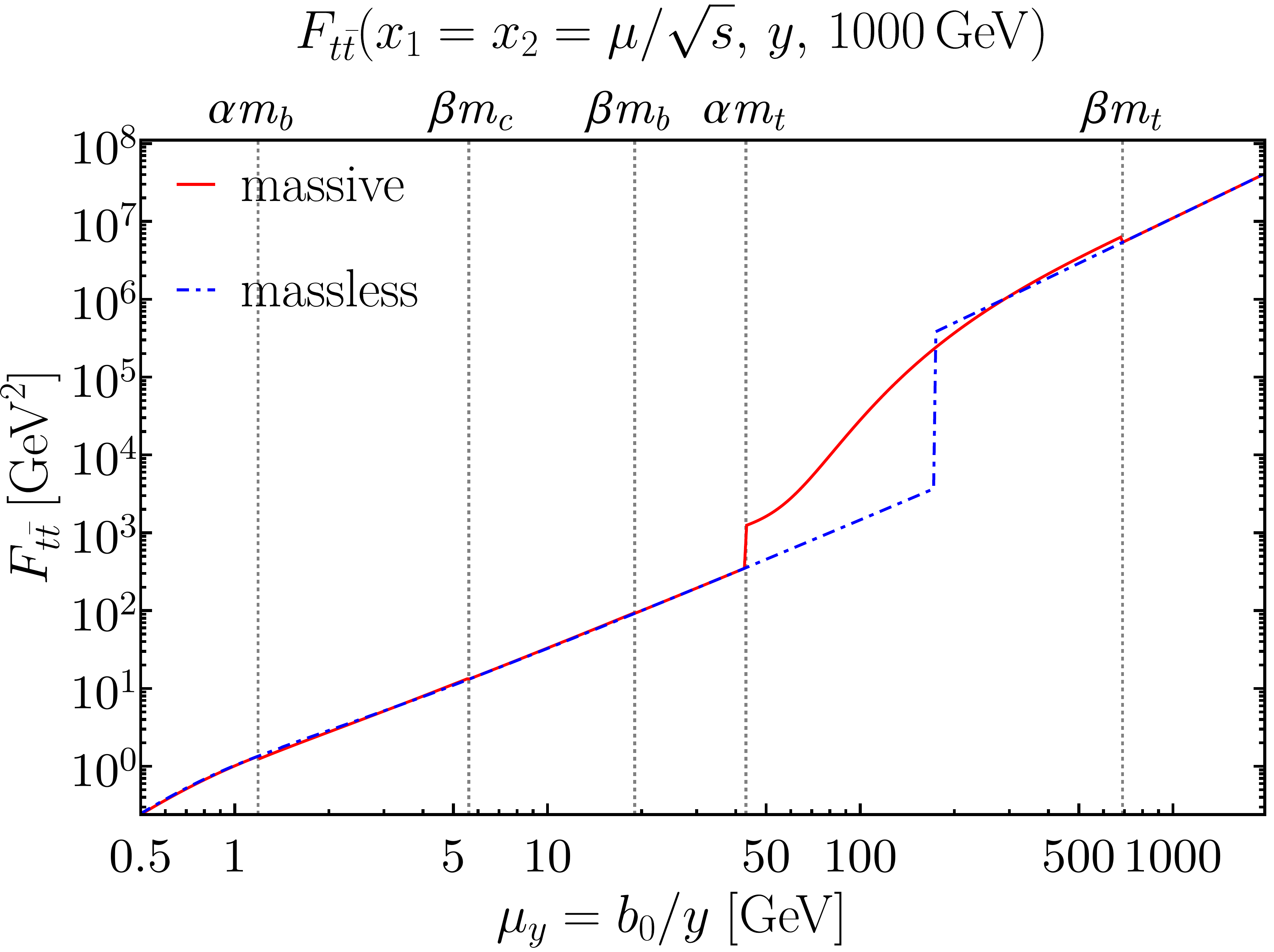}
      }
      \\
      \subfigure[\label{subfig:Fgt-massive-2}${g t}, \; 1/\alpha = \beta = 2$]{
         \includegraphics[width=0.475\linewidth, trim=0 0 0 50, clip]{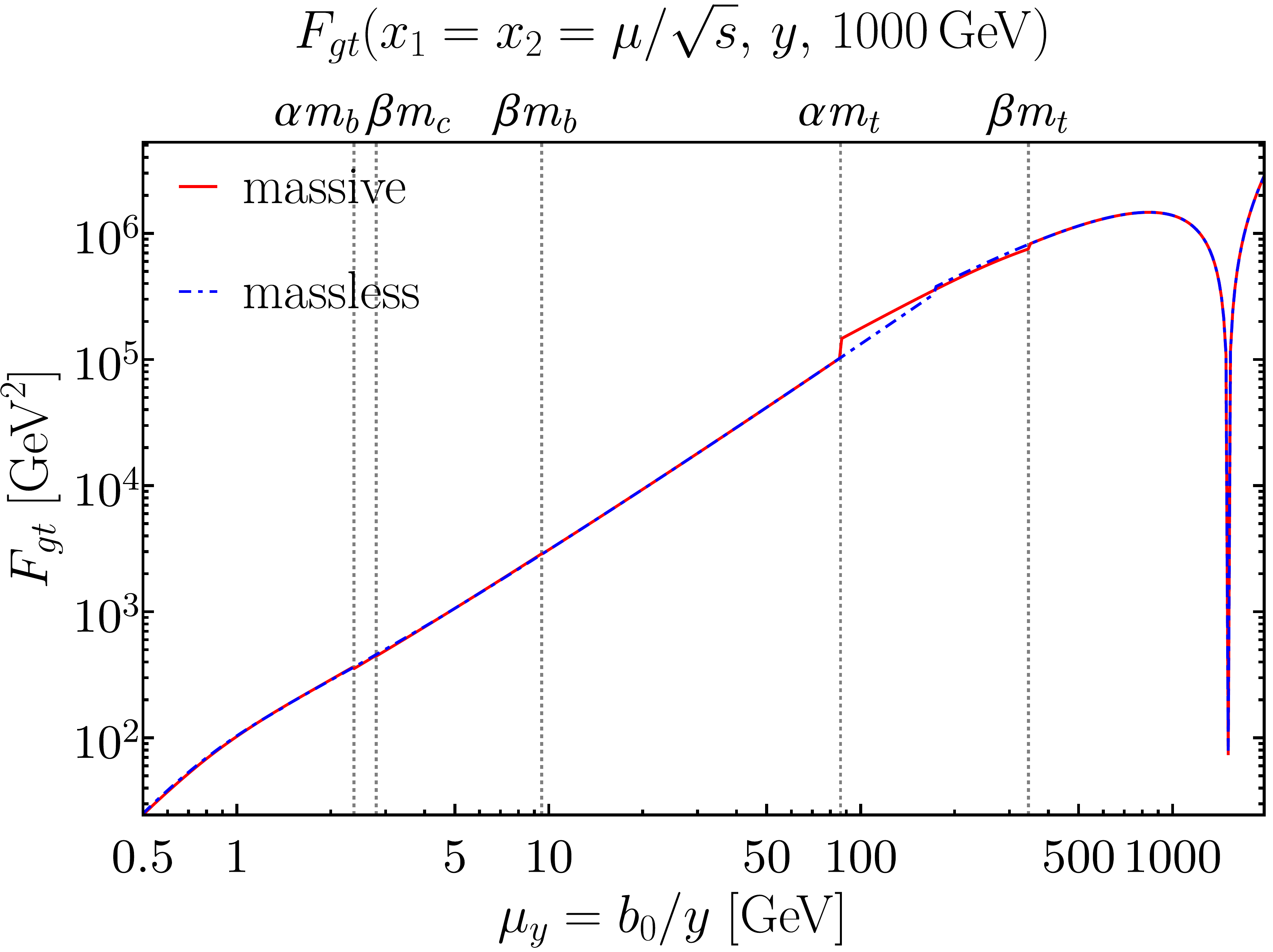}
      }
      \subfigure[\label{subfig:Fgt-massive-4}${g t}, \; 1/\alpha = \beta = 4$]{
         \includegraphics[width=0.475\linewidth, trim=0 0 0 50, clip]{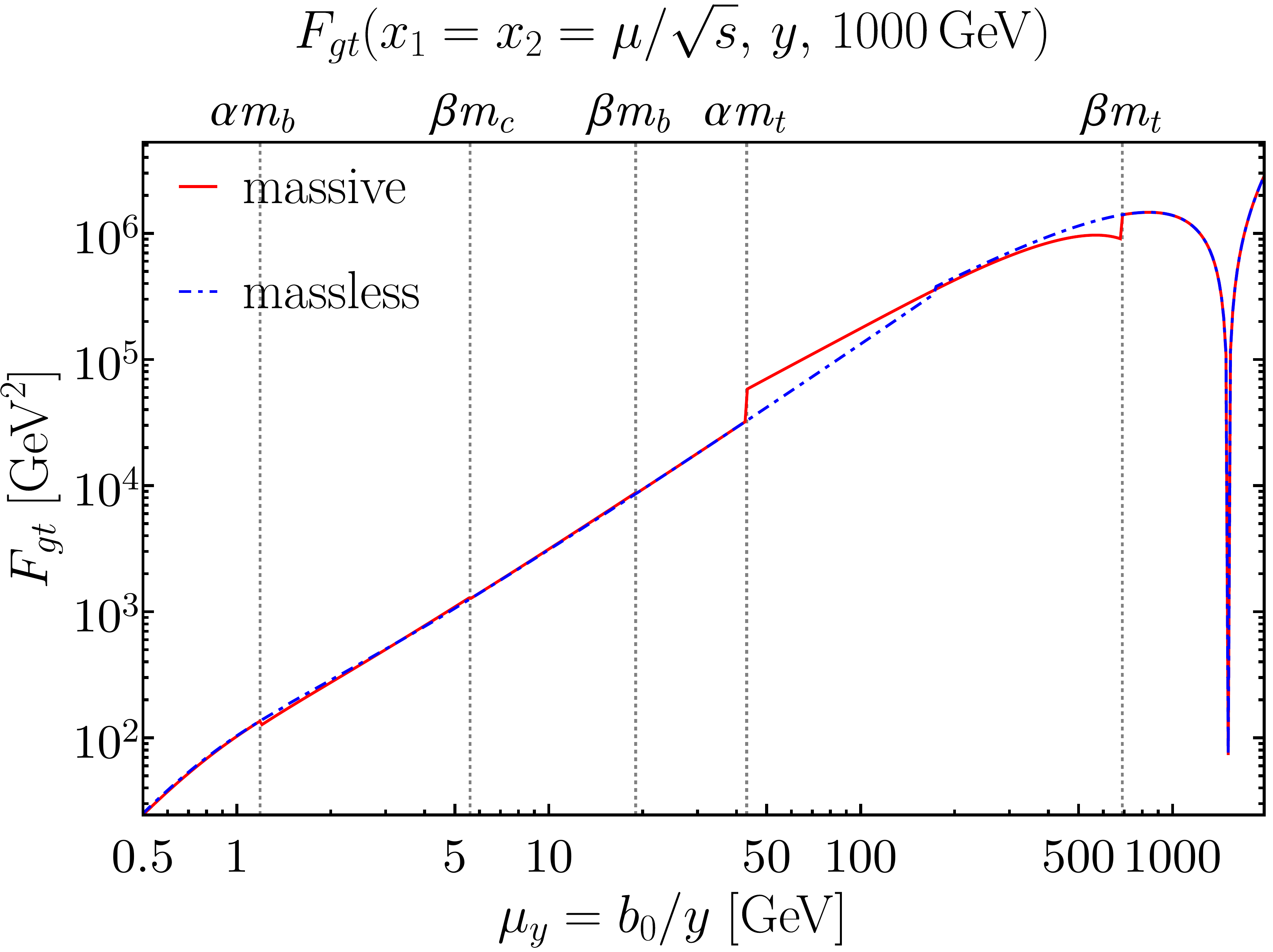}
      }
   \end{center}
   \caption{\label{fig:F-ttbar-gt-massive} $\nf = 6$ splitting DPDs at $\mu = 1 \tev$ for the setting \protect\eqref{eq:ttbar-setting}.  Solid lines are for the massive scheme with $1 / \alpha = \beta$, and dashed lines are for the massless scheme with $\gamma = 1$.  The momentum fractions are $x_1 = x_2 = 10^{-2}$ according to \protect\eqref{eq:mom-fracs-central-rap}.}
\end{figure}

\begin{figure}[t!]
   \begin{center}
      \subfigure[\label{subfig:Fbbbar-massive-24-ttbar}${b \bbar}$]{
         \includegraphics[width=0.475\linewidth, trim=0 0 0 50, clip]{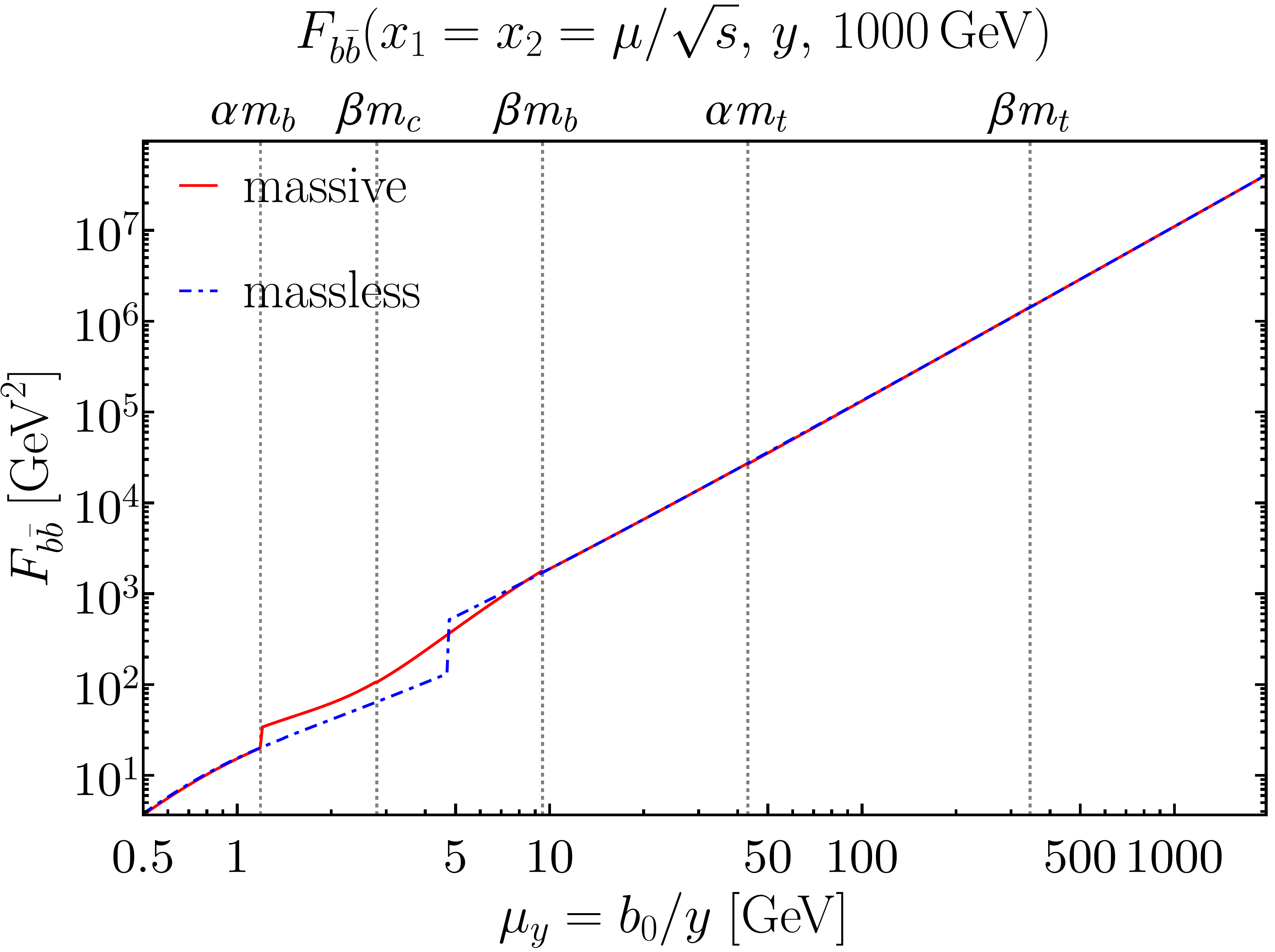}
      }
      \subfigure[\label{subfig:Fttbar-massive-24}${t \tbar}$]{
         \includegraphics[width=0.475\linewidth, trim=0 0 0 50, clip]{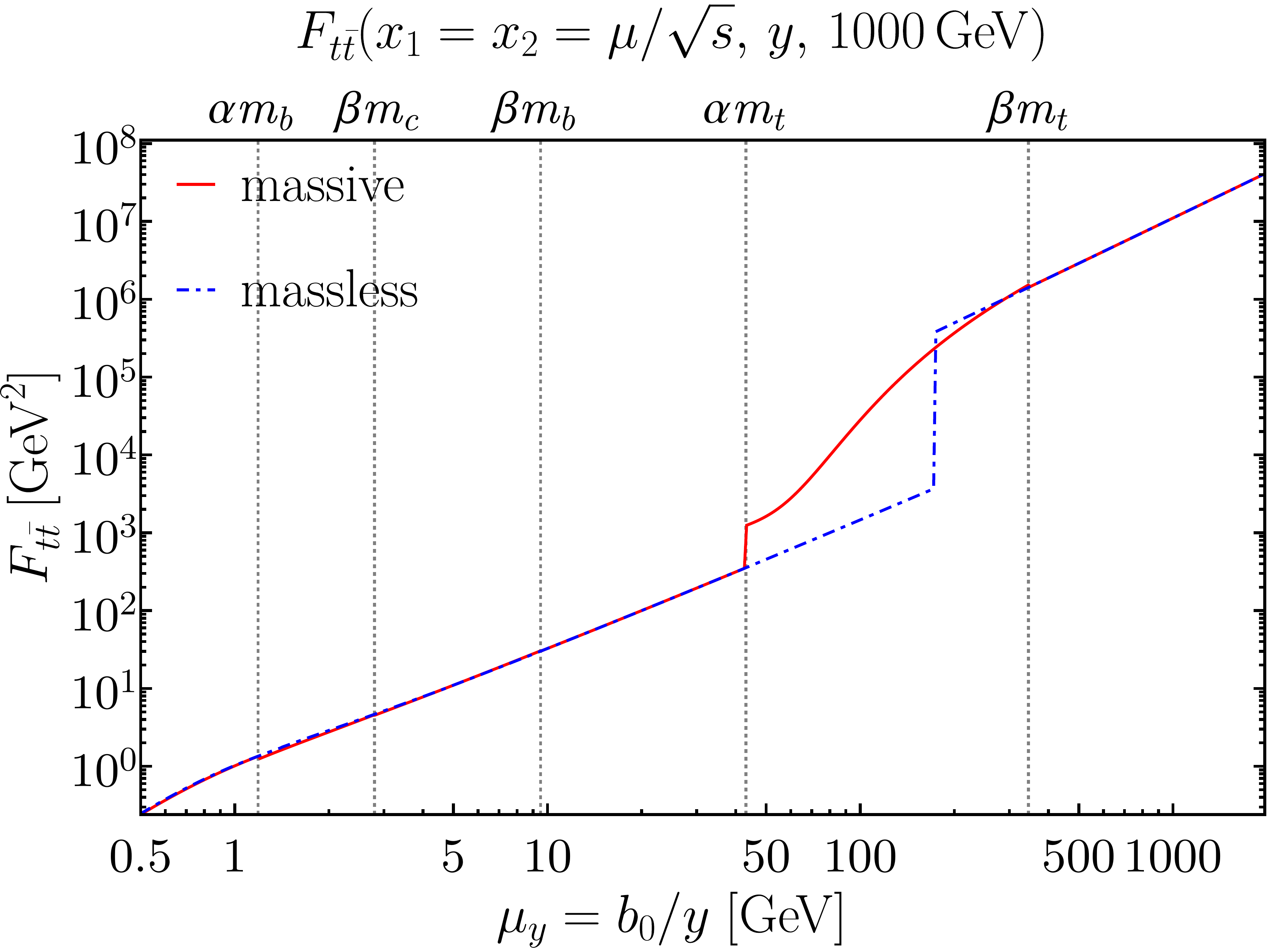}
      }
      \\
      \subfigure[\label{subfig:Fcb-massive-24-ttbar}${c b}$]{
         \includegraphics[width=0.475\linewidth, trim=0 0 0 50, clip]{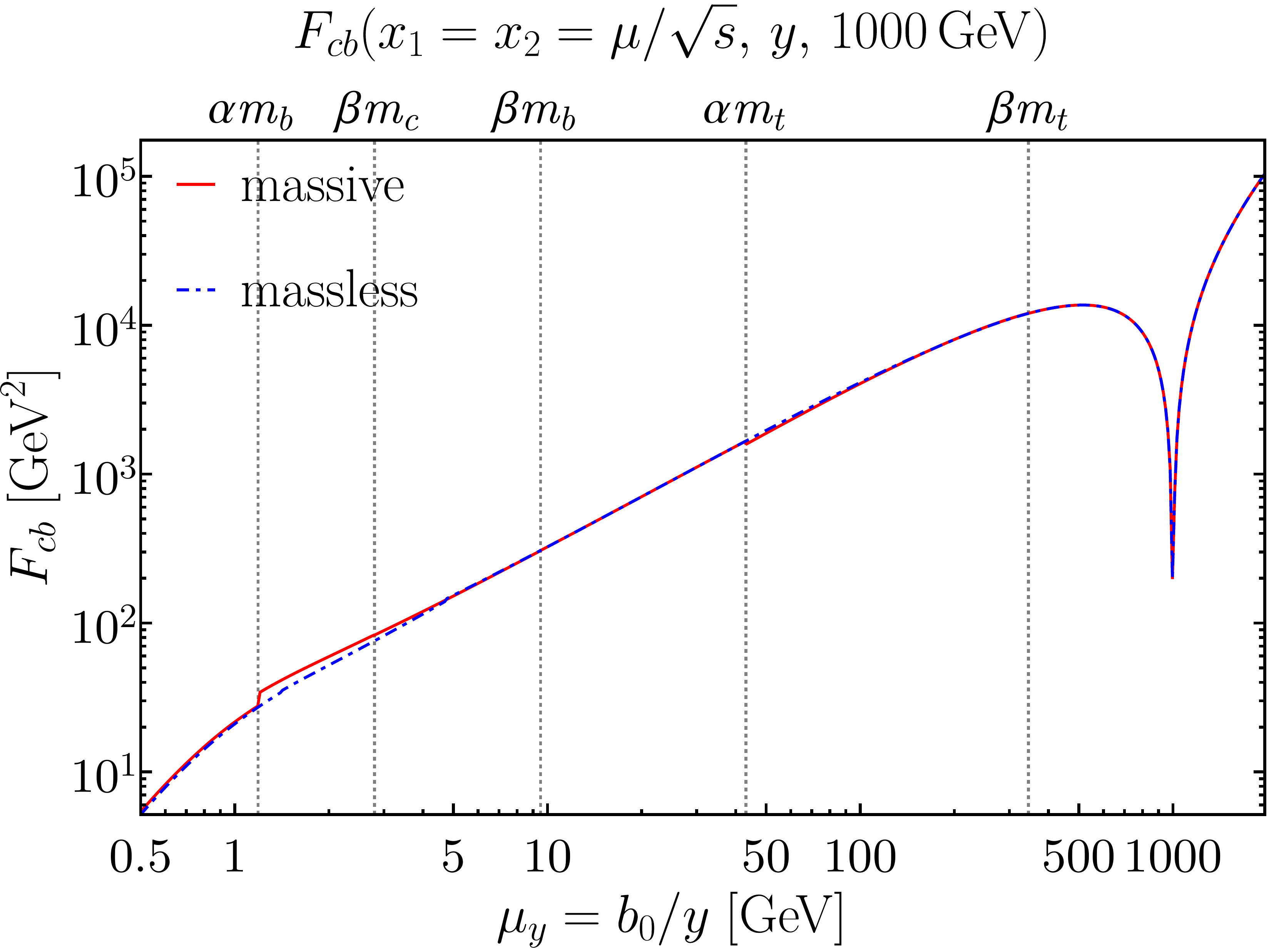}
      }
      \subfigure[\label{subfig:Fgt-massive-24}${g t}$]{
         \includegraphics[width=0.475\linewidth, trim=0 0 0 50, clip]{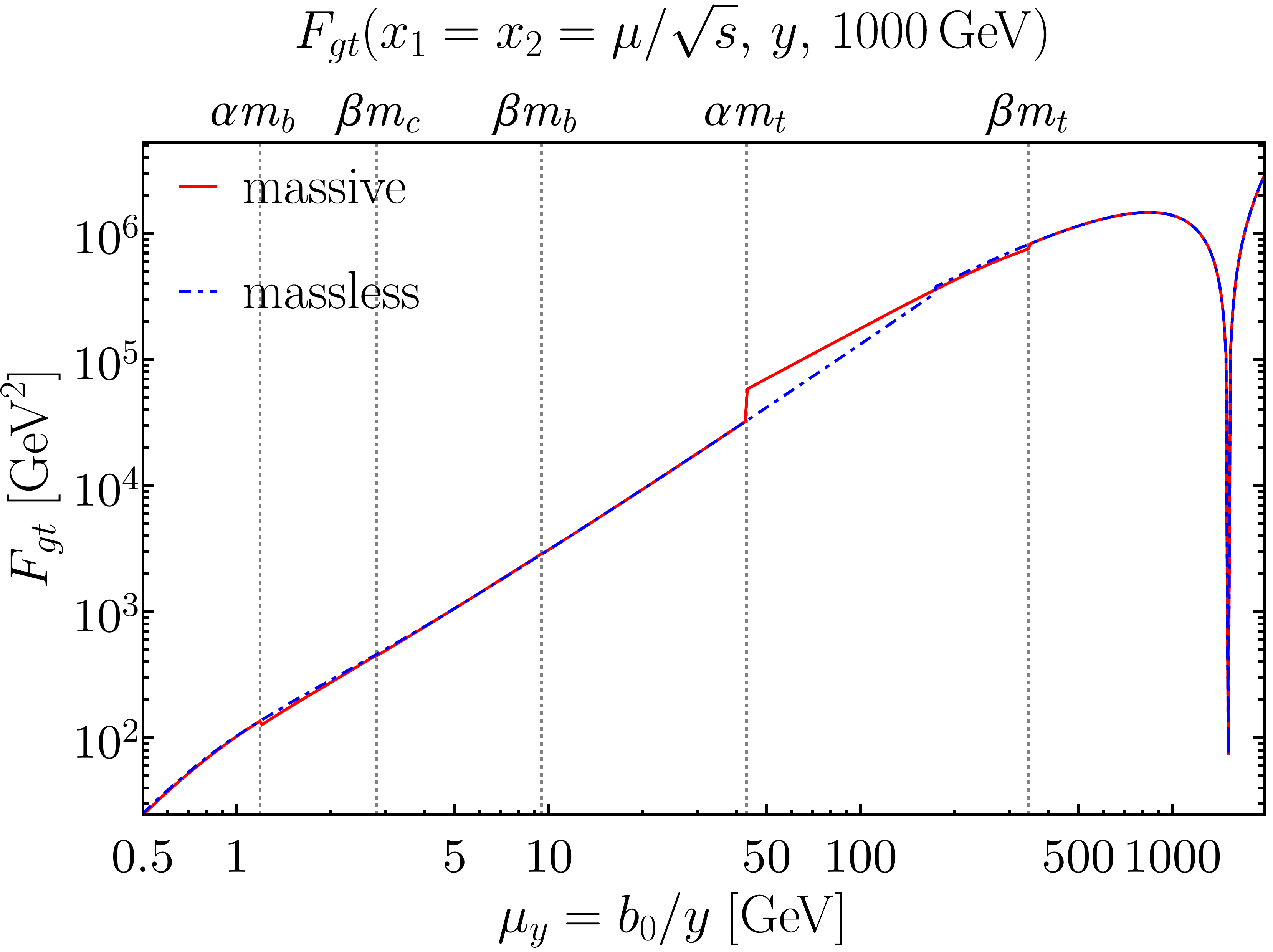}
      }
   \end{center}
   \caption{\label{fig:F-ttbar-gt-massive-24} $\nf = 6$ splitting DPDs at $\mu = 1 \tev$ for the setting \protect\eqref{eq:ttbar-setting}.  Solid lines are for the massive scheme of section \ref{sec:two-massive} with our preferred values $\alpha = 1/4$ and $\beta = 2$, and dashed lines are for the massless scheme with $\gamma = 1$.}
\end{figure}

To see whether the same parameters give a good description in other situations, we show in \fig{\ref{fig:F-ttbar-gt-massive}} the $t \tbar$ and $g t$ distributions for the $t \tbar$ production setting \eqref{eq:ttbar-setting}.
The pattern of discontinuities in the massive scheme is the same as the one we see in \fig{\ref{fig:F-alpha-beta-comp}} for $b$ quarks, confirming our above choice of parameters.  Remarkably, the discontinuity of the $t \tbar$ distribution amounts to almost two orders of magnitude at $\mu_y = \gamma \ms m_t$ in the massless scheme, and to more than one order of magnitude in the massive scheme with $\alpha = 1/2$.  DPDs for our preferred scheme parameters are shown in \fig{\ref{fig:F-ttbar-gt-massive-24}}.  It is remarkable that the discontinuities of the $b \bbar$ and $c b$ distributions at $\alpha \ms m_b$ are not entirely washed out by evolution to the much larger final scale $1 \tev$.
%
%
\subsection{Double parton luminosities}
\label{sec:LO-DPD-lumis}
\begin{figure}[!t]
   \begin{center}
      \subfigure[\label{subfig:Lccbarbbbar-jets}${c\cbar b\bbar}$]{
         \includegraphics[width=0.475\linewidth, trim=0 0 0 50, clip]{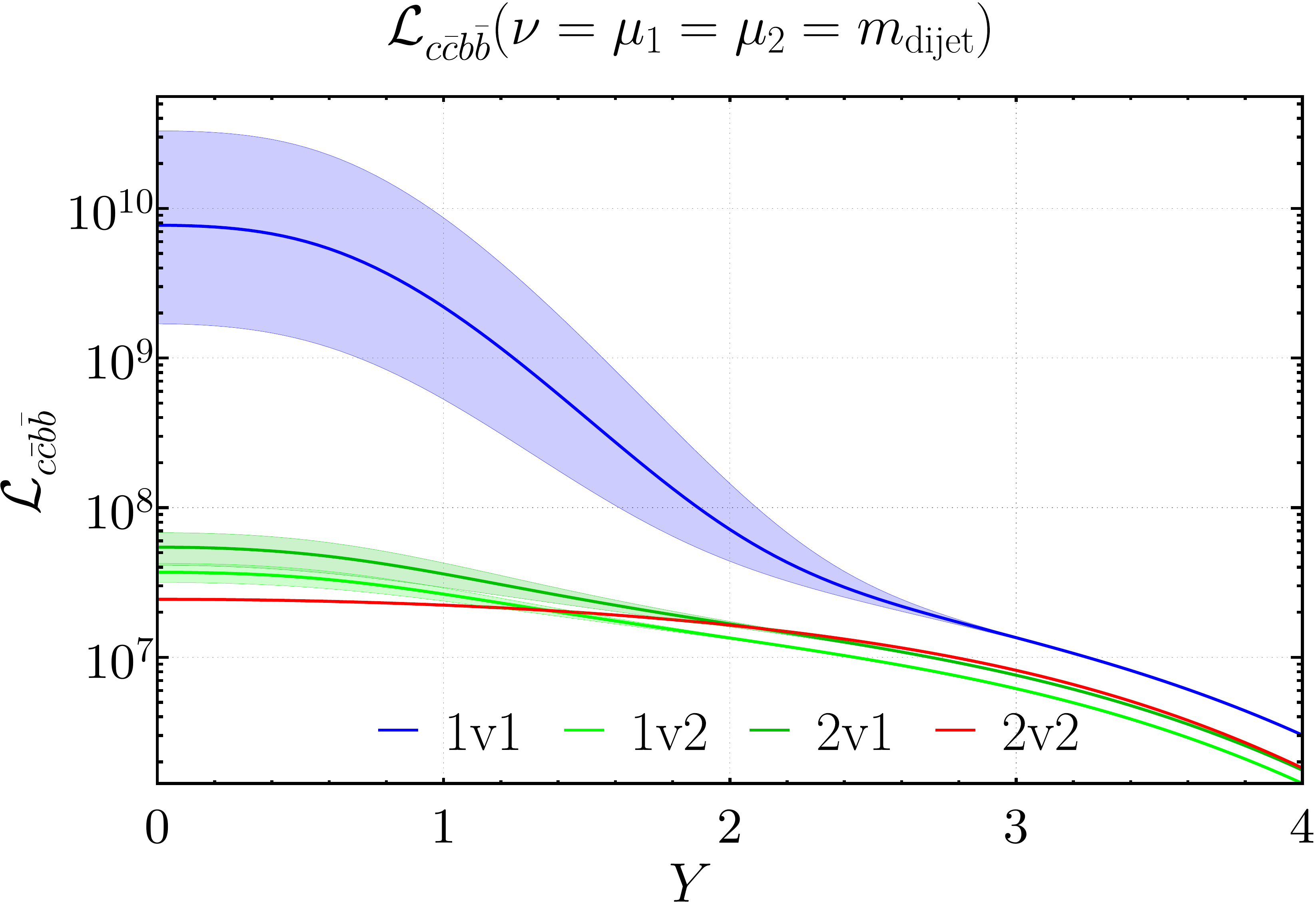}
      }
      \subfigure[\label{subfig:Lcbgg-jets}${c b g g}$]{
         \includegraphics[width=0.475\linewidth, trim=0 0 0 50, clip]{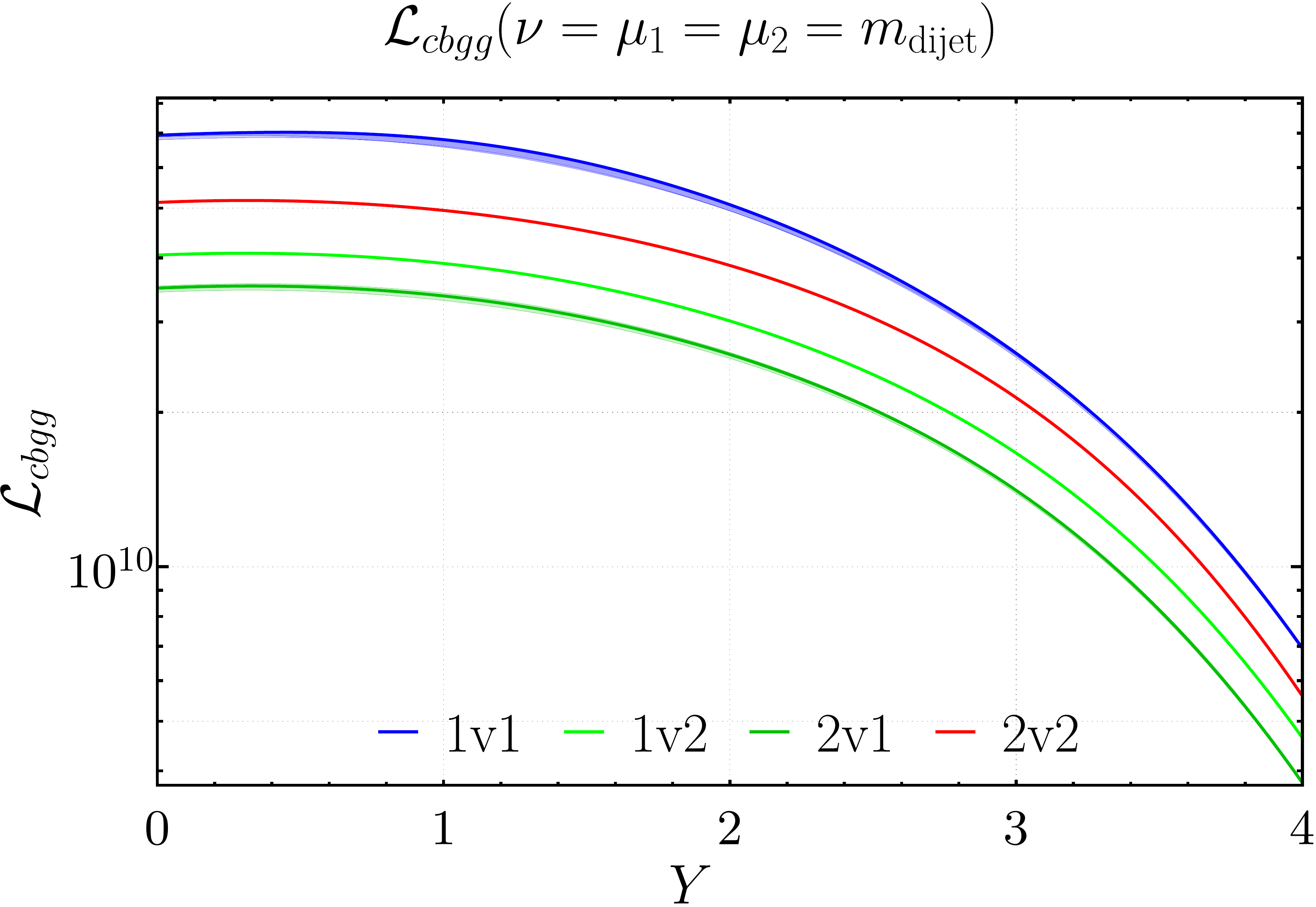}
      }
      \\
      \subfigure[\label{subfig:Lgcgb-jets}${g c b g}$]{
         \includegraphics[width=0.475\linewidth, trim=0 0 0 50, clip]{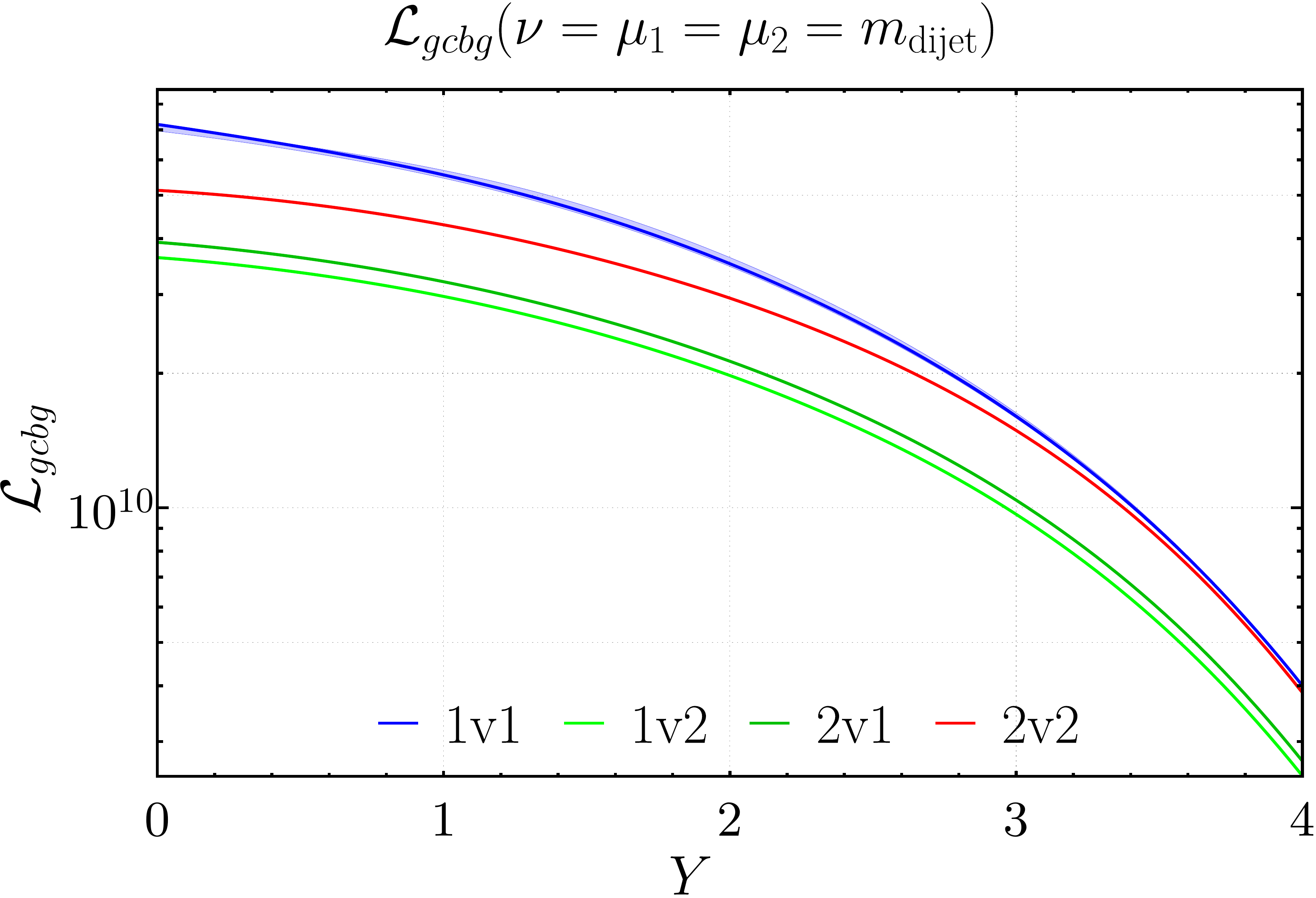}
      }
      \subfigure[\label{subfig:Lgggg-jets}${g g g g}$]{
         \includegraphics[width=0.475\linewidth, trim=0 0 0 50, clip]{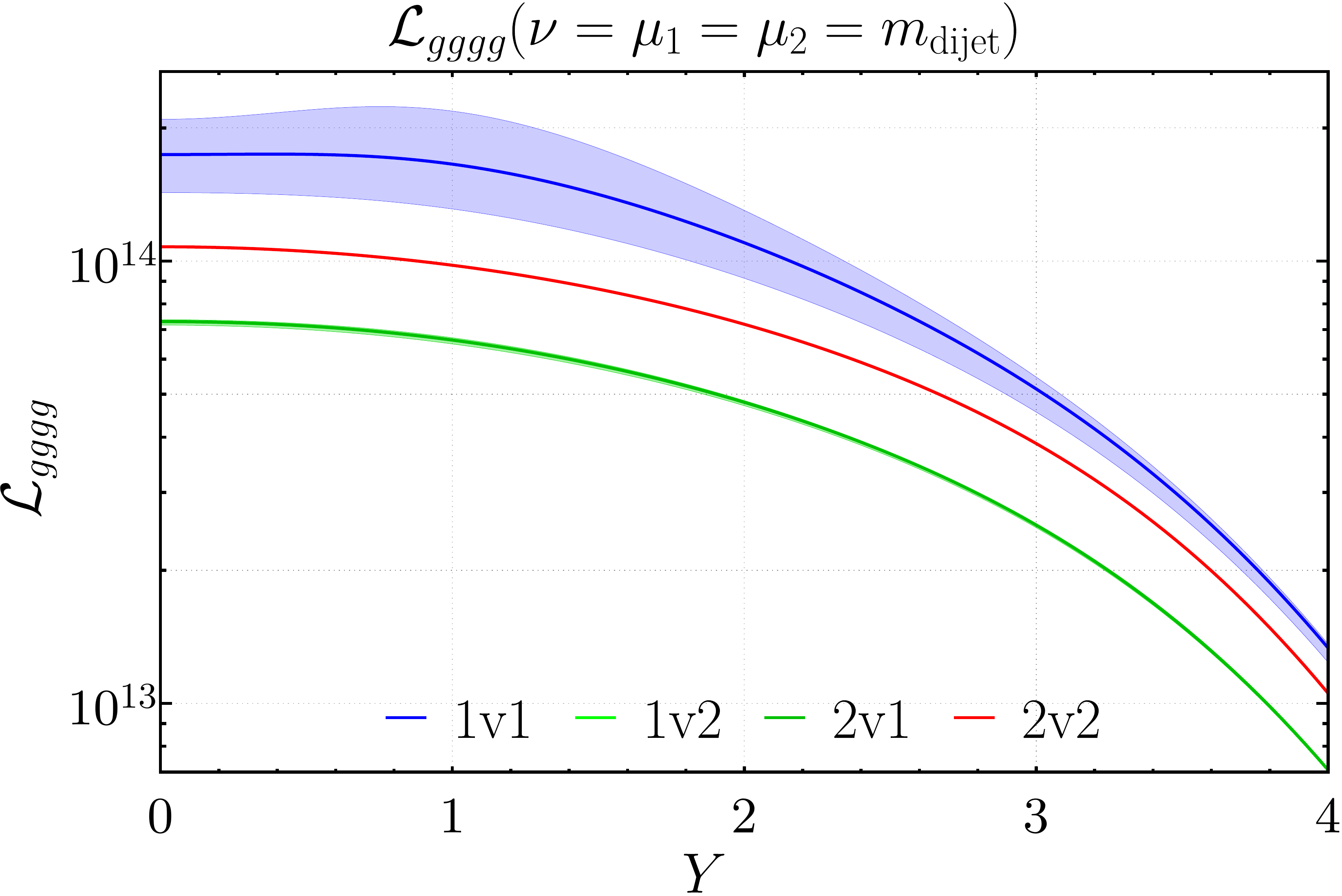}
      }
   \end{center}
   \caption{\label{fig:lumis-jets-contribs} Double parton luminosities at $\mu = 25 \gev$ for the dijet production setting, computed in the massive scheme of section \ref{sec:two-massive} with $\alpha = 1/4$ and $\beta = 2$. The momentum fractions in the DPDs are specified by \eqs{\protect\eqref{eq:dijet-setting}} and \eqref{eq:mom-fracs-opposite-rap}.  Central values correspond to a cutoff parameter $\nu = \mu$, and bands to the variation of $\nu$ between $\mu/2$ and $2 \mu$.}
\end{figure}

\begin{figure}
   \begin{center}
      \subfigure[\label{subfig:Lttbartbart-ttbar}${t \tbar \ms \tbar t}$]{
         \includegraphics[width=0.475\linewidth, trim=0 0 0 50, clip]{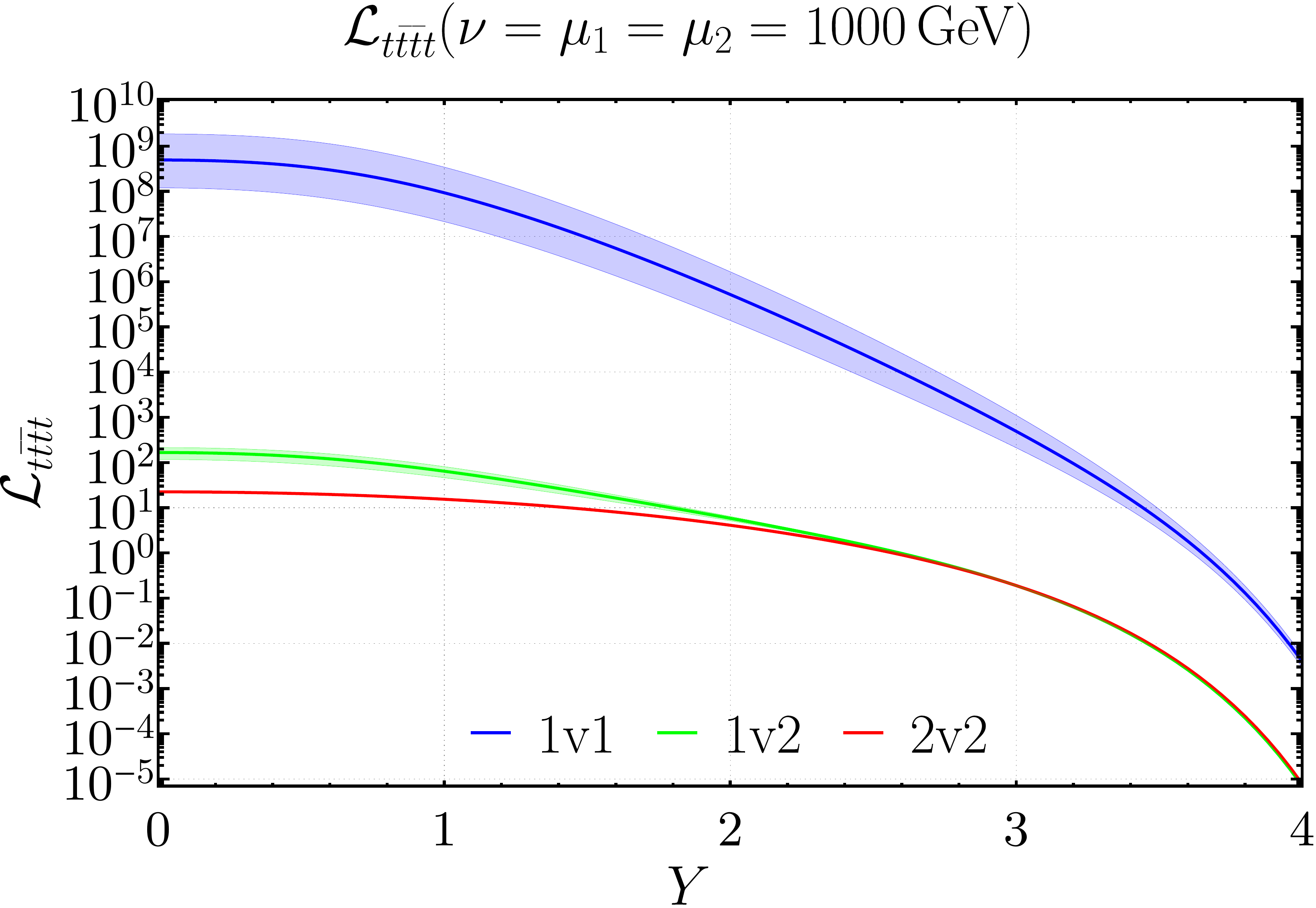}
      }
      \subfigure[\label{subfig:Lgtgtbar-ttbar}${g t g \tbar}$]{
         \includegraphics[width=0.475\linewidth, trim=0 0 0 50, clip]{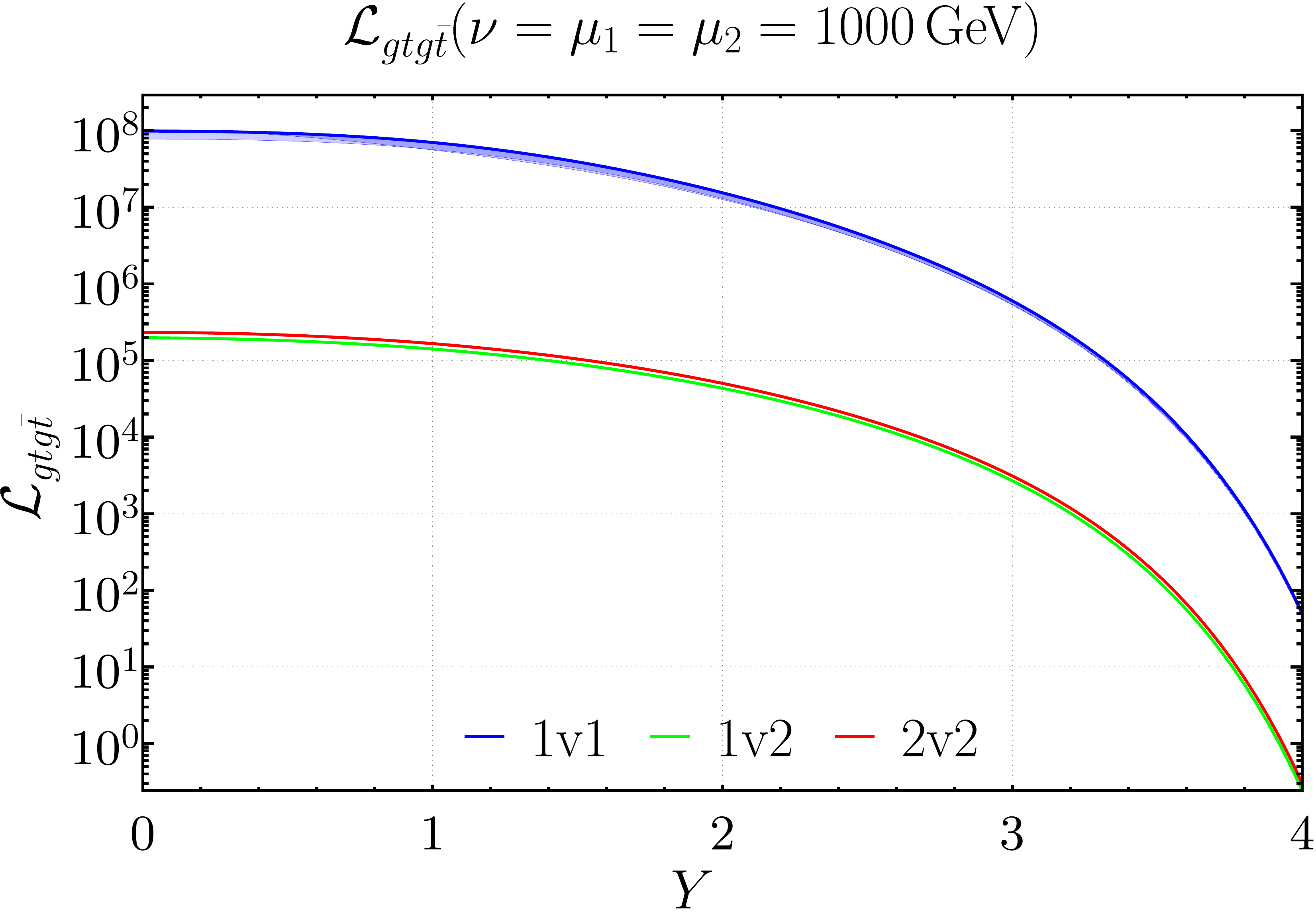}
      }
   \end{center}
   \caption{\label{fig:lumis-ttbar-contribs} As \fig{\protect\ref{fig:lumis-jets-contribs}} but at $\mu = 1 \tev$ for the $t \tbar$ production setting \protect\eqref{eq:ttbar-setting}.  The 2v1 luminosities are not shown here, because they are almost indistinguishable from the 1v2 ones.}
\end{figure}

So far our focus has been on the DPDs and their $y$ dependence.  This dependence is not directly observable in DPS processes, where DPDs only appear in integrals over $y$.  As already mentioned in the introduction, the relevant quantities are double parton luminosities
\begin{align}
   \label{eq:DPD-lumi}
   &  \mathcal{L}_{a_1 a_2 b_1 b_2}
      (x_{1a}, x_{2a}, x_{1b}, x_{2b}; \mu, \nu)
   \nonumber \\[0.2em]
   & \qquad
      = \int \mathrm{d}^2 y \; \theta(y - b_0 / \nu) \,
      F^{\nf}_{a_1 a_2}(x_{1a}, x_{2a}, y; \mu) \,
      F^{\nf}_{b_1 b_2}(x_{1b}, x_{2b}, y; \mu) \,,
\end{align}
where we express the lower cutoff on $y$ in terms of a momentum scale $\nu$ following \cite{Diehl:2017kgu}.  We always set $\nu = \mu$ in this section, except for \figs{\ref{fig:lumis-jets-contribs}} and \ref{fig:lumis-ttbar-contribs}.

We evaluate these luminosities at parton momentum fractions that correspond to a kinematic setting where the system produced in the first and second hard scattering has rapidity $Y$ and $-Y$, respectively.  For equal invariant mass $Q$ of both systems, this corresponds to
\begin{align}
   \label{eq:mom-fracs-opposite-rap}
   x_{1 a} &= x_{2 b} = \frac{Q}{\sqrt{s}} \, \exp (Y) \,,
   &
   x_{2 a} &= x_{1 b} = \frac{Q}{\sqrt{s}} \, \exp (-Y) \,.
\end{align}

At this point, we recall that the splitting DPDs $F^{\text{spl}}$ discussed so far are the leading contribution in a small-distance expansion that contains also an ``intrinsic''  contribution $F^{\text{intr}}$.  While subleading in $y$, this contribution can be important for parton combinations where the splitting DPD is small, and it is generally growing more strongly when the parton momentum fractions decrease.

Following section 9.2.1 of \cite{Diehl:2017kgu}, we model the intrinsic contribution as
\begin{align}
   \label{eq:int-DPD}
      F_{a_1 a_2}^{\text{intr}}(x_1,x_2,y; \mu_{0})
   &= \frac{1}{4 \pi h_{a_1 a_2}}
      \exp \left[\frac{-y^2}{4 h_{a_1 a_2}}\right] \;
      \frac{(1 - x_1 - x_2)^2}{(1 - x_1)^2 \ms (1 - x_2)^2} \;
      f_{a_1}(x_1; \mu_{0}) \, f_{a_2}(x_2; \mu_{0})
\end{align}
with $\mu_{0} = 1 \gev$.  The normalised Gaussian with the widths given in equation \eqref{eq:damping-parameters} is used to model the $y$ dependence of the DPDs, and the $x_1$ and $x_2$ dependent prefactor ensures a sensible behaviour of the DPDs as the momentum fractions approach the kinematic threshold $x_1 + x_2 \to 1$.  The ansatz \eqref{eq:int-DPD} is made for $\nf = 3$ active flavours, and flavour matching is used to obtain the DPDs for higher $\nf$.

We note that the distinction between a splitting and an intrinsic part of the DPD is unambiguously defined only in the small $y$ limit; for large $y$ we simply define our DPD model as the sum of the regularised splitting form \eqref{eq:small-y-DPD-LO-mod} and the intrinsic term \eqref{eq:int-DPD} (after evolving them to a common scale).
We emphasise that this model has not been tuned against data and is likely too simplistic, but we estimate that it contains enough realistic features for the studies that will follow.

The decomposition $F^{\text{spl}} + F^{\text{intr}}$ induces a decomposition of the  double parton luminosity \eqref{eq:DPD-lumi} into four contributions:
\begin{align}
   \label{eq:DPD-lumi-combinations}
      \mathcal{L}_{a_1 a_2 b_1 b_2}^{\text{1v1}}
   &= \int \mathrm{d}^2 y \,
      F_{a_1 a_2}^{\text{spl}} \, F_{b_1 b_2}^{\text{spl}} \,,
   && \mathcal{L}_{a_1 a_2 b_1 b_2}^{\text{2v2}}
    = \int \mathrm{d}^2 y \,
      F_{a_1 a_2}^{\text{intr}} \, F_{b_1 b_2}^{\text{intr}} \,,
   \nonumber \\
      \mathcal{L}_{a_1 a_2 b_1 b_2}^{\text{1v2}}
   &= \int \mathrm{d}^2 y \,
      F_{a_1 a_2}^{\text{spl}} \, F_{b_1 b_2}^{\text{intr}} \,,
   && \mathcal{L}_{a_1 a_2 b_1 b_2}^{\text{2v1}}
    = \int \mathrm{d}^2 y \,
      F_{a_1 a_2}^{\text{intr}} \, F_{b_1 b_2}^{\text{spl}} \,,
\end{align}
where the arguments and integration boundaries are the same as in equation \eqref{eq:DPD-lumi}.  The superscripts ``1v1'', ``1v2'' etc.\ follow the nomenclature introduced in \cite{Gaunt:2012dd}.

In \fig{\ref{fig:lumis-jets-contribs}}, we show the different contributions to $\mathcal{L}_{c \cbar b \smash{\bbar}}$, $\mathcal{L}_{c b g g}$, $\mathcal{L}_{g c b g}$, and $\mathcal{L}_{g g g g}$ for the dijet production setting \eqref{eq:dijet-setting}.
We see that the 1v1 and 2v2 and contributions are of comparable size, except for $\mathcal{L}_{c\cbar b \smash{\bbar}}$, where the 1v1 contribution is dominant for small and intermediate values of the rapidity $Y$.  The luminosity for four gluons is by far the largest one.
\rev{To which extent this channel dominates the production of two dijets containing heavy flavours depends of course on the relative size of the relevant parton-level cross sections and jet functions in given kinematics.  To study this question is beyond the scope of the present work.}

For the $t \tbar$ production setting \eqref{eq:ttbar-setting}, the mass scheme dependence is most pronounced for DPDs containing top quarks, so that we expect significant mass effects for the luminosities $\mathcal{L}_{t \tbar \ms \tbar t}$ and $\mathcal{L}_{g t g \tbar}$ shown in \fig{\ref{fig:lumis-ttbar-contribs}}.  We see that for both channels, the 1v1 contribution dominates in the full $Y$ range considered.


At this point, we recall that the luminosities shown here need to be combined with luminosities that correspond to the double counting subtraction terms in the overall cross section.  The importance of these terms, which we will not investigate in this work, can be estimated by varying the lower cut-off of the $y$ integration.  This is because the dependence on this cutoff approximately cancels between DPS and the subtraction terms, up to perturbative orders that are beyond the accuracy of the calculation.  A detailed discussion is given in \sects{6.2} and 9.2 of \cite{Diehl:2017kgu}.  The largest variation with $\nu$ seen in \figs{\ref{fig:lumis-jets-contribs}} and \ref{fig:lumis-ttbar-contribs} appears in the 1v1 contributions to $c \cbar b \bbar$, $gggg$, and $t \tbar \ms \tbar t$ at low $Y$.  One can hence expect the double counting subtraction to be most important in these cases.

Given the very different $y$ dependence of $F^{\text{spl}}$ and $F^{\text{intr}}$, the dominant region of $y$ in the partial luminosities \eqref{eq:DPD-lumi-combinations} in general differs appreciably between the 1v1 and the 1v2 or 2v1 terms.  We can therefore expect corresponding differences in the sensitivity of these terms to details of the heavy-flavour treatment.  The 2v2 contribution is of course independent of how heavy quarks are treated in the splitting DPDs.
%
%
\subsubsection{Dependence on the scheme parameters}
\label{sec:DPD-lumis-params}
We now study how the double parton luminosities depend on the scheme parameters --- $\alpha$ and $\beta$ in the massive scheme and $\gamma$ in the massless one.  To this end, we consider the ratios
\begin{align}
   \label{eq:lumi-ratios}
   r(\beta)
      &= \frac{\mathcal{L}_{\text{massive}}(1/\alpha = \beta)}{
         \mathcal{L}_{\text{massive}}(\alpha = 1/4, \beta = 2)} \,,
   &
   r(\gamma)
      &= \frac{\mathcal{L}_{\text{massless}}(\gamma)}{
         \mathcal{L}_{\text{massive}}(\alpha = 1/4, \beta = 2)} \,,
\end{align}
whose denominator is evaluated with our preferred values in the massive scheme (see \figs{\ref{fig:lumis-jets-contribs}} and \ref{fig:lumis-ttbar-contribs}).  Within the massive scheme, we plot $r (\beta)$ for a range of values $\beta=2, 3, 4$ where logarithms in the massive splitting formula remain of moderate size.  For the massless scheme, we plot $r (\gamma)$ with $\gamma = 1/2, 1, 2$.  We discard values $\gamma < 1/2$, for which quarks as treated as massless in the splitting at scales much below their mass, as well as values $\gamma > 2$, for which direct $g\to Q \Qbar$ splitting is omitted at splitting scales much larger than $\mQ$.

\begin{figure}[p]
   \begin{center}
      \subfigure[\label{subfig:Lccbarbbbar-jets-1v1}${c\cbar b\bbar}, \, {\text{1v1}}$]{
         \includegraphics[width=0.475\linewidth, trim=0 0 0 55, clip]{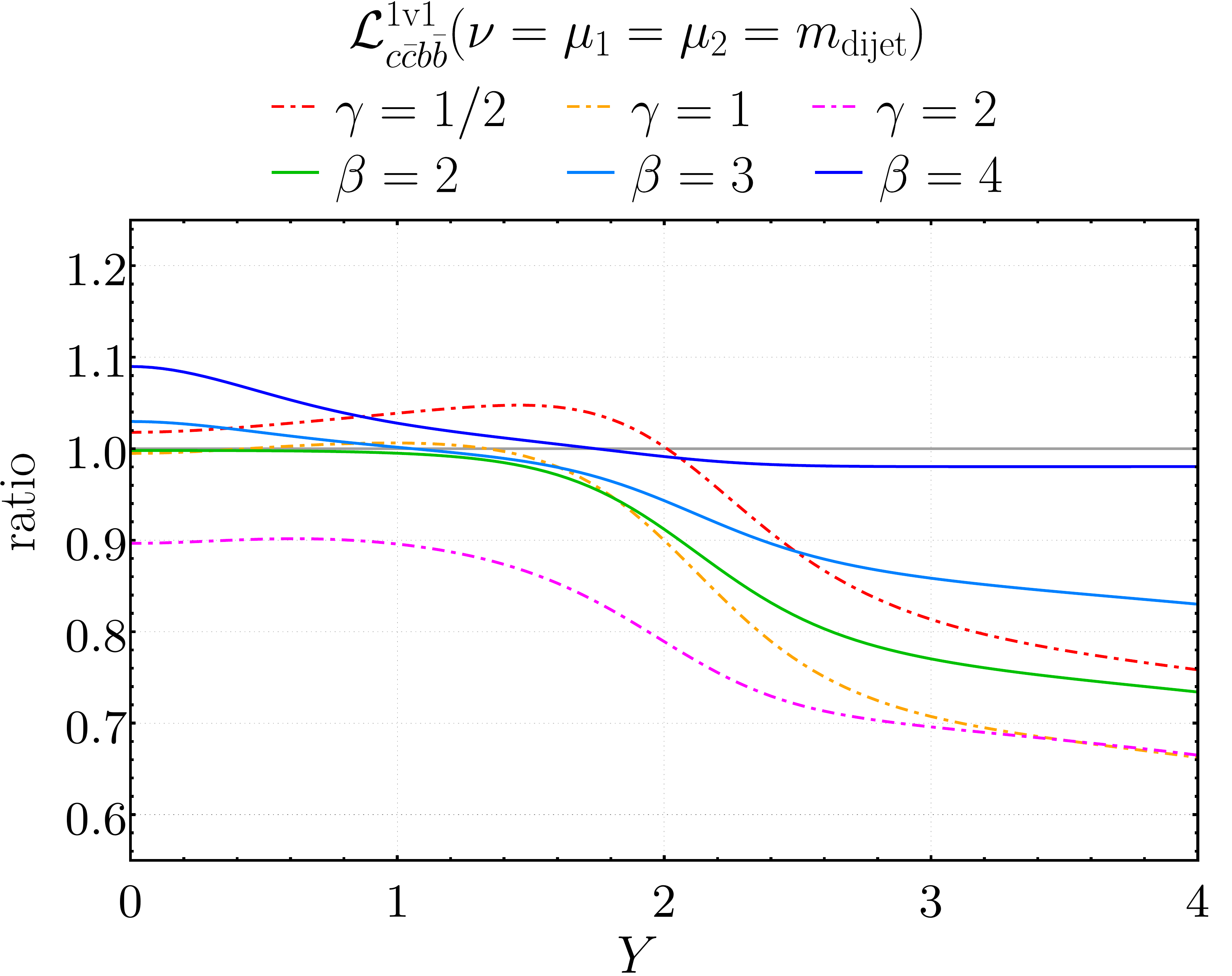}
      }
      \subfigure[\label{subfig:Lccbarbbbar-jets-1v2}${c\cbar b\bbar}, \, {\text{1v2}}$]{
         \includegraphics[width=0.475\linewidth, trim=0 0 0 55, clip]{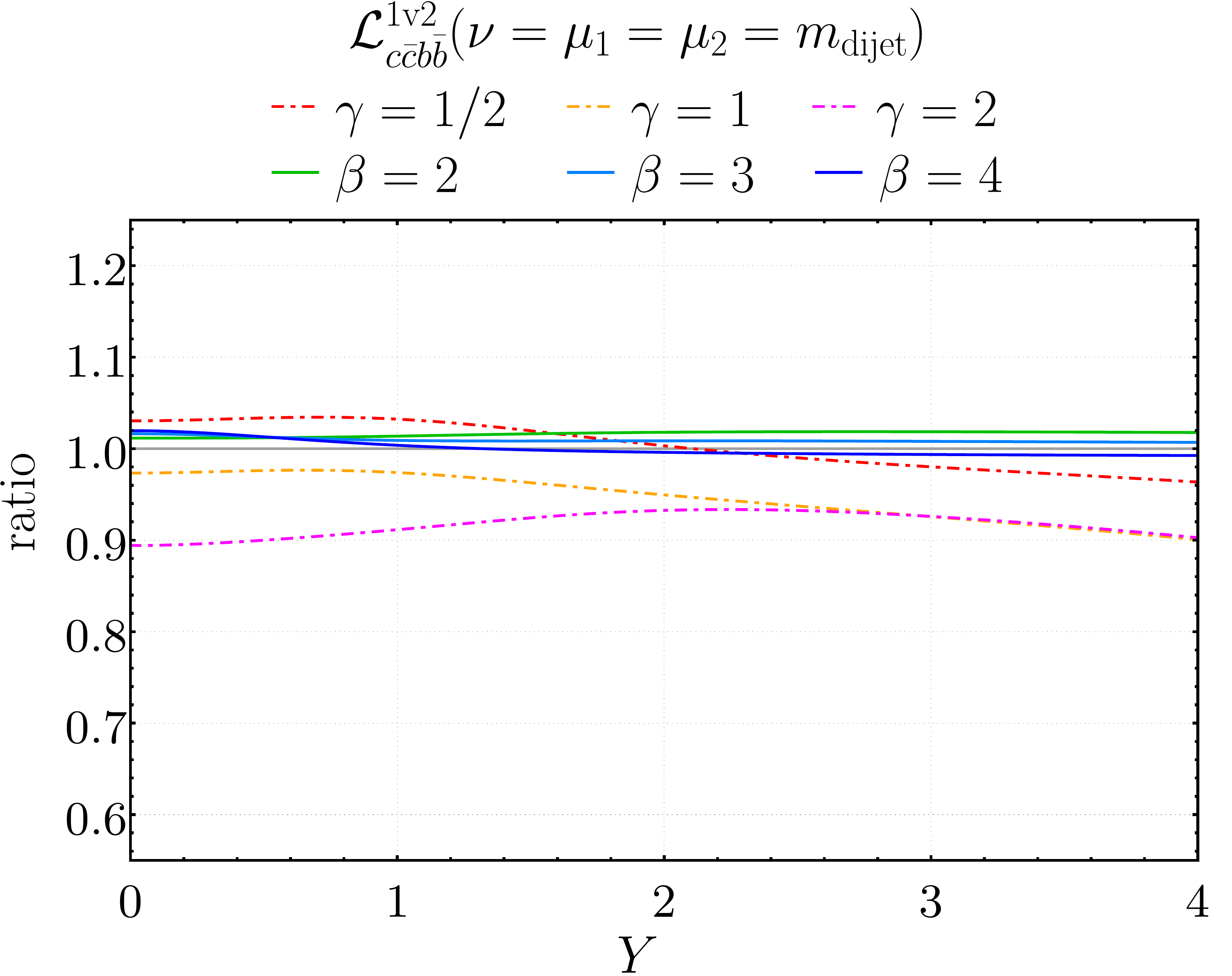}
      }
      \\
      \subfigure[\label{subfig:Lccbarbbbar-jets-2v1}${c\cbar b\bbar}, \, {\text{2v1}}$]{
         \includegraphics[width=0.475\linewidth, trim=0 0 0 55, clip]{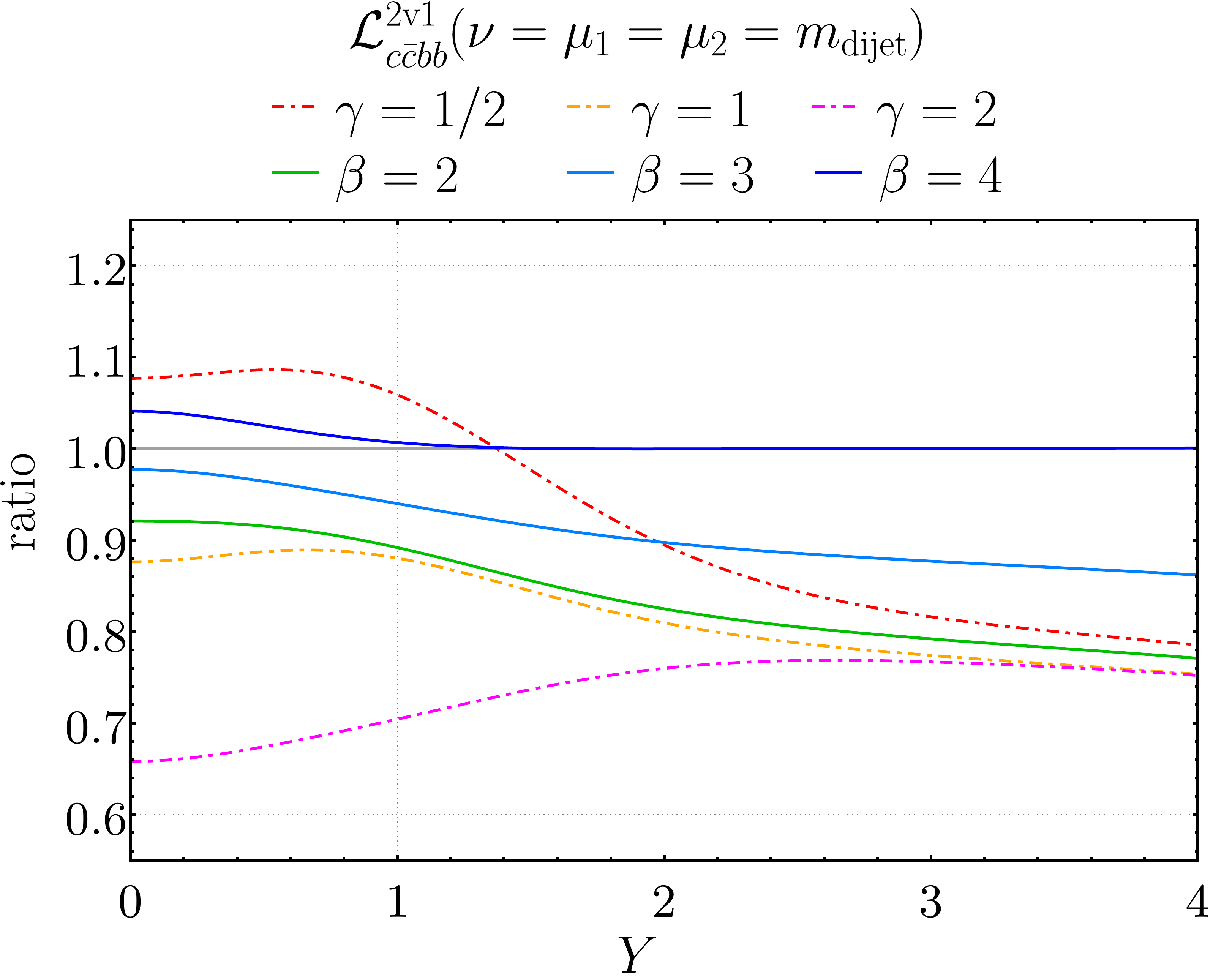}
      }
   \end{center}
   \caption{\label{fig:Lccbarbbbar-jets} Double parton luminosity ratios \protect\eqref{eq:lumi-ratios} for $c \cbar b \bbar$ in the dijet production setting \protect\eqref{eq:dijet-setting} with $\nu = \mu = 25 \gev$.  Solid (dashed) lines correspond to the numerator computed in the (massless) scheme.}
   \begin{center}
      \subfigure[\label{subfig:Lcbgg-jets-1v1}${c b g g}, \, {\text{1v1}}$]{
         \includegraphics[width=0.475\linewidth, trim=0 0 0 55, clip]{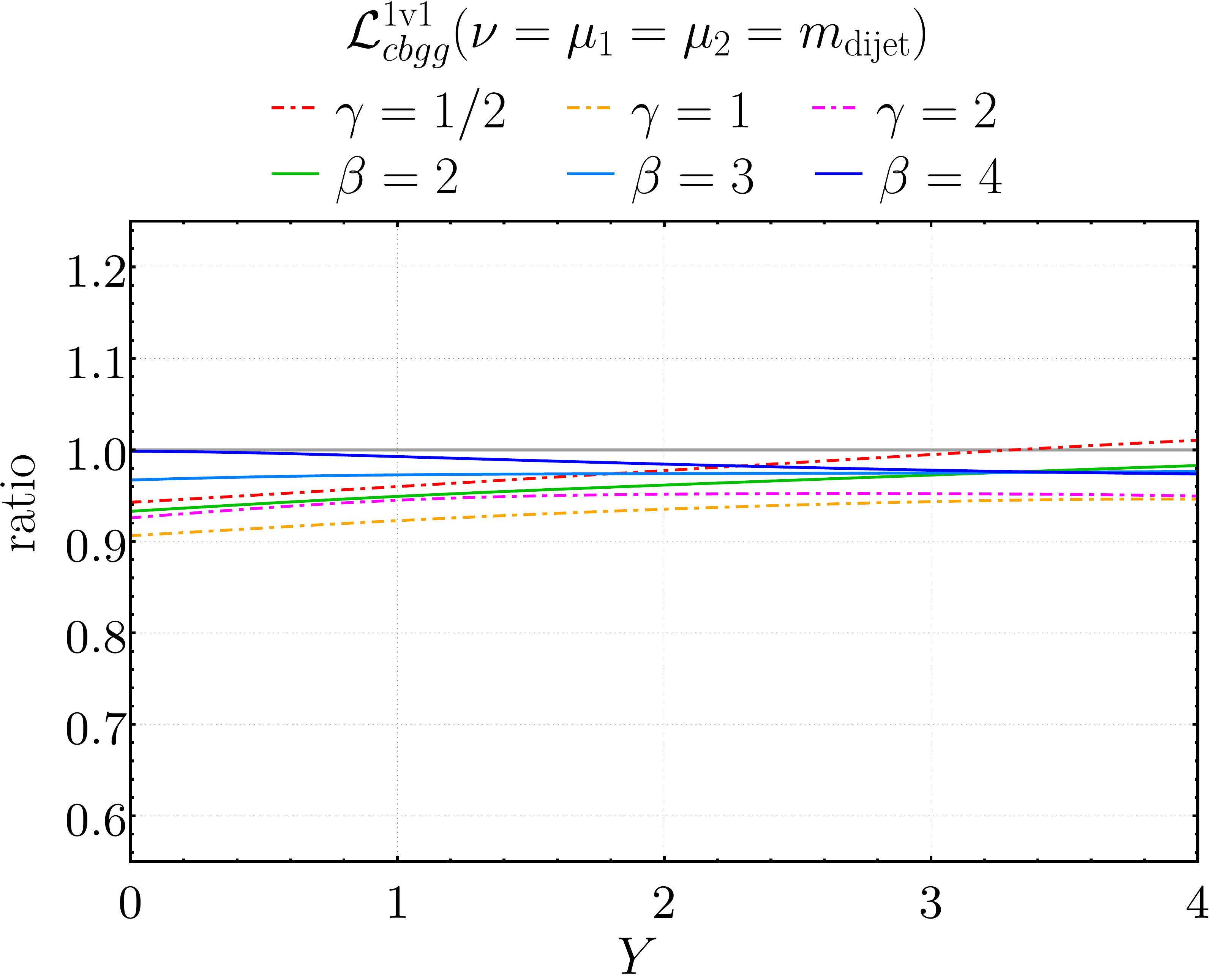}
      }
      \subfigure[\label{subfig:Lcbgg-jets-1v2}${c b g g}, \, {\text{1v2}}$]{
         \includegraphics[width=0.475\linewidth, trim=0 0 0 55, clip]{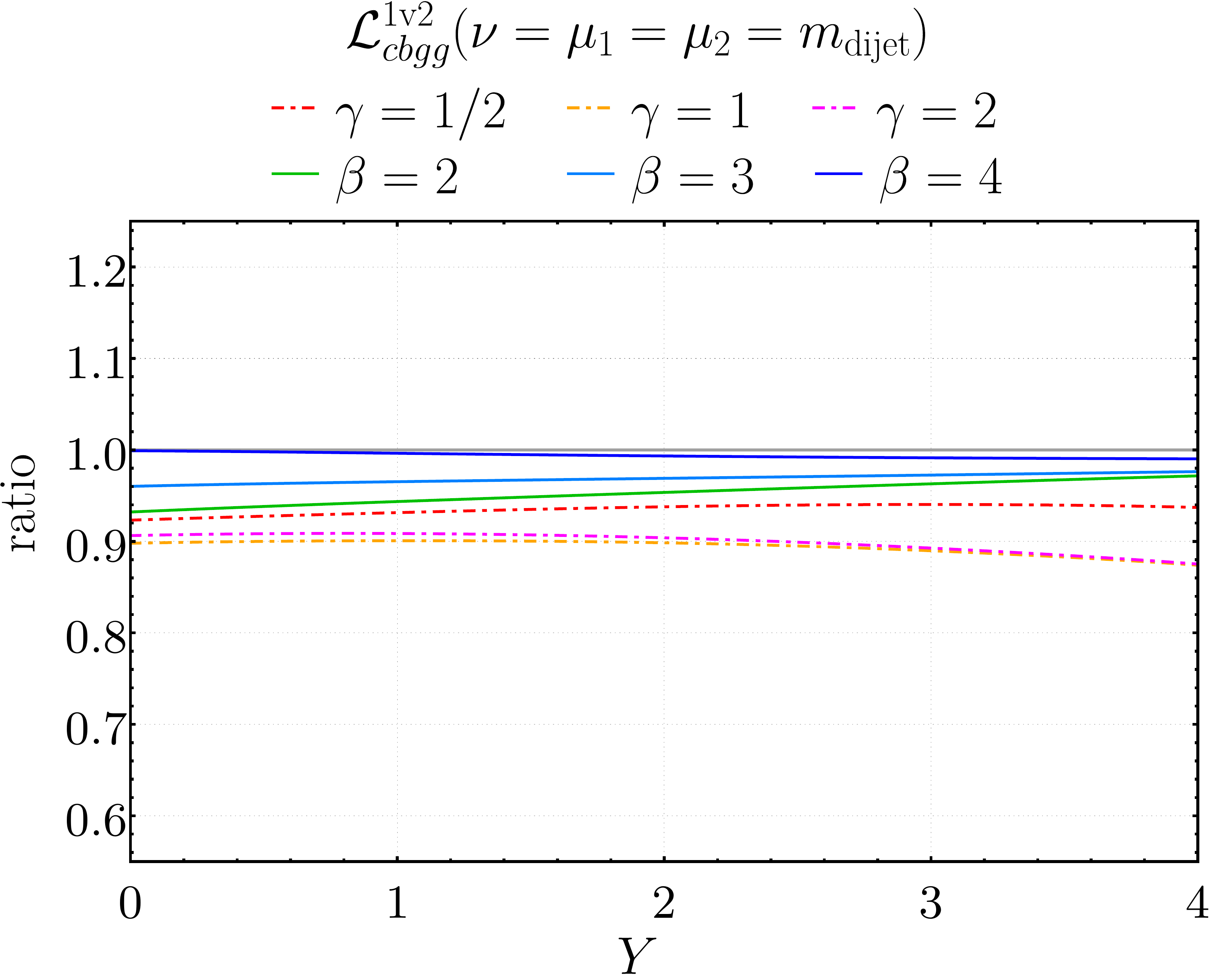}
      }
   \end{center}
   \caption{\label{fig:Lcbgg-jets} As \fig{\protect\ref{fig:Lccbarbbbar-jets}}, but for the $c b g g$ luminosity.  The ratios for the 2v1 contribution deviate from unity by at most 4\% and are not shown here.}
\end{figure}

In \fig{\ref{fig:Lccbarbbbar-jets}} we show ratios for the different contributions to $\mathcal{L}_{c \cbar b \smash{\bbar}}$ in the dijet production setting.  Generally, deviations from unity are larger in the massless scheme than in the massive one (note however that different parameters are varied in the two cases).  Furthermore, the ratios show a pronounced $Y$ dependence in many cases.  If the $b \bbar$ pair is produced by splitting (\figs{\ref{subfig:Lccbarbbbar-jets-1v1}} and \ref{subfig:Lccbarbbbar-jets-2v1}), the largest deviations are slightly below 30\% in the massive scheme and slightly above in the massless one.  Deviations are smaller if only the $c \cbar$ pair originates from splitting (\fig{\ref{subfig:Lccbarbbbar-jets-1v2}}).  This is in line with the weak scheme dependence we observed for $F_{c \cbar}$ in \fig{\ref{subfig:Fccbar-massive-24}}.

We now turn to the channels in \fig{\ref{fig:lumis-jets-contribs}} for which the heavy flavours are not directly produced by $1\to 2$ splitting.  An example is the $c b g g$ channel shown in \fig{\ref{fig:Lcbgg-jets}}.  Again, the deviations of the ratios from unity are somewhat larger in the massless scheme than in the massive one, but they remain below 15\% in both cases.  Deviations in the $g c b g$ luminosity are not larger than those shown for $c b g g$.  In the $g g g g$ channel, we find a parameter deviations of at most 6\%, which originate in the small but visible scheme dependence we observed for $F_{g g}$ in \fig{\ref{fig:F-alpha-beta-comp}}.

\begin{figure}[t!]
   \begin{center}
      \subfigure[\label{subfig:Lttbartbart-ttbar-1v1}${t \tbar \ms \tbar t}, \, {\text{1v1}}$]{
         \includegraphics[width=0.475\linewidth, trim=0 0 0 55, clip]{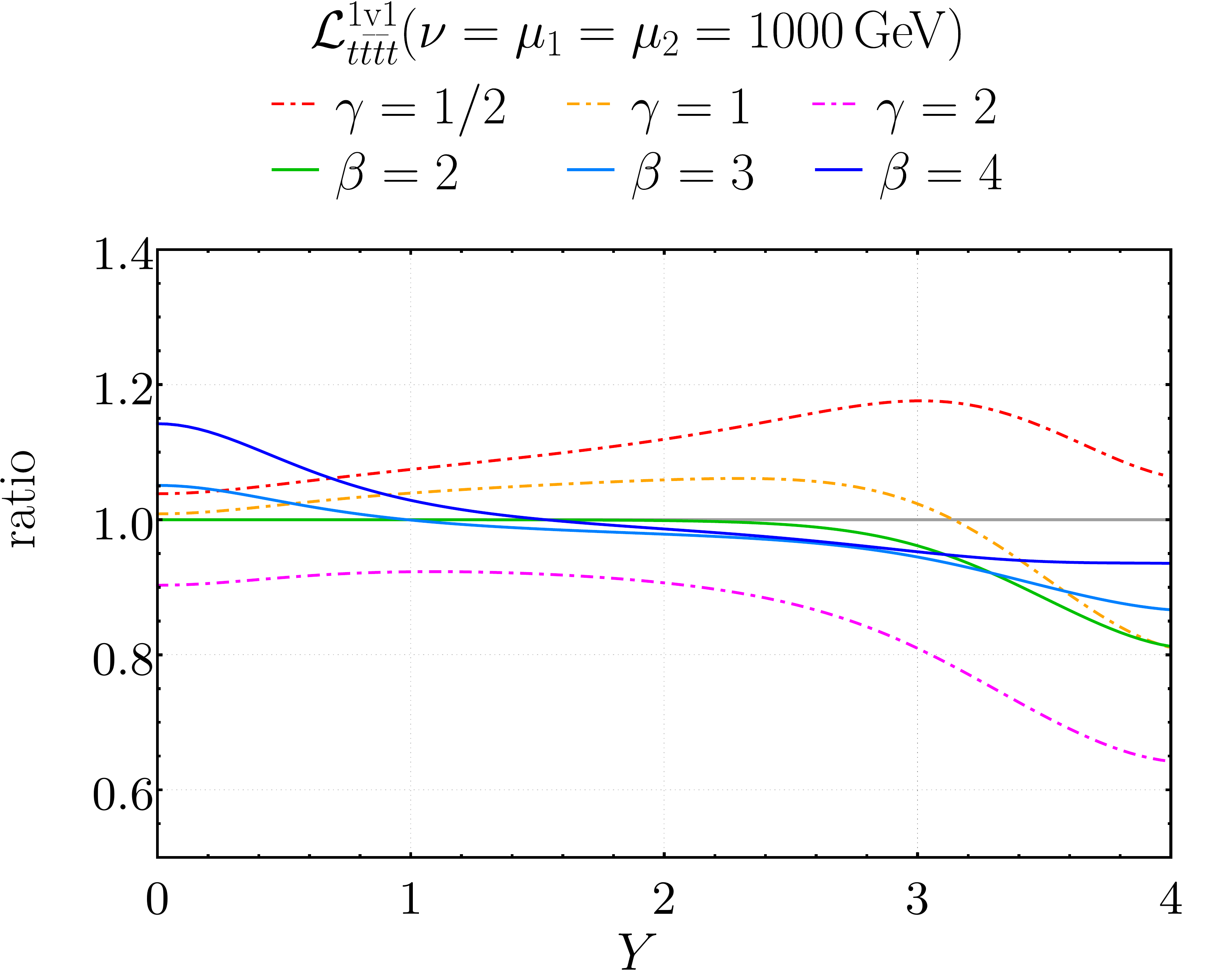}
      }
      \subfigure[\label{subfig:Lgtgtbar-ttbar-1v2}${t \tbar \ms \tbar t}, \, {\text{1v2}}$]{
         \includegraphics[width=0.475\linewidth, trim=0 0 0 55, clip]{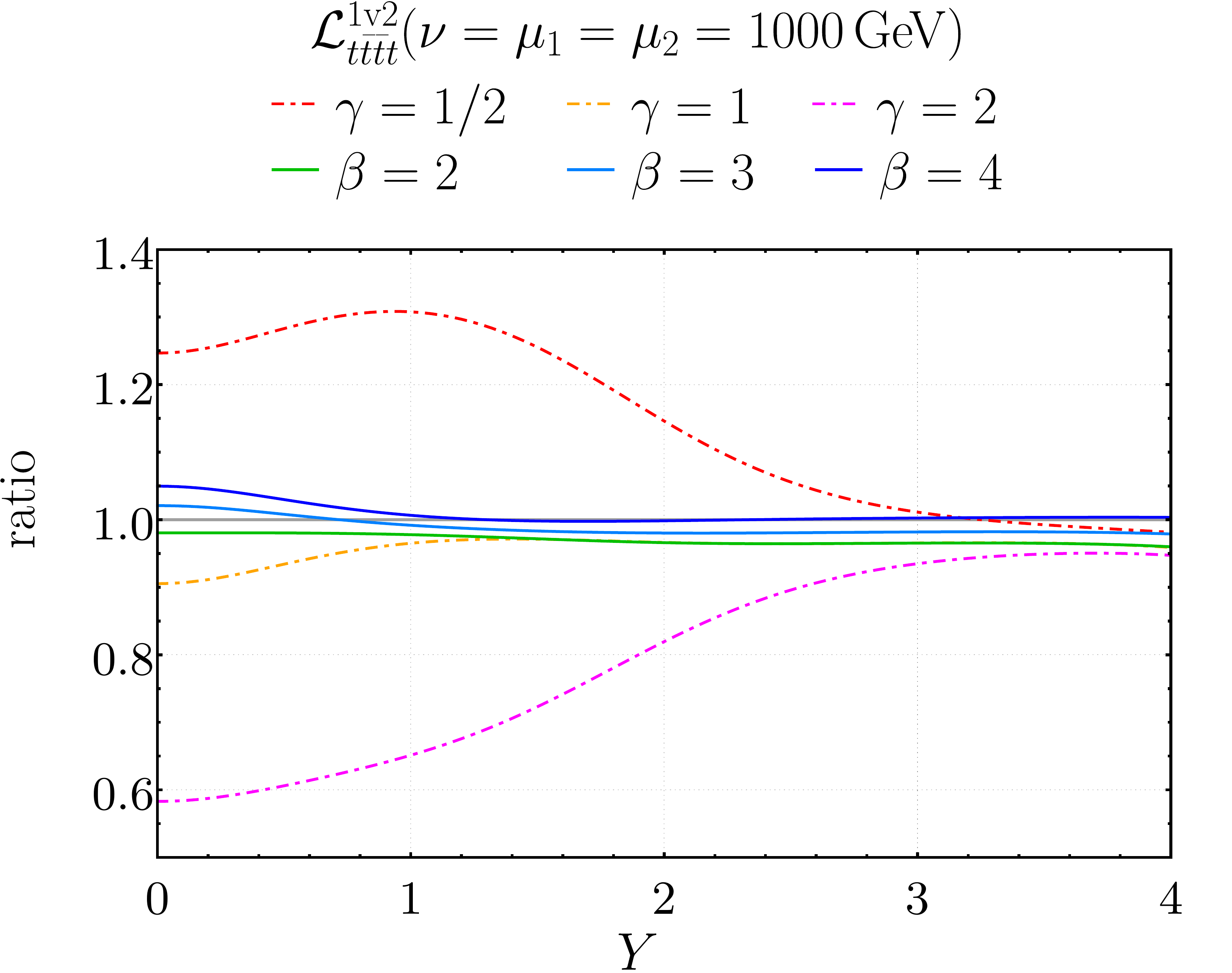}
      }
      \subfigure[\label{subfig:Lgtgtbar-ttbar-1v1}${g t g \tbar}, \, {\text{1v1}}$]{
         \includegraphics[width=0.475\linewidth, trim=0 0 0 55, clip]{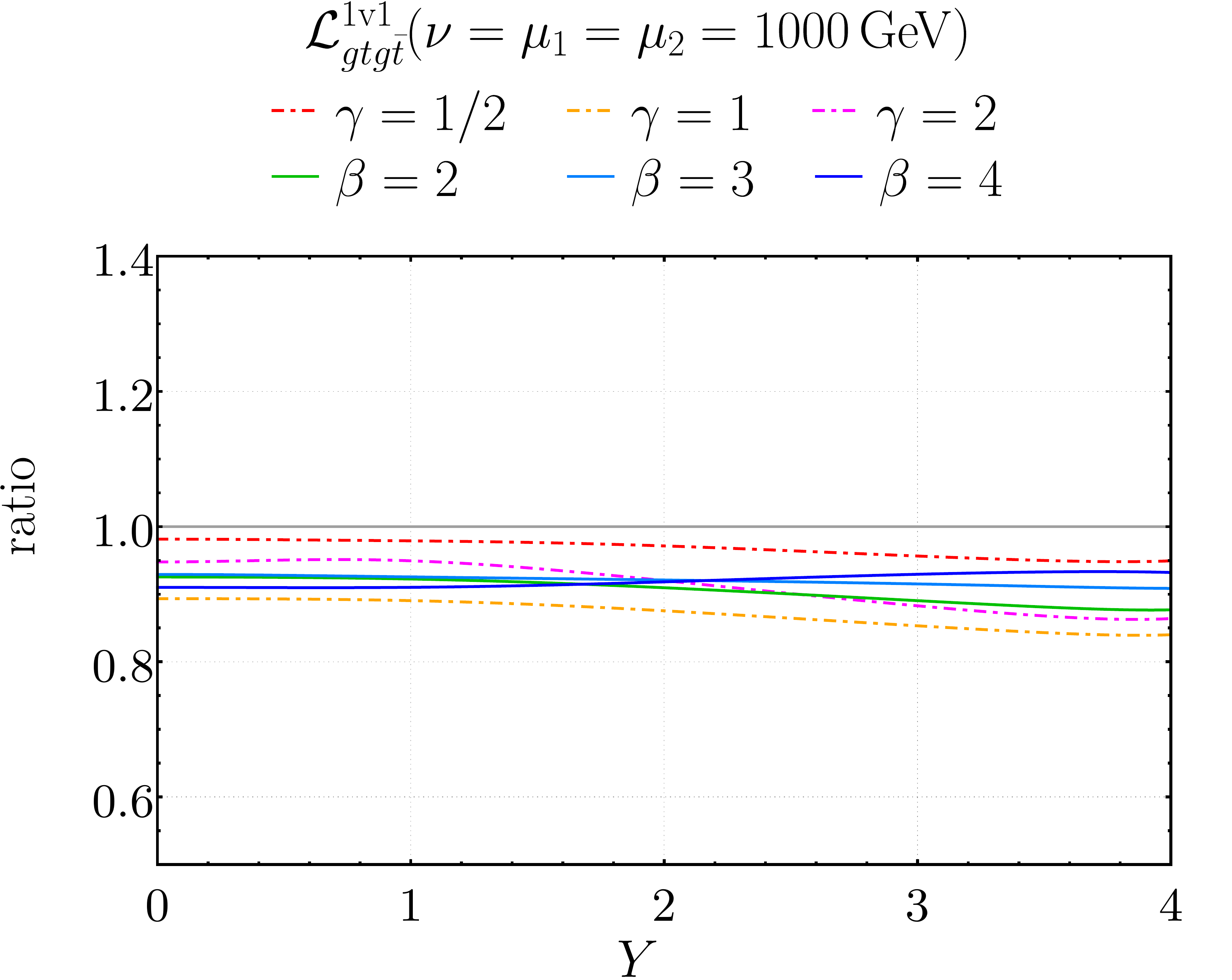}
      }
   \end{center}
   \caption{\label{fig:Lttbartbart-ttbar} As \fig{\protect\ref{fig:Lccbarbbbar-jets}}, but for selected contributions to the $t \tbar \ms \tbar t$ and $g t g \tbar$ luminosities at $\mu = 1 \tev$ for the setting \protect\eqref{eq:ttbar-setting}.  The 1v2 and 2v1 contributions to $t \tbar \ms \tbar t$ are identical to each other.  Deviations of the ratio from unity are tiny for the 1v2 and 2v1 contributions to $g t g \tbar$ and not shown here.}
\end{figure}%

Plots for the $t \tbar$ production setting \eqref{eq:ttbar-setting} are shown in \fig{\ref{fig:Lttbartbart-ttbar}}.  Deviations for the $t \tbar \ms \tbar t$ luminosity reach at most 20\% in the massless and at most 40\% in the massless scheme.  For the $g t g \tbar$ channel, effects are smaller in either scheme, and for $g g g g$ they do not exceed $1\%$.

In both settings, we thus find that the largest scheme dependence arises for parton combinations that can directly be produced by $1\to 2$ splitting in both DPDs.  Results in the massive scheme depend on $\alpha$ and $\beta$ at the level of at most 30\%.

One may ask whether there is a choice of $\gamma$ in the massless scheme that typically gives the best agreement with the more realistic results in the massive scheme.  Whilst for $c \cbar b \bbar$ luminosities in \fig{\ref{fig:lumis-ttbar-contribs}}, ratios closest to unity are obtained for $\gamma = 1/2$, the preferred value for the ${g t g \tbar}$ luminosities in \fig{\ref{fig:lumis-ttbar-contribs}} is between $1/2$ and $1$.  Values below 1/2 or above 2 would not improve the global agreement.  For $1/2 \le \gamma \le 1$ the luminosities in the massless scheme approximately reproduce the ones in the massive scheme with $\alpha = 1/4$ and $\beta = 2$.  Typical deviations reach 30\% in either direction and can depend strongly on $Y$, i.e.\ on the momentum fractions in the DPDs.
%
%
\subsubsection{Splitting scale variation}
\label{sec:DPD-lumis-split-scale}
To estimate the importance of higher-order corrections in the DPD splitting formulae \eqref{eq:small-y-DPD} and \eqref{eq:small-y-DPD-massive}, one may vary the scale $\mu_{\text{split}}$ at which they are evaluated before the DPD is evolved to the final scale $\mu$ (with flavour matching at an intermediate scale if appropriate).  It is customary to vary the renormalisation scale in a fixed-order formula by a factor of 2 around its central value.  We modify this to
\begin{align}
   \label{eq:mu-split-variation}
   \min\bigl( \mu_{\min} , \mu_{y^*} / 2 \bigr)
      \le \mu_{\text{split}}
      \le 2 \mu_{y^*} \,,
\end{align}
which avoids renormalisation scales below $\mu_{\text{min}} = 1 \gev$.  This results in an asymmetric scale variation for $\mu_{y^*} < 2 \mu_{\text{min}}$, which translates to $y > 0.57 \gev^{-1}$ with our choice of $y^*(y)$.
%
%
In the following, we study the impact of this scale variation on double parton luminosities in the massive scheme with $\alpha = 1/4$ and $\beta = 2$.

\begin{figure}[t!]
   \begin{center}
      \subfigure[\label{subfig:Lccbarbbbar-jets-scale}${c \cbar b \bbar}$]{
         \includegraphics[width=0.475\linewidth, trim=0 0 0 50, clip]{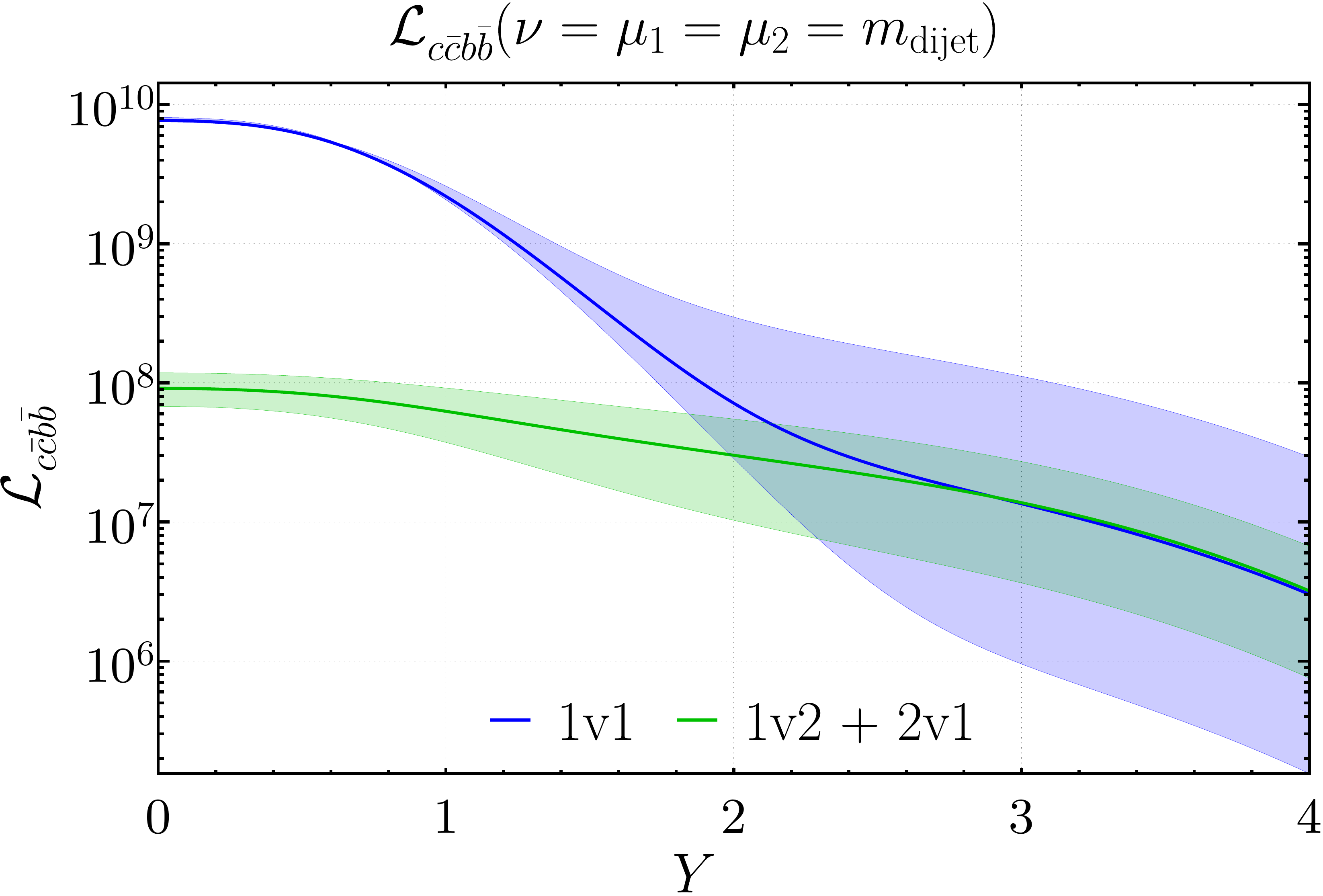}
      }
      \subfigure[\label{subfig:Lcbgg-jets-scale}${c b g g}$]{
         \includegraphics[width=0.475\linewidth, trim=0 0 0 50, clip]{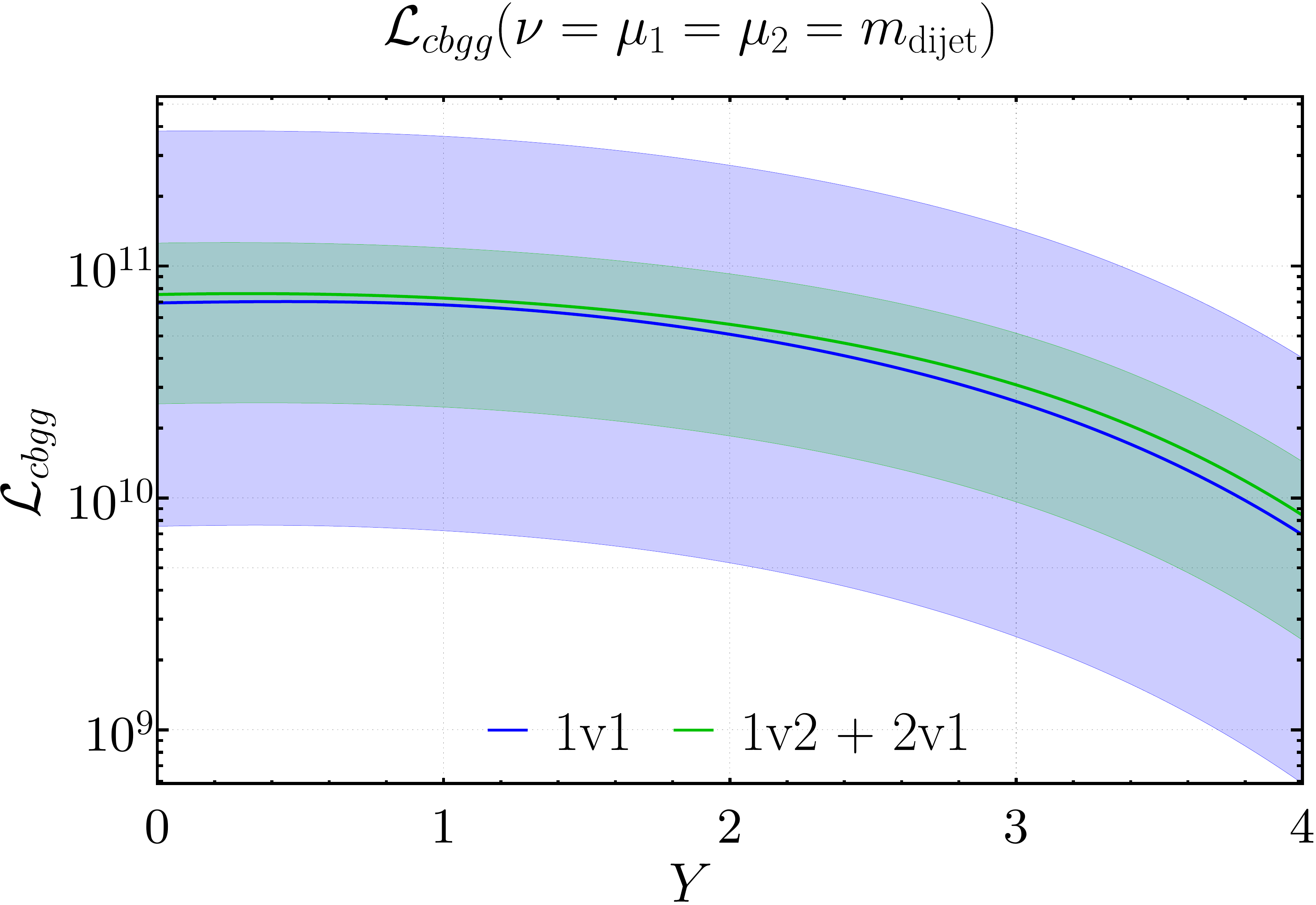}
      }
      \\
      \subfigure[\label{subfig:Lgcbg-jets-scale}${g c b g}$]{
         \includegraphics[width=0.475\linewidth, trim=0 0 0 50, clip]{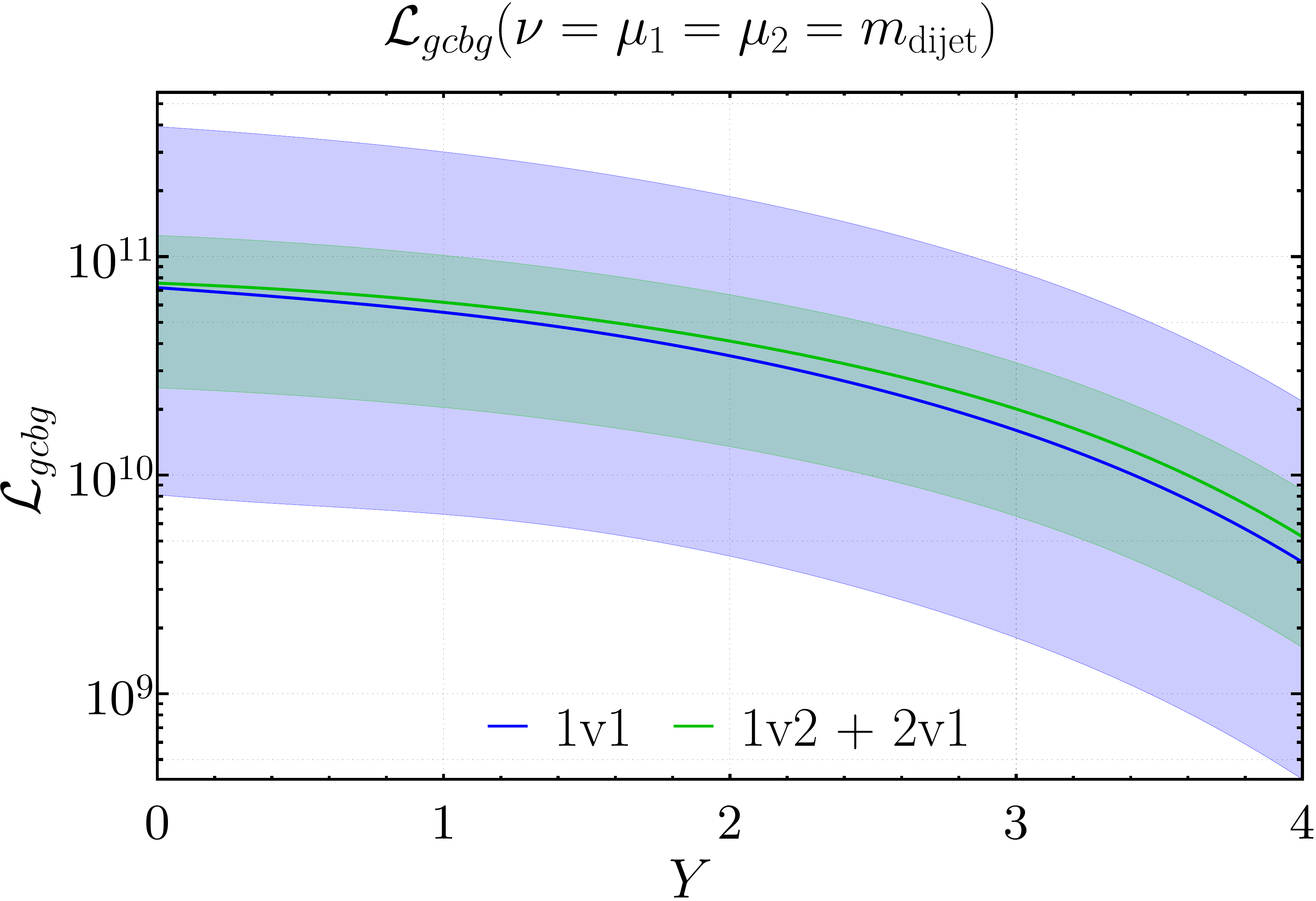}
      }
      \subfigure[\label{subfig:Lgggg-jets-scale}${g g g g}$]{
         \includegraphics[width=0.475\linewidth, trim=0 0 0 50, clip]{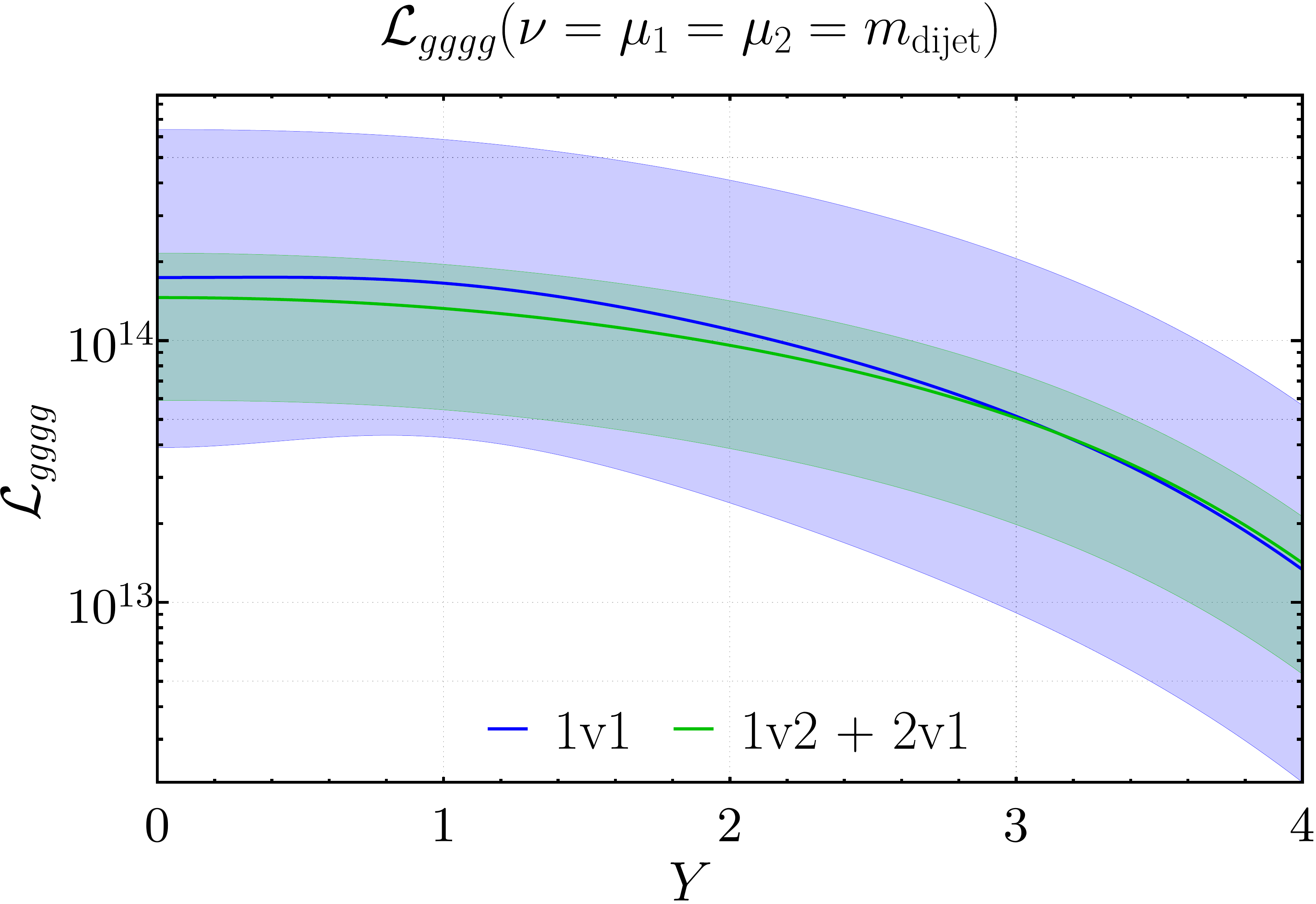}
      }
   \end{center}
   \caption{\label{fig:lumis-jets-scale} Splitting scale variation of the double parton luminosities shown in \fig{\protect\ref{fig:lumis-jets-contribs}} for the dijet production setting.  Central curves are for $\mu_{\text{split}} = \mu_{y^*}$, and bands correspond to the variation specified in \protect\eqref{eq:mu-split-variation}.  Notice that the 1v2 and 2v1 contributions have been added.}
   \begin{center}
      \subfigure[\label{subfig:Lttbartbart-ttbar-scale}${t \tbar \ms \tbar t}$]{
         \includegraphics[width=0.475\linewidth, trim=0 0 0 50, clip]{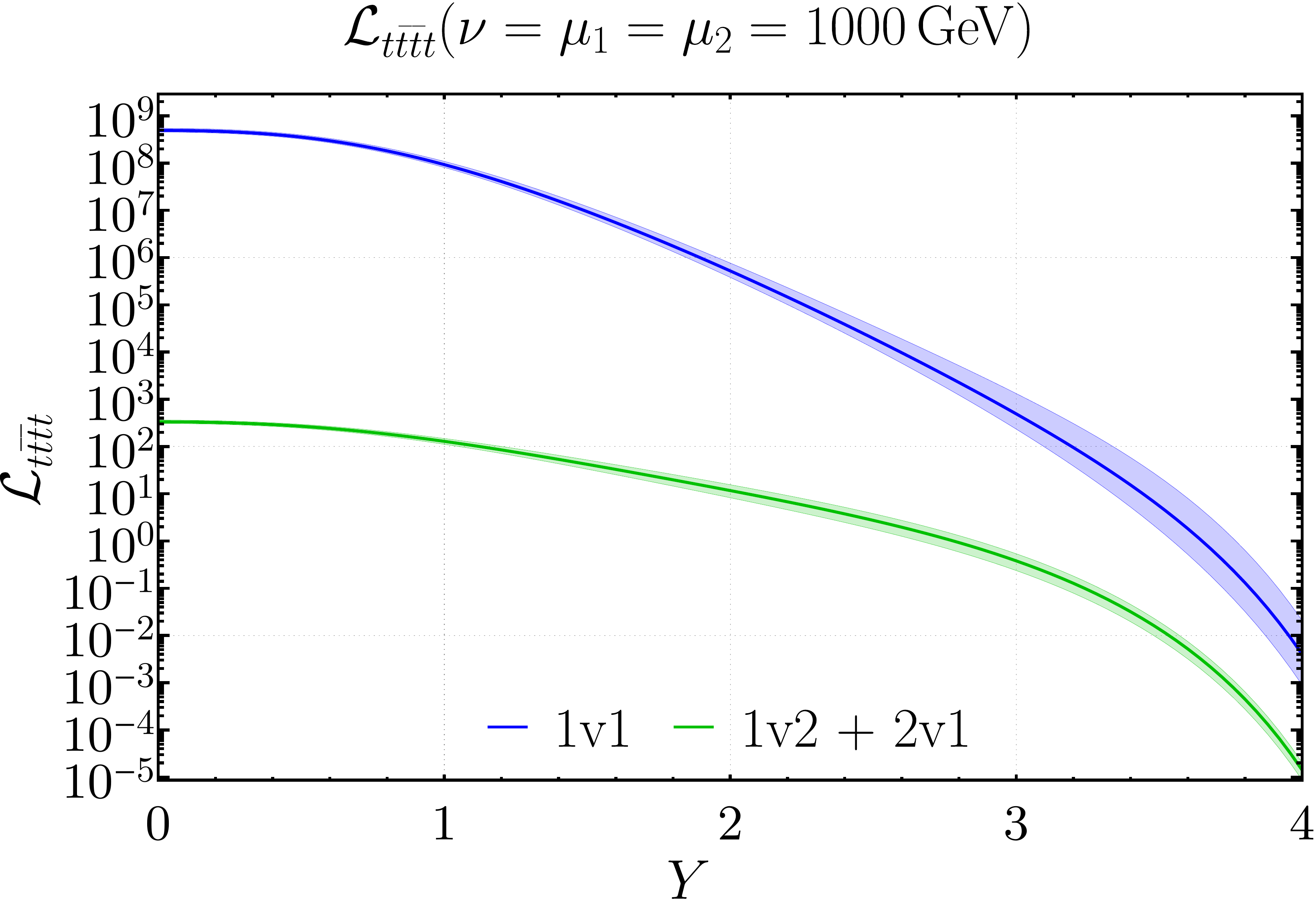}
      }
      \subfigure[\label{subfig:Lgtgtbar-ttbar-scale}${g t g \tbar}$]{
         \includegraphics[width=0.475\linewidth, trim=0 0 0 50, clip]{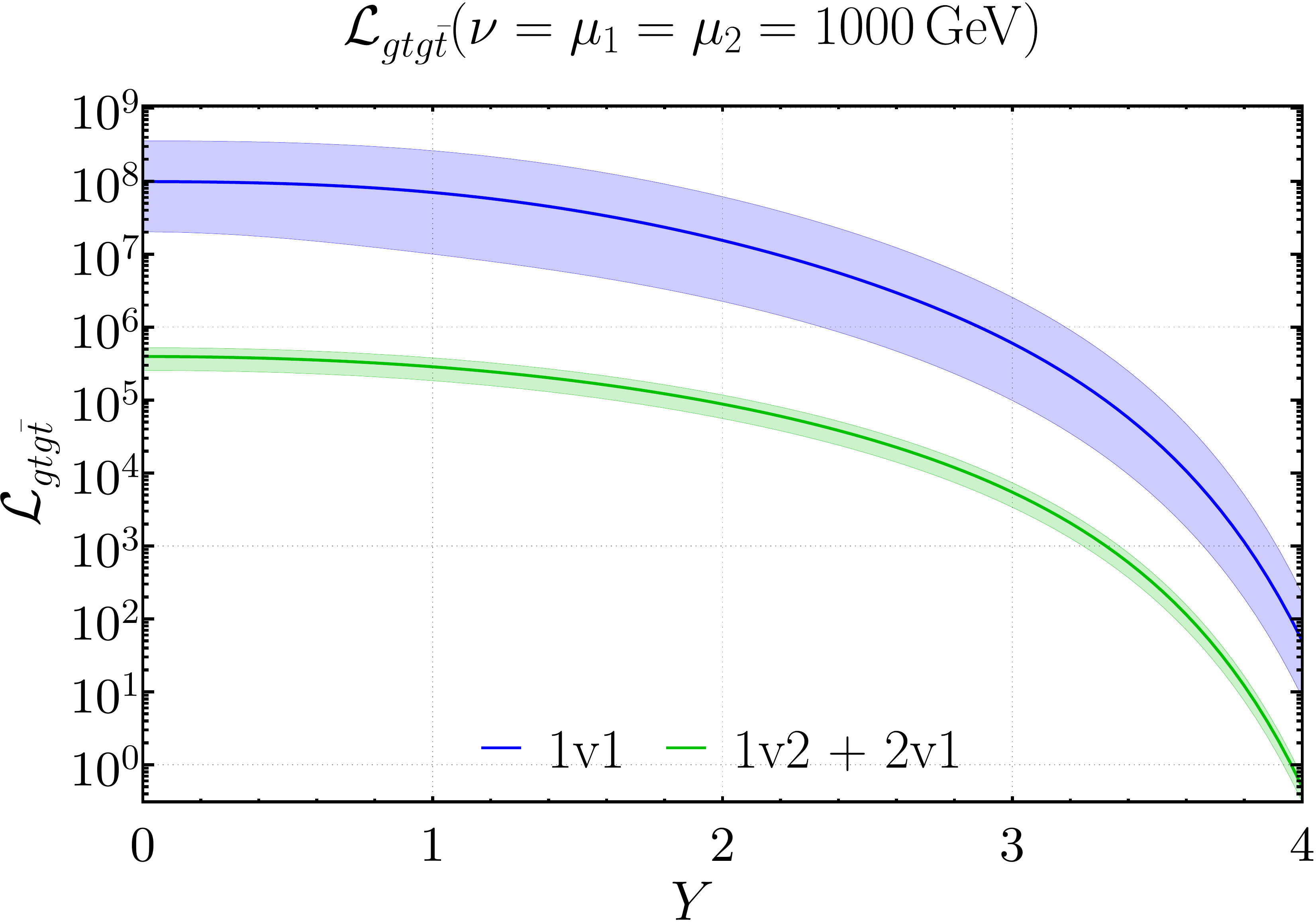}
      }
   \end{center}
   \caption{\label{fig:lumis-ttbar-scale} As \fig{\protect\ref{fig:lumis-jets-scale}}, but for the luminosities shown in \fig{\protect\ref{fig:lumis-ttbar-contribs}} for the $t \tbar$ production setting.}
\end{figure}

Starting with the dijet production setting, we show in \fig{\ref{fig:lumis-jets-scale}} the splitting scale dependence for the same luminosities that were given in \fig{\ref{fig:lumis-jets-contribs}}.  The variation of the 1v1 terms is far greater than for the 1v2 and 2v1 terms, which is not surprising because the former involve two splitting DPDs and the latter only one.  We also observe that the bands are more asymmetric for the sum of the 1v2 and 2v1 terms than for 1v1.  This is because the former receive important contributions from the region $y > 0.57 \gev$ where the scale variation \eqref{eq:mu-split-variation} becomes asymmetric.  In the channels ${c b g g}$, ${g c b g}$, and ${g g g g}$, the scale variation is of considerable size and depends rather weakly on $Y$.  By contrast, the scale variation of for $c \cbar b \bbar$ is very weak at low $Y$ but large at high $Y$.  An explanation for this remarkable feature is given in \app{\ref{app:add-numerics}}.

The splitting scale dependence of double parton luminosities in the $t\tbar$ production setting is shown in \fig{\ref{fig:lumis-ttbar-scale}} and shows similarities to the cases just discussed.  In the 1v1 terms, we find a large scale variation for ${g t g \tbar}$ at all $Y$ and for ${t \tbar \ms \tbar t}$ at high $Y$.

%
%
\subsubsection{Variation of the flavour matching scale}
\label{sec:DPD-lumis-match-scale}
Apart from the DPD splitting scale $\mu_{\text{split}}$, the evaluation of DPDs in the schemes of \sect{\ref{sec:schemes}} involve a second scale choice, namely for the scale $\mu_Q$ at which flavour matching is performed for DPDs, PDFs, and $\alpha_s$.  It is therefore natural to study the impact of varying this scale as well.  In analogy to \eqref{eq:mu-split-variation}, we vary the matching scale in the interval
\begin{align}
   \label{eq:mu-Q-variation}
   \min( \mu_{\text{min}}, \mQ / 2 ) \le \mu_Q \le 2 \mQ
   & &
   \text{ for } Q = c, b, t,
\end{align}
where the minimum-prescription on the l.h.s.\ is relevant for the charm quark mass.

In the following, we also compare results with flavour matching evaluated either at LO or at NLO.  For the default choice $\mu_Q = \mQ$, this makes no difference, since all one-loop flavour matching kernels are zero at that point.  For different scale choices, this is no longer the case.  We note that computing $A^{Q}$ at NLO but $V$ and $V^{Q}$ at LO corresponds to order $\alpha_s$ in all kernels appearing in the factorised graphs of \fig{\ref{fig:splitting-regions}}.  Of course, the overall accuracy remains at LO with such a hybrid choice.

Notice that not only the computation of the splitting DPDs involves flavour matching, but also the evaluation of the intrinsic part $F^{\text{intr}}$.  The latter is initialised for $\nf = 3$ flavours for all $y$ and thus requires several steps of DPD flavour matching in our settings with $\nf = 5$ or $6$ for dijet or $t\tbar$ production, respectively.

\begin{figure}[!t]
   \begin{center}
      \subfigure[\label{subfig:Lccbarbbbar-jets-match-LO}${c \cbar b \bbar}$ with LO flavour matching]{
         \includegraphics[width=0.475\linewidth, trim=0 0 0 50, clip]{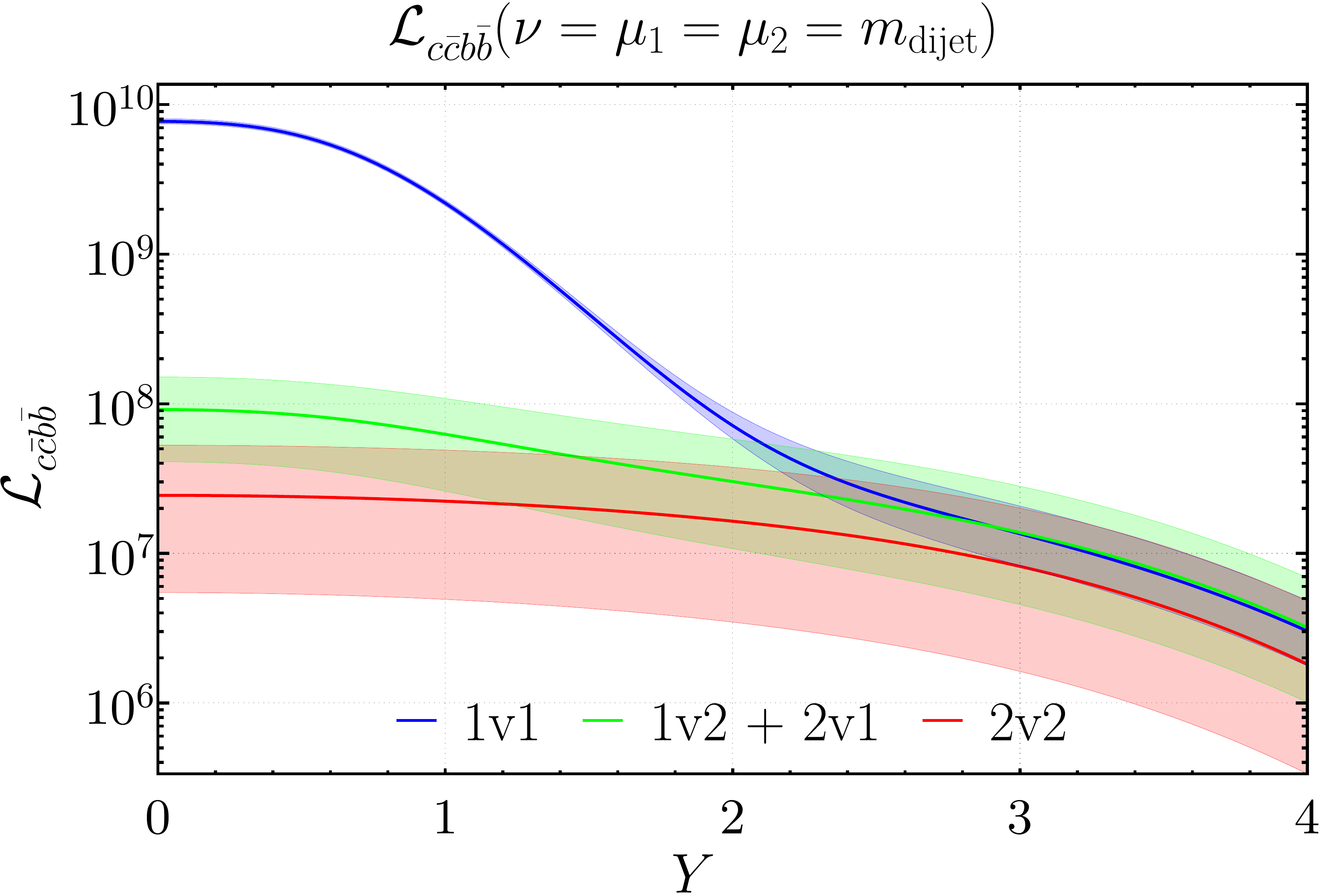}
      }
      \subfigure[\label{subfig:Lccbarbbbar-jets-match-NLO}${c \cbar b \bbar}$ with NLO flavour matching]{
         \includegraphics[width=0.475\linewidth, trim=0 0 0 50, clip]{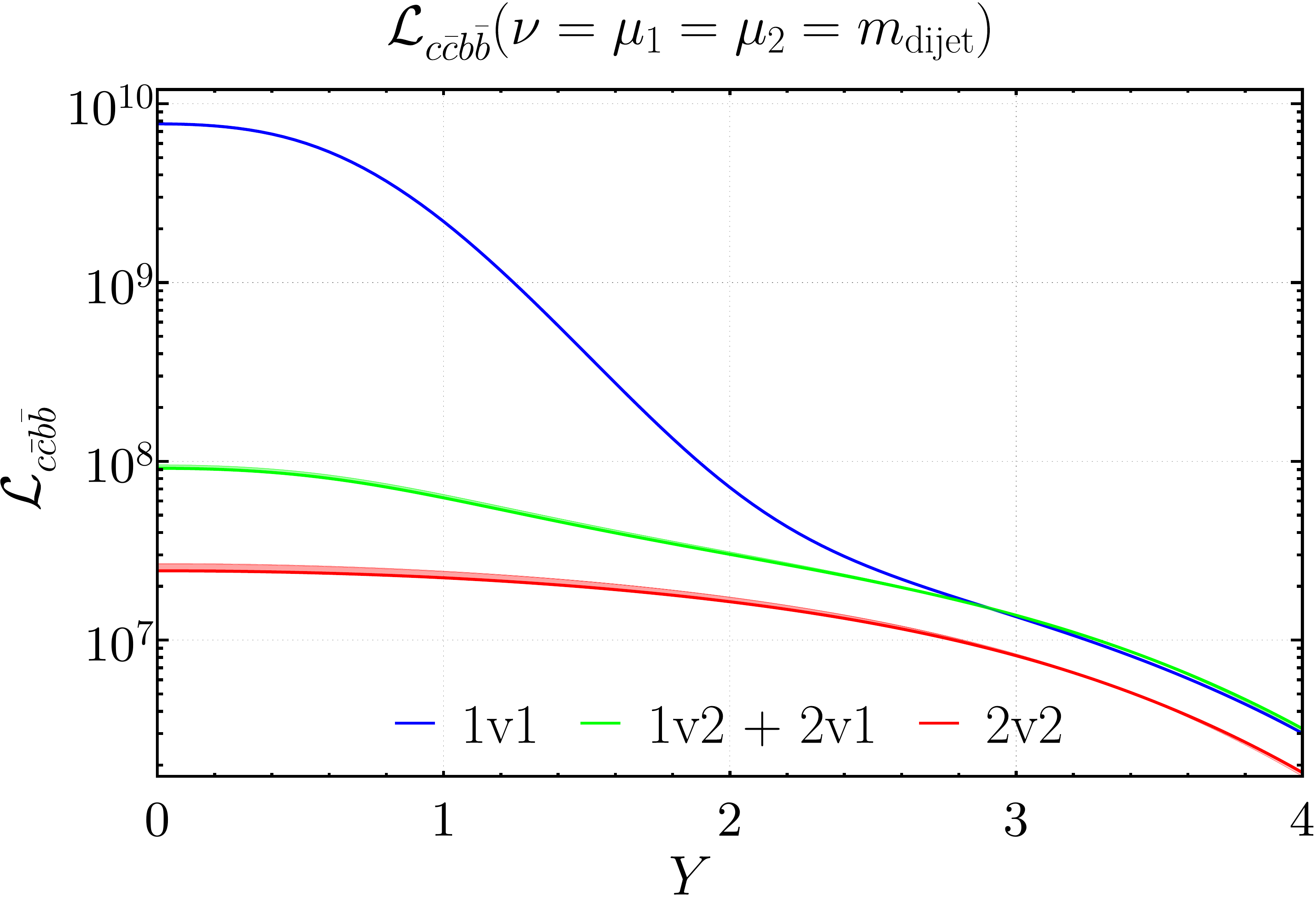}
      }
      \\
      \subfigure[\label{subfig:Lcbgg-jets-match-LO}${c b g g}$ with LO flavour matching]{
         \includegraphics[width=0.475\linewidth, trim=0 0 0 50, clip]{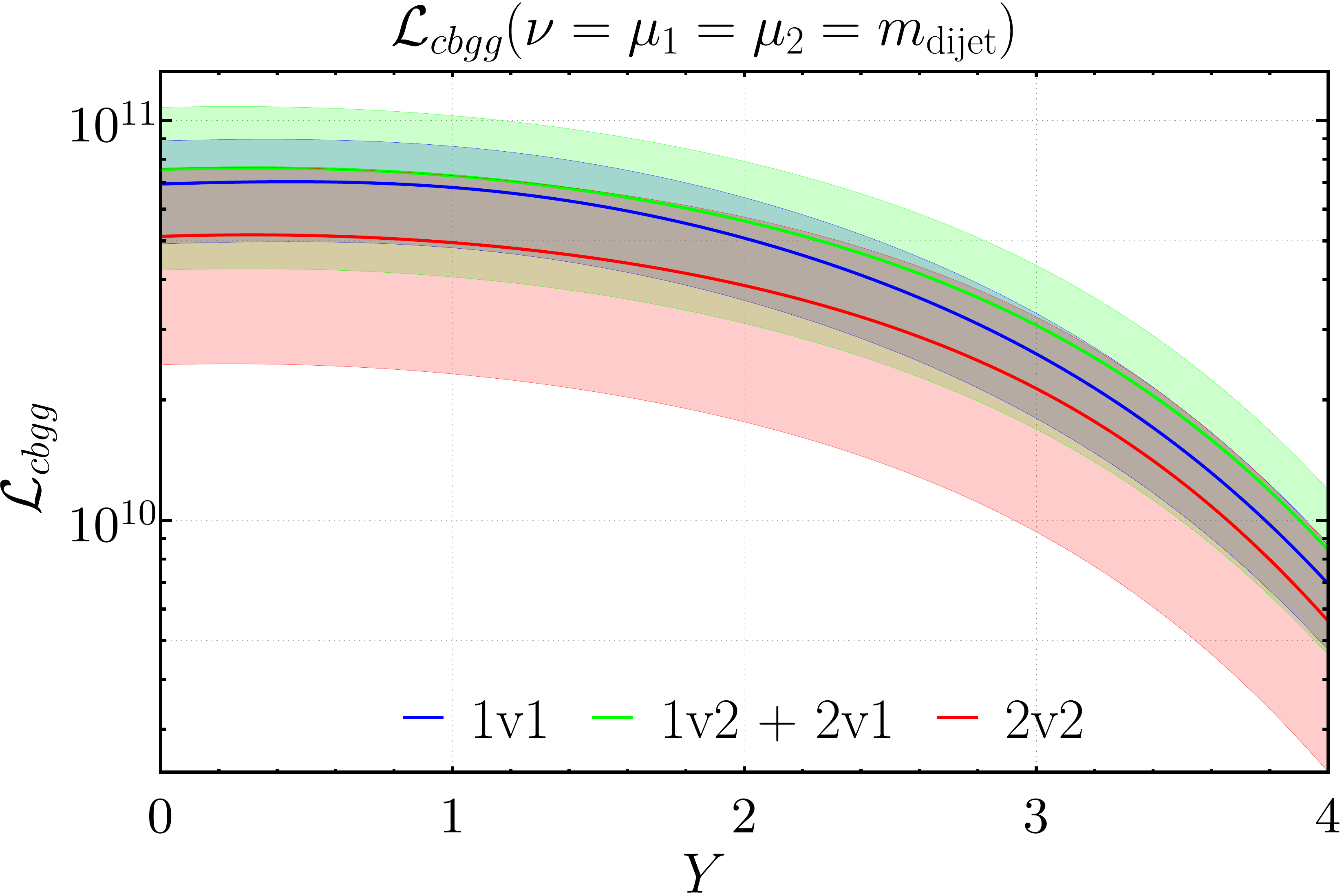}
      }
      \subfigure[\label{subfig:Lcbgg-jets-match-NLO}${c b g g}$ with NLO flavour matching]{
         \includegraphics[width=0.475\linewidth, trim=0 0 0 50, clip]{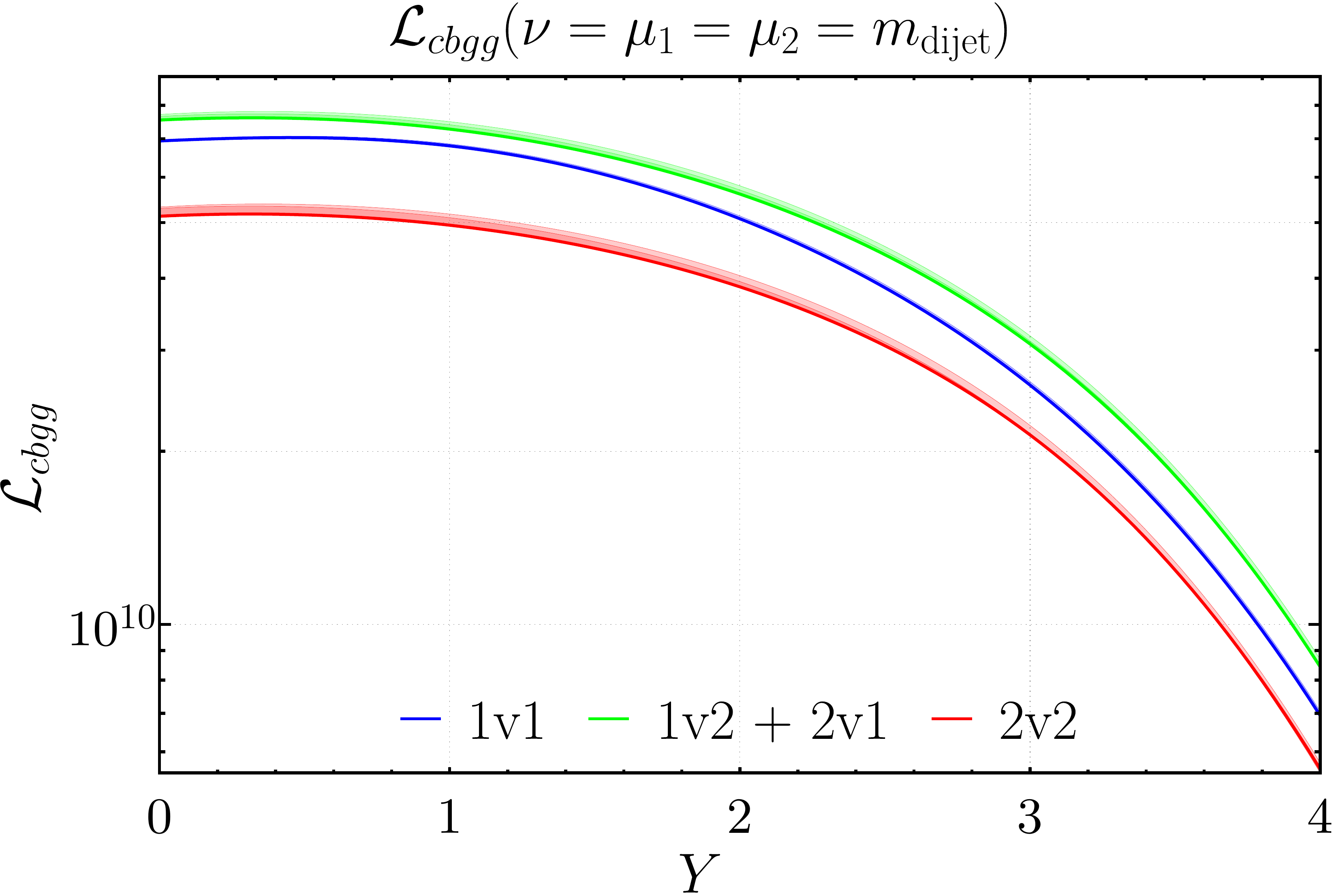}
      }
      \\
      \subfigure[\label{subfig:Lgggg-jets-match-LO}${g g g g}$ with LO flavour matching]{
         \includegraphics[width=0.475\linewidth, trim=0 0 0 48, clip]{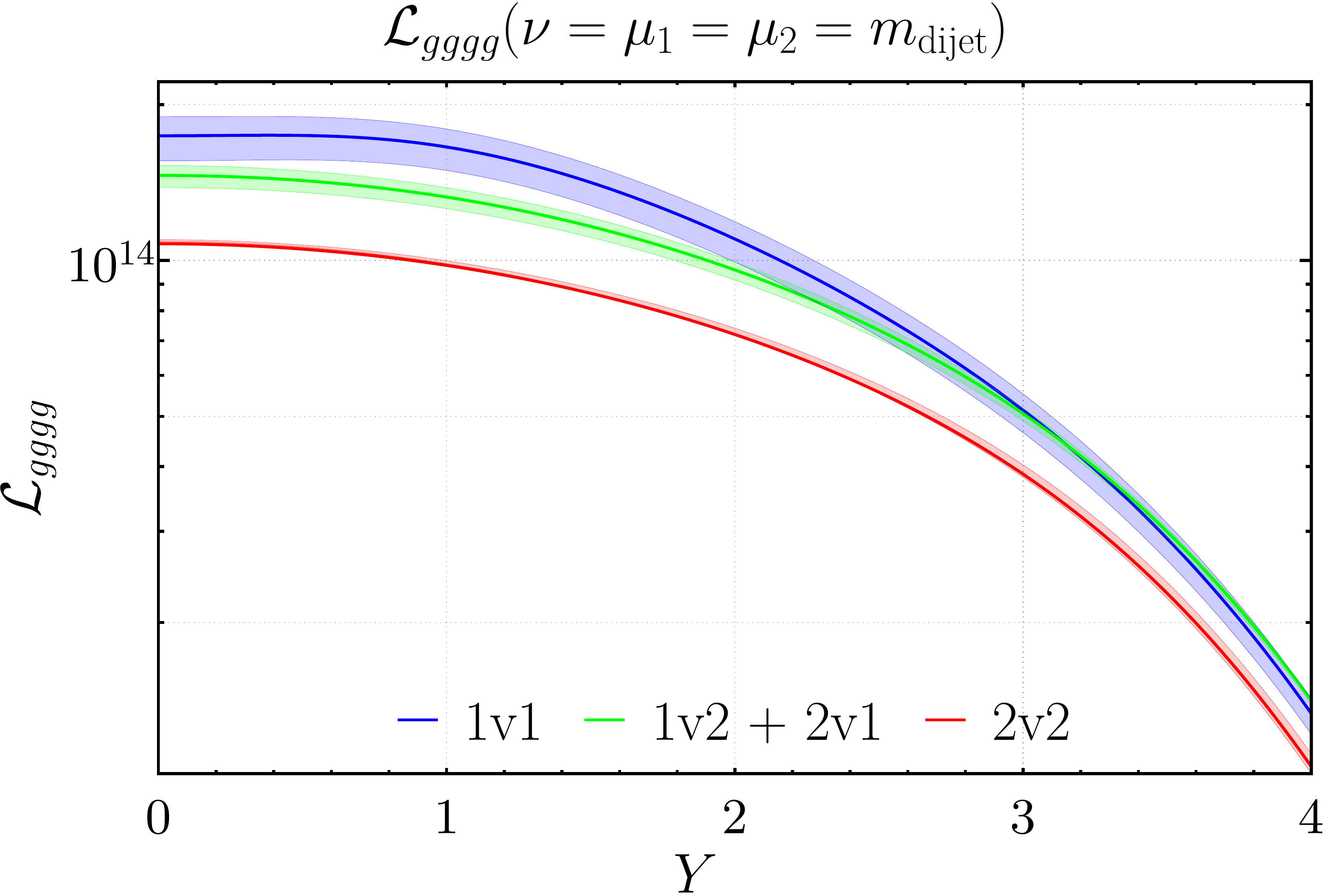}
      }
      \subfigure[\label{subfig:Lgggg-jets-match-NLO}${g g g g}$ with NLO flavour matching]{
         \includegraphics[width=0.475\linewidth, trim=0 0 0 48, clip]{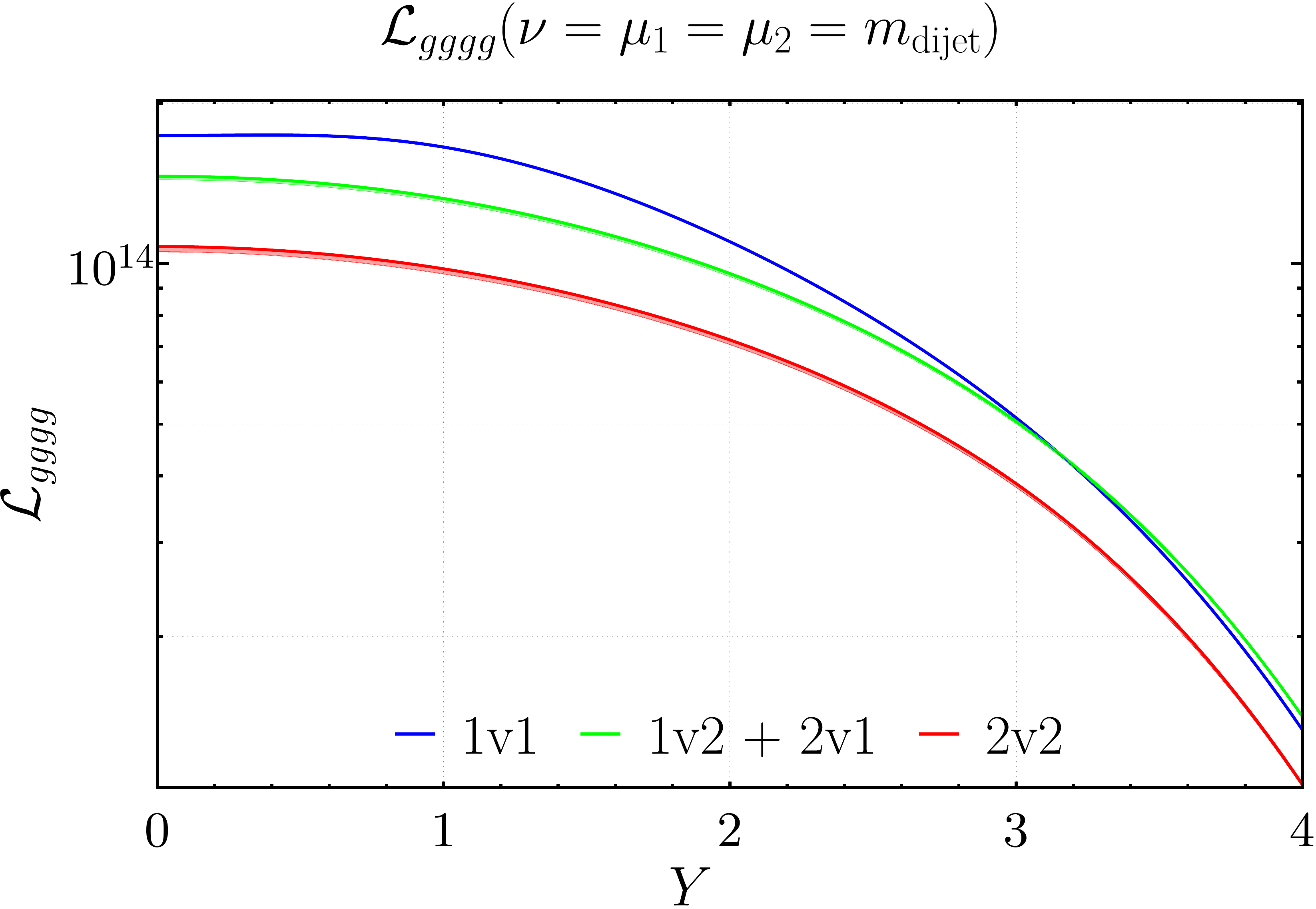}
      }
   \end{center}
   \caption{\label{fig:lumis-jets-match}  Dependence on the flavour matching scale of the double parton luminosities shown in \fig{\protect\ref{fig:lumis-jets-contribs}} for the dijet production setting.  Central curves are for $\mu_{Q} = \mQ$, and bands correspond to the variation specified in \protect\eqref{eq:mu-Q-variation}.  Flavour matching (of DPDs, PDFs, and $\alpha_s$) is performed at LO in the panels on the left and at NLO in the panels on the right.}
\end{figure}

The matching scale dependence of luminosities in the dijet production setting is shown in \fig{\ref{fig:lumis-jets-match}}.  With LO flavour matching we find large scale uncertainties, except in $\mathcal{L}^{\text{1v1}}_{c \cbar b \bbar}$ at low or intermediate $Y$ and in the $g g g g$ channel.  Not surprisingly, the effects are largest for terms where the observed heavy partons can only be produced by flavour matching, such as the 2v2 contributions to $c \cbar b \bbar$ and $c b g g$.  The scale dependence is greatly reduced when flavour matching is performed at NLO, with small variations in all channels and in the full $Y$ range.

\begin{figure}[!t]
   \begin{center}
      \subfigure[\label{subfig:Lttbartbart-ttbar-match-LO}${t \tbar \ms \tbar t}$ with LO flavour matching]{
         \includegraphics[width=0.475\linewidth, trim=0 0 0 50, clip]{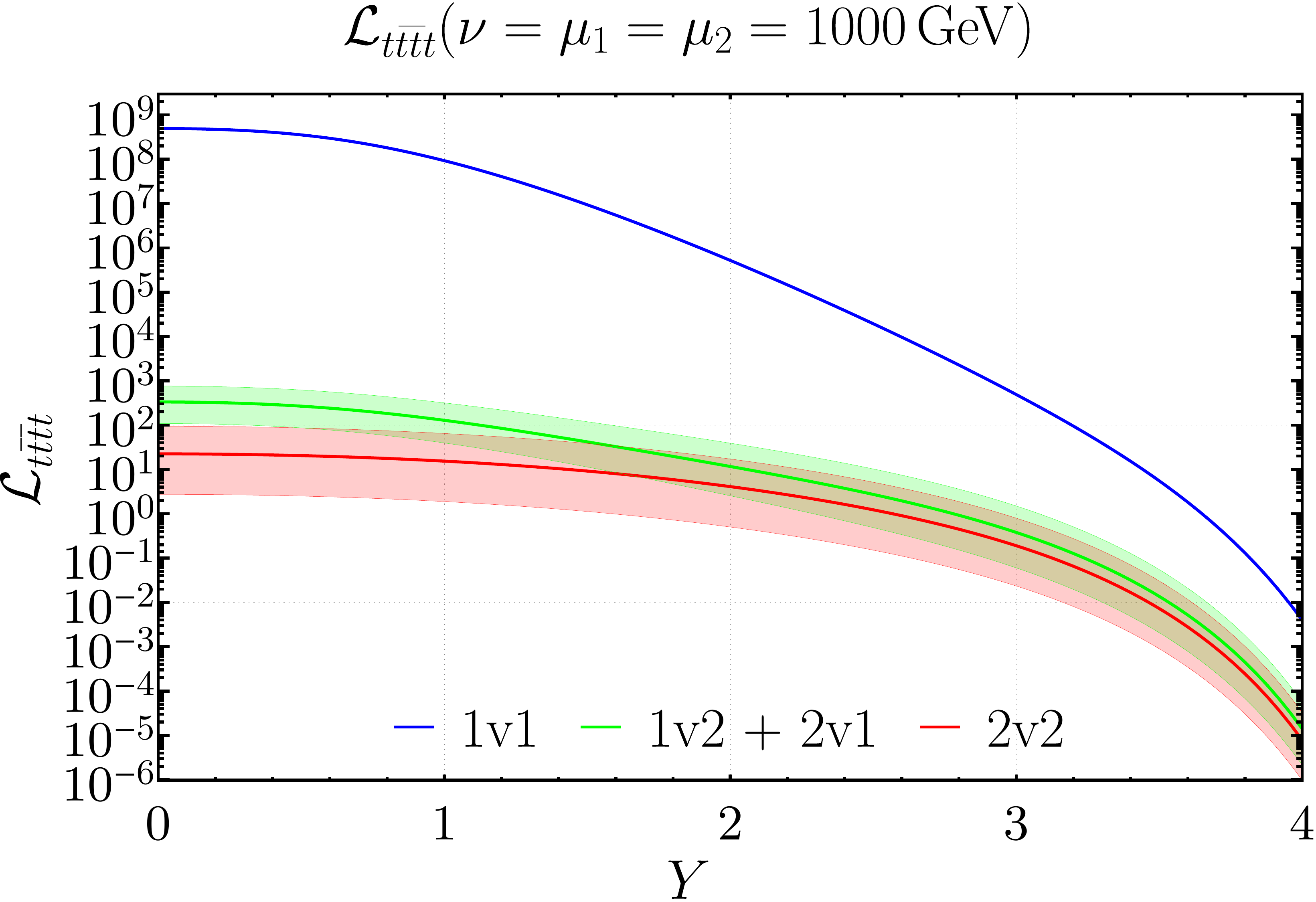}
      }
      \subfigure[\label{subfig:Lttbartbart-ttbar-match-NLO}${t \tbar \ms \tbar t}$ with NLO flavour matching]{
         \includegraphics[width=0.475\linewidth, trim=0 0 0 50, clip]{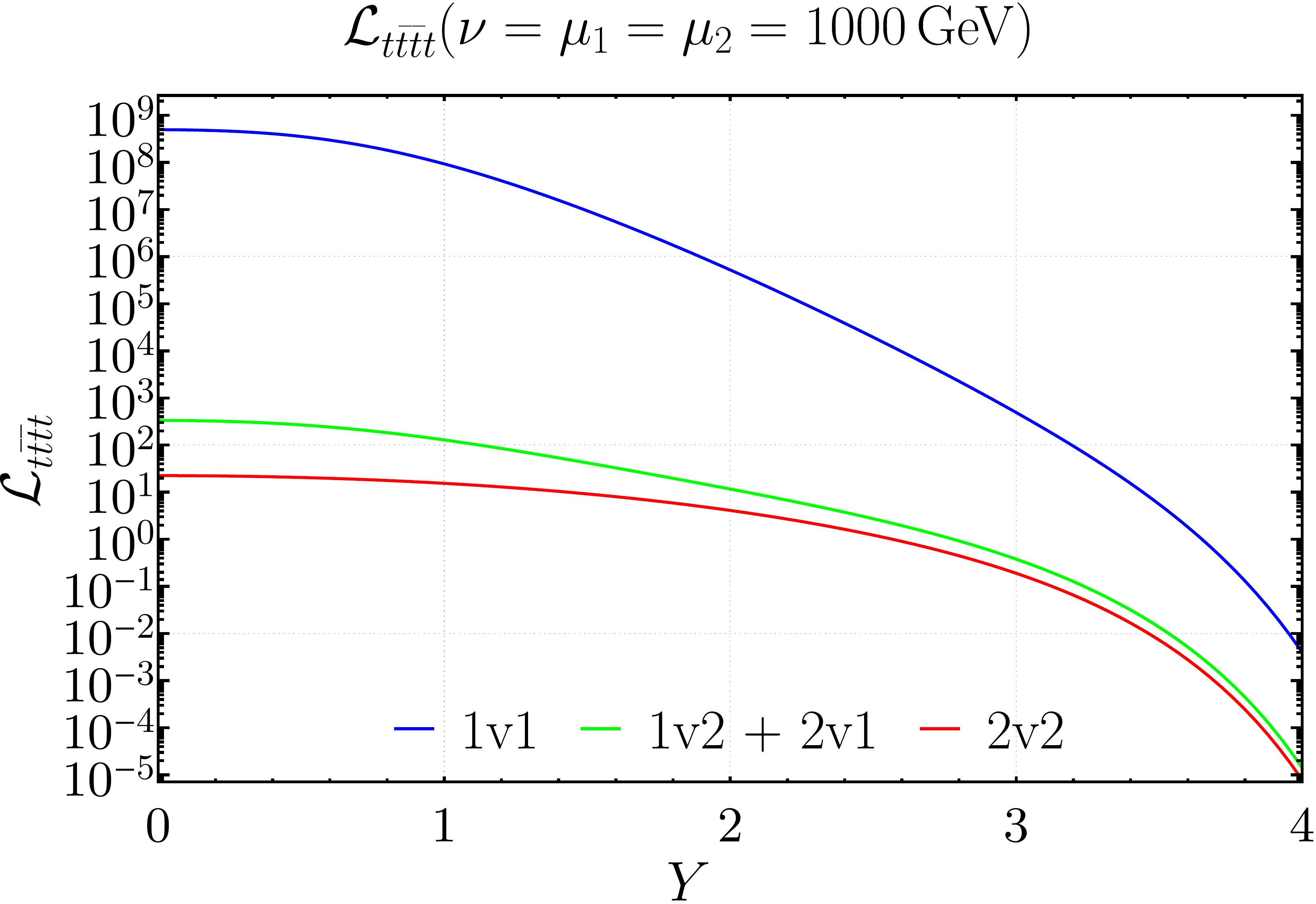}
      }
      \\
      \subfigure[\label{subfig:Lgtgtbar-ttbar-match-LO}${g t g \tbar}$ with LO flavour matching]{
         \includegraphics[width=0.475\linewidth, trim=0 0 0 50, clip]{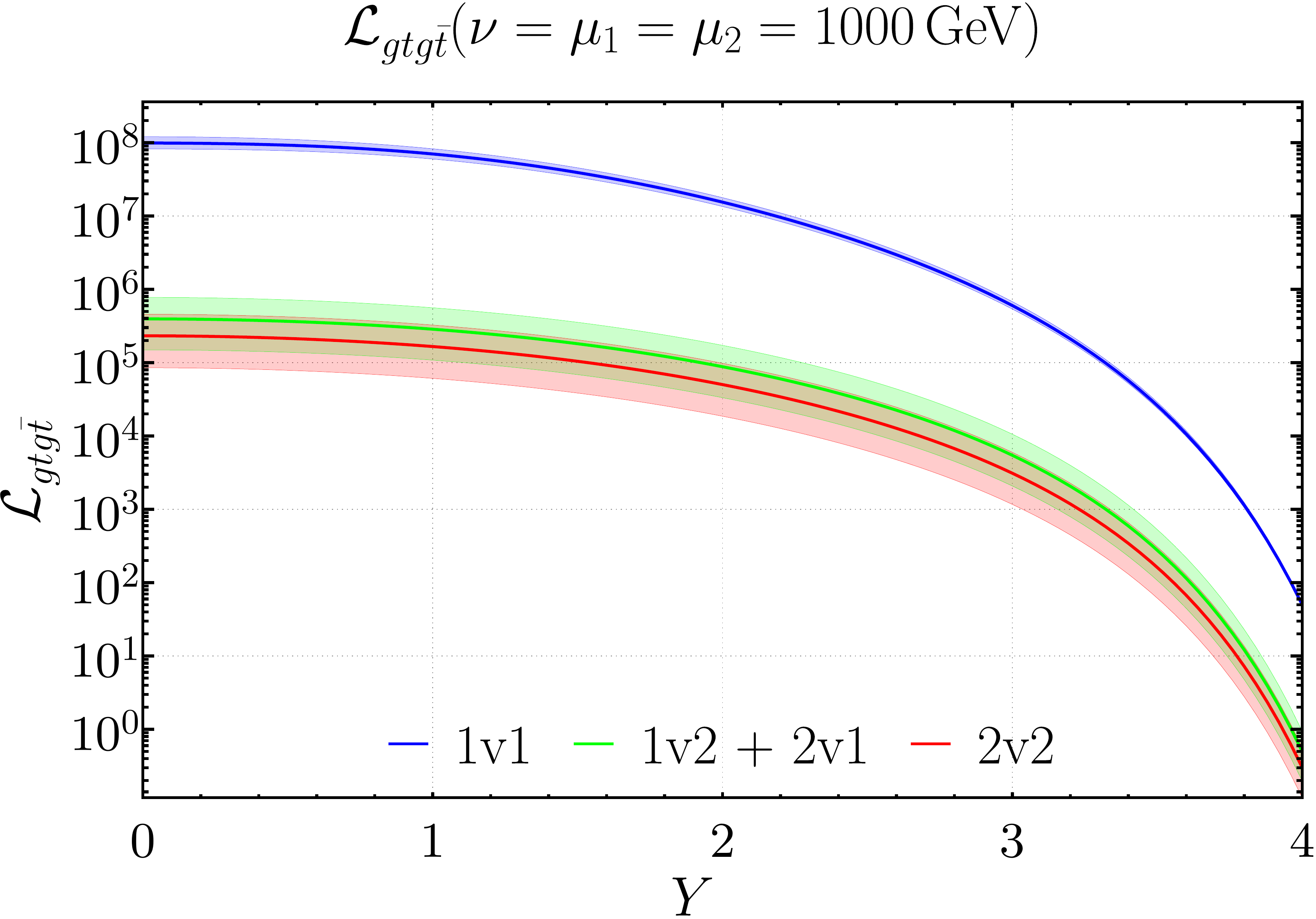}
      }
      \subfigure[\label{subfig:Lgtgtbar-ttbar-match-NLO}${g t g \tbar}$ with NLO flavour matching]{
         \includegraphics[width=0.475\linewidth, trim=0 0 0 50, clip]{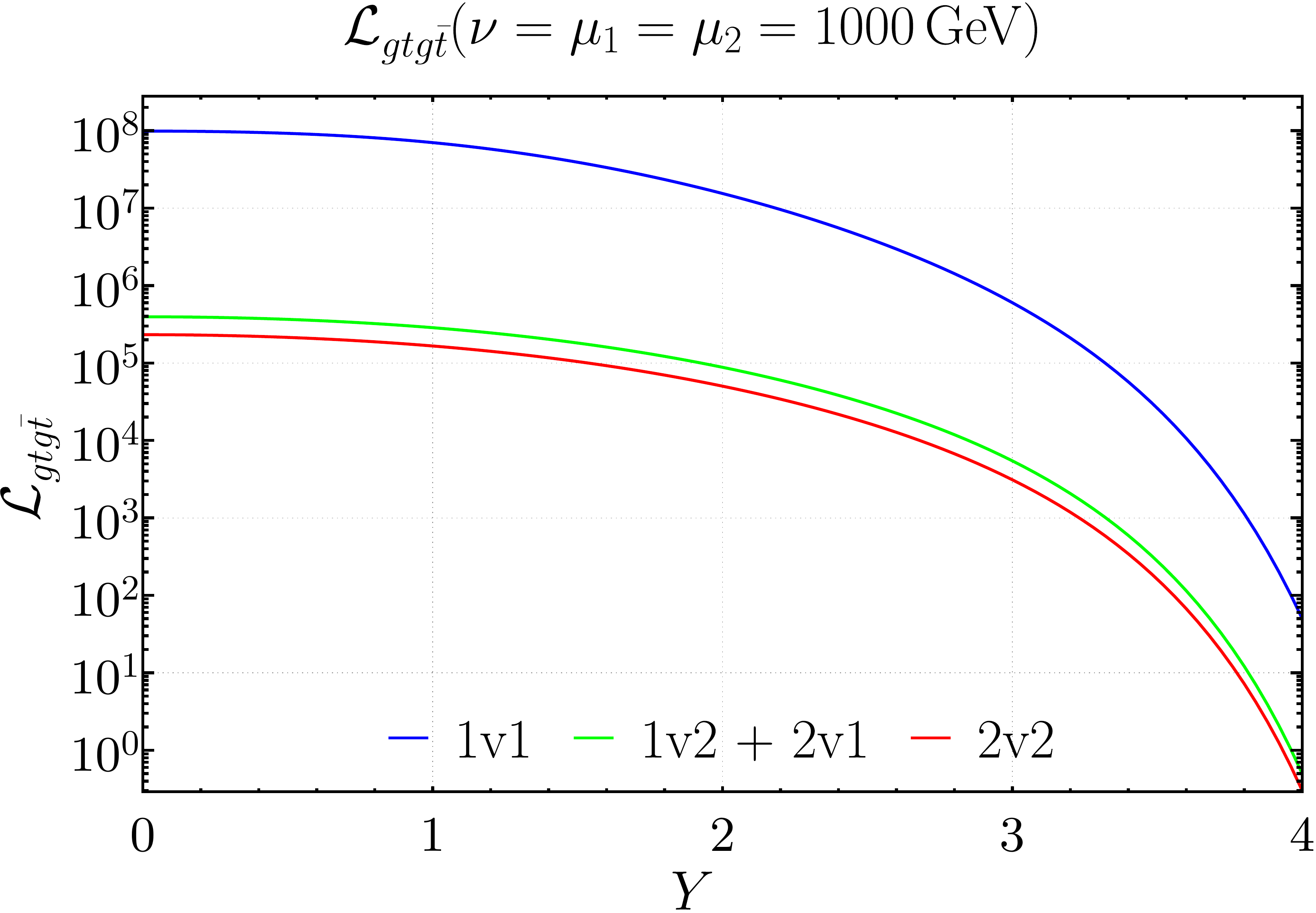}
      }
   \end{center}
   \caption{\label{fig:lumis-ttbar-match} As \fig{\protect\ref{fig:lumis-jets-match}}, but for the luminosities shown in \fig{\protect\ref{fig:lumis-ttbar-contribs}} for the $t \tbar$ production setting.}
\end{figure}

Corresponding plots for the $t\tbar$ production setting can be seen in \fig{\ref{fig:lumis-ttbar-match}} and show a very similar picture.  At LO, the scale variation for the 2v2 terms is huge, in particular for the $t \tbar \ms \tbar t$ channel, whereas only small variations are found in all cases at NLO.

Since the scale variation at NLO is very small, we expect that higher-order graphs associated with flavour matching will provide only small corrections to the results obtained with $\mu_Q = \mQ$ (at LO or NLO).  This is in stark contrast to the higher-order corrections to DPD splitting, where based on the scale variation exercise, we must expect very substantial corrections to the leading-order results.

\FloatBarrier

%
\graphicspath{{Figures/NLO_kernels/}}
%
%
\section{Massive splitting kernels at NLO}
\label{sec:NLO-kernels-constraints}
The large variation of LO splitting DPDs reported in the previous section provides a strong motivation for evaluating the splitting kernels at NLO accuracy, i.e.\ at two loops.  For massless kernels and unpolarised partons, this has been achieved in \cite{Diehl:2019rdh, Diehl:2021wpp}, and the extension of these calculations to polarisation is rather straightforward.  By contrast, the massive splitting kernels are only known at LO (see \eqs{\eqref{eq:VQQbarg-LO}} and \eqref{eq:VQ-light-LO}).  To compute them at NLO is well beyond the scope of the present work.  Instead, we derive their limiting behaviour for small and large $y$, as well as their dependence on $\nf$ and on the renormalisation scale $\mu$.  In a later section, we will also show which constraints follow from the DPD sum rules \cite{Gaunt:2009re, Diehl:2018kgr} for unpolarised partons.

From now on it is understood that PDFs $f^{\nf}$, DPDs $F^{\nf}$, the splitting kernels $V^{\nf}$ and $V^{Q, \nf}$, and the flavour matching kernels $A^{Q, \nf}$ are to be evaluated at the following arguments:
\begin{align}
   \label{eq:arguments-fnF}
      f^{\nf}_{a_0}
   &  \equiv f^{\nf}_{a_0}(x; \mu) \,,
   \\
   \label{eq:arguments-FnF}
      F^{\nf}_{a_1 a_2}
   &  \equiv F^{\nf}_{a_1 a_2}(x_1, x_2, y; \mu) \,,
   \\
   \label{eq:arguments-VQ}
      V^{Q, \nf}_{a_1 a_2, a_0}
   &  \equiv V^{Q, \nf}_{a_1 a_2, a_0}(z_1, z_2, y, \mQ; \mu) \,,
   \\
   \label{eq:arguments-VnF}
      V^{\nf}_{a_1 a_2, a_0}
   &  \equiv V^{\nf}_{a_1 a_2, a_0}(z_1, z_2, y; \mu) \,,
   \\
   \label{eq:arguments-A}
      A^{Q, \nf}_{a_1 a_0}
   &  \equiv A^{Q, \nf}_{a_1 a_0}(z, \mQ; \mu) \,,
\end{align}
unless indicated otherwise.  Notice our convention to use momentum fraction arguments $x$, $x_1$, $x_2$ in distributions and $z$, $z_1$, $z_2$ in kernels.
%
%
\subsection{Two-loop splitting graphs with massive quarks}
\label{sec:two-loop-graphs}
\begin{figure}[p]
   \begin{center}
      \subfigure[\label{subfig:gqqbar}]{
         \includegraphics[width=0.12\linewidth]{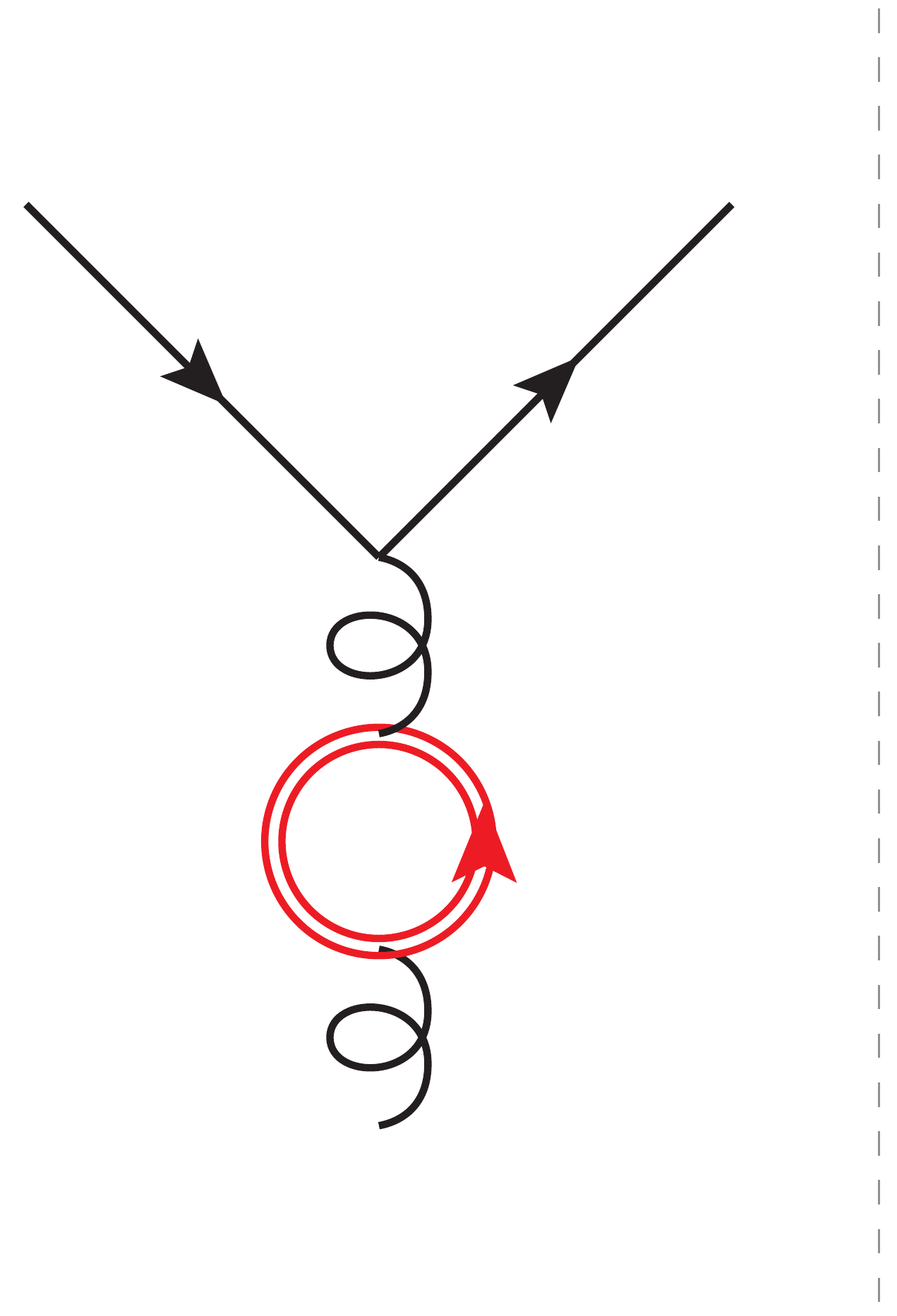}
      }
      \hspace{0.06\linewidth}
      \subfigure[\label{subfig:qqg}]{
         \includegraphics[width=0.12\linewidth]{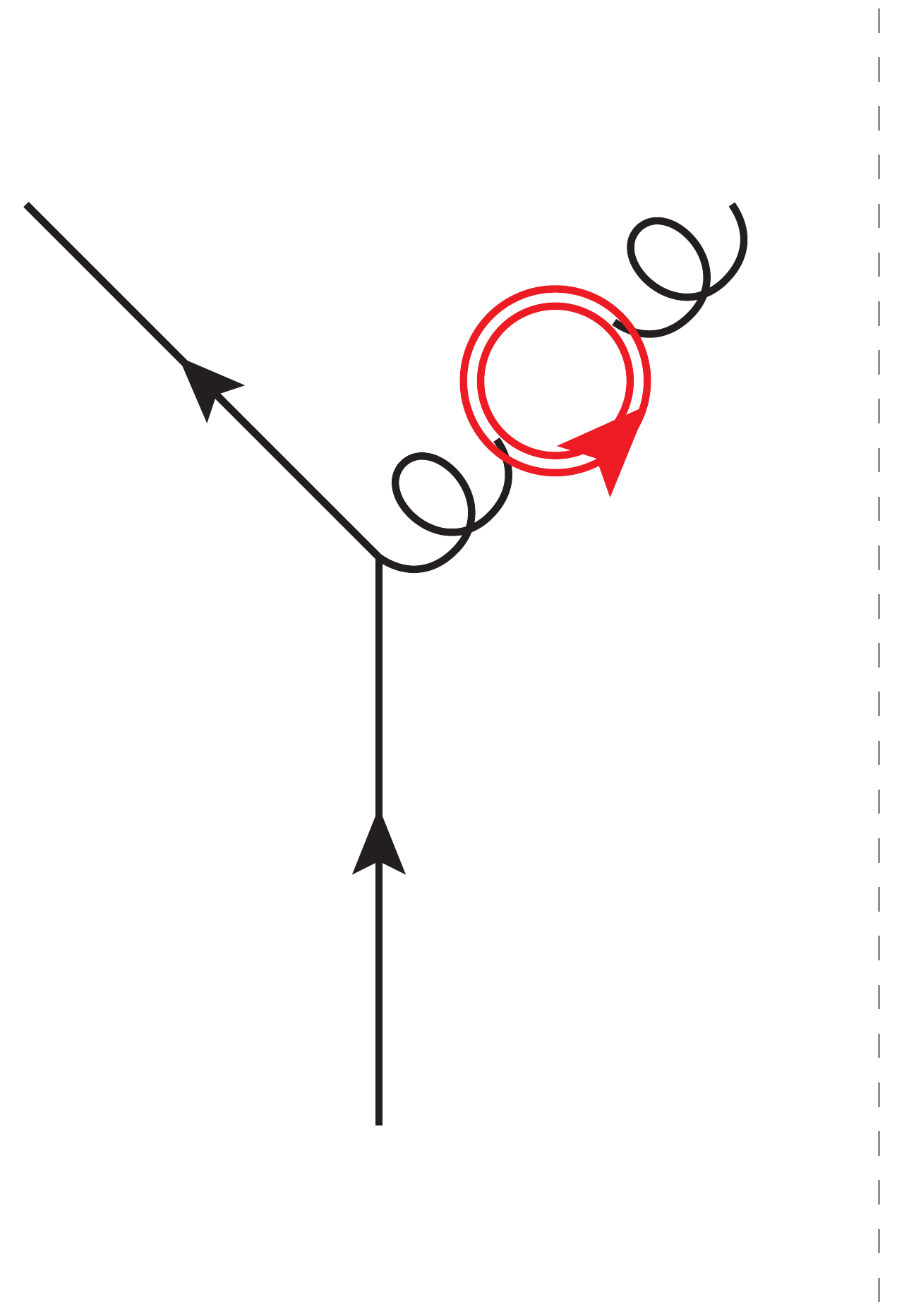}
      }
      \hspace{0.06\linewidth}
      \subfigure[\label{subfig:ggg-1}]{
         \includegraphics[width=0.12\linewidth]{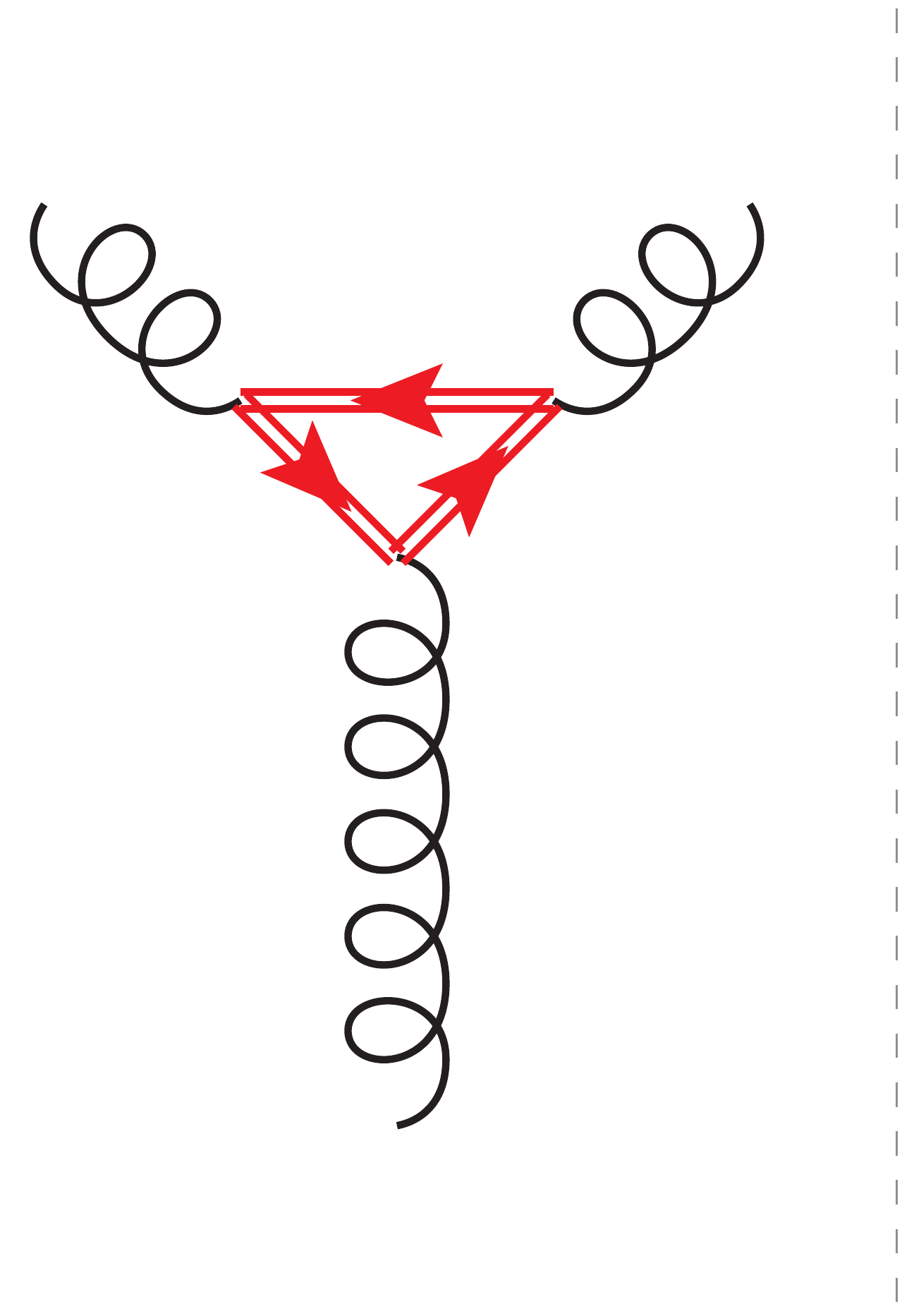}
      }
      \hspace{0.06\linewidth}
      \subfigure[\label{subfig:ggg-2}]{
         \includegraphics[width=0.12\linewidth]{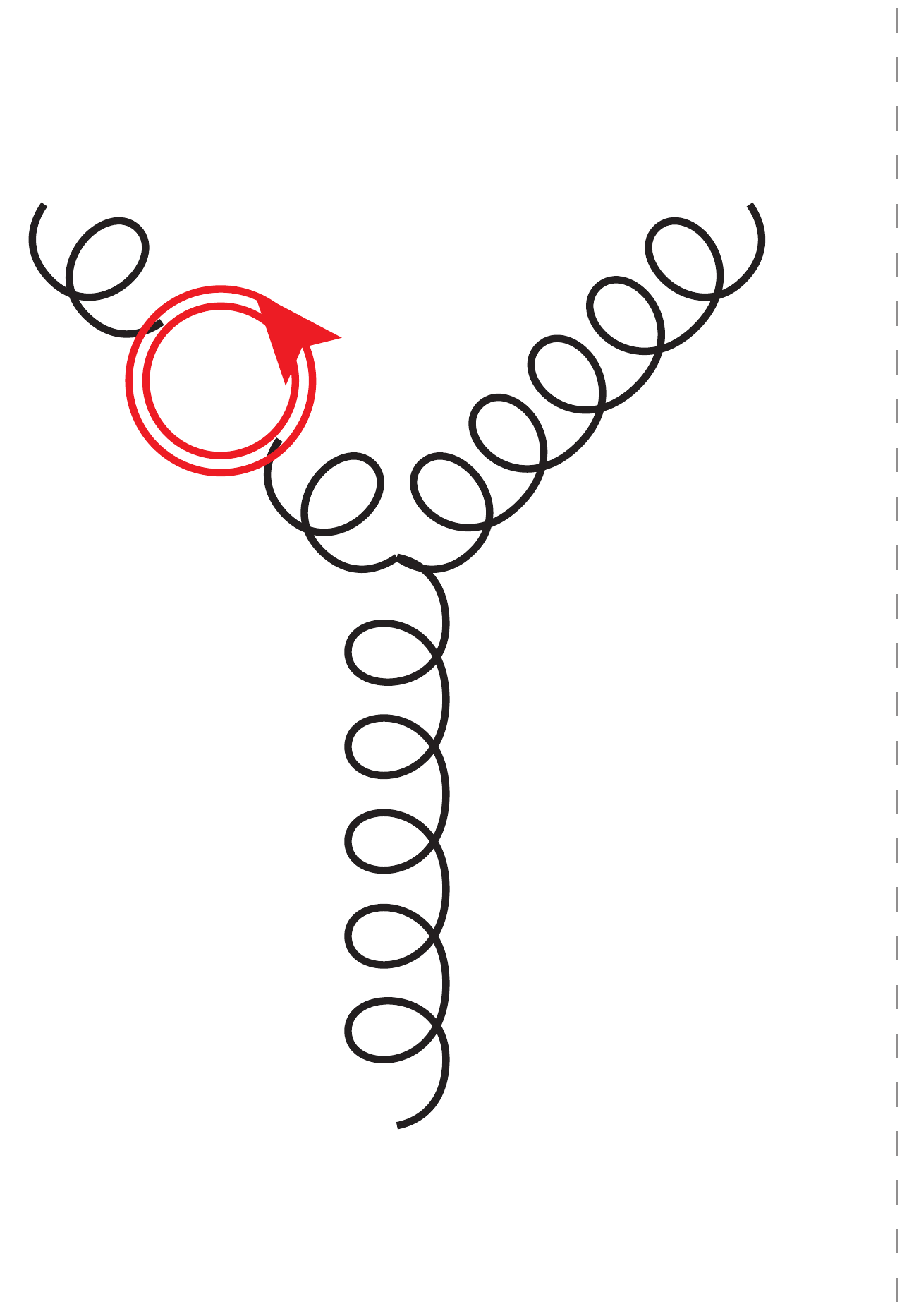}
      }
      \hspace{0.06\linewidth}
      \subfigure[\label{subfig:ggg-3}]{
         \includegraphics[width=0.12\linewidth]{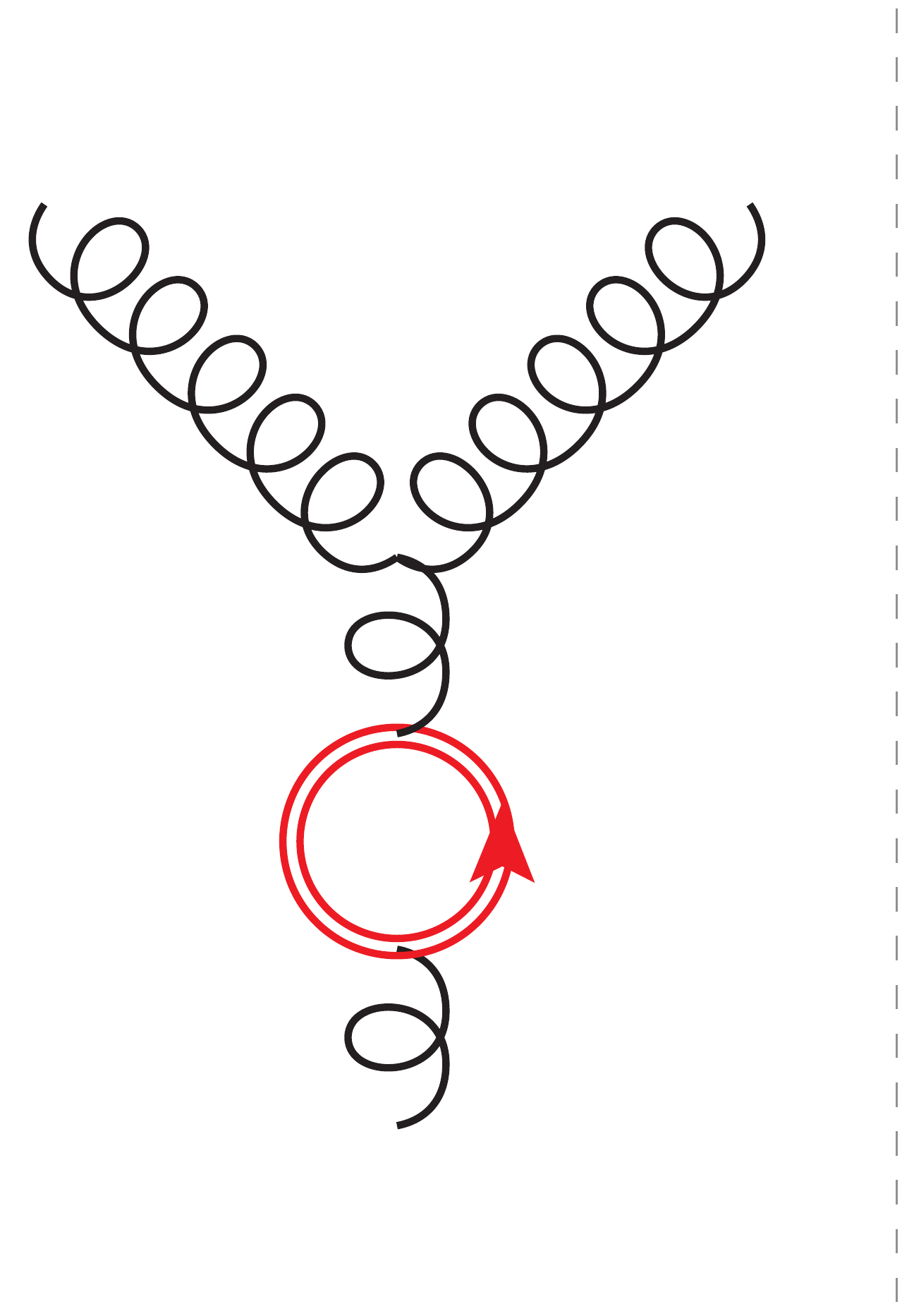}
      }
      \vspace{0.5cm}\\
      \subfigure[\label{subfig:gQQbar-5}]{
         \includegraphics[width=0.12\linewidth]{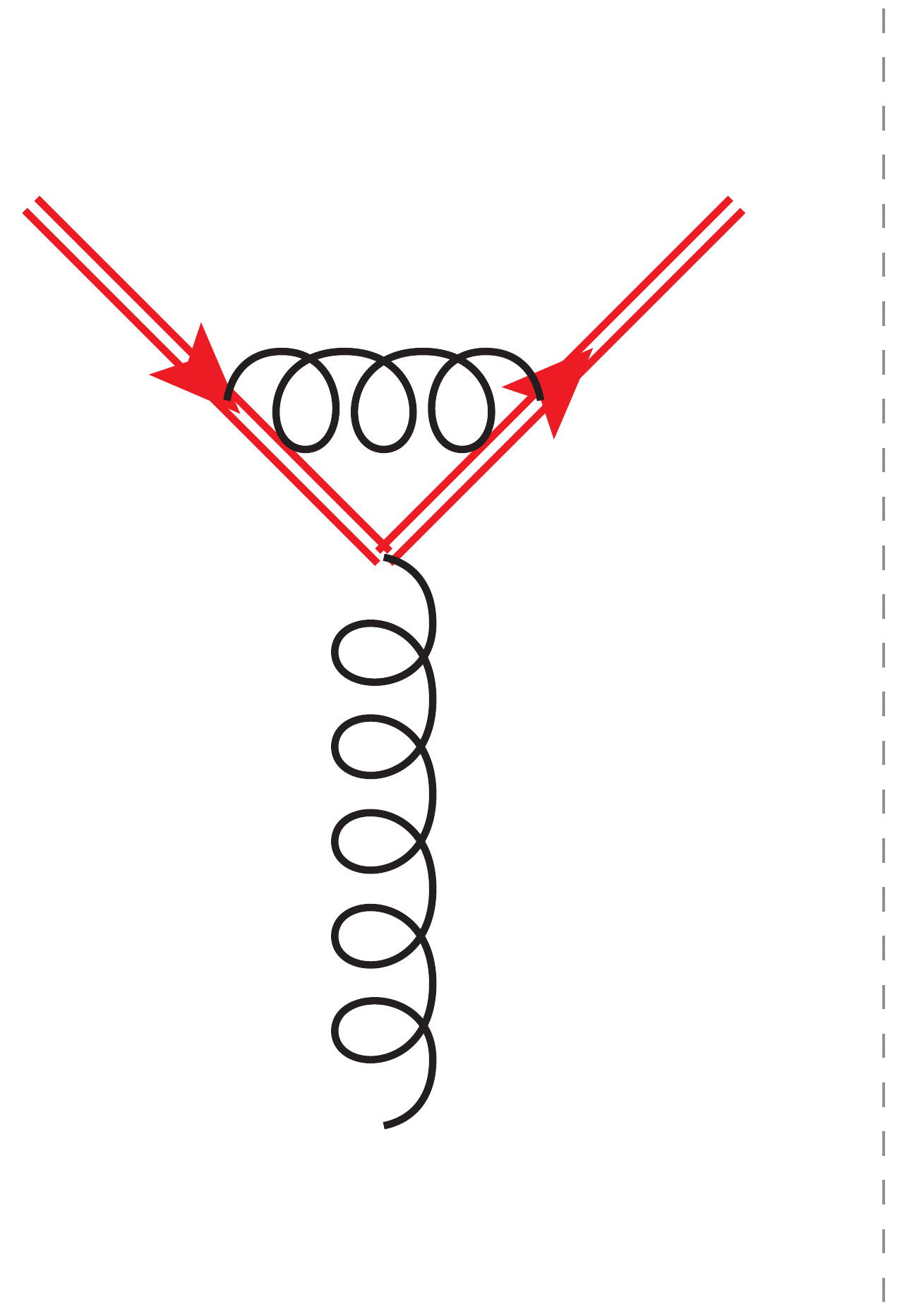}
      }
      \hspace{0.06\linewidth}
      \subfigure[\label{subfig:gQQbar-6}]{
         \includegraphics[width=0.12\linewidth]{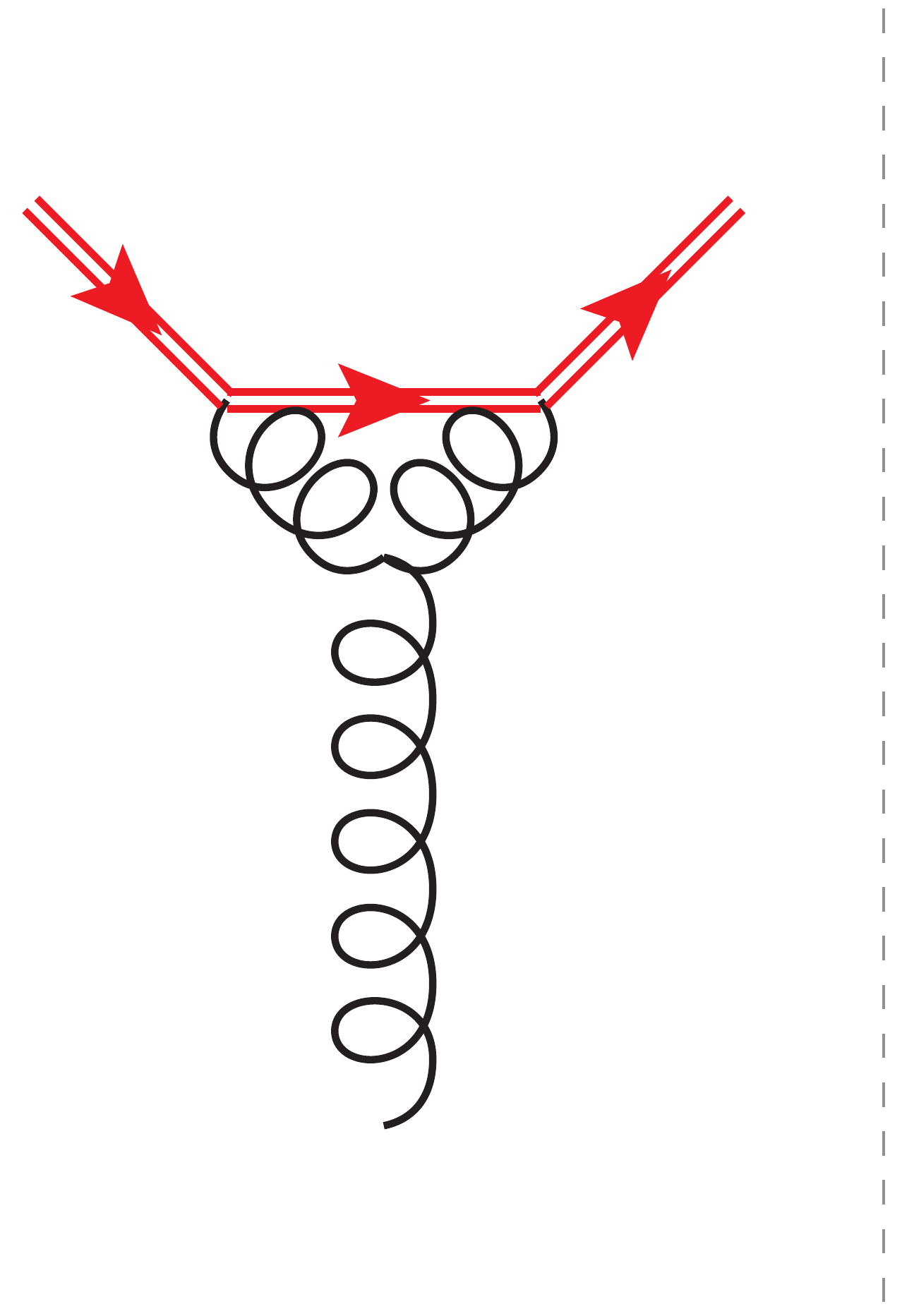}
      }
      \hspace{0.06\linewidth}
      \subfigure[\label{subfig:gQQbar-7}]{
         \includegraphics[width=0.12\linewidth]{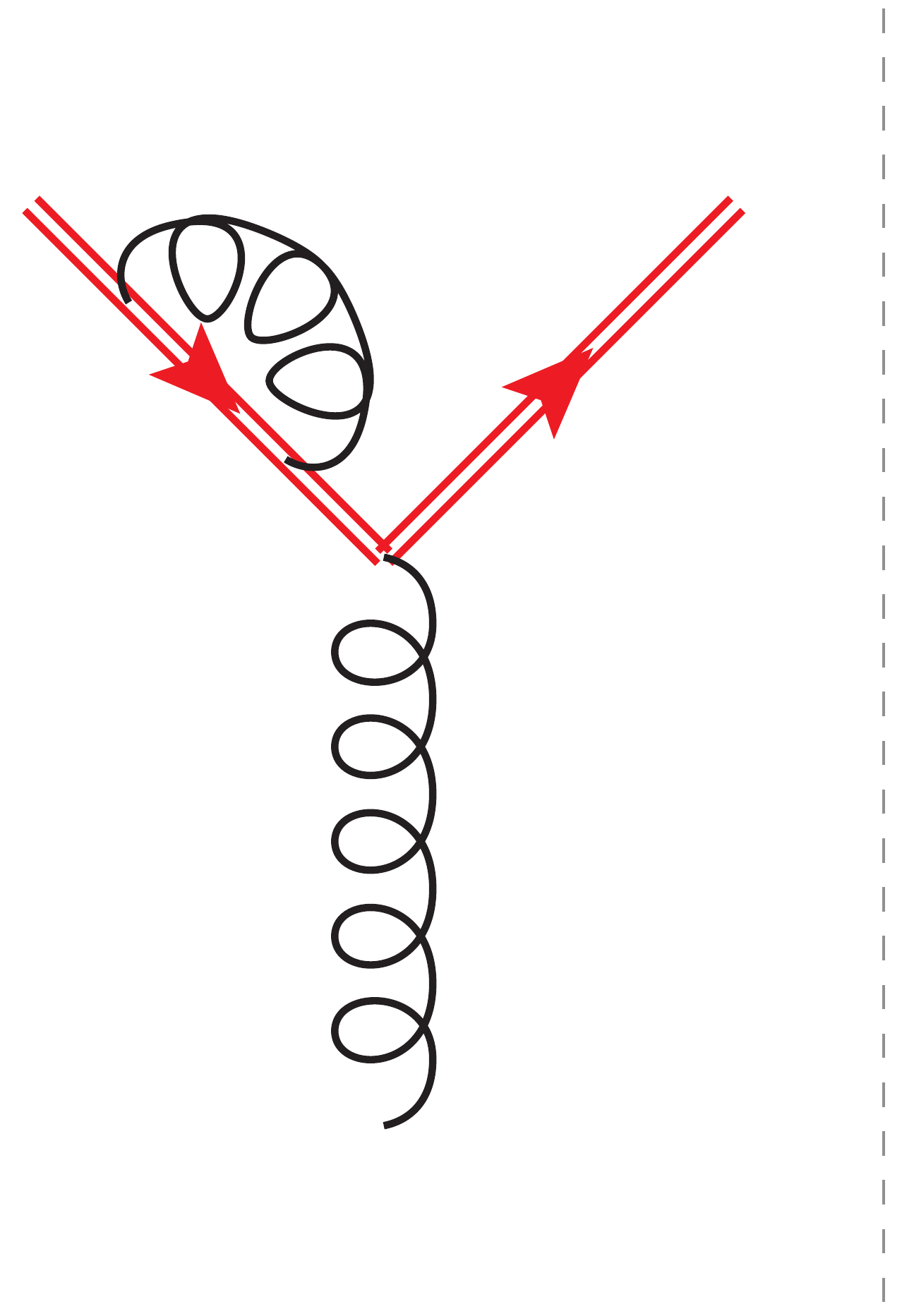}
      }
      \hspace{0.06\linewidth}
      \subfigure[\label{subfig:gQQbar-8}]{
         \includegraphics[width=0.12\linewidth]{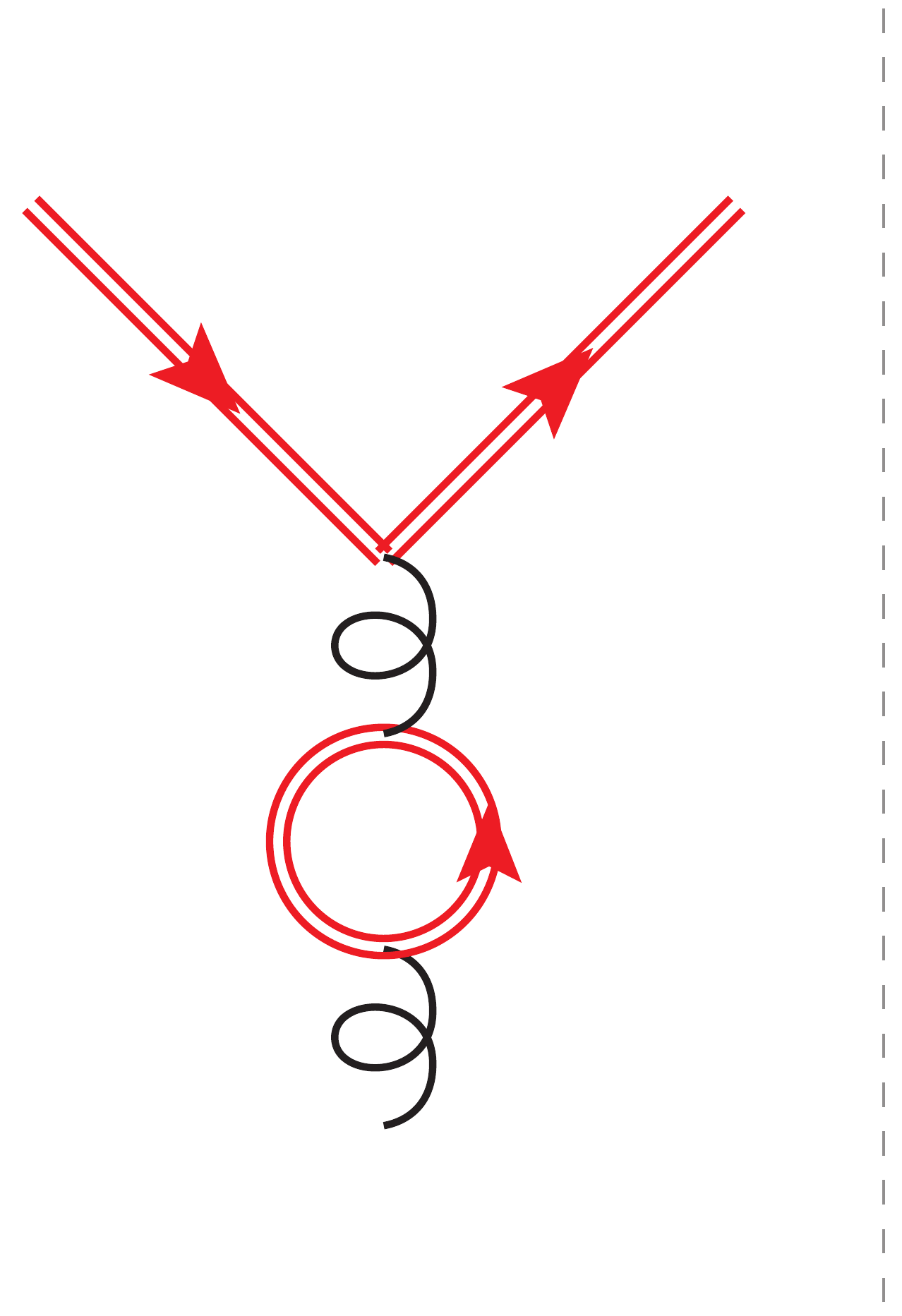}
      }
      \caption[Virtual graphs for LO channels]
      {\label{fig:virt-LO} Virtual NLO diagrams for the amplitude that contain heavy-quark lines, indicated by red double lines. The tree graphs in the complex conjugate amplitude are not shown.  In Feynman gauge, graphs with eikonal lines are to be added, as specified in \fig{4} of \protect\cite{Diehl:2019rdh}.}
   \end{center}
   \begin{center}
      \subfigure[\label{subfig:gQQbar-1}]{
         \includegraphics[width=0.22\linewidth]{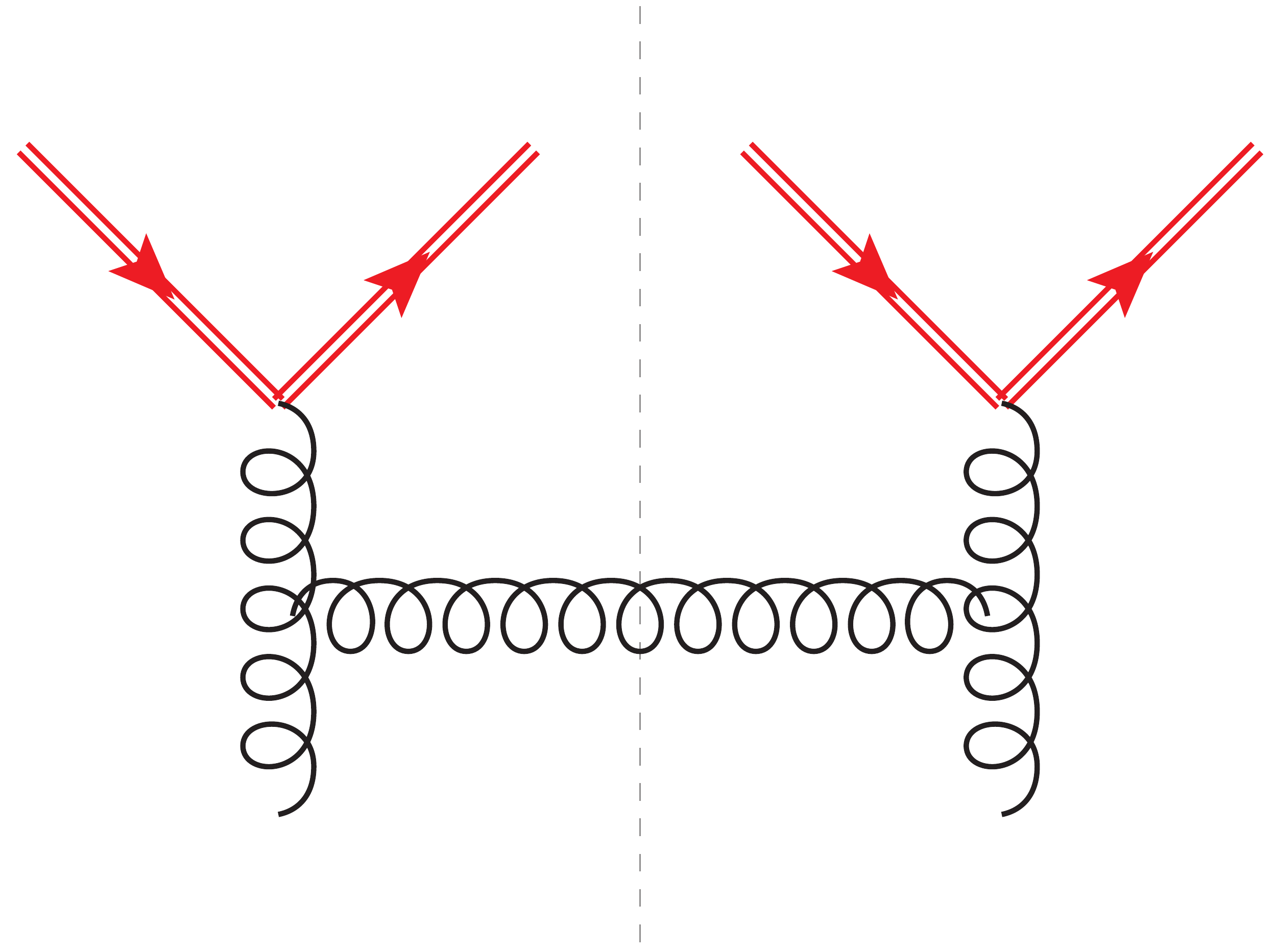}
      }
      \subfigure[\label{subfig:gQQbar-2}]{
         \includegraphics[width=0.22\linewidth]{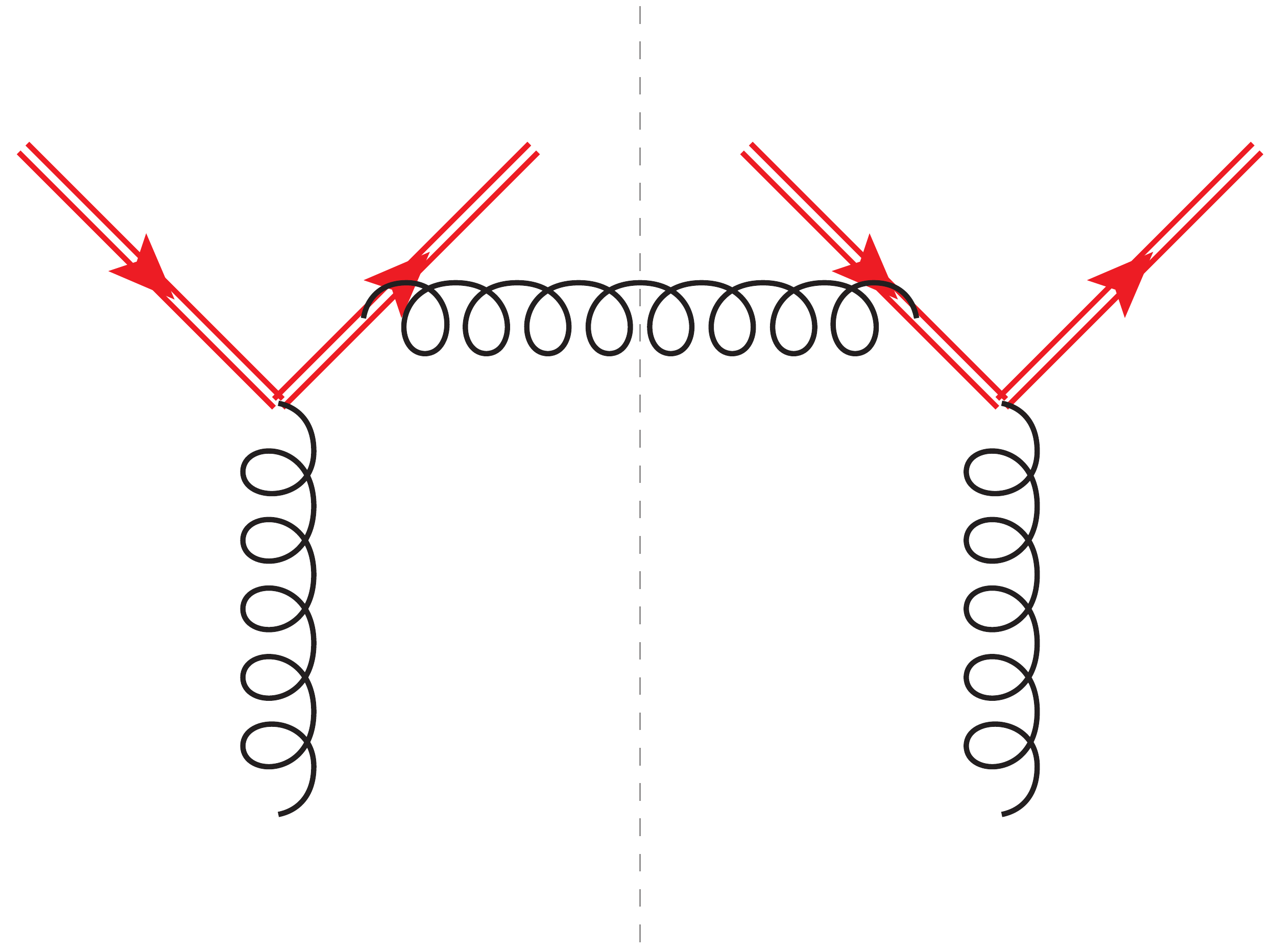}
      }
      \subfigure[\label{subfig:gQQbar-3}]{
         \includegraphics[width=0.22\linewidth]{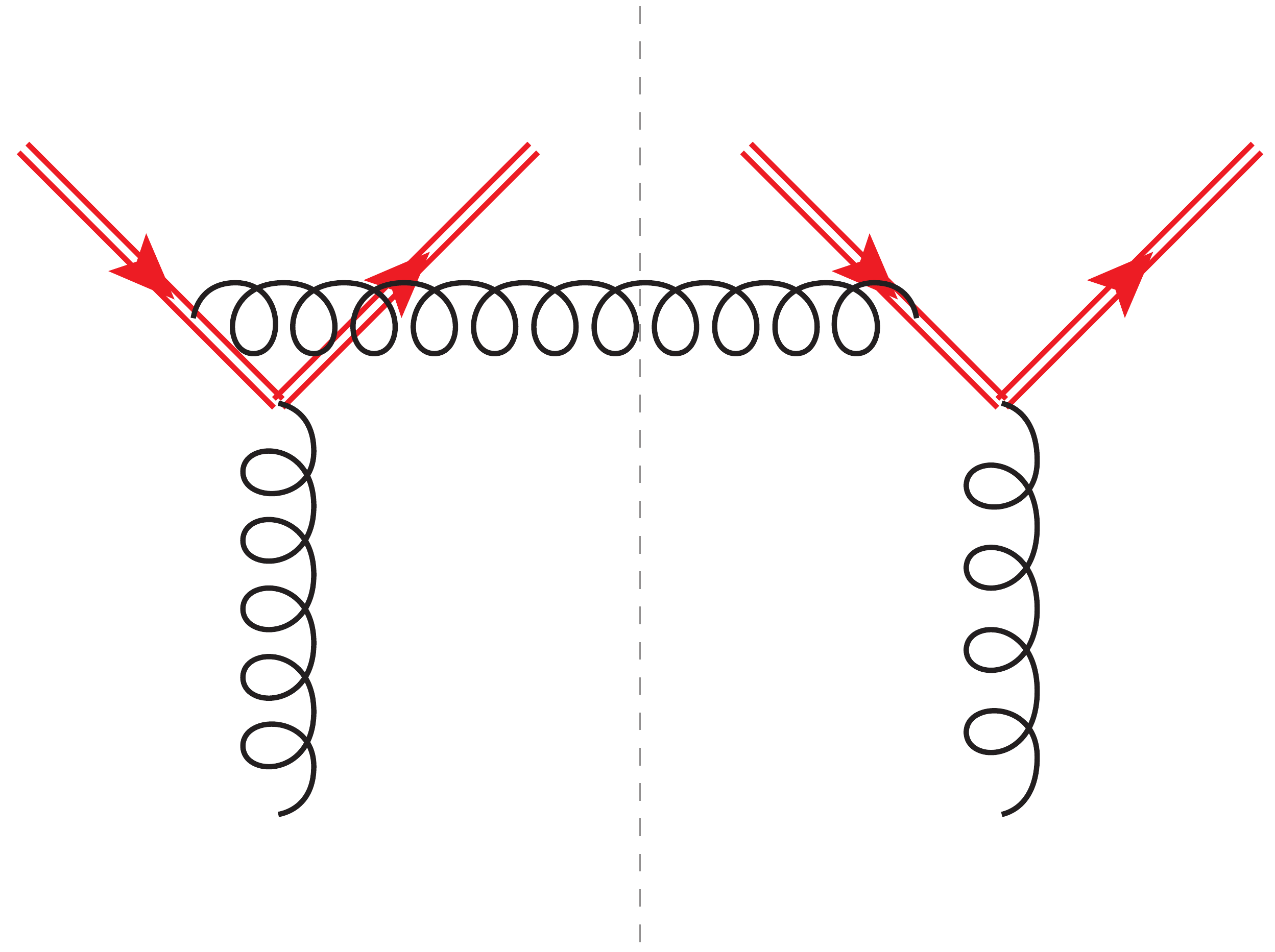}
      }
      \subfigure[\label{subfig:gQQbar-4}]{
         \includegraphics[width=0.22\linewidth]{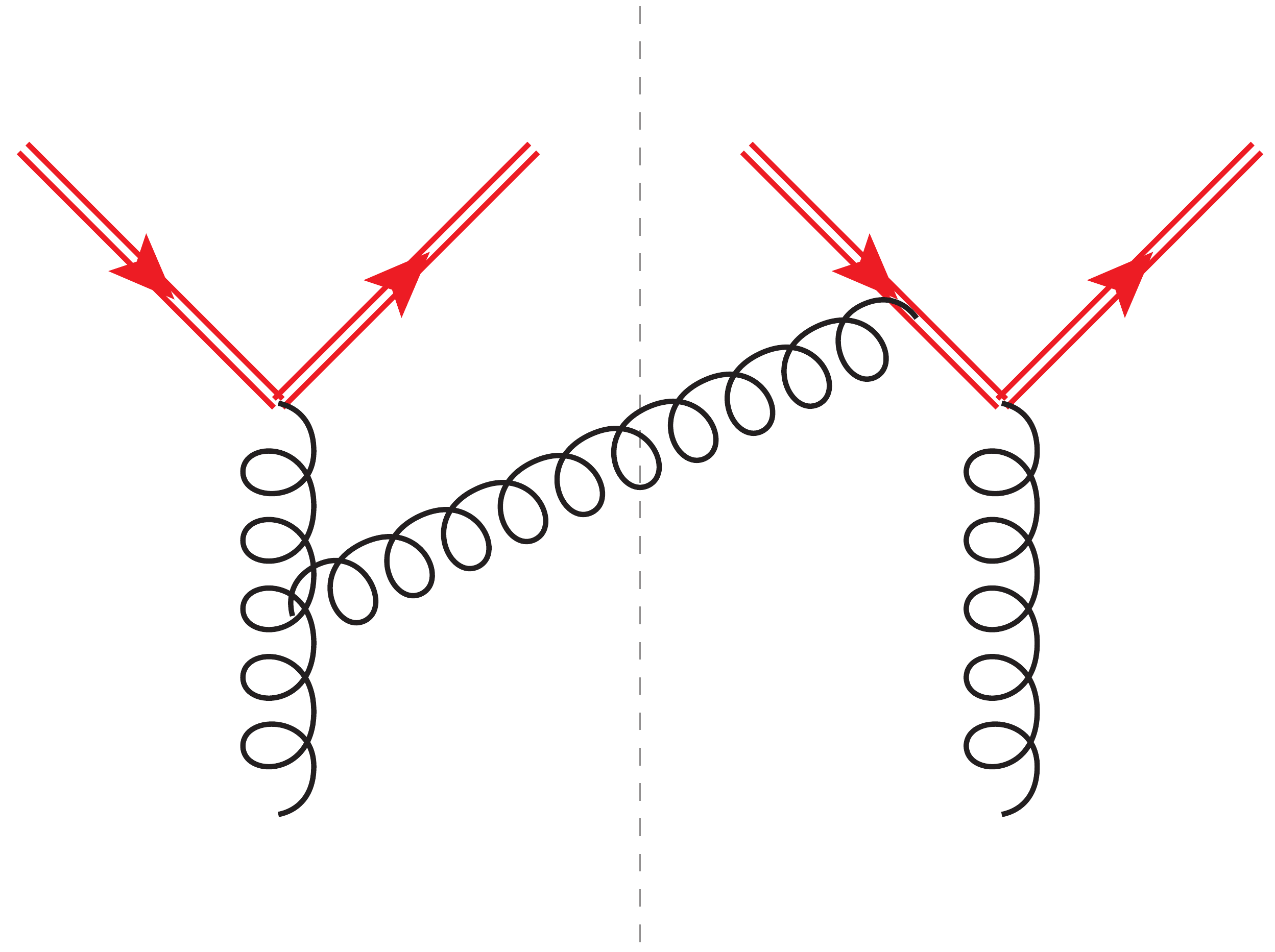}
      }
      \vspace{1cm}\\
      \subfigure[\label{subfig:gQg-1}]{
         \includegraphics[width=0.22\linewidth]{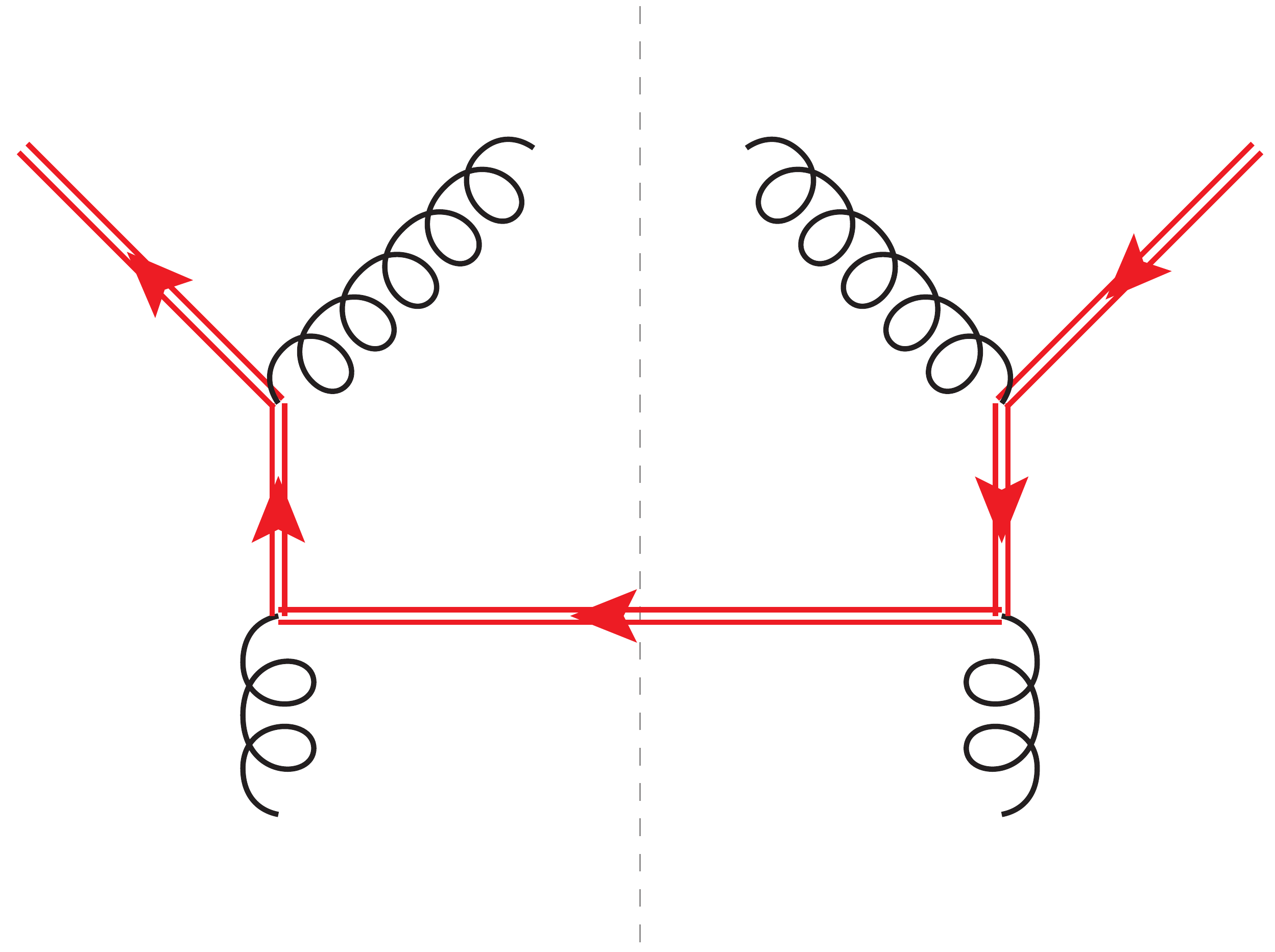}
      }
      \subfigure[\label{subfig:gQg-2}]{
         \includegraphics[width=0.22\linewidth]{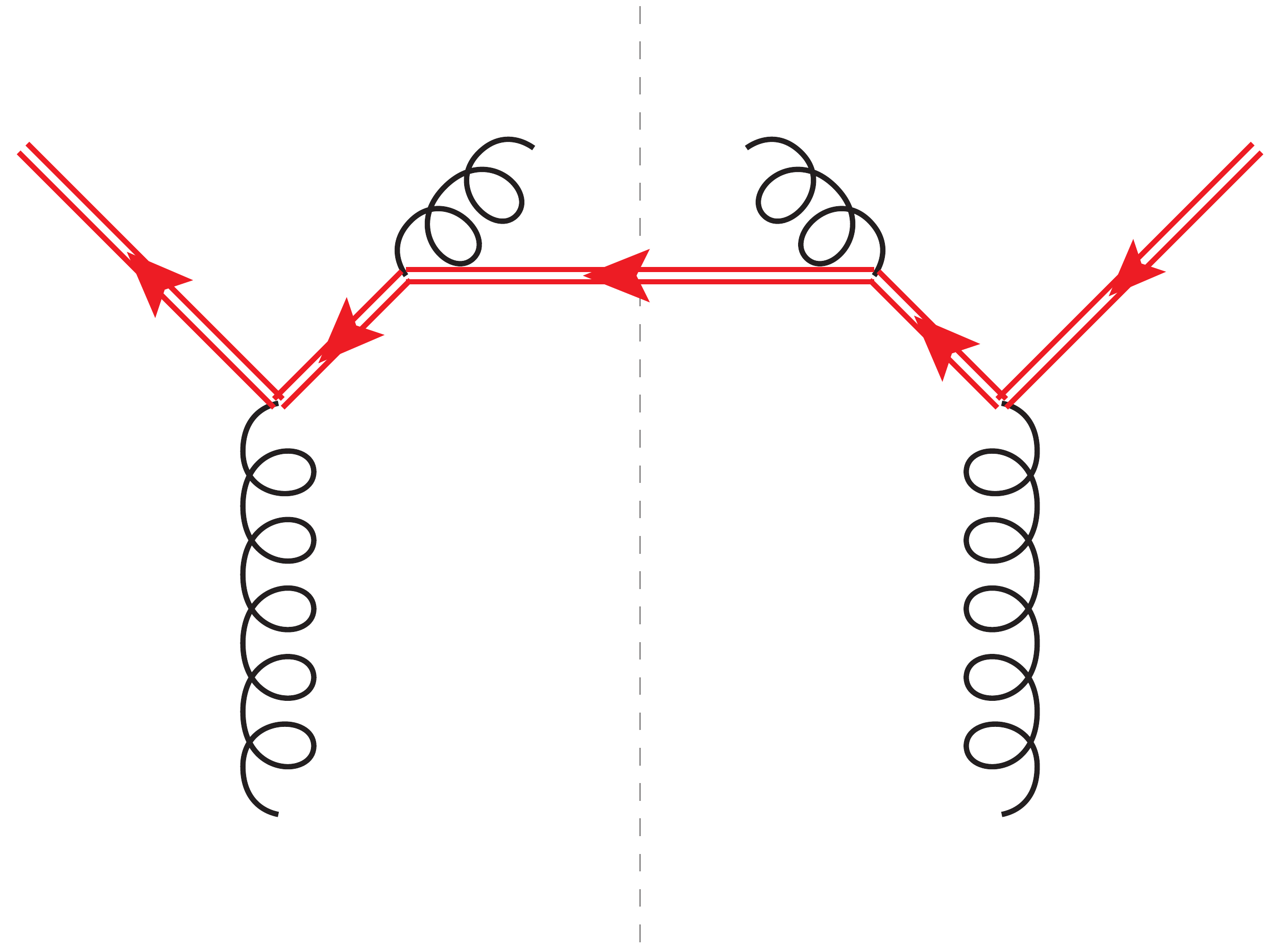}
      }
      \subfigure[\label{subfig:gQg-3}]{
         \includegraphics[width=0.22\linewidth]{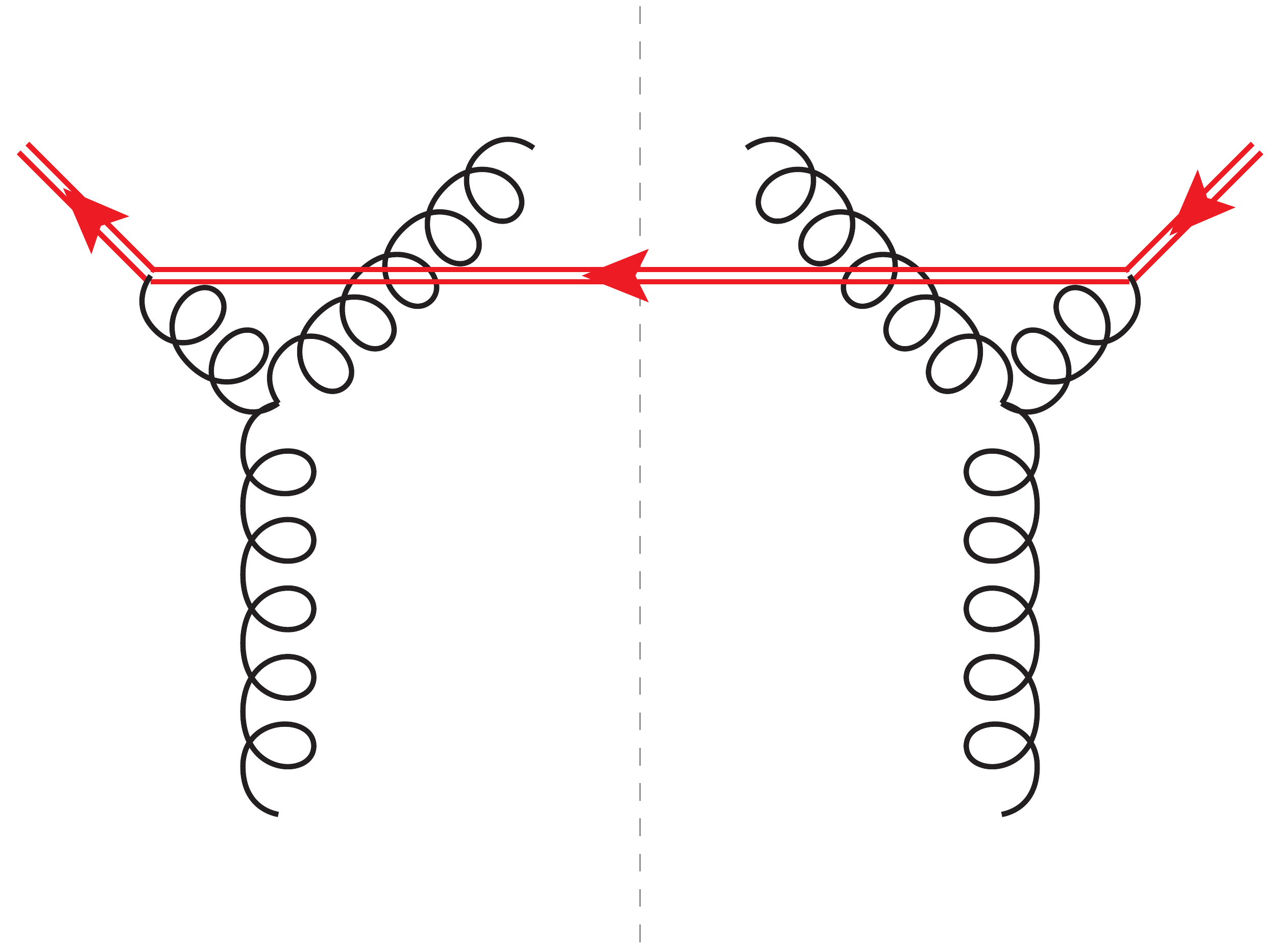}
      }
      \subfigure[\label{subfig:gQg-4}]{
         \includegraphics[width=0.22\linewidth]{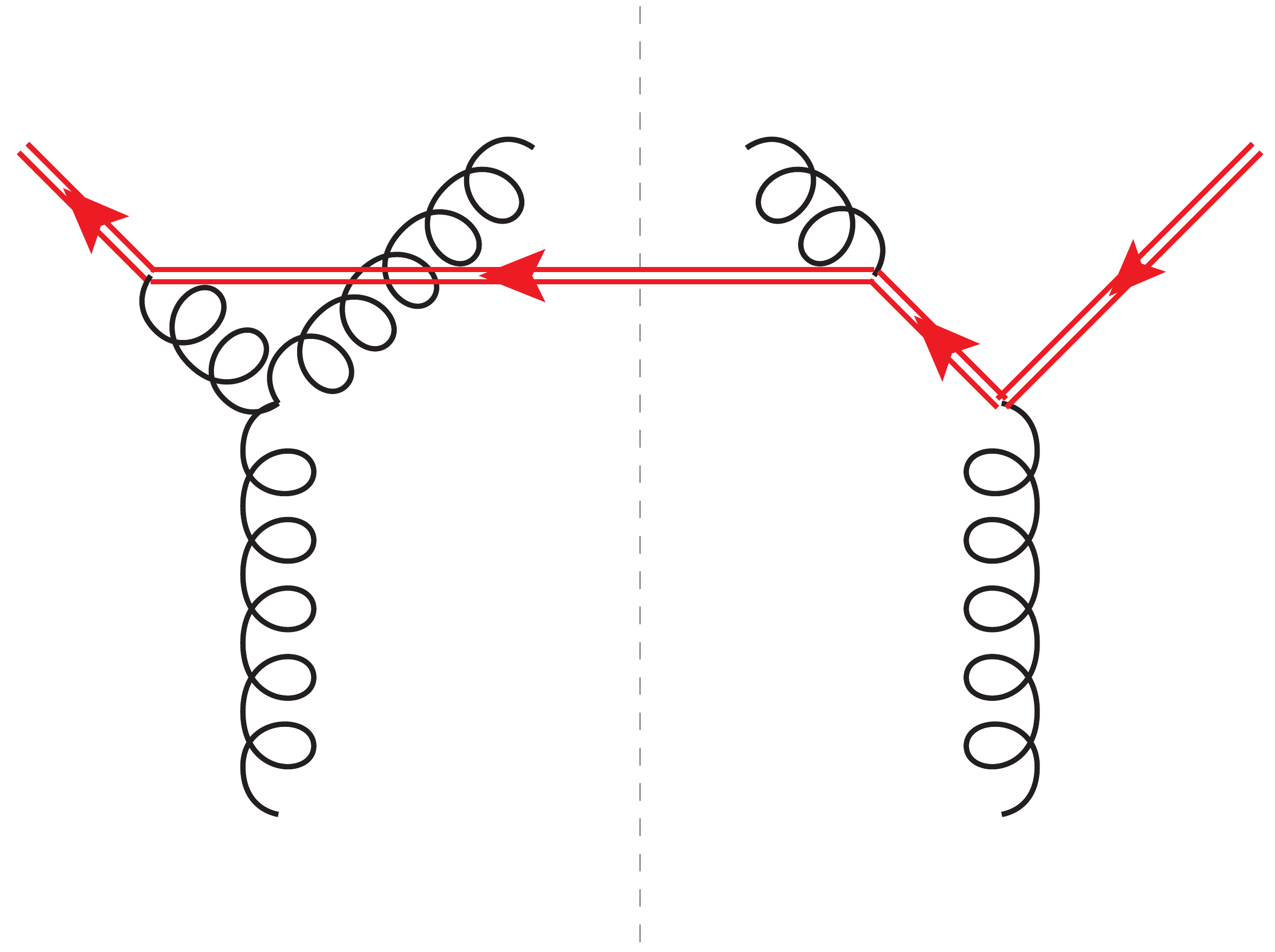}
      }
      \vspace{1cm}\\
      \subfigure[\label{subfig:gQg-5}]{
         \includegraphics[width=0.22\linewidth]{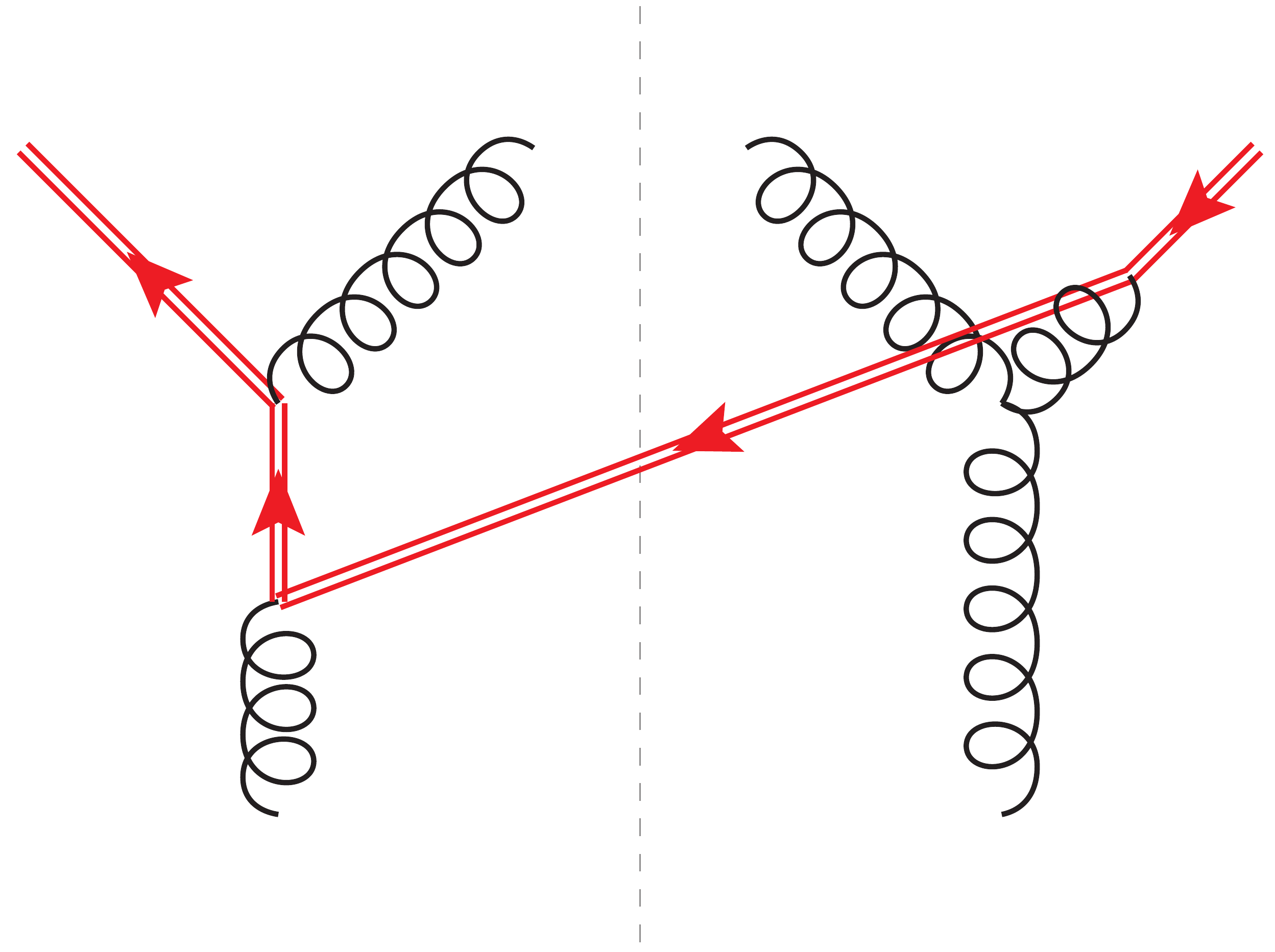}
      }
      \subfigure[\label{subfig:gQg-6}]{
         \includegraphics[width=0.22\linewidth]{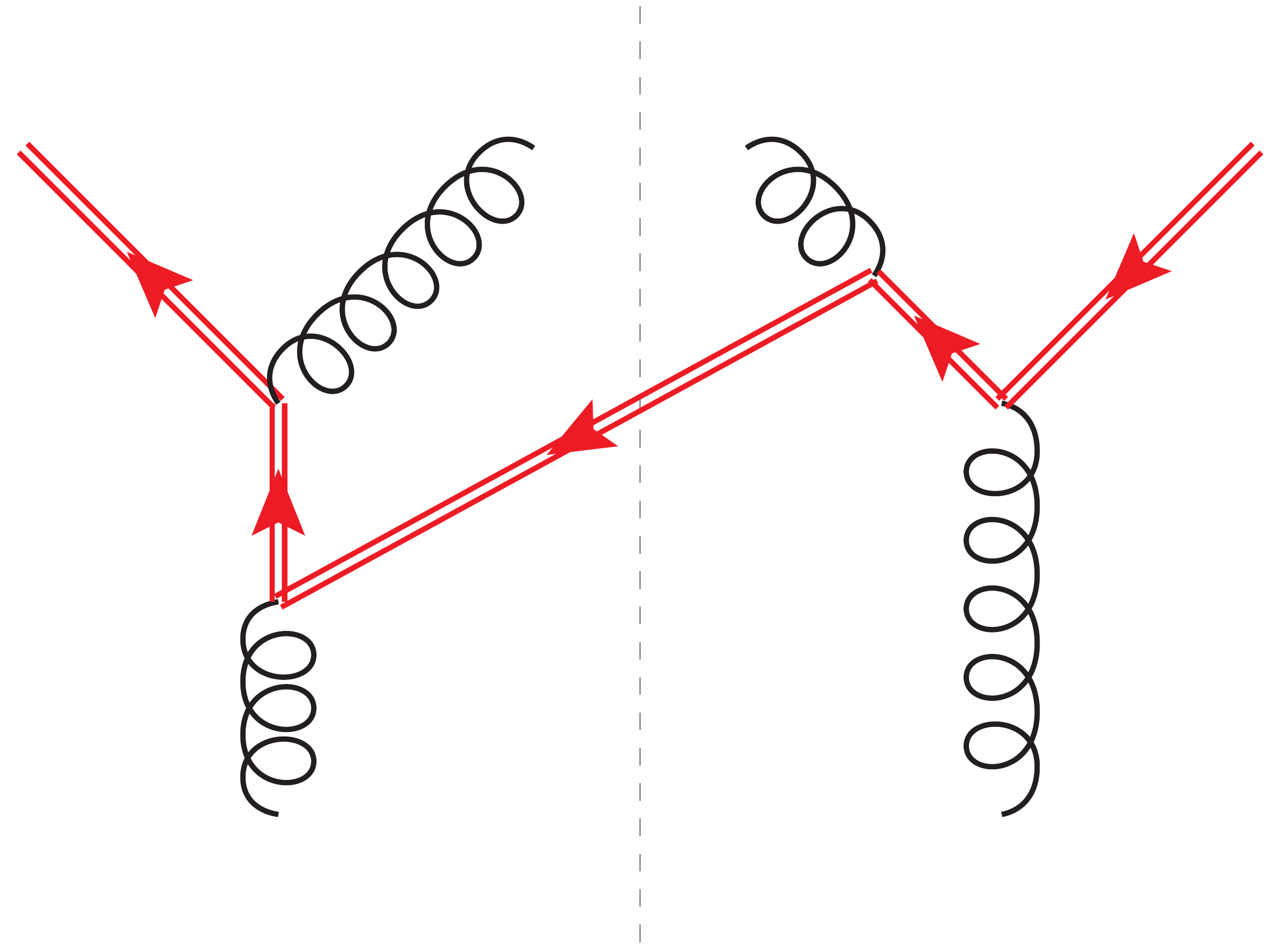}
      }
      \subfigure[\label{subfig:qQQbar}]{
         \includegraphics[width=0.22\linewidth]{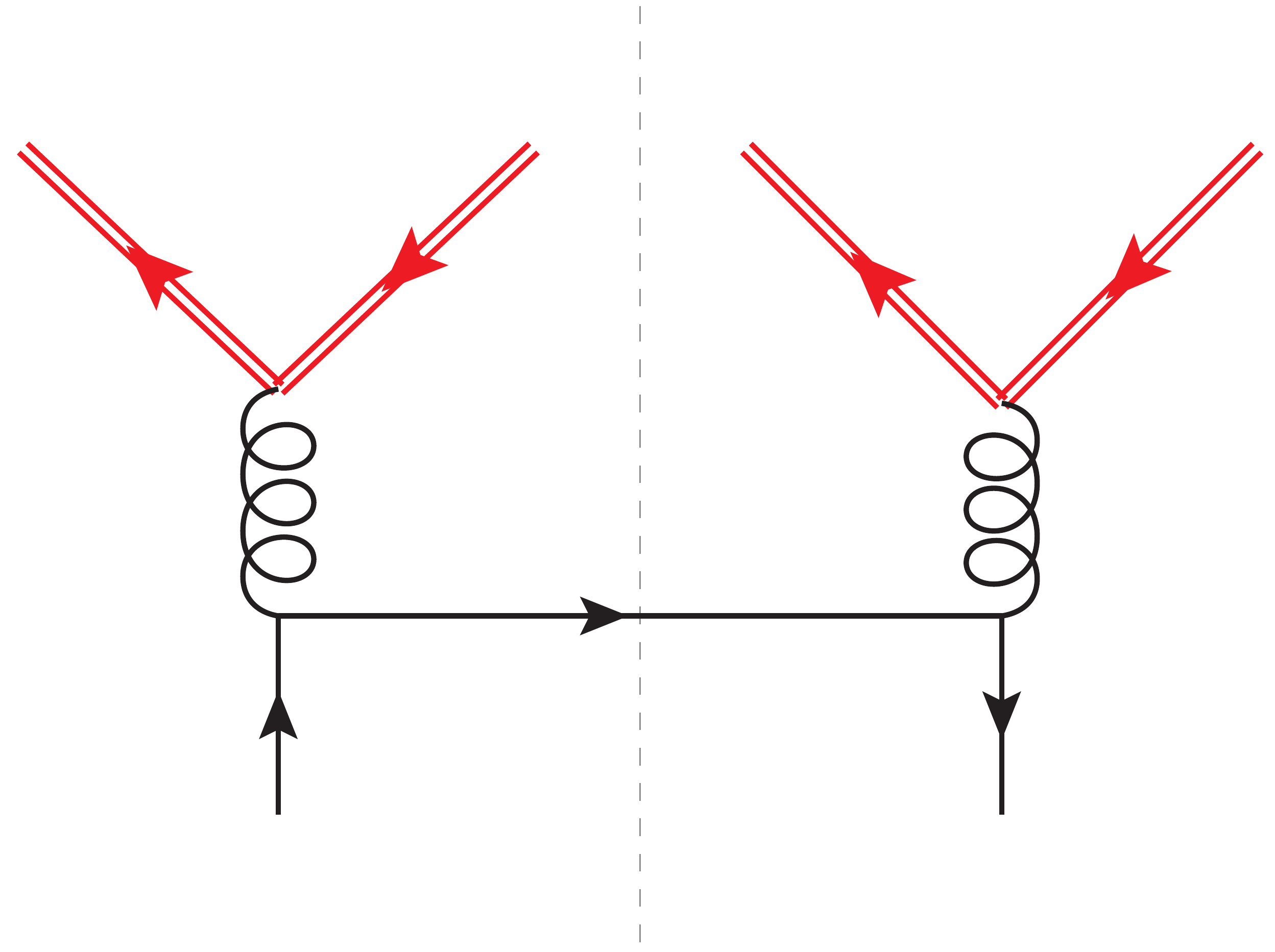}
      }
      \subfigure[\label{subfig:qqQ}]{
         \includegraphics[width=0.22\linewidth]{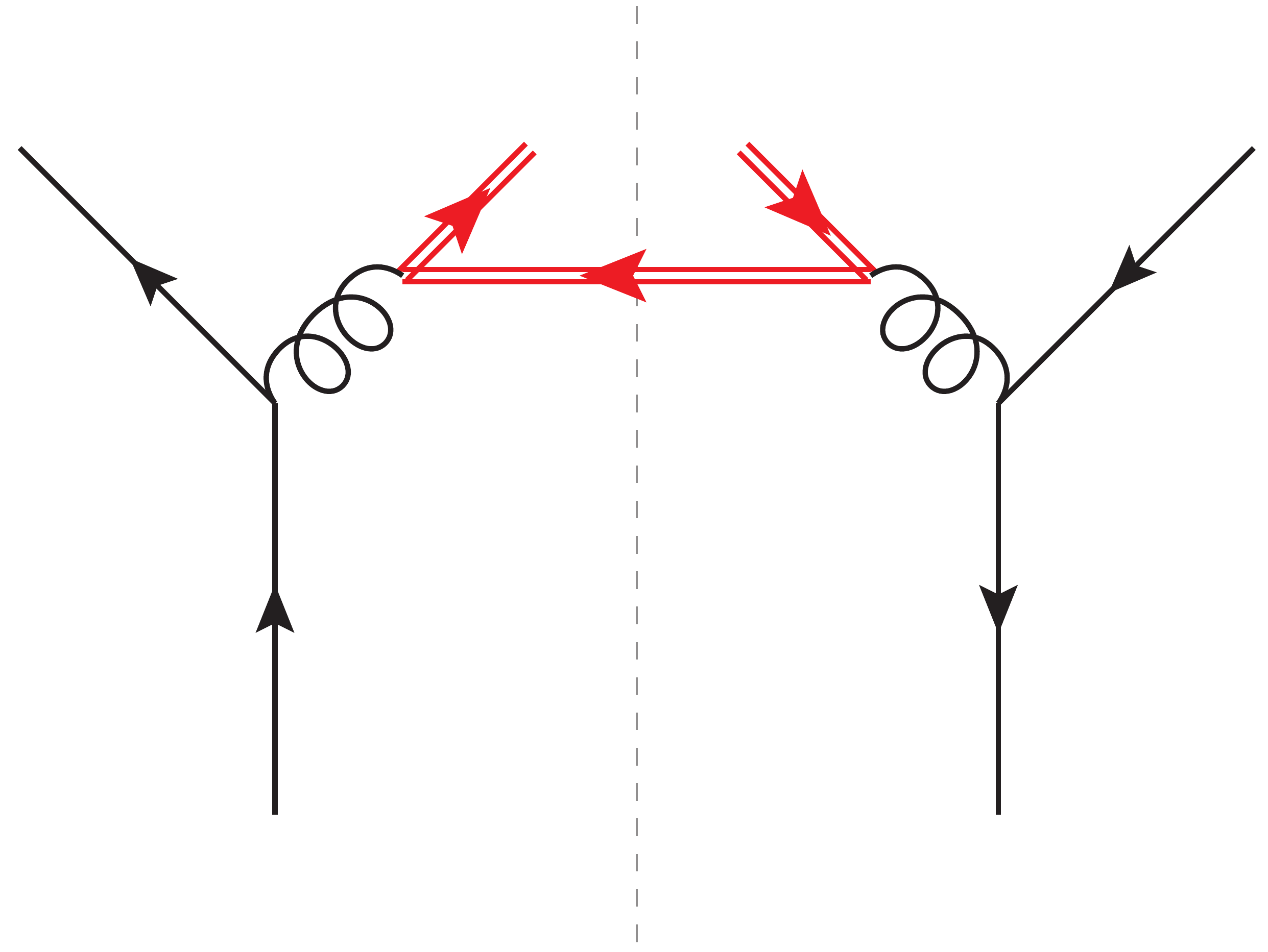}
      }
      \caption[Real graphs for NLO channels]
      {\label{fig:real-NLO} As \fig{\protect\ref{fig:virt-LO}}, but for real NLO diagrams with heavy quarks.}
   \end{center}
\end{figure}

Some properties of the NLO massive splitting kernels follow directly from the Feynman graphs for the associated splitting process.  Let us briefly review these and point out in which ones massive quark lines appear.  We distinguish between ``LO channels'' with splitting processes that are already possible at LO ($g \to q \qbar$, $q \to q g$, $g \to g g$, and $g \to Q \Qbar$) and ``NLO channels'', which first appear at two-loop order ($q \to Q \Qbar$, $g \to Q g$, $q \to Q q$, $q \to \Qbar q$, and corresponding channels with $q$ or $q'$ instead of $Q$).  Recall that we denote light flavours by $q$ and $q'$ and heavy ones by $Q$.  In detail, we have
\begin{enumerate}[itemsep=1ex]
  \item $\boldsymbol{g \to q \qbar}$. The graphs without heavy-quark lines are given in \figs{1(h)-1(k) and 3(f)-3(h)} of reference \cite{Diehl:2019rdh}.  In addition, there are virtual graphs with a heavy-quark loop on one of the incoming gluons, as shown in \fig{\ref{subfig:gqqbar}}.  In the massless case this diagram does not contribute, because incoming partons are on-shell, and scaleless integrals vanish in dimensional regularisation.
  \item $\boldsymbol{q \to q g}$. Besides the diagrams with massless quarks in \figs{1(l)-1(q)} and 3(i)-3(m) of \cite{Diehl:2019rdh}), there are virtual graphs with a heavy-quark loop on one of the outgoing gluon lines, cf.~\fig{\ref{subfig:qqg}}.
  \item $\boldsymbol{g \to g g}$. In addition to the massless graphs in \figs{1(a)-1(g)} and 3(a)-3(m) of \cite{Diehl:2019rdh})), there are virtual graphs with a heavy-quark loop in a vertex or propagator correction to the tree-level graph, as shown in \figs{\ref{subfig:ggg-1}} to \ref{subfig:ggg-3}.
  \item $\boldsymbol{g \to Q \Qbar}$. The diagrams for this channel correspond to the ones for the massless $g \to q \qbar$ case, with all quark lines replaced by massive quark lines.  They are shown in \figs{\ref{subfig:gQQbar-5}} to \ref{subfig:gQQbar-7} and \ref{subfig:gQQbar-1} to \ref{subfig:gQQbar-4}. In addition, there are virtual graphs with a heavy-quark loop on one of the incoming gluon lines, as shown in \fig{\ref{subfig:gQQbar-8}}.
  \item $\boldsymbol{q \to Q \Qbar}$. The diagram for this channel is obtained from the one for the massless $q \to q' \qbar'$ kernel by replacing all $q'$ quark lines by massive lines, as depicted in figure \ref{subfig:qQQbar}.
  \item $\boldsymbol{g \to Q g}$. The diagrams for this channel correspond to the ones for the massless $g \to q g$ kernel, again with all quark lines replaced by massive ones.  They are shown in \figs{\ref{subfig:gQg-1}} to \ref{subfig:gQg-6}.
  \item $\boldsymbol{q \to Q q}$. The diagram for this channel is obtained from the one for the massless $q \to q' q$ kernel by replacing all $q'$ quark lines by massive lines, shown in figure \ref{subfig:qqQ}.
\end{enumerate}

The two-loop diagrams for ${q \to \Qbar q}$ and ${q \to Q q}$ are related by charge reversal of the heavy-quark line, which results in
\begin{align}
   \label{eq:VQ-qQbarq}
   V^{Q (2)}_{\ms \smash{\Qbar}\vphantom{Q} q, q}
      &= V^{Q (2)}_{Q q, q}
   \,.
\end{align}
The two-loop graphs for NLO channels with observed light partons do not involve any heavy-quark lines, so that we have $V^{Q (2)}_{a_1 a_2, a_0} = V^{(2)}_{a_1 a_2, a_0}$ for the corresponding parton combinations.
%
%
\subsection{Reminders about massless kernels}
\label{sec:two-loop-massless}
To begin with, let us recall two properties of the massless two-loop splitting kernels derived in \cite{Diehl:2018kgr,Diehl:2021wpp}.
\begin{enumerate}
\item The part $V^{\nf [2,1]}$ of a two-loop kernel that is multiplied by $\log(\mu^2 / \mu_y^2)$ can readily be deduced by taking the $\mu$ derivative of the splitting formula \eqref{eq:small-y-DPD} and using the evolution equations for the DPD, PDF, and $\as(\mu)$.  One obtains
\begin{align}
   \label{eq:V-kernel-21}
V^{\nf [2,1]}_{a_1 a_2, a_0}
   &= \sum_{b_1} P_{a_1 b_1}^{\nf (0)} \conv{1} V^{(1)}_{b_1 a_2, a_0}
      + \sum_{b_2} P_{a_2 b_2}^{\nf (0)} \conv{2} V^{(1)}_{a_1 b_2, a_0}
   \nonumber\\
   & \quad
   {} - \sum_{b_0} V^{(1)}_{a_1 a_2, b_0} \conv{12} P_{b_0 a_0}^{\nf (0)}
      + \frac{\beta_0^{\nf}}{2} V^{(1)}_{a_1 a_2, a_0}\,.
\end{align}
\item The $\nf$ dependence of the NLO kernels enters only via quark loops in virtual graphs.  This affects the two channels $q\to q g$ and $g\to g g$, whereas the NLO kernels for all other channels are $\nf$ independent.  The $\nf$ dependence naturally appears via $\beta_0^{\nf}$, see \eqs{(4.17)} and (4.23) in \cite{Diehl:2021wpp}, and we write
\begin{align}
   \label{eq:V-kernel-nf-dep}
   V^{\nf [2, \ell\ms]}_{a_1 a_2, a_0}
      &= \widetilde{V}^{[2, \ell\ms]}_{a_1 a_2, a_0}
         + \beta_0^{\nf} \ms V^{\beta \ms [2, \ell\ms]}_{a_1 a_2, a_0}
   & \text{ for $(a_1 a_2, a_0) = (q g, q)$ and $(g g, g)$,}
\intertext{where $\ell = 0,1$.  Since $V^{\beta \ms [2, \ell\ms]}$ originates from virtual graphs, it is proportional to $\delta(1 - z_1 - z_2)$, just as the LO splitting kernels in \eqref{eq:DPD-split-LO}.  Using \eqref{eq:V-kernel-21} and the $\nf$ dependence of the LO DGLAP kernels, we can derive that}
   \label{eq:V-beta-21}
   V^{\beta \ms [2,1]}_{a_1 a_2, a_0}
      &= V^{(1)}_{a_1 a_2, a_0}
   & \text{ for $(a_1 a_2, a_0) = (q g, q)$ and $(g g, g)$.}
\end{align}
We note that in the expression of \eqref{eq:V-kernel-21} for $(a_1 a_2, a_0) = (q \qbar, g)$ the $\nf$ dependence cancels between the third and fourth terms, owing to the relation \eqref{eq:Pgg-LO}.
\end{enumerate}
The preceding relations also hold if the produced partons $a_1$ and $a_2$ are polarised.  The first two DGLAP kernels in \eqref{eq:V-kernel-21} are then polarised, whilst the third one is unpolarised since it refers to the incoming parton $a_0$.
%
%
\subsection{Limiting behaviour for small or large distances}
\label{sec:NLO-kernels-limits}
In the limits of small or large $y$ --- corresponding to large or small momentum scales $\mu_y$, respectively --- the massive splitting kernels can be expressed in terms of massless splitting kernels and flavour matching kernels.  This follows from the consistency between the different factorisation regimes in \fig{\ref{fig:splitting-regions}}.  If in the intermediate regime (where the massive splitting kernels are used) one takes $\mu_y \gg \mQ$, then one must recover the factorisation regime in which the heavy flavour $Q$ is treated as massless in the $1\to 2$ splitting process, whereas for $\mu_y \ll \mQ$ one must recover the regime in which $Q$ decouples in the $1\to 2$ splitting.  Comparing the DPD splitting formulae for the three regimes (see \sect{\ref{sec:one-massive}}) we thus find the following.

For $\mu_y \gg \mQ$ (i.e.\ in the small-$y$ limit) the massive kernels can be expressed by $\nf + 1$ flavour massless kernels, convoluted with a flavour matching kernel:
\begin{align}
   \label{eq:Limit-small-y}
      V^{Q, \nf}_{a_1 a_2, a_0}
   &  \, \overset{\mu_y \gg \mQ}{\longrightarrow} \,
      \sum_{b_0} V^{\nf + 1}_{a_1 a_2, b_0} \conv{12} A^{Q, \nf}_{b_0 a_0} \,.
\end{align}
This corresponds to the right panel in \fig{\ref{fig:splitting-regions}}, where one has first a flavour matching in the PDF and then the $1\to 2$ splitting for $\nf + 1$ massless flavours.  The $\mathcal{O}(a_s^2)$ term in equation \eqref{eq:Limit-small-y} reads
\begin{align}
   \label{eq:Limit-small-y-NLO}
      V^{Q, \nf (2)}_{a_1 a_2, a_0}
   &  \, \overset{\mu_y \gg \mQ}{\longrightarrow} \,
      \delta^{\nf}_{a_0 \ms l} \; V^{\nf + 1 (2)}_{a_1 a_2, a_0\phantom{b}}
      + \sum_{b_0} V^{(1)}_{a_1 a_2, b_0}
         \conv{12} A_{b_0 a_0}^{Q (1)}
\end{align}
with $\delta^{\nf}_{a_0 \ms l}$ defined in \eqref{eq:delta-light-def}.

For $\mu_y \ll \mQ$ (i.e.\ in the large-$y$ limit) one has
\begin{align}
   \label{eq:Limit-large-y}
      V^{Q, \nf}_{a_1 a_2, a_0}
   &  \, \overset{\mu_y \ll \mQ}{\longrightarrow} \,
      \sum_{b_1, b_2} A^{Q, \nf}_{a_1 b_1} \conv{1} A^{Q, \nf}_{a_2 b_2}
      \conv{2} V^{\nf}_{b_1 b_2, a_0} \,.
\end{align}
This corresponds to the left panel in \fig{\ref{fig:splitting-regions}}, where one has a $1\to 2$ splitting process for $\nf$ flavours and then DPD flavour matching to $\nf + 1$ flavours.  The $\mathcal{O}(a_s^2)$ kernel reads
\begin{align}
   \label{eq:Limit-large-y-NLO}
      V^{Q, \nf (2)}_{a_1 a_2, a_0}
   &  \, \overset{\mu_y \ll \mQ}{\longrightarrow} \,
      V^{\nf (2)}_{a_1 a_2, a_0}  +
      \sum_{b_1} A_{a_1 b_1}^{Q (1)} \conv{1} V^{(1)}_{b_1 a_2, a_0} +
      \sum_{b_2} A_{a_2 b_2}^{Q (1)} \conv{2} V^{(1)}_{a_1 b_2, a_0} +
      A_{\alpha\phantom{b}}^{Q (1)} \, V^{(1)}_{a_1 a_2, a_0\phantom{b}}
   \,,
\end{align}
where the term with $A_{\alpha}$ appears because $V^{Q, \nf}$ is expanded in $a_s^{\nf + 1}$ and $V^{\nf}$ in $a_s^{\nf}$.
%
%
\subsection{Scale dependence}
\label{sec:NLO-kernels-RGE}
Truncated at NLO, the massive splitting \eqref{eq:small-y-DPD-massive} formula reads
\begin{align}
   \label{eq:massive-splitting-DPD}
      F^{\nf + 1}_{a_1 a_2}
   &= \frac{1}{\pi y^2} \sum_{a_0}
      \left(
         a_s^{\nf + 1} \ms V^{Q (1)}_{a_1 a_2, a_0} \conv{12} f^{\nf}_{a_0} +
         \bigl( a_s^{\nf + 1} \bigr)^2 \, V^{Q, \nf (2)}_{a_1 a_2, a_0}
            \conv{12} f^{\nf}_{a_0}
      \right)
      + \mathcal{O}(a_s^3) \,.
\end{align}
Inserting \eqref{eq:massive-splitting-DPD} on the l.h.s.\ of the double DGLAP equation \eqref{eq:RGE-1} gives
\begin{align}
   \label{eq:RGE-2}
   &  a_s^{\nf + 1} \sum_{a_0}
      \left(
         \frac{\text{d}}{\text{d} \log \mu^2} V^{Q (1)}_{a_1 a_2, a_0}
      \right)
      \conv{12} f^{\nf}_{a_0}
   + (a_s^{\nf + 1})^2 \sum_{a_0}
      \Biggl[
         \left(
            \frac{\text{d}}{\text{d} \log \mu^2} V^{Q (2)}_{a_1 a_2, a_0}
         \right)
         \conv{12} f^{\nf}_{a_0}
   \nonumber \\[0.2em]
   & \quad
         - \frac{\beta_0^{\nf + 1}}{2} \, V^{Q (1)}_{a_1 a_2, a_0}
         \conv{12} f^{\nf}_{a_0}
         + V^{Q (1)}_{a_1 a_2, b_0} \conv{12}
         \left(
            P_{b_0 a_0}^{\nf (0)} \otimes f^{\nf}_{a_0}
         \right)
      \Biggr]
      + \mathcal{O}(a_s^3)\,,
\end{align}
whilst for the r.h.s.\ of \eqref{eq:RGE-1} one obtains
\begin{align}
   \label{eq:RGE-3}
   &  \bigl( a_s^{\nf + 1} \bigr)^2
      \left[
         \sum_{b_1, a_0}
         P_{a_1 b_1}^{\nf + 1 (0)} \conv{1} V^{Q (1)}_{b_1 a_2, a_0}
         \conv{12} f^{\nf}_{a_0}
         + \sum_{b_2, a_0}
         P_{a_2 b_2}^{\nf + 1 (0)} \conv{2} V^{Q (1)}_{a_1 b_2, a_0} \!
         \conv{12} f^{\nf}_{a_0}
      \right]
      + \mathcal{O}(a_s^3)
   \,.
\end{align}
The prefactor $1 /(\pi y^2)$ has been omitted in both equations.  Using the relation \eqref{eq:conv-12-relation} for the convolutions with three factors, one derives
\begin{align}
   \label{eq:RGE-4}
   \frac{\text{d}}{\text{d} \log \mu^2} \, V^{Q (1)}_{a_1 a_2, a_0}
      &= 0 \,,
   &
   \frac{\text{d}}{\text{d} \log \mu^2} \, V^{Q, \nf (2)}_{a_1 a_2, a_0}
      &= v^{\nf,\, \text{RGE}}_{a_1 a_2, a_0}
\end{align}
with
\begin{align}
   \label{eq:RGE-5}
   v^{\nf,\, \text{RGE}}_{a_1 a_2, a_0}
      &= \sum_{b_1} P_{a_1 b_1}^{\nf + 1 (0)} \conv{1} V^{Q (1)}_{b_1 a_2, a_0}
         + \sum_{b_2} P_{a_2 b_2}^{\nf + 1 (0)}
            \conv{2} V^{Q (1)}_{a_1 b_2, a_0}
   \nonumber\\
   & \quad {}
      - \sum_{b_0} V^{Q (1)}_{a_1 a_2, b_0} \conv{12} P_{b_0 a_0}^{\nf (0)}
      + \frac{\beta_0^{\nf + 1}}{2} \, V^{Q (1)}_{a_1 a_2, a_0} \,.
\end{align}
The first equation in \eqref{eq:RGE-4} confirms the $\mu$ independence of the LO kernels given in \eqref{eq:VQQbarg-LO} and \eqref{eq:VQ-light-LO}. Taking into account the $\nf$ dependence of the LO DGLAP kernels, one obtains the explicit expressions
\begin{align}
   \label{eq:RGE-NLO-qqbarg}
      v^{\text{RGE}}_{q \qbar, g}
   &= V^{[2,1]}_{q \qbar, g}
      + \frac{\Delta\beta_0}{2} \, V^{(1)}_{q \qbar, g}\,,
   \\
   \label{eq:RGE-NLO-qgq}
      v^{\nf,\,\text{RGE}}_{q g, q}
   &= V^{\nf [2,1]}_{q g, q}
      + \Delta\beta_0 \, V^{(1)}_{q g, q}\,,
   \\
   \label{eq:RGE-NLO-ggg}
      v^{\nf,\,\text{RGE}}_{g g, g}
   &= V^{\nf [2,1]}_{g g, g}
   + \frac{3\ms \Delta\beta_0}{2} \, V^{(1)}_{g g, g}\,,
   \\
   \label{eq:RGE-NLO-QQbarg}
      v^{\text{RGE}}_{Q \Qbar, g}
   &= P_{q q}^{(0)} \conv{1} V^{Q (1)}_{Q \Qbar, g} +
      P_{q q}^{(0)} \conv{2} V^{Q (1)}_{Q \Qbar, g} -
      V^{Q (1)}_{Q \Qbar, g} \conv{12} \widetilde{P}_{g g}^{(0)} +
      \frac{\Delta\beta_0}{2} \, V^{Q (1)}_{Q \Qbar, g}\,,
   \\
   \label{eq:RGE-NLO-QQbarq}
      v^{\text{RGE}}_{Q \Qbar, q}
   &= {}- V^{Q (1)}_{Q \Qbar, g} \conv{12} P_{g q}^{(0)}\,,
   \\
   \label{eq:RGE-NLO-Qgg}
      v^{\text{RGE}}_{Q g, g}
   &= P_{q g}^{(0)} \conv{1} V^{(1)}_{g g, g} +
      P_{g q}^{(0)} \conv{2} V^{Q (1)}_{Q \Qbar, g}\,,
   \\
   \label{eq:RGE-NLO-Qqq}
      v^{\text{RGE}}_{Q q, q}
   &= P_{q g}^{(0)} \conv{1} V^{(1)}_{g q, q}\,.
\end{align}
In \eqref{eq:RGE-NLO-qqbarg} to \eqref{eq:RGE-NLO-ggg}, we used the expression \eqref{eq:V-kernel-21} of $V^{\nf [2,1]}$, and in \eqref{eq:RGE-NLO-QQbarg} we made use of the explicit form \eqref{eq:Pgg-LO} of $P^{\nf (0)}_{g g}$.  In accordance with the inspection of the two-loop Feynman graphs, we find that the only kernels with an $\nf$ dependence are those for $q\to  q g$ and $g\to g g$.

The analogues of \eqs{\eqref{eq:RGE-NLO-qqbarg}} to \eqref{eq:RGE-NLO-Qqq} for polarised partons $a_1$ and $a_2$ are obtained by replacing the one-loop $V$ kernels on the r.h.s.\ with their polarised counterparts, as well as the LO DGLAP kernels in the convolutions $\otimes_{1}$ and $\otimes_{2}$.

\rev{As a cross check, let us verify that the form \eqref{eq:RGE-5} has the
correct behaviour at small and at large $y$.  To this end, we take the scale
derivative of the limiting relations \eqref{eq:Limit-small-y-NLO} and
\eqref{eq:Limit-large-y-NLO}.  Using $\text{d} V^{\nf (1)}_{a_1 a_2, a_0} /
\text{d} \log \mu^2 = 0$ and $\text{d} V^{\nf (2)}_{a_1 a_2, a_0} / \text{d}
\log \mu^2 = V^{\nf [2,1]}_{a_1 a_2, a_0}$, together with the relations
\eqref{eq:Aalpha-1}, \eqref{eq:delta-beta0}, \eqref{eq:A-NLO}, and \eqref{eq:V-kernel-21}, we obtain}
\begin{align}
   \label{eq:RGE-small-y}
   v^{\nf,\, \text{RGE}}_{a_1 a_2, a_0}
   \, \overset{\mu_y \gg \mQ}{\longrightarrow} \,
   \delta^{\nf}_{a_0 \ms l} \,
   &
   \biggl[ \,
   \sum_{b_1} P_{a_1 b_1}^{\nf + 1 (0)} \conv{1} V^{(1)}_{b_1 a_2, a_0}
   + \sum_{b_2} P_{a_2 b_2}^{\nf + 1 (0)} \conv{2} V^{(1)}_{a_1 b_2, a_0}
   \nonumber\\
   & {}
   - \sum_{b_0} V^{(1)}_{a_1 a_2, b_0} \conv{12} P_{b_0 a_0}^{\nf (0)}
   + \frac{\beta_0^{\nf + 1}}{2} \, V^{(1)}_{a_1 a_2, a_0}
   \biggr]
   \,,
   \\
   \label{eq:RGE-large-y}
   v^{\nf,\, \text{RGE}}_{a_1 a_2, a_0}
   \, \overset{\mu_y \ll \mQ}{\longrightarrow} \,
   \delta^{\nf}_{a_0 \ms l} \,
   &
   \biggl[ \,
   \delta^{\nf}_{a_2 \ms l} \,
   \sum_{b_1} \delta^{\nf}_{b_1 l} \,
      P_{a_1 b_1}^{\nf + 1 (0)} \conv{1} V^{(1)}_{b_1 a_2, a_0}
   + \delta^{\nf}_{a_1 l} \,
   \sum_{b_2}  \delta^{\nf}_{b_2 \ms l} \,
      P_{a_2 b_2}^{\nf + 1 (0)} \conv{2} V^{(1)}_{a_1 b_2, a_0}
   \nonumber\\
   & {}
   - \delta^{\nf}_{a_1 l}\, \delta^{\nf}_{a_2 \ms l} \,
     \sum_{b_0} V^{(1)}_{a_1 a_2, b_0} \conv{12} P_{b_0 a_0}^{\nf (0)}
   + \delta^{\nf}_{a_1 \ms l}\, \delta^{\nf}_{a_2 \ms l} \,
     \frac{\beta_0^{\nf + 1}}{2} \, V^{(1)}_{a_1 a_2, a_0}
   \biggr]
   \,.
\end{align}
\rev{This is consistent with \eqref{eq:RGE-5} if}
\begin{align}
   \label{eq:VQ-small-y}
   V^{Q (1)}_{b_1 b_2, b_0}
   &
   \, \overset{\mu_y \gg \mQ}{\longrightarrow} \,
   \delta_{b_0 l}^{\nf} \, V^{(1)}_{b_1 b_2, b_0}
   \,,
   \\[0.2em]
   \label{eq:VQ-large-y}
   V^{Q (1)}_{b_1 b_2, b_0}
   &
   \, \overset{\mu_y \ll \mQ}{\longrightarrow} \,
   \delta_{b_0 l}^{\nf} \, \delta_{b_1 l}^{\nf} \,
   \delta_{b_2 \ms l}^{\nf} \, V^{(1)}_{b_1 b_2, b_0}
\end{align}
\rev{Given the relations \eqref{eq:VQ-light-LO} to \eqref{eq:VQQbarg-LO-small-y} and the fact that at LO there is no splitting from a light parton to a light and a heavy one, we see that \eqref{eq:VQ-small-y} and \eqref{eq:VQ-large-y} are fulfilled.}
%
%
\subsection{An explicit parametrisation}
\label{sec:NLO-kernels-param}
We now give an explicit parametrisation of the massive two-loop kernels that incorporates their small- and large-$y$ limits as well as their dependence on $\mu$.  It reads
\begin{align}
   \label{eq:Vm-ansatz}
      V^{Q, \nf (2)}_{a_1 a_2, a_0}
   &= V^{\nf [2,0]}_{a_1 a_2, a_0}
      + V^{\nf [2,1]}_{a_1 a_2, a_0} \log \frac{\mQ^2}{\mu_y^2}
      + V^{I}_{a_1 a_2, a_0}(z_1, z_2, y \ms \mQ)
   \nonumber\\
   &  \quad
      + k_{11}(y \ms \mQ)
      \left(
         V^{\nf + 1 [2,0]}_{a_1 a_2, a_0} - V^{\nf [2,0]}_{a_1 a_2, a_0}
      \right)
      - k_{02}(y \ms \mQ)
      \left(
         V^{\nf + 1 [2,1]}_{a_1 a_2, a_0} - V^{\nf [2,1]}_{a_1 a_2, a_0}
      \right)
   \nonumber\\
   &  \quad
      + \log \frac{\mu^2}{\mQ^2} \,
         v^{\nf, \text{RGE}}_{a_1 a_2, a_0}(z_1, z_2)
\end{align}
with $v^{\nf, \text{RGE}}$ given in \eqref{eq:RGE-5}, where we abbreviate
\begin{align}
   \label{eq:kij}
      k_{i j}(w)
   &= w^2 K_{i}(w) K_{j}(w)\,.
\end{align}
The function $V^{I}_{a_1 a_2, a_0}$ is $\mu$ independent and interpolates between the small and large $y$ limits of the kernels with the properties
\begin{align}
   \label{eq:VI-limits}
      V^{I}_{a_1 a_2, a_0}(z_1, z_2, y \ms \mQ)
      \overset{y \to 0}{\longrightarrow} 0 \,,
   && V^{I}_{a_1 a_2, a_0}(z_1, z_2, y \ms \mQ)
      \overset{y \to \infty}{\longrightarrow} 0 \,.
\end{align}
For dimensional reasons, it depends on the product of $y$ and $\mQ$.  The functions $k_{11}(y \ms \mQ)$ and $k_{02}(y \ms \mQ)$ in \eqref{eq:Vm-ansatz} can be replaced with other functions that have the same limiting behaviour for $y\to 0$ and $y\to \infty$, provided that one makes a corresponding change in $V^{I}$.
Note that we do not allow for an $\nf$ dependence of $V^{I}$.  Indeed, the $\nf$ dependence of the two-loop kernels originates from graphs with massless quark loops, and the contribution of these graphs is fully contained in the first two terms and in the last term of \eqref{eq:Vm-ansatz}.

To verify that the parametrisation \eqref{eq:Vm-ansatz} has the correct limits for small and for large $y$, we recall that the modified Bessel functions $K_i$ satisfy
\begin{align}
   \label{eq:limits-Bessel}
   K_0(w)
      &  \overset{w \to 0}{\longrightarrow} \log \frac{b_0}{w} \,,
   &
   K_1(w)
      &  \overset{w \to 0}{\longrightarrow} \frac{1}{w} \,,
   &
   K_2(w)
      &  \overset{w \to 0}{\longrightarrow} \frac{2}{w^{2}} \,,
   \nonumber \\
   K_i(w)
      &  \overset{w \to \infty}{\longrightarrow} 0 \,.
\end{align}
As a consequence, one has
\begin{align}
   \label{eq:limits-kij}
   k_{11}(y \ms \mQ)
      & \overset{y \to 0}{\longrightarrow} 1 \,,
   &
   k_{02}(y \ms \mQ)
      & \overset{y \to 0}{\longrightarrow} \log \frac{\mu_y^2}{\mQ^2} \,,
   \nonumber \\
   k_{i j}(y \ms \mQ)
      & \overset{y \to \infty}{\longrightarrow} 0 \,.
\end{align}
Along with the limiting behaviour \eqref{eq:VI-limits} of $V^{I}_{a_1 a_2, a_0}$, one readily finds that
\begin{align}
   \label{eq:Vm-ansatz-limits}
      V^{Q (2)}_{a_1 a_2, a_0}(z_1, z_2, y, \mQ; \mQ)
      \overset{y \to 0}{\longrightarrow}
   &  \;V^{\nf + 1 [2,0]}_{a_1 a_2, a_0} +
      V^{\nf + 1 [2,1]}_{a_1 a_2, a_0} \, \log \frac{\mQ^2}{\mu_y^2}
    = V^{\nf + 1 (2)}_{a_1 a_2, a_0}(z_1, z_2, y; \mQ) \,,
   \nonumber\\
      V^{Q (2)}_{a_1 a_2, a_0}(z_1, z_2, y, \mQ; \mQ)
      \overset{y \to \infty}{\longrightarrow}
   &  \;V^{\nf [2,0]}_{a_1 a_2, a_0} +
      V^{\nf [2,1]}_{a_1 a_2, a_0} \, \log \frac{\mQ^2}{\mu_y^2}
    = V^{\nf (2)}_{a_1 a_2, a_0}(z_1, z_2, y; \mQ) \,.
\end{align}
This is agrees with \eqref{eq:Limit-small-y-NLO} and \eqref{eq:Limit-large-y-NLO}, given that for $\mu = \mQ$, the LO matching kernels $A^{Q (1)}$ in these equations are zero.
\rev{Since the $\mu$ dependence of the parametrisation \eqref{eq:Vm-ansatz}
resides entirely in its last term, the correct limiting behaviour at any other scale follows from the discussion at the end of the previous subsection.}

The explicit form of our parametrisation \eqref{eq:Vm-ansatz} for the different channels reads
\begin{align}
   \label{eq:Vm-qqbarg-explicit}
V^{Q (2)}_{q \qbar, g}
   &= V^{[2,0]}_{q \qbar, g} +
      V^{[2,1]}_{q \qbar, g} \log \frac{\mu^2}{\mu_y^2} +
      V^{I}_{q \qbar, g} +
      \frac{\Delta\beta_0}{2} \, \log \frac{\mu^2}{\mQ^2} \,
      V^{(1)}_{q \qbar, g} \,,
   \\[0.3em]
   \label{eq:Vm-qgq-explicit}
V^{Q, \nf (2)}_{q g, q}
   &= V^{\nf [2,0]}_{q g, q}
      + V^{\nf [2,1]}_{q g, q} \, \log \frac{\mu^2}{\mu_y^2}
      + V^{I}_{q g, q}
      + k_{11}(y \ms \mQ) \, \Delta\beta_0 \, V^{\beta \ms [2,0]}_{q g, q}
   \nonumber\\
   &  \quad
      + \biggl[ \log \frac{\mu^2}{\mQ^2} - k_{02}(y \ms \mQ) \biggr] \, \Delta\beta_0 \, V^{(1)}_{q g, q} \,,
   \\[0.3em]
   \label{eq:Vm-ggg-explicit}
V^{Q, \nf (2)}_{g g, g}
   &= V^{\nf [2,0]}_{g g, g}
      + V^{\nf [2,1]}_{g g, g} \log \frac{\mu^2}{\mu_y^2}
      + V^{I}_{g g, g}
      + k_{11}(y \ms \mQ) \, \Delta\beta_0 \, V^{\beta \ms [2,0]}_{g g, g}
   \nonumber\\
   &  \quad
      + \biggl[ \frac{3}{2} \log \frac{\mu^2}{\mQ^2}
         - k_{02}(y \ms \mQ) \biggr] \, \Delta\beta_0 \, V^{(1)}_{g g, g} \,,
   \\[0.3em]
   \label{eq:Vm-QQbarg-explicit}
V^{Q (2)}_{Q \Qbar, g}
   &= V^{I}_{Q \Qbar, g}
      + k_{11}(y \ms \mQ) \, V^{[2,0]}_{q \qbar, g}
      - k_{02}(y \ms \mQ) \, V^{[2,1]}_{q \qbar, g}
   \nonumber\\
   &  \quad
      + \log \frac{\mu^2}{\mQ^2} \,
      \biggl[
         P_{q q}^{(0)} \conv{1} V^{Q (1)}_{Q \Qbar, g}
         + P_{q q}^{(0)} \conv{2} V^{Q (1)}_{Q \Qbar, g}
         - V^{Q (1)}_{Q \Qbar, g} \conv{12} \widetilde{P}_{g g}^{(0)}
         + \frac{\Delta\beta_0}{2} \, V^{Q (1)}_{Q \Qbar, g}
      \biggr] \,,
   \\[0.3em]
   \label{eq:Vm-QQbarq-explicit}
V^{Q (2)}_{Q \Qbar, q}
   &= V^{I}_{Q \Qbar, q}
      + k_{11}(y \ms \mQ) \, V^{[2,0]}_{q' \qbar', q}
      + k_{02}(y \ms \mQ) \, V^{(1)}_{q \qbar, g} \conv{12} P_{g q}^{(0)}
      - \log \frac{\mu^2}{\mQ^2} \, V^{Q (1)}_{Q \Qbar, g}
         \conv{12} P_{g q}^{(0)} \,,
   \\[0.3em]
   \label{eq:Vm-Qgg-explicit}
V^{Q (2)}_{Q g, g}
   &= V^{I}_{Q g, g}
      + k_{11}(y \ms \mQ) \, V^{[2,0]}_{q g, g}
      - k_{02}(y \ms \mQ) \,
         \biggl[
            P_{q g}^{(0)} \conv{1} V^{(1)}_{g g, g} +
            P_{g q}^{(0)} \conv{2} V^{(1)}_{q \qbar, g}
            - V^{(1)}_{q g, q} \conv{12} P^{(0)}_{q g}
         \biggr]
   \nonumber\\
   &  \quad
      + \log \frac{\mu^2}{\mQ^2} \,
      \biggl[
         P_{q g}^{(0)} \conv{1} V^{(1)}_{g g, g} +
         P_{g q}^{(0)} \conv{2} V^{Q (1)}_{Q \Qbar, g}
      \biggr] \,,
   \\[0.3em]
   \label{eq:Vm-Qqq-explicit}
V^{Q (2)}_{Q q, q}
   &= V^{I}_{Q q, q}
      + k_{11}(y \ms \mQ) \, V^{[2,0]}_{q' q, q}
      - k_{02}(y \ms \mQ) \, P^{(0)}_{q g} \conv{1} V^{(1)}_{g q, q}
      + \log \frac{\mu^2}{\mQ^2} \, P_{q g}^{(0)} \conv{1} V^{(1)}_{g q, q}
      \,,
\end{align}
where $\Delta\beta_0$ is defined in \eqref{eq:delta-beta0} and $V^{\beta \ms [2,0]}$ in \eqref{eq:V-kernel-nf-dep}.  In \eqs{\eqref{eq:Vm-QQbarq-explicit}} to \eqref{eq:Vm-Qqq-explicit} we have inserted the explicit forms \eqref{eq:V-kernel-21} of $V^{[2,1]}$ for the relevant channel.
The analogues of \eqref{eq:Vm-qqbarg-explicit} to \eqref{eq:Vm-Qqq-explicit} for polarised partons are obtained by replacing the $V$ kernels on the r.h.s.\ with their polarised counterparts, as well as the DGLAP kernels in the convolutions $\otimes_{1}$ and $\otimes_{2}$.

The interpolating function for $g\to g g$ splitting receives contributions only from virtual graphs with a massive quark loop (see \figs{\ref{subfig:ggg-1}} to \ref{subfig:ggg-3}).  It therefore has the form
\begin{align}
   \label{eq:VI-ggg-delta}
   V^{I}_{g g, g}(z_1, z_2, y\ms \mQ)
      &= \delta(1 - z_1 - z_2) \, V^{I}_{g g, g}(z_1, y \ms \mQ)
   \,,
\end{align}
where the function on the r.h.s.\ depends on one instead of two momentum fractions.
In \sect{\ref{sec:Vqqbarg-massive-NLO}} we will show that the interpolating function for $g\to q \qbar$ is zero, i.e.\
\begin{align}
   \label{eq:VI-qqbarg-zero}
   V^{I}_{q \qbar, g} &= 0
\end{align}
with corresponding relations for longitudinal or transverse quark polarisation.

We already noted that for NLO channels with light observed partons, the massive two-loop kernels are equal to their massless counterparts.  Correspondingly, one finds $V^{I}_{a_1 a_2, a_0} = 0$ for the relevant parton combinations.
%
%
\subsection{Large logarithms}
\label{sec:NLO-kernels-large-logs}
For both $\mu_y \ll \mQ$ and $\mu_y \gg \mQ$, the massive splitting kernels involve two physical scales of very different size.  The expressions just derived allow us to explicitly investigate the large logarithms that appear at NLO in these limits.
Notice that according to \eqref{eq:limits-kij}, the function $k_{02}(y \ms \mQ)$ counts as a large logarithm in the limit of small $y$, whereas this is not the case for $k_{11}(y \ms \mQ)$.  Both functions vanish for $\mu_y \ll \mQ$.

From the explicit forms \eqref{eq:Vm-qqbarg-explicit} to \eqref{eq:Vm-Qqq-explicit} it is clear that no choice of $\mu$ can remove all large logarithms, not even for light observed partons.  In the following we will focus on the scale choice $\mu \sim \mu_y$.  This provides an element of continuity between the massive regime and the two regimes with massless DPD splitting kernels in \fig{\ref{fig:scheme-one-massive}} --- for the latter, $\mu_y$ is the only relevant scale in the $1\to 2$ splitting process.  Moreover, the numerical study in \cite{Diehl:2021wpp} showed that the logarithmic part $V^{\nf [2,1]}_{g g, g}$ of the gluon splitting kernel gives rise to large NLO corrections when the splitting formula is evaluated at scales away from $\mu_y$.  In fact, this kernel contains large logarithms in several kinematic limits.

We now discuss the different parton channels in turn.
\begin{itemize}
\item \textbf{LO channels with observed light partons.}  With the choice $\mu \sim \mu_y$ the expressions for all three channels $g \to q \qbar$, $q \to q g$, and $g \to g g$ receive logarithmic contributions of the type
\begin{align}
   \label{eq:large-log-beta0}
   k\, \frac{\Delta\beta_0}{2} \, \log \frac{\mu_y^2}{\mQ^2} \,
      V^{(1)}_{a_1 a_2, a_0}
\end{align}
with $0\le k \le 1$ for $\mu_y \gg \mQ$ and $1 \le k \le 3$ for $\mu_y \ll \mQ$.  There is hence no kinematic enhancement of the NLO correction relative to the LO result.  The prefactor of the logarithm is always negative and at most of size $3 \ms |\Delta\beta_0| / 2 = 1$.  Taking for instance $\mu_y = m_b / 4$, we have $a_s^{5}(\mu_y) \approx 0.066$ and $a_s^5(\mu_y) \, \log (\mu_y^2 / m_b^2) \approx -0.18$.
\item \textbf{Channels with an observed heavy-quark pair.}  In the large-$y$ limit, the NLO kernels $V_{Q \smash{\Qbar}, g}^{Q (2)}$ and $V_{Q \smash{\Qbar}, q}^{Q (2)}$ go to zero, as does the LO kernel $V_{Q \smash{\Qbar}, g}^{Q (1)}$.  At two-loop accuracy, the $Q \Qbar$ distribution hence decouples for $\mu_y \ll \mQ$.

To analyse the small-$y$ limit, we use the explicit form of $V^{[2,1]}_{q \qbar, g}$ in equation \eqref{eq:V-kernel-21}, together with the limiting behaviour
\begin{align}
   \label{eq:VQQbarg-small-y-log}
   V^{Q (1)}_{Q \Qbar, g}
      & \overset{\mu_y \gg \mQ}{\longrightarrow}  V^{(1)}_{q \qbar, g}
\end{align}
and the limit of $k_{02}$ given in \eqref{eq:limits-kij}.  This yields
\begin{align}
   V_{Q \Qbar, g}^{Q (2)}
   &  \overset{\mu_y \gg \mQ}{\longrightarrow}
      \log \frac{\mu^2}{\mu_y^2} \,
      \biggl[ P_{q q}^{(0)} \conv{1} V^{(1)}_{q \qbar, g}
            + P_{q q}^{(0)} \conv{2}
            - V^{(1)}_{q \qbar, g} \conv{12} \widetilde{P}_{g g}^{(0)}
      \biggr]
      + \frac{\Delta\beta_0}{2} \, \log \frac{\mu^2}{\mQ^2} \,
         V^{(1)}_{q \qbar, g} + \cdots \,,
   \\[0.3em]
   V_{Q \Qbar, q}^{Q (2)}
   &  \overset{\mu_y \gg \mQ}{\longrightarrow}
      -\log \frac{\mu^2}{\mu_y^2} \,
      V^{(1)}_{q \qbar, g} \conv{12} P_{g q}^{(0)} + \cdots \,,
\end{align}
where the ellipsis denotes non-logarithmic contributions from $V^I_{} + V^{[2,0]}$.  With the choice $\mu \sim \mu_y$ one thus finds that $V_{Q \smash{\Qbar}, g}^{Q (2)}$ has the same large logarithm as its counterpart $V_{q \qbar, g}^{Q (2)}$ for light quarks, whilst $V_{Q \smash{\Qbar}, q}^{Q (2)}$ has no large logarithm at all.

\item \textbf{Channels with one observed heavy quark.}  In the large-$y$ limit, one has
\begin{align}
   \label{eq:VQgg-large-y-log}
   V^{Q (2)}_{Q g, g}
      & \overset{\mu_y \ll \mQ}{\longrightarrow}
      \log \frac{\mu^2}{\mQ^2} P_{q g}^{(0)} \conv{1} V^{(1)}_{g g, g}
      = A_{Q g}^{Q (1)}(\mQ; \mu) \conv{1} V^{(1)}_{g g, g} \,,
   \\[0.3em]
   \label{eq:VQqq-large-y-log}
   V^{Q (2)}_{Q q, q}
      & \overset{\mu_y \ll \mQ}{\longrightarrow}
      \log \frac{\mu^2}{\mQ^2} \, P_{q g}^{(0)} \conv{1} V^{(1)}_{g q, q}
      = A_{Q g}^{Q (1)}(\mQ; \mu) \conv{1} V^{(1)}_{g q, q} \,,
\end{align}
so that the production of one heavy quark does \emph{not} decouple in the limit $\mu_y \ll \mQ$.  At this point we must remember that we defined the DPDs in the $\msbar$ renormalisation scheme, which does \emph{not} yield heavy-quark decoupling for scales much smaller than the quark mass, see e.g.~\cite{Chetyrkin:1997sg}.
We note in passing that at order $\as^3$, non-decoupling in the limit $\mu_y \ll \mQ$ will also occur in the channel $g \to Q \Qbar$, because the three-loop kernel contains a term $A^{Q (1)}_{Q g} \conv{1} A^{Q (1)}_{Q g} \conv{2} V^{(1)}_{g g, g}$ according to \eqref{eq:Limit-large-y}.

For the scale choice $\mu \sim \mu_y$, the two-loop kernels in  \eqref{eq:VQgg-large-y-log} and \eqref{eq:VQqq-large-y-log} yield splitting DPDs $F_{Q g}$ and $F_{Q q}$ that are negative and enhanced by a large logarithm $\log (\mu_y^2 / \mQ^2)$.  This is similar to the negative values of the heavy-quark PDF that one obtains when flavour matching at scales $\mu \ll \mQ$.  Both in the DPD and in the PDF case, subsequent DGLAP evolution up to the scale $\mQ$ yields a distribution close to zero, provided that $a_s \log(\mQ^2 / \mu^2)$ is sufficiently small.  Indeed, the large logarithm in \eqref{eq:VQgg-large-y-log} and \eqref{eq:VQqq-large-y-log} is then cancelled by a corresponding term with $\log(\mQ^2 / \mu^2)$ from DGLAP evolution, and further evolution terms come with at least one more factor of $a_s \log(\mQ^2 / \mu^2)$.

Let us now turn to the small-$y$ limit.  In this case, we have
\begin{align}
V^{Q (2)}_{Q g, g}
   & \overset{\mu_y \gg \mQ}{\longrightarrow}
      \log \frac{\mu^2}{\mu_y^2} \,
      \biggl[
         P_{q g}^{(0)} \conv{1} V^{(1)}_{g g, g} +
         P_{g q}^{(0)} \conv{2} V^{(1)}_{q \qbar, g}
      \biggr]
      + \log \frac{\mu_y^2}{\mQ^2} \,
         V^{(1)}_{q g, q} \conv{12} P_{q g}^{(0)} + \cdots \,,
   \\[0.3em]
V^{Q (2)}_{Q q, q}
   & \overset{\mu_y \gg \mQ}{\longrightarrow}
      \log \frac{\mu^2}{\mu_y^2} \,
         P_{q g}^{(0)} \conv{1} V^{(1)}_{g q, q} + \cdots
   \,,
\end{align}
where the ellipsis again denotes non-logarithmic contributions.
For the choice $\mu \sim \mu_y$, a large logarithm remains only in the $g \to Q g$ channel, which can be rewritten as
\begin{align}
   \label{eq:VQgg-small-y-log}
   V^{Q (2)}_{Q g, g}
      & \overset{\mu_y \gg \mQ}{\longrightarrow}
         V^{(1)}_{q g, q} \conv{12} A_{Q g}^{Q (1)}(\mQ; \mu_y) + \cdots
&
   & \text{ for $\mu \sim \mu_y$.}
\end{align}
This is just the expression of the graph in \fig{\ref{subfig:Fgb-massless-scheme}}.  We recall that in \sect{\ref{sec:massive-DPDs}}, this was identified as a large missing NLO contribution when the massive splitting scheme is evaluated at LO.  It led us to choose a rather low value for the upper limit $\beta \ms \mQ$ of the $\mu_y$ region where massive splitting kernels are used.  One might expect that using massive splitting kernels at NLO will mitigate the problem and possibly allow for using larger $\beta$.  Of course, this should be confirmed by a numerical study, which is beyond the scope of this work.
\end{itemize}

The preceding discussion, and in particular the relations \eqref{eq:VQQbarg-small-y-log} to \eqref{eq:VQgg-small-y-log} are readily generalised to the case of polarised partons.
%
%
%
\section{NLO splitting kernels with two heavy flavours}
\label{sec:NLO-kernels-constraints-two-flavours}
In the previous section, we analysed the massive NLO $1 \to 2$ splitting kernels for one heavy flavour $Q$ with mass $\mQ$.  We now extend our analysis to kernels for two massive flavours, which appear in the scheme we laid out in \sect{\ref{sec:two-massive}}.  The corresponding LO kernels have been given in \sect{\ref{sec:massive-kernels}}.
In analogy to \eqs{\eqref{eq:arguments-fnF}} to \eqref{eq:arguments-A}, it is understood that the massive kernels are taken at arguments
\begin{align}
   \label{eq:arguments-Vcb}
      V^{c b}_{a_1 a_2, a_0}
   &  \equiv V^{c b}_{a_1 a_2, a_0}(z_1, z_2, y, m_c, m_b; \mu)
\end{align}
unless specified otherwise.
%
%
\subsection{Expressions in terms of kernels with one heavy flavour}
\label{sec:NLO-kernels-two-flavours-from-one-flavour}
Inspection of the relevant Feynman graphs allows us to obtain the kernels for massive charm and bottom from the ones for a single heavy flavour.  We discuss the different types of parton channels in turn.

\paragraph{LO channels with observed light partons.} For the LO channels $g \to g g$, $g \to q \qbar$, and $q \to q g$, heavy-quark effects arise solely from closed quark loops, as shown in figure \ref{fig:real-NLO}. Each of these loop graphs is evaluated once for charm and once for bottom, and there are no graphs that involve both heavy flavours.

We now decompose the kernels for three light flavours and one heavy flavour $Q$ as
\begin{align}
   \label{eq:LO-channels-one-heavy-decomp}
      V_{a_1 a_2, a_0}^{Q, 3 (2)}(\mQ)
   &= V_{a_1 a_2, a_0}^{3 (2)} +
      \Delta V_{a_1 a_2, a_0}^{Q (2)}(\mQ)\,,
\end{align}
where $V_{a_1 a_2, a_0}^{3 (2)}$ is the massless three-flavour NLO kernel and $\Delta V_{a_1 a_2, a_0}^{Q (2)}(\mQ)$ contains the contributions from diagrams with massive quark loops (see \figs{ \ref{subfig:gqqbar}} to \ref{subfig:ggg-3}). Each term on the r.h.s.\ contains the ultraviolet counterterms in the $\msbar$ scheme for the flavours running in the quark loops.  The sum of counterterms for the kernel \eqref{eq:LO-channels-one-heavy-decomp} hence corresponds to renormalising the strong coupling for $3 + 1$ flavours, in agreement with \eqref{eq:VmQ-expanded}.

According to our discussion of the relevant Feynman graphs, the kernel for three light and two heavy flavours is then given by
\begin{align}
   \label{eq:LO-channels-two-heavy-decomp}
      V_{a_1 a_2, a_0}^{c b (2)}(m_c, m_b)
   &= V_{a_1 a_2, a_0}^{3 (2)} +
      \Delta V_{a_1 a_2, a_0}^{Q (2)}(m_c) +
      \Delta V_{a_1 a_2, a_0}^{Q (2)}(m_b)\,.
\end{align}
The sum of the kernels on the r.h.s.\ contains the ultraviolet counterterms for $3 + 1 + 1$ flavours, which corresponds to the use of $a_s^5$ in the perturbative expansion \eqref{eq:Vcs-expanded} of $V^{c b}$.

Combining the two previous equations, we obtain
\begin{align}
   \label{eq:LO-channels-one-to-two-1}
      V_{a_1 a_2, a_0}^{c b (2)}(m_c, m_b)
   &= V_{a_1 a_2, a_0}^{Q, 3 (2)}(m_c) +
      V_{a_1 a_2, a_0}^{Q, 3 (2)}(m_b) -
      V_{a_1 a_2, a_0}^{3 (2)} \,.
\end{align}
Here and in the following, we explicitly indicate the dependence of kernels on the two quark masses, but omit the arguments $z_1, z_2, y$ and $\mu$.
\paragraph{LO channels with observed heavy partons.} In the NLO kernels for $g\to c \cbar$ and $g\to b \bbar$, the two massive quarks enter on a different footing.
For definiteness, let us consider the kernel for $g \to c \cbar$, from which the one for $g \to b \bbar$ is readily obtained by interchanging the roles of $c$ and $b$ quarks, i.e.
\begin{align}
   \label{eq:Vccbarg-Vbbbarg-relation}
      V^{c b (2)}_{\smash{b \bbar, g} \phantom{|}}(m_c, m_b)
   &= V^{c b (2)}_{c \cbar, g}(m_b, m_c) \,.
\end{align}
The kernel $V^{c b (2)}_{c \cbar, g}$ differs from $V^{c (2)}_{c \cbar, g}$, where no $b$ quarks appear, by the contribution from the diagram in figure \ref{subfig:gQQbar-8} with charm in the upper quark lines and bottom in the quark loop.  Denoting the contribution of this graph and its ultraviolet counterterm by $v_{c \cbar, g}^{c b}$, we have
\begin{align}
   \label{eq:LO-channels-one-to-two-heavy-1}
      V_{c \cbar, g}^{c b (2)}(m_c, m_b)
   &= V_{c \cbar, g}^{c (2)}(m_c) +
      v_{c \cbar, g}^{c b}(m_c, m_b)\,.
\end{align}
In \sect{\ref{sec:Vqqbarg-massive-NLO}} we will derive the expression for the kernel resulting from \fig{\ref{subfig:gqqbar}}, where the upper lines are for a light quark.  The derivation is trivially extended to the case where that quark is massive and yields
\begin{align}
   \label{eq:Vcb-gQQbar-two-flavours}
   v_{c \cbar, g}^{c b}(m_c, m_b)
      &= \frac{\Delta \beta_0}{2} \, \log \frac{\mu^2}{m_b^2} \;
         V^{c (1)}_{c \cbar, g}(m_c)
       = V^{c (1)}_{c \cbar, g}(m_c) \conv{12} A^{b (1)}_{g g}(m_b)
      \,.
\end{align}
\paragraph{NLO channels with observed heavy partons.} These are the channels $q \to Q \Qbar$, $g \to Q g$, and $q \to Q q$ with $Q = c, \, b$.    The relevant graphs involve massive lines only for the observed flavour $Q$, as seen in the second and third rows of \fig{\ref{fig:real-NLO}}.  They are hence not sensitive to the presence of a second heavy quark and read
\begin{align}
   \label{eq:NLO-channels-one-to-two-flavours}
      V_{a_1 a_2, a_0}^{c b (2)}(m_c)
   &= V_{a_1 a_2, a_0}^{c (2)}(m_c)
   &
   \text{ for } (a_1 a_2, a_0) = (c \cbar, g), (c g, g), (c q, q)
\end{align}
for produced charm.  For produced bottom, one has to replace the flavour labels and the quark mass.
\paragraph{NLO channels with observed light partons.} Since the two-loop graphs for these channels do not involve any heavy-quark lines, we have $V^{c b (2)}_{a_1 a_2, a_0} = V^{(2)}_{a_1 a_2, a_0}$ for the corresponding parton combinations, just as in the case of a single heavy flavour.

In the following subsections, we will verify that \eqref{eq:LO-channels-one-to-two-1}, \eqref{eq:Vcb-gQQbar-two-flavours}, and \eqref{eq:NLO-channels-one-to-two-flavours} have the correct limits for small or large $y$ and the correct dependence on the renormalisation scale.
%
%
\subsection{Limiting behaviour for small or large distances}
\label{sec:NLO-kernels-limits-two-flavours}
For small or large $y$, the massive splitting kernels with two heavy flavours can be reduced to the convolution of massless splitting kernels with flavour matching kernels, just like in the one-flavour case (\sect{\ref{sec:NLO-kernels-limits}}).  This follows from the consistency between the different factorisation regimes in \fig{\ref{fig:splitting-regions}} for the case that the heavy-quark subprocesses contain two heavy flavours instead of one.

\paragraph{PDF matching for two heavy flavours.}  Instead of successively matching a three-flavour PDF first to four and then to five active flavours, one may directly match
\begin{align}
   \label{eq:two-flavour-PDF-matching}
   f_{a_1}^{5}(x; \mu)
      &= \sum_{a_0} A_{a_1 a_0}^{c b}(m_c, m_b; \mu) \otimes f_{a_0}^{3}(\mu)
\end{align}
with a perturbative expansion of the kernel in $a_s^{5}$:
\begin{align}
   \label{eq:Acb-general-form}
A^{c b}_{a_1 a_0}(z, m_c, m_b; \mu)
   &= \sum_{k = 0}^\infty \bigl[ a_{s}^{5}(\mu) \bigr]^k \,
   A_{a_1 a_0}^{c b (k)}(z, m_c, m_b; \mu)
   \,.
\end{align}
For brevity, we do not indicate explicitly that $A^{c b}$ assumes the presence of 3 massless quark flavours.
At tree-level, we have
\begin{align}
A^{c b (0)}_{a_1 a_0}
   &= A_{a_1 a_0}^{Q (0)}
    = \delta^{3}_{a_1 \ms l} \, \delta^{}_{a_1 a_0}\, \delta(1-z)
\end{align}
with $\delta^{3}_{a_1 \ms l}$ defined in \eqref{eq:delta-light-def}, whereas at one-loop order the matching coefficients are additive:
\begin{align}
   \label{eq:Acb-NLO}
   A^{c b (1)}_{a_1 a_0}(m_c, m_b)
   &= A^{c (1)}_{a_1 a_0}(m_c)
      + \delta^{3}_{a_0 \ms l} \, A^{b (1)}_{a_1 a_0}(m_b)
   \,.
\end{align}
Results for $A^{c b}$ up to three loops can be found in reference \cite{Ablinger:2017xml}.  Up to two loops, the coefficients can be obtained from the single-flavour case, as shown in \cite{Blumlein:2018jfm}.

In full analogy, one can perform two-flavour matching for the strong coupling,
\begin{align}
   \label{eq:a-cb-matching-expanded}
      a_s^{3}(\mu)
   &= \sum_{k = 0}^{\infty} \bigl[ a_s^{5}(\mu) \bigr]^{k + 1} \,
      A^{c b (k)}_{\alpha}(m_c, m_b; \mu) \,,
\end{align}
with coefficients
\begin{align}
   \label{eq:Aalpha-cb-1}
   A^{c b (0)}_{\alpha} &= 1 \,,
&
   A^{c b (1)}_{\alpha}(m_c, m_b; \mu)
      &= A^{c (1)}_{\alpha}(m_c; \mu) + A^{b (1)}_{\alpha}(m_b; \mu)
\end{align}
at LO and NLO.

\paragraph{General limiting behaviour.}
For small $y$, or more precisely for $\mu_y \gg m_b$, one finds that the two-flavour kernels can be expressed as
\begin{align}
   \label{eq:Limit-small-y-two-flavours}
      V^{c b}_{a_1 a_2, a_0}
   &  \,\overset{y \to 0}{\longrightarrow}\,
      \sum_{b_0} V^{5}_{a_1 a_2, b_0} \conv{12} A^{c b}_{b_0 a_0}
   \,.
\end{align}
Expanding this in the five-flavour strong coupling $a_s^{5}$ gives
\begin{align}
   \label{eq:Limit-small-y-NLO-two-flavours}
      V^{c b (2)}_{a_1 a_2, a_0}
   &  \,\overset{y \to 0}{\longrightarrow}\,
      \delta^{3}_{a_0 \ms l} V^{5 (2)}_{a_1 a_2, a_0}
      + \sum_{b_0} V^{5 (1)}_{a_1 a_2, b_0} \conv{12} A_{b_0 a_0}^{c b (1)}
\end{align}
for the NLO coefficient.
In the opposite limit of large $y$, or more precisely for $\mu_y \ll m_c$, the two-flavour kernels may be expressed as
\begin{align}
   \label{eq:Limit-large-y-two-flavours}
      V^{c b}_{a_1 a_2, a_0}
   &  \,\overset{y \to \infty}{\longrightarrow}\,
      A^{c b}_{a_1 b_1} \conv{1} A^{c b}_{a_2 b_2} \conv{2}
      V^{3}_{b_1 b_2, a_0} \,,
\end{align}
where the $\mathcal{O}(a_s^2)$ term reads
\begin{align}
   \label{eq:Limit-large-y-NLO-two-flavours}
      V^{c b (2)}_{a_1 a_2, a_0}
   &  \,\overset{y \to \infty}{\longrightarrow}\,
      V^{3 (2)}_{a_1 a_2, a_0} +
      \sum_{b_1} A_{a_1 b_1}^{c b (1)} \conv{1} V^{3 (1)}_{b_1 a_2, a_0} +
      \sum_{b_2} A_{a_2 b_2}^{c b (1)} \conv{2} V^{3 (1)}_{a_1 b_2, a_0} +
      A_{\alpha}^{c b (1)} \, V^{3 (1)}_{a_1 a_2, a_0} \,.
\end{align}
We now show that the kernels derived in section \ref{sec:NLO-kernels-two-flavours-from-one-flavour} correctly fulfil these limits, again considering separately the different groups of parton channels.

\paragraph{LO channels with observed light partons.}  To have the same small-$y$ limit on both sides of the relation \eqref{eq:LO-channels-one-to-two-1}, we need
\begin{align}
   \label{eq:Limit-small-y-LO-channels-light}
   &  V^{5 (2)}_{a_1 a_2, a_0} + \sum_{b_0} V^{(1)}_{a_1 a_2, b_0}
      \conv{12} A^{c b (1)}_{b_0 a_0}
   \nonumber\\
   & \quad
      \overset{!}{=}
      \biggl[
      V^{4 (2)}_{a_1 a_2, a_0} + \sum_{b_0} V^{(1)}_{a_1 a_2, b_0} \conv{12}
         A^{c (1)}_{b_0 a_0}
      \biggr]
      + \biggl[
      V^{4 (2)}_{a_1 a_2, a_0} + \sum_{b_0} V^{(1)}_{a_1 a_2, b_0} \conv{12}
         A^{b (1)}_{b_0 a_0}
      \biggl]
      - V^{3 (2)}_{a_1 a_2, a_0} \,,
\end{align}
where we have used \eqref{eq:Limit-small-y-NLO} and \eqref{eq:Limit-small-y-NLO-two-flavours}.
To see that this is the case, we use \eqref{eq:Acb-NLO} and the relation
\begin{align}
   V^{5 (2)}_{a_1 a_2, a_0}
   &= 2 V^{4 (2)}_{a_1 a_2, a_0} - V^{3 (2)}_{a_1 a_2, a_0} \,,
\end{align}
which holds because $V^{\nf (2)}$ depends linearly on $\nf$.

For \eqref{eq:LO-channels-one-to-two-1} to have the correct large-$y$ limit, we need
\begin{align}
   \label{eq:Limit-large-y-LO-channels-light}
   &  V^{3 (2)}_{a_1 a_2, a_0} +
      \sum_{b_1} A^{c b (1)}_{a_1 b_1} \conv{1} V^{(1)}_{b_1 a_2, a_0} +
      \sum_{b_2} A^{c b (1)}_{a_2 b_2} \conv{1} V^{(1)}_{a_1 b_2, a_0}
      + A^{c b (1)}_{\alpha} \, V^{(1)}_{a_1 a_2, a_0}
   \nonumber\\
   & \quad
      \, \overset{!}{=} \,
      V^{3 (2)}_{a_1 a_2, a_0} + \sum_{b_1}
      \left(
         A^{c (1)}_{a_1 b_1} +
         A^{b (1)}_{a_1 b_1}
      \right) \conv{1} V^{(1)}_{b_1 a_2, a_0}
      + \sum_{b_2}
      \left(
         A^{c (1)}_{a_2 b_2} +
         A^{b (1)}_{a_2 b_2}
      \right) \conv{1} V^{(1)}_{a_1 b_2, a_0}
   \nonumber \\
   & \quad
      + \left(
         A^{c (1)}_{\alpha} + A^{b (1)}_{\alpha}
        \right)
        V^{(1)}_{a_1 a_2, a_0}
\end{align}
according to \eqref{eq:Limit-large-y-NLO} and \eqref{eq:Limit-large-y-NLO-two-flavours}.  Given \eqref{eq:Acb-NLO} and \eqref{eq:Aalpha-cb-1}, this is indeed fulfilled.
\paragraph{LO channels with observed heavy partons.}  According to \eqref{eq:Limit-small-y-NLO} and \eqref{eq:Limit-small-y-NLO-two-flavours}, the consistency of relation \eqref{eq:LO-channels-one-to-two-heavy-1} in the small-$y$ limit requires
\begin{align}
   \label{eq:Limit-small-y-LO-channels-heavy}
   V^{(2)}_{c \cbar, g} + V^{(1)}_{c \cbar, g}
      \conv{12} A^{c b (1)}_{g g}
   & \, \overset{!}{=} \,
     V^{(2)}_{c \cbar, g} + V^{(1)}_{c \cbar, g}
      \conv{12}
      \left(
         A^{c (1)}_{g g} +
         A^{b (1)}_{g g}
      \right) \,,
\end{align}
which is satisfied because of \eqref{eq:Acb-NLO}.
In the large-$y$ limit, both sides of equation \eqref{eq:LO-channels-one-to-two-heavy-1} tend to zero according to equations \eqref{eq:Limit-large-y-NLO} and \eqref{eq:Limit-large-y-NLO-two-flavours}.  An analogous analysis applies to $V^{c b (2)}_{\smash{b \bbar, g}\phantom{|}}.$
\paragraph{NLO channels with observed heavy partons.}  For definiteness, we consider again the case of observed charm quarks.

The small-$y$ behaviour given in \eqref{eq:Limit-small-y-NLO} and \eqref{eq:Limit-small-y-NLO-two-flavours} implies that
\begin{align}
   \label{eq:NLO-channels-1-small-y}
   V^{c b (2)}_{c \cbar, q} \text{ and } V^{c (2)}_{c \cbar, q}
      & \;\overset{y \to 0}{\longrightarrow}\,
         V^{(2)}_{q' \qbar'\!, q} \,,
   &
   V^{c b (2)}_{c q, q} \text{ and } V^{c (2)}_{c q, q}
      & \;\overset{y \to 0}{\longrightarrow}\,
         V^{(2)}_{q' q, q} \,,
\end{align}
whilst
\begin{align}
   \label{eq:NLO-channels-2-small-y}
   V^{c b (2)}_{c g, g}
      & \,\overset{y \to 0}{\longrightarrow}\,
         V^{(2)}_{q g, g} + V^{(1)}_{q g, q} \conv{12} A^{c b (1)}_{c g} \,,
   &
   V^{c (2)}_{c g, g}
      & \,\overset{y \to 0}{\longrightarrow}\,
         V^{(2)}_{q g, g} + V^{(1)}_{q g, q} \conv{12} A^{c (1)}_{c g} \,.
\end{align}
The consistency of the relation \eqref{eq:NLO-channels-one-to-two-flavours} with \eqref{eq:NLO-channels-1-small-y} is evident, whilst for \eqref{eq:NLO-channels-2-small-y} it requires
\begin{align}
   \label{eq:A-cg-relation}
   A^{c b (1)}_{c g} & \overset{!}{=} A^{c (1)}_{c g}
   \,.
\end{align}
This holds due to \eqref{eq:Acb-NLO} and the fact that $A^{b (1)}_{c g} = 0$.

In the large-$y$ limit, both $V^{c b (2)}_{c \cbar, q}$ and $V^{c (2)}_{c \cbar, q}$ tend to zero, whilst
\begin{align}
   V_{c a}^{c b (2)}
      & \,\overset{y \to \infty}{\longrightarrow}\,
        A^{c b (2)}_{c g} \conv{1} V^{(1)}_{g a, a} \,,
   &
   V_{c a}^{c (2)}
      & \,\overset{y \to \infty}{\longrightarrow}\,
        A^{c (2)}_{c g} \conv{1} V^{(1)}_{g a, a}
&
   & \text{ for $a = g, q$}
\end{align}
according to \eqref{eq:Limit-large-y-NLO} and \eqref{eq:Limit-large-y-NLO-two-flavours}.  The consistency of \eqref{eq:NLO-channels-one-to-two-flavours} is again ensured by the relation \eqref{eq:A-cg-relation}.
%
%
\subsection{Scale dependence}
\label{sec:NLO-kernels-RGE-two-flavours}
The analysis in section \ref{sec:NLO-kernels-RGE} can be extended to the case of two heavy flavours in a straightforward manner. Starting from
\begin{align}
   \label{eq:massive-splitting-DPD-two-flavours}
      F^{5}_{a_1 a_2}
   &= \frac{1}{\pi y^2} \sum_{a_0}
      \left(
         a_s^5 \, V^{c b (1)}_{a_1 a_2, a_0} \conv{12} f^{3}_{a_0} +
         \bigl( a_s^5 \bigr)^2 \;
         V^{c b (2)}_{a_1 a_2, a_0} \conv{12} f^{3}_{a_0}
      \right)
      + \mathcal{O}(a_s^3)
\end{align}
and adapting the steps that lead from \eqref{eq:massive-splitting-DPD} to \eqref{eq:RGE-4} and \eqref{eq:RGE-5}, one finds the following scale dependence of the massive splitting kernels with two heavy flavours:
\begin{align}
   \label{eq:RGE-4-two-flavours}
      \frac{\text{d}}{\text{d} \log \mu^2} \, V^{c b (1)}_{a_1 a_2, a_0}
   &= 0 \,,
   \\
   \label{eq:RGE-5-two-flavours}
      \frac{\text{d}}{\text{d} \log \mu^2} \, V^{c b (2)}_{a_1 a_2, a_0}
   &= \sum_{b_1} P_{a_1 b_1}^{5 (0)} \conv{1} V^{c b (1)}_{b_1 a_2, a_0} +
      \sum_{b_2} P_{a_2 b_2}^{5 (0)} \conv{2} V^{c b (1)}_{a_1 b_2, a_0}
   \nonumber\\
   & \quad
      - V^{c b (1)}_{a_1 a_2, b_0} \conv{12} P_{b_0 a_0}^{3 (0)}
      + \frac{\beta_0^{5}}{2} \, V^{c b (1)}_{a_1 a_2, a_0}\,.
\end{align}
We now show that this is fulfilled by the two-flavour kernels in \eqref{eq:LO-channels-one-to-two-1}, \eqref{eq:Vcb-gQQbar-two-flavours}, and \eqref{eq:NLO-channels-one-to-two-flavours}.
\paragraph{LO channels with observed light partons.} The scale dependence of the two-loop kernels for one heavy flavour is given by \eqref{eq:RGE-5}, and for the massless kernel with three flavours one has
\begin{align}
   \label{eq:RGE-5-massless}
   \frac{\text{d}}{\text{d} \log \mu^2} \, V^{3 (2)}_{a_1 a_2, a_0}
   &= V^{3 [2,1]}_{a_1 a_2, a_0}
\end{align}
with the r.h.s.\ given in equation \eqref{eq:V-kernel-21}.  Inserting this on the r.h.s.\ of the relation \eqref{eq:LO-channels-one-to-two-1}, one obtains
\begin{align}
   \label{eq:eq:LO-channels-one-to-two-RGE-rhs-1}
   \frac{\text{d}}{\text{d} \log \mu^2} \, V^{c b (2)}_{a_1 a_2, a_0}
   &=
      \sum_{b_1} \left(
         2 P_{a_1 b_1}^{4 (0)} -
         P_{a_1 b_1}^{3 (0)}
      \right)
      \conv{1} V^{(1)}_{b_1 a_2, a_0} +
      \sum_{b_2} \left(
         2 P_{a_2 b_2}^{4 (0)} -
         P_{a_2 b_2}^{3 (0)}
      \right)
      \conv{2} V^{(1)}_{a_1 b_2, a_0}
\nonumber \\
   & \quad {}
      - \sum_{b_0} V^{(1)}_{a_1 a_2, b_0} \conv{12} P_{b_0 a_0}^{3 (0)}
      + \frac{1}{2} \bigl( 2 \beta_0^{4} -  \beta_0^{3} \bigr) \,
        V^{(1)}_{a_1 a_2, a_0} \,.
\end{align}
Here we have used that for observed light partons the kernels $V^{Q (1)}$ on the r.h.s.\ of \eqref{eq:RGE-5} are equal to their massless counterparts $V^{(1)}$.  Since $P^{\nf (1)}$ and $\beta_0^{\nf}$ are linear functions of $\nf$, we have
\begin{align}
   \label{eq:Pab-beta0-relation}
   2 P_{a b}^{4 (0)} - P_{a b}^{3 (0)}
   &= P_{a b}^{5 (0)} \,,
&
   2 \beta_0^{4 (0)} - \beta_0^{3 (0)}
   &= \beta_0^{5 (0)} \,,
\end{align}
so that \eqn{\eqref{eq:eq:LO-channels-one-to-two-RGE-rhs-1}} reduces to
\begin{align}
   \label{eq:LO-channels-one-to-two-RGE-rhs-2}
   \frac{\text{d}}{\text{d} \log \mu^2} \, V^{c b (2)}_{a_1 a_2, a_0}
   &=  \sum_{b_1} P_{a_1 b_1}^{5 (0)} \conv{1} V^{(1)}_{b_1 a_2, a_0} +
       \sum_{b_2} P_{a_2 b_2}^{5 (0)} \conv{2} V^{(1)}_{a_1 b_2, a_0}
\nonumber \\
   & \quad {}
      - \sum_{b_0} V^{(1)}_{a_1 a_2, b_0} \conv{12} P_{b_0 a_0}^{3 (0)} +
      \frac{\beta_0^{5}}{2} \, V^{(1)}_{a_1 a_2, a_0}\,.
\end{align}
This is consistent with \eqref{eq:RGE-5-two-flavours}, because the LO kernels $V^{c b (1)}$ are equal to $V^{(1)}$ for observed massless partons.
\paragraph{LO channels with observed heavy partons.} Using \eqref{eq:RGE-5} and \eqref{eq:RGE-5-two-flavours} together with the fact that the DGLAP kernel $P^{(0)}_{q q}$ does not depend on the number of active flavours, one finds that $V_{c \cbar, g}^{c b (2)}$ in \eqref{eq:LO-channels-one-to-two-heavy-1} has the correct scale behaviour.
\paragraph{NLO channels with observed heavy partons.}  For NLO channels, the kernels $V^{Q (1)}$  and $V^{c b (1)}$ on the r.h.s.\ of \eqref{eq:RGE-5} and \eqref{eq:RGE-5-two-flavours} are zero, leaving only terms with flavour non-diagonal DGLAP kernels, which are independent of $\nf$.  This ensures the correct scale dependence of the kernels $V_{a_1 a_2, a_0}^{c b (2)}$ in \eqref{eq:NLO-channels-one-to-two-flavours}.
%
%
%
\subsection{Explicit parametrisation}
\label{sec:NLO-kernels-param-two-flavours}
An explicit parametrisation of the kernels $V^{c b (2)}$ is readily obtained from the relations in \sect{\ref{sec:NLO-kernels-two-flavours-from-one-flavour}} and the parametrisation of $V^{Q (2)}$ in \sect{\ref{sec:NLO-kernels-param}}.  For the LO channels, this gives
\begin{align}
   \label{eq:Vcb-qqbarg-explicit}
V^{c b (2)}_{q \qbar, g}
   &= V^{[2,0]}_{q \qbar, g}
      + V^{[2,1]}_{q \qbar, g} \log \frac{\mu^2}{\mu_y^2}
      + V^{I}_{q \qbar, g}(m_c) + V^{I}_{q \qbar, g}(m_b)
      + L(\mu, m_c, m_b) \, \frac{\Delta\beta_0}{2} \,
        V^{(1)}_{q \qbar, g} \,,
   \\[0.3em]
   \label{eq:Vcb-qgq-explicit}
V^{c b (2)}_{q g, q}
   &= V^{3 \ms [2,0]}_{q g, q}
      + V^{3 \ms [2,1]}_{q g, q} \, \log \frac{\mu^2}{\mu_y^2}
      + V^{I}_{q g, q}(m_c) + V^{I}_{q g, q}(m_b)
      + \tilde{k}_{11}(y, m_c, m_b) \,
        \Delta\beta_0 \, V^{\beta \ms [2,0]}_{q g, q}
   \nonumber\\
   &  \quad
      + \Bigl[ L(\mu, m_c, m_b) - \tilde{k}_{02}(y, m_c, m_b) \Bigr] \,
        \Delta\beta_0 \, V^{(1)}_{q g, q} \,,
   \\[0.3em]
   \label{eq:Vcb-ggg-explicit}
V^{c b (2)}_{g g, g}
   &= V^{3 \ms [2,0]}_{g g, g}
      + V^{3 \ms [2,1]}_{g g, g} \log \frac{\mu^2}{\mu_y^2}
      + V^{I}_{g g, g}(m_c) + V^{I}_{g g, g}(m_b)
      + \tilde{k}_{11}(y, m_c, m_b) \, \Delta\beta_0 \,
        V^{\beta \ms [2,0]}_{g g, g}
   \nonumber\\
   &  \quad
      + \biggl[ \frac{3}{2} \ms L(\mu, m_c, m_b)
         - \tilde{k}_{02}(y, m_c, m_b) \biggr] \,
            \Delta\beta_0 \, V^{(1)}_{g g, g} \,,
   \\[0.3em]
   \label{eq:Vcb-ccbarg-explicit}
V^{c b (2)}_{c \cbar, g}
   &= V^{I}_{c \cbar, g}(m_c)
      + k_{11}(y \ms m_c) \, V^{[2,0]}_{q \qbar, g}
      - k_{02}(y \ms m_c) \, V^{[2,1]}_{q \qbar, g}
   \nonumber\\
   &  \quad
      + \log \frac{\mu^2}{m_c^2} \,
      \biggl[
         P_{q q}^{(0)} \conv{1} V^{c (1)}_{c \cbar, g}(m_c)
         + P_{q q}^{(0)} \conv{2} V^{c (1)}_{c \cbar, g}(m_c)
         - V^{c (1)}_{c \cbar, g}(m_c)
            \conv{12} \widetilde{P}_{g g}^{(0)}
      \biggr]
   \nonumber \\
   &  \quad
      + L(\mu, m_c, m_b) \,
        \frac{\Delta\beta_0}{2} \, V^{c, (1)}_{c \cbar, g}(m_c) \,,
   \\[0.3em]
   \label{eq:Vcb-bbbarg-explicit}
V^{c b (2)}_{b \bbar, g}
   &= V^{I}_{b \bbar, g}(m_b)
      + k_{11}(y \ms m_b) \, V^{[2,0]}_{q \qbar, g}
      - k_{02}(y \ms m_b) \, V^{[2,1]}_{q \qbar, g}
   \nonumber\\
   &  \quad
      + \log \frac{\mu^2}{m_b^2} \,
      \biggl[
         P_{q q}^{(0)} \conv{1} V^{b (1)}_{b \bbar, g}(m_b)
         + P_{q q}^{(0)} \conv{2} V^{b (1)}_{b \bbar, g}(m_b)
         - V^{b (1)}_{b \bbar, g}(m_b)
            \conv{12} \widetilde{P}_{g g}^{(0)}
      \biggr]
   \nonumber \\
   &  \quad
      + L(\mu, m_c, m_b) \,
        \frac{\Delta\beta_0}{2} \, V^{b, (1)}_{b \bbar, g}(m_b)
   \,,
\end{align}
where we abbreviated
\begin{align}
   \label{eq:Lcb-def}
   L(\mu, m_c, m_b)
      &= \log \frac{\mu^2}{m_c^2} + \log \frac{\mu^2}{m_b^2} \,,
   \\
   \label{eq:kij-tilde-def}
   \tilde{k}_{i j}(y, m_c, m_b)
      &= k_{i j}(y \ms m_c) + k_{i j}(y \ms m_b)
   \,.
\end{align}
For definiteness, we have indicated all dependence on the heavy-quark masses and also given the expression for $V^{c b (2)}_{b \bbar, g}$.

According to \eqref{eq:NLO-channels-one-to-two-flavours}, the kernels $V^{c b (2)}$ for NLO channels are directly obtained from \eqref{eq:Vm-QQbarq-explicit} to \eqref{eq:Vm-Qqq-explicit} by appropriate substitution of parton labels and masses.

\paragraph{Large logarithms.} The analysis of logarithms in \eqref{eq:Vcb-qqbarg-explicit} to \eqref{eq:Vcb-bbbarg-explicit} is very similar to the one in \sect{\ref{sec:NLO-kernels-large-logs}}, with similar conclusions regarding the large logarithms that remain if one chooses $\mu \sim \mu_y$.  The most notable change is that for the NLO corrections from massive quark loops, which appear in all LO channels, one needs to replace $\log (\mu_y^2 / \mQ^2)$ in \eqref{eq:large-log-beta0} with the sum of logarithms $\log(\mu_y^2 / m_c^2) + \log (\mu_y^2 / m_b^2)$.

%
\graphicspath{{Figures/NLO_kernels/}}
%
%
\section{DPD sum rules and massive splitting kernels}
\label{sec:sum-rules}
In this section we will show that the number and momentum sum rules for unpolarised DPDs \cite{Gaunt:2009re, Diehl:2018kgr} imply corresponding sum rules for the massive splitting kernels $V^{Q}$.  These sum rules are easily verified for the LO results given in \eqref{eq:VQQbarg-LO} and \eqref{eq:VQ-light-LO}.  More importantly, they provide valuable constraints on the interpolating terms $V^{I}$ in our parametrisation \eqref{eq:Vm-ansatz} of the NLO kernels.  Throughout this section, it is understood that all partons are unpolarised unless stated otherwise.  As in \sect{\ref{sec:NLO-kernels-constraints}}, we consider $\nf$ light flavours along with one heavy flavour $Q$.

The DPD sum rules do not directly apply to the distributions $F(x_1, x_2, y)$ we have considered so far.  They require an integral of these distributions over all distances $\tvec{y}$ in the transverse plane, which due to the $1/y^2$ behaviour from perturbative splitting is logarithmically divergent at small $y$ if carried out naively.  In \cite{Diehl:2017kgu, Diehl:2018kgr} it was shown that the sum rules hold if this splitting singularity is renormalised in the $\msbar$ scheme (with the integral over $\tvec{y}$ carried out in $2 - 2\epsilon$ dimensions before modified minimal subtraction of the resulting pole at $\epsilon = 0$).  Moreover, a matching formula was derived that connects these distributions $F^{\msbar}(x_1, x_2)$ to the integral of $F(x_1, x_2, y)$ over $\tvec{y}$ in two dimensions, evaluated with a lower cutoff on $y$.  The corresponding matching kernels were computed in \cite{Diehl:2018kgr} at LO and in \cite{Diehl:2019rdh} at NLO accuracy.  For DPDs with $\nf + 1$ active flavours, the matching relation reads
\begin{align}
   \label{eq:MSbarDPD-1}
      F_{a_1 a_2}^{\nf + 1, \msbar}(x_1, x_2; \mu)
   &= \int\limits_{b_0/\nu}^\infty \!\! \text{d}^2 y \,
      F^{\nf + 1}_{a_1 a_2}(x_1, x_2, y; \mu)
      + \sum_{a_0}
         U^{\nf + 1}_{a_1 a_2, a_0}(\nu, \mu)
         \conv{12} f^{\nf + 1}_{a_0}(\mu)
   \,,
\end{align}
where we use the notation
\begin{align}
   \label{eq:y-integral-notation}
   \int\limits_{y_0}^{y_1} \! \text{d}^2 y
   &= 2 \pi \int\limits_{y_0}^{y_1} \! \text{d}y \, y
\end{align}
for an integral over a ring in the $\tvec{y}$ plane.  The $\nu$ dependence cancels between the two terms on the r.h.s.\ of \eqref{eq:MSbarDPD-1}, up to power corrections in $\Lambda / \nu$, where $\Lambda$ is a hadronic scale.  We therefore require $\nu \gg \Lambda$.
In line with the convention in \eqref{eq:arguments-FnF}, we will suppress arguments and write
\begin{align}
   \label{eq:arguments-FMSbar}
      F^{\nf, \msbar}_{a_1 a_2}
   &  \equiv F^{\nf, \msbar}_{a_1 a_2}(x_1, x_2; \mu)
\end{align}
in the following.  The perturbative expansion of the matching kernels $U$ reads
\begin{align}
   \label{eq:U-general-form}
      U^{\nf}_{a_1 a_2, a_0}(z_1, z_2; \nu, \mu)
   &= \sum_{k = 1}^\infty \bigl[ a_s^{\nf}(\mu) \bigr]^k \;
      U^{\nf (k)}_{a_1 a_2, a_0}(z_1, z_2; \nu, \mu)
\intertext{with}
   \label{eq:U-logarithms}
   U^{\nf (k)}_{a_1 a_2, a_0}(z_1, z_2; \nu, \mu)
   & = \sum_{\ell = 0}^{k - 1} \log^{\ms \ell \bs} \frac{\mu^2}{\nu^2} \;
      U^{\nf \ms [k, \ell\ms]}_{a_1 a_2, a_0}(z_1, z_2)
   \,.
\end{align}
In the following derivations, the kernels $U$ will be needed at different values of $\nu$ but always the same $\mu$.  Using that the $\nu$ dependence appears only via the ratio $\mu/\nu$ in \eqref{eq:U-logarithms}, we write
\begin{align}
   \label{eq:arguments-UnF}
      U^{\nf}_{a_1 a_2, a_0} (\mu/\nu)
   &  \equiv U^{\nf}_{a_1 a_2, a_0}(z_1, z_2; \nu, \mu) \,,
\end{align}
where the separate dependence on $\mu$, due to the factors of $a_s(\mu)$ in \eqref{eq:U-general-form}, is suppressed on the l.h.s.
The coefficients $U^{\nf [k, \ell\ms]}$ are related to $V^{\nf \ms [k, \ell - 1]}$ via
\begin{align}
   \label{eq:relation-U-V}
      U^{\nf \ms [k, \ell\ms]}_{a_1 a_2, a_0}
   &= \frac{1}{\ell} \, V^{\nf \ms [k, \ell - 1]}_{a_1 a_2, a_0}
   && \text{for $\ell \ge 1$.}
\end{align}
Details about the $U$ kernels are given in equations (79), (116) and (140) of reference \cite{Diehl:2019rdh}.  In the following, we need that
the coefficients $U^{[1, 0]}$ are $\nf$ independent, whilst $U^{\nf \ms [2,0]}$ has the same type of $\nf$ dependence as $V^{\nf \ms [2,0]}$.  In analogy to \eqref{eq:V-kernel-nf-dep}, we can hence write
\begin{align}
   \label{eq:U-kernel-nf-dep}
   U^{\nf [2, 0]}_{a_1 a_2, a_0}
      &= \widetilde{U}^{[2, 0]}_{a_1 a_2, a_0}
         + \beta_0^{\nf} \ms U^{\beta \ms [2, 0]}_{a_1 a_2, a_0}
   & \text{ for $(a_1 a_2, a_0) = (q g, q)$ and $(g g, g)$,}
\end{align}
where $U^{\beta \ms [2, 0]}$ is proportional to $\delta(1 - z_1 - z_2)$.
The kernels in the matching equation \eqref{eq:MSbarDPD-1} have been computed only for massless quarks.  They are associated with physics at momentum scales above $\nu$, so that in our setting with a heavy flavour $Q$ we can use them if $\nu \gg \mQ$.  This is ensured by taking
\begin{align}
   \label{eq:nu-choice}
\nu &= \beta \ms \mQ
&
\text{ with $\beta \gg 1$.}
\end{align}

For DPDs with $\nf + 1$ active flavours, the momentum sum rule reads
\begin{align}
   \label{eq:momsum-1}
      \sum_{a_2} \int\limits_2 X_2 \,
      F_{a_1 a_2}^{\nf + 1, \msbar}
   &= (1 - X)\, f^{\nf + 1}_{a_1} \,,
\end{align}
and the number sum rule is given by
\begin{align}
   \label{eq:numsum-1}
      \int\limits_2
      F^{\nf + 1, \msbar}_{a_1 a_{2 v}}
   &= \bigl( N_{a_{2 v}} + \delta_{a_1 \abar_2} - \delta_{a_1 a_2} \bigr) \,
      f^{\nf + 1}_{a_1} \,,
\end{align}
where $N_{a_{2 v}}$ is the number of valence quarks with flavour $a_2$ in the hadron.  The parton label $a_{2 v}$ denotes the difference between $a_2$ and $\abar_2$, i.e.\ we write
$F_{a_1 a_{2 v}} = F_{a_1 a_2} - F_{a_1 \abar_2}$
for DPDs and likewise for various kernels appearing below.  We also use the shorthand notation introduced in reference \cite{Diehl:2018kgr}, where integrals over momentum fractions are denoted by
\begin{align}
   \label{eq:mom-frac-int}
      \int C
   &= \int \text{d} x \, C(x)
   \,,
   && \int\limits_2 D
    = \int \text{d} x_2 \, D(x_1, x_2)
\end{align}
for functions of one or two momentum fractions. The multiplication with a power of a momentum fraction is indicated by operators $X^n$ and $X_2^n$, which act as
\begin{align}
   \label{eq:mom-frac-mult}
      (X^n C)(x)
   &= x^n C(x) \,,
   && (X_2^n D)(x_1, x_2) = x_2^n \, D(x_1, x_2) \,.
\end{align}
The integration boundaries in \eqref{eq:momsum-1} and \eqref{eq:numsum-1} are readily inferred from the support properties of the DPDs, and one finds that the $x_2$ integrations run from $0$ to $1 - x_1$ in both sum rules. It is furthermore understood that the left- and right-hand sides of \eqref{eq:momsum-1} and \eqref{eq:numsum-1} still depend on the momentum fraction $x_1$, which is not written out explicitly.

In the following calculations, we will use the identities
\begin{align}
   \label{eq:int-conv-2}
   \int\limits_2 X_2^n \ms (C \conv{2} D)
   &= \biggl( \int X^n  C \biggr) \,
      \Biggl( \ms \int\limits_2 X_2^n  D \Biggr) \,,
\\
   \label{eq:int-conv-12}
   \int\limits_2 X_2^n \ms (D \conv{12} C)
   &= \Biggl( \ms \int\limits_2 X_2^n D \Biggr) \conv{} (X^n C)
\end{align}
derived in \sect{6.1} of \cite{Diehl:2018kgr}, where it is understood that the overall expressions depend on $x_1$ in both cases.
%
%
\subsection{Sum rules for the massive splitting kernels}
\label{sec:NLO-kernels-sumrules}
The sum rules just spelled out involve an integral over DPDs from the perturbative to the nonperturbative region of $y$.  To make contact with the DPD splitting kernels, we split the integration over $y$ into the two intervals $[y_{\beta}, y_{\alpha}]$ and $[y_\alpha, \infty]$, where
\begin{align}
   \label{eq:y-alpha-beta-def}
y_{\beta} &= b_0 / (\beta \ms \mQ) \,,
&
y_{\alpha} &= b_0 / (\alpha \ms \mQ)
\end{align}
with $\beta \gg 1$ and $\alpha \ll 1$, and with $\alpha \ms \mQ$ being in the perturbative region.\footnote{%
These conditions on $\alpha$ cannot be satisfied for $\mQ = m_c$.  Since the aim of this section is to derive constraints on the perturbative kernels $V^{Q}$ for a generic heavy flavour $Q$, we can consider a value $\mQ$ that is as large as needed for our arguments to hold.}
The lower limit $y_{\beta}$ is equal to $\nu / b_0$ with $\nu$ given in \eqref{eq:nu-choice}.  A number of integrals over $y$ that will be used in the following is collected in \tab{\ref{tab:sumrule-integrals}}.
\begin{table}[t]
   \centering
   \begin{tabular}{c c c c c c}
      \toprule
      $f(y)$
      & $1$
      & $\log ( \mQ^2 / \mu_y^2 )$
      & $k_{11}(y \ms \mQ)$
      & $k_{02}(y \ms \mQ)$
      & $k_{00}(y \ms \mQ)$
      \\[0.5em]
      $I[f]$
      & $-2 \log\alpha + 2 \log\beta$
      & $2 \log^2 \alpha - 2 \log^2 \beta$
      & $- 1 + 2 \log\beta$
      & $1 + 2 \log^2\beta$
      & $1$
      \\
      \bottomrule
   \end{tabular}
   \caption{\label{tab:sumrule-integrals} Integrals $I[f] = \int_{y_{\beta}}^{y_{\alpha}} \text{d}^2 y \; f(y) \big/ (\pi y^2)$ of functions $f(y)$ that appear in our parametrisation of the splitting kernels.  The integration limits are defined in \protect\eqref{eq:y-alpha-beta-def}, and the limits $\alpha \ll 1$ and $\beta \gg 1$ have been taken.}
\end{table}

The $y$ integral on the r.h.s.\ of the matching equation \eqref{eq:MSbarDPD-1} can now be written as
\begin{align}
   \label{eq:MSbarDPD-2}
      \int\limits_{y_\beta}^\infty \text{d}^2 y \,
      F^{\nf + 1}_{a_1 a_2}
   &= \int\limits_{y_\beta}^{y_\alpha}
      \text{d}^2 y \frac{1}{\pi y^2} \sum_{a_0} V^{Q, \nf}_{a_1 a_2, a_0}
      \conv{12} f^{\nf}_{a_0}
      + \sum_{b_1, b_2}  A^{Q, \nf}_{a_1 b_1}
      \conv{1} A^{Q, \nf}_{a_2 b_2} \conv{2}
      \int\limits_{y_\alpha}^\infty
      \text{d}^2 y \, F^{\nf}_{b_1 b_2}
      \,.
\end{align}
For $y_\beta \le y \le y_\alpha$, we have used the massive DPD splitting formula, whereas for $y > y_{\alpha}$ we used DPD flavour matching to relate $F^{\nf + 1}$ to $F^{\nf}$.  This is in line with the scheme we presented in \sect{\ref{sec:one-massive}}, except that for large $y$ one cannot compute $F^{\nf}$ using a splitting formula.  We now use the analogue of the matching formula \eqref{eq:MSbarDPD-1} for $\nf$ flavours and take $\nu = \alpha \ms \mQ$, which satisfies $\nu \gg \Lambda$ by assumption.  This gives
\begin{align}
   \label{eq:MSbarDPD-3}
      \int_{y_{\alpha}}^\infty \!\! \text{d}^2 y \, F^{\nf}_{b_1 b_2}
   &= F_{b_1 b_2}^{\nf, \msbar}
      - \sum_{a_0} U^{\nf}_{b_1 b_2, a_0}\Bigl(\frac{\mu}{\alpha \ms \mQ}\Bigr)
      \conv{12} f^{\nf}_{a_0} \,.
\end{align}
Inserting \eqref{eq:MSbarDPD-2} and \eqref{eq:MSbarDPD-3} into \eqref{eq:MSbarDPD-1}, we obtain our master formula
\begin{align}
   \label{eq:MSbarDPD-4}
   &   F_{a_1 a_2}^{\nf + 1, \msbar}
   \nonumber \\[0.3em]
   & \quad
      = \sum_{a_0} \Biggl[
      \int_{y_{\beta}}^{y_{\alpha}} \!\! \text{d}^2 y
      \frac{1}{\pi y^2} \,
         V^{Q, \nf}_{a_1 a_2, a_0} \conv{12} f^{\nf}_{a_0}
      + \sum_{b_0}
         U^{\nf + 1}_{a_1 a_2, b_0}(r_\beta)
         \conv{12} A^{Q, \nf}_{b_0 a_0} \Biggr]
      \conv{12} f^{\nf}_{a_0}
   \nonumber\\[0.2em]
   & \qquad
      + \sum_{b_1, b_2} A^{Q\, \nf}_{a_1 b_1}
      \conv{1} A^{Q, \nf}_{a_2 b_2} \conv{2}
      \biggl[
        F_{b_1 b_2}^{\nf, \msbar}
        - \sum_{a_0}
        U^{\nf}_{b_1 b_2, a_0}(r_\alpha)
            \conv{12} f^{\nf}_{a_0}
      \biggr] \,,
\end{align}
where for the term going with $U^{\nf + 1}$ we have expressed $f^{\nf + 1}$ in terms of $f^{\nf}$ and used the relation \eqref{eq:conv-12-relation} for a triple convolution.  For the argument $\mu / \nu$ of the $U$ kernels (see \eqref{eq:arguments-UnF}), we have introduced the notation
\begin{align}
   \label{eq:r-alpha-beta-def}
r_{\beta} &= \mu / (\beta \ms \mQ) \,,
&
r_{\alpha} &= \mu / (\alpha \ms \mQ)
\end{align}
in analogy to \eqref{eq:y-alpha-beta-def}.
Using the DPD sum rules for both $F^{\nf + 1, \msbar}$ and $F^{\nf, \msbar}$ in \eqref{eq:MSbarDPD-4}, we will obtain sum rules for the massive splitting kernels $V^{Q, \nf}$.
%
%
\subsubsection{Momentum sum rule}
When inserting \eqref{eq:MSbarDPD-4} into the momentum sum rule \eqref{eq:momsum-1} for a DPD with $\nf + 1$ active flavours, one needs the expression
\begin{align}
   \label{eq:momsum-aux}
   M^{\nf}_{b_1}(x_1)
   &= \sum_{a_2, b_2} \int\limits_2 X_2 \, \Bigl( A^{Q, \nf}_{a_2 b_2}
      \conv{2} D^{\nf}_{b_1 b_2} \Bigr)
    = \sum_{a_2, b_2} \Biggl( \int X A^{Q, \nf}_{a_2 b_2} \Biggr) \,
      \Biggl( \ms \int\limits_2 X_2 D^{\nf}_{b_1 b_2} \Biggr)
   \,,
\end{align}
where $D^{\nf}$ is the expression in square brackets on the last line of \eqref{eq:MSbarDPD-4}, and where we used \eqref{eq:int-conv-2} in the last step.  We now use the momentum sum rule for the PDF matching kernels,
\begin{align}
   \sum_{a_2} \int\limits X A^{Q, \nf}_{a_2 b_2}
   &= \delta^{\nf}_{b_2 \ms l} \,,
\end{align}
which ensures the consistency of the momentum sum rules for PDFs with $\nf$ and $\nf + 1$ flavours and can readily be checked for the LO and NLO expressions \eqref{eq:A-LO} and \eqref{eq:A-NLO}.  This gives
\begin{align}
   \label{eq:momsum-aux-2}
   M^{\nf}_{b_1}(x_1)
   &= \sum_{b_2} \int\limits_2 X_2 \, D^{\nf}_{b_1 b_2}
    = \sum_{b_2} \int\limits_2 X_2 \, F_{b_1 b_2}^{\nf, \msbar}
        - \sum_{b_2, a_0} \int\limits_2 X_2 \,
            \biggl( U^{\nf}_{b_1 b_2, a_0}(r_\alpha)
            \conv{12} f^{\nf}_{a_0} \biggr)
   \,,
\end{align}
where we could drop $\delta^{\nf}_{b_2 \ms l}$ because $D^{\nf}$ is constructed from quantities with $\nf$ flavours.  The first term on the r.h.s.\ of \eqref{eq:momsum-aux-2} is equal to $(1 - X) \, f^{\nf}_{b_1}$ according to the DPD momentum sum rule for $\nf$ flavours.  Putting everything together and using the relations \eqref{eq:int-conv-2} and \eqref{eq:int-conv-12} for integrals over convolution products, we obtain
\begin{align}
\label{eq:momsum-2}
   & \sum_{a_2}
      \int\limits_2 X_2 \, F_{a_1 a_2}^{\nf + 1, \msbar}
   = \sum_{a_0, a_2} \Biggl( \ms
         \int\limits_2 X_2 \int_{y_{\beta}}^{y_{\alpha}} \!\! \text{d}^2 y
         \frac{1}{\pi y^2} V^{Q}_{a_1 a_2, a_0}
      \Biggr)
      \conv{} (X f^{\nf}_{a_0})
   \nonumber \\
   & \quad
      + \sum_{a_0, a_2, b_0} \Biggl( \ms
         \int\limits_2 X_2 \,
         U^{\nf + 1}_{a_1 a_2, b_0}(r_\beta)
         \Biggr)
      \conv{} \Bigl( X A^{Q, \nf}_{b_0 a_0} \Bigr)
      \conv{} (X f^{\nf}_{a_0})
   \nonumber \\
   & \quad
     + \sum_{b_1} A^{Q, \nf}_{a_1 b_1} \conv{1} (1 - X) \ms f^{\nf}_{b_1}
     - \sum_{a_0, b_1, b_2} A^{Q, \nf}_{a_1 b_1} \conv{1}
      \Biggl( \ms
         \int\limits_2 X_2 \,
         U^{\nf}_{b_1 b_2, a_0}(r_\alpha)
      \Biggr) \conv{} (X f^{\nf}_{a_0}) \,.
\end{align}
The r.h.s.\ of the DPD momentum sum rule \eqref{eq:momsum-1} for $\nf + 1$ flavours can be expressed in terms of an $\nf$ flavour PDF as
\begin{align}
   \label{eq:momsum-3}
      \sum_{a_0} (1 - X) (A^{Q}_{a_1 a_0} \otimes f^{\nf}_{a_0})
   &= \sum_{a_0} A^{Q}_{a_1 a_0} \otimes f^{\nf}_{a_0}
      - \sum_{a_0} (X A^{Q}_{a_1 a_0}) \otimes (X f^{\nf}_{a_0})
\end{align}
and is to be evaluated at momentum fraction $x_1$.  Using that \eqref{eq:momsum-2} and \eqref{eq:momsum-3} must be equal for any set of PDFs $f^{\nf}$, we obtain the desired momentum sum rule for the heavy-quark kernels $\smash{V^{Q}_{a_1 a_2, a_0}}$:
\begin{align}
   \label{eq:momsum-7}
   &  \sum_{a_2} \int\limits_2 X_2 \int_{y_{\beta}}^{y_{\alpha}}
      \!\! \text{d}^2 y \frac{1}{\pi y^2} \,
      V^{Q}_{a_1 a_2, a_0}
   = (1 - X) \ms A^{Q, \nf}_{a_1 a_0}
   \nonumber \\
   & \qquad {}
   + \sum_{b_1, a_2} A^{Q, \nf}_{a_1 b_1} \otimes
      \Biggl( \ms
         \int\limits_2 X_2 \, U^{\nf}_{b_1 a_2, a_0}(r_\alpha)
      \Biggr)
   - \sum_{a_2, b_0} \,
      \Biggl( \ms
         \int\limits_2 X_2 \, U^{\nf + 1}_{a_1 a_2, b_0}(r_\beta)
      \Biggr)
      \conv{} \Bigl( X A^{Q, \nf}_{b_0 a_0} \Bigr)
   \,.
\end{align}

At LO, the only heavy-quark kernel is $V^{Q (1)}_{Q \smash{\Qbar}, g}$, which appears in the sum rule for $a_0 = g$ and $a_1 = Q$.  The term of order $a_s^{\nf + 1}$ in \eqref{eq:momsum-7} reads
\begin{align}
   \label{eq:momsum-7-LO}
      \int\limits_2 X_2
      \int_{y_{\beta}}^{y_{\alpha}} \!\! \text{d}^2 y \frac{1}{\pi y^2} \,
      V^{Q (1)}_{Q \Qbar, g}
   &= (1 - X) \ms A^{Q (1)}_{Q g} -
   \int\limits_2 X_2 \, U^{(1)}_{q \qbar, g}(r_\beta)
\end{align}
for this channel.  With the explicit form \eqref{eq:AQg1} of $A^{Q (1)}_{Q g}$ and with
\begin{align}
   \label{eq:Uqqbarg1}
      U^{(1)}_{q \qbar, g}\Bigl( \frac{\mu}{\nu} \Bigr)
   &= \delta(1 - z_1 - z_2) \, T_F
      \left[
         (z_1^2 + z_2^2) \log\frac{\mu^2}{\nu^2} - 2 z_1 z_2 \ms
      \right] \,,
\end{align}
the r.h.s.\ of \eqref{eq:momsum-7-LO} reduces to
\begin{align}
      2 \, T_F \, (1 - z_1)
      \Bigl[
         \bigl( z_1^2 + (1 - z_1)^2 \ms \bigr) \log \alpha
         + (1 - z_1) z_1 \ms
      \Bigr] \,.
\end{align}
For the l.h.s.\ of \eqn{\eqref{eq:momsum-7-LO}} one obtains the same expression, using the result \eqref{eq:VQQbarg-LO} for the LO splitting kernel and the relevant integral in \tab{\ref{tab:sumrule-integrals}}.  This confirms the validity of the momentum sum rules for the massive splitting kernels at LO.

The term of order $(a_s^{\nf + 1})^{\ms 2}$ in \eqn{\eqref{eq:momsum-7}} is given by
\begin{align}
   \label{eq:momsum-8}
   &  \sum_{a_2} \int\limits_2 X_2 \int_{y_{\beta}}^{y_{\alpha}}
      \!\! \text{d}^2 y \, V^{Q, \nf (2)}_{a_1 a_2, a_0}
   = (1 - X) \ms A_{a_1 a_0}^{Q (2)}
   \nonumber \\
   & \quad {}
      + \sum_{a_2} \int\limits_2 X_2 \,
         \Bigl[\ms
            U^{\nf (2)}_{a_1 a_2, a_0}(r_\alpha)
            - U^{\nf + 1 (2)}_{a_1 a_2, a_0}(r_\beta)
         \Bigr]
      + A_\alpha^{(1)} \, \sum_{a_2}
         \int\limits_2 X_2 \, U^{(1)}_{a_1 a_2, a_0}(r_\alpha)
   \nonumber\\
   & \quad {}
      + \sum_{b_1, a_2} A_{a_1 b_1}^{Q (1)} \conv{1}
      \Biggl( \ms
         \int\limits_2 X_2 \, U^{(1)}_{b_1 a_2, a_0}(r_\alpha)
      \Biggr)
      - \sum_{a_2, b_0} \,
         \Biggl( \ms
            \int\limits_2 X_2 \, U^{(1)}_{a_1 a_2, b_0}(r_\beta)
         \Biggr)
         \conv{} \Bigl( X A_{b_0 a_0}^{Q (1)} \Bigr)
\end{align}
where we have rewritten the perturbative expansion \eqref{eq:U-general-form} of $U^{\nf}$ in terms of $a_s^{\nf + 1}$ using the matching equation \eqref{eq:a-matching-expanded} for the strong coupling.  Since all quantities on its r.h.s.\ are known, \eqref{eq:momsum-8} provides a non-trivial constraint on the massive two-loop splitting kernels.
%
%
\subsubsection{Number sum rule}

In analogy to \eqref{eq:momsum-aux}, we now need the expression
\begin{align}
   \label{eq:numsum-aux}
   N^{\nf}_{b_1 a_{2 v}}
   &= \sum_{b_2} \int\limits_2  A^{Q, \nf}_{a_{2 v}\ms b_2}
      \conv{2} D^{\nf}_{b_1 b_2}
    = \sum_{b_2} \Biggl( \int A^{Q, \nf}_{a_{2 v}\ms b_2} \Biggr) \,
      \Biggl( \ms \int\limits_2 D^{\nf}_{b_1 b_2} \Biggr)
\end{align}
where $D^{\nf}$ denotes again the expression in square brackets on the last line of \eqref{eq:MSbarDPD-4}.  The number sum rule for the PDF matching kernels reads
\begin{align}
   \int\limits A^{Q, \nf}_{a_{2 v}\ms b_2}
   &= \delta^{\nf}_{b_2 \ms l} \,
      \bigl( \delta_{a_2\ms b_2} - \delta_{\abar_2 b_2} \bigr) \,,
\end{align}
which is easily verified for the LO and NLO expressions \eqref{eq:A-LO} and \eqref{eq:A-NLO}.  Inserting this in \eqref{eq:numsum-aux} and performing the sum over $b_2$, we get
\begin{align}
   N^{\nf}_{b_1 a_{2 v}}(x_1)
   &= \int\limits_2 D^{\nf}_{b_1 a_{2 v}}
    = \int\limits_2 F^{\nf, \msbar}_{b_1 a_{2 v}}
       - \sum_{a_0} \int\limits_2 \,
            \biggl( U^{\nf}_{b_1 a_{2 v}, a_0}(r_\alpha)
            \conv{12} f^{\nf}_{a_0} \biggr)
   \,.
\end{align}
The first term on the r.h.s.\ is equal to $\bigl( N_{a_{2 v}} + \delta_{b_1 \abar_2} - \delta_{b_1 a_2} \bigr) \, f^{\nf}_{b_1}$ by virtue of the DPD number sum rule for $\nf$ flavours.  With this, we obtain
\begin{align}
\label{eq:numsum-2}
   & \int\limits_2 F_{a_1 a_{2 v}}^{\nf + 1, \msbar}
   = \sum_{a_0} \Biggl( \ms
         \int\limits_2 \int_{y_{\beta}}^{y_{\alpha}} \!\!
         \frac{1}{\pi y^2} V^{Q}_{a_1 a_{2 v}, a_0}
      \Biggr)
      \conv{} f^{\nf}_{a_0}
   \nonumber \\
   & \quad
      + \sum_{a_0, b_0} \Biggl( \ms
         \int\limits_2 \,
         U^{\nf + 1}_{a_1 a_{2 v}, b_0}(r_\beta)
         \Biggr)
      \conv{} A^{Q, \nf}_{b_0 a_0}
      \conv{} f^{\nf}_{a_0}
   \nonumber \\
   & \quad
     + \sum_{b_1}
      \bigl( N_{a_{2 v}} + \delta_{b_1 \abar_2} - \delta_{b_1 a_2} \bigr) \,
      A^{Q, \nf}_{a_1 b_1} \conv{1}
      f^{\nf}_{b_1}
     - \sum_{a_0, b_1} A^{Q, \nf}_{a_1 b_1} \conv{1}
      \Biggl( \ms
         \int\limits_2
         U^{\nf}_{b_1 a_{2 v}, a_0}(r_\alpha)
      \Biggr) \conv{} f^{\nf}_{a_0} \,.
\end{align}
Expressing the r.h.s.\ of the number sum rule \eqref{eq:numsum-1} for $\nf + 1$ flavours in terms of a $\nf$ flavour PDF, we get
\begin{align}
  \label{eq:numsum-5}
    \bigl( N_{a_{2 v}} + \delta_{a_1 \abar_2} - \delta_{a_1 a_2} \bigr) \,
    \sum_{a_0}
    A^{Q, \nf}_{a_1 a_0} \otimes f^{\nf}_{a_0} \,.
\end{align}
Using that \eqref{eq:numsum-2} and \eqref{eq:numsum-5} must be equal, we obtain the number sum rule for the massive splitting kernels:
\begin{align}
   \label{eq:numsum-7}
   &  \int\limits_2 \int_{y_{\beta}}^{y_{\alpha}} \!\!
      \text{d}^2 y
      \frac{1}{\pi y^2} \, V^{Q, \nf}_{a_1 {a_2}_v, a_0}
   = \bigl( \delta_{a_1 \abar_2} - \delta_{a_1 a_2}
          - \delta_{a_2 \abar_0} + \delta_{a_2 a_0} \bigr) \,
      A^{Q, \nf}_{a_1 a_0}
   \nonumber\\
   &  \qquad
      + \sum_{b_1} A^{Q, \nf}_{a_1 b_1} \otimes
      \Biggl( \ms
         \int\limits_2 U^{\nf}_{b_1 {a_2}_v, a_0}(r_\alpha)
      \Biggr)
      - \sum_{b_2} \,
         \Biggl( \ms
         \int\limits_2 U^{\nf + 1}_{a_1 {a_2}_v, b_0}(r_\beta)
         \Biggr)
         \conv{} A^{Q, \nf}_{b_0 a_0}
   \,.
\end{align}

Let us explicitly verify this at LO for the kernel $V^{Q (1)}_{Q Q_v, g} = {}- V^{Q (1)}_{Q \smash{\Qbar}, g}$.  In this case, the term of order $a_s^{\nf + 1}$ in \eqref{eq:numsum-7} reads
\begin{align}
   \label{eq:numsum-7-LO}
      \int\limits_2 \int_{y_{\beta}}^{y_{\alpha}} \!\!
      \text{d}^2 y
      \frac{1}{\pi y^2} \, V^{Q (1)}_{Q \Qbar, g}
   &= A^{Q (1)}_{Q g} -
      \int\limits_2 U^{\nf + 1 (1)}_{q \qbar, g}(r_\beta) \,.
\end{align}
Using the expression \eqref{eq:VQQbarg-LO} of the massive splitting kernel together with \eqref{eq:AQg1} and \eqref{eq:Uqqbarg1}, one finds that both sides of this equation evaluate to
\begin{align}
      -2 \ms T_F
      \Bigl[
         \big( z_1^2 + z_2^2 \bigr) \log \alpha
         - z_1 z_2
      \Bigr] \,.
\end{align}
The term of order $(a_s^{\nf + 1})^{\ms 2}$ in \eqn{\eqref{eq:numsum-7}} reads
\begin{align}
   \label{eq:numsum-8}
   &  \int\limits_2 \int_{y_{\beta}}^{y_{\alpha}} \!\!
      \text{d}^2 y
      \frac{1}{\pi y^2} \, V^{Q, \nf (2)}_{a_1 {a_2}_v, a_0}
   = \bigl( \delta_{a_1 \abar_2} - \delta_{a_1 a_2}
          - \delta_{a_2 \abar_0} + \delta_{a_2 a_0} \bigr) \,
      A^{Q (2)}_{a_1 a_0}
   \nonumber\\
   &  \qquad
   + \int\limits_2
      \Bigl[\ms
         U^{\nf (2)}_{a_1 {a_2}_v, a_0}(r_\alpha)
         - U^{\nf + 1 (2)}_{a_1 {a_2}_v, a_0}(r_\beta)
      \ms\Bigr]
   + A^{(1)}_{\alpha}
         \int\limits_2 U^{(1)}_{a_1 {a_2}_v, a_0}(r_\alpha)
   \nonumber\\
   &  \qquad {}
   + \sum_{b_1} A^{Q (1)}_{a_1 b_1} \otimes
      \Biggl( \ms
         \int\limits_2 U^{(1)}_{b_1 {a_2}_v, a_0}(r_\alpha)
      \Biggr)
   - \sum_{b_2} \,
         \Biggl( \ms
            \int\limits_2 U^{(1)}_{a_1 {a_2}_v, b_0}(r_\beta)
         \Biggr)
         \conv{} A^{Q (1)}_{b_0 a_0}
\end{align}
and provides another constraint on the massive NLO splitting kernels.
%
%
\subsection{Sum rules for the interpolating function \texorpdfstring{$V^{I}$}{VnFI}}
\label{sec:interpolation-sumrules}%

Let us now rewrite the sum rules \eqref{eq:momsum-8} and \eqref{eq:numsum-8} for the two-loop kernels in terms of the interpolating term $V^{I}$ in our parametrisation \eqref{eq:Vm-ansatz}.  To this end, it is sufficient to evaluate these sum rules at the scale $\mu = m_Q$, where they simplify considerably.  Using \eqref{eq:Vm-ansatz} and the integrals in \tab{\ref{tab:sumrule-integrals}}, we find that the $y$ integral needed in the sum rules reads
\begin{align}
   \label{eq:Vm-yint}
   &  \int_{y_\beta}^{y_\alpha} \!\! \text{d}^2 y
      \frac{1}{\pi y^2} \, V^{Q, \nf (2)}_{a_1 a_2, a_0}(\mu = \mQ)
   =  (1 - 2 \log\alpha) \, V^{\nf [2,0]}_{a_1 a_2, a_0}
    + (1 + 2 \log^2\alpha) \, V^{\nf [2,1]}_{a_1 a_2, a_0}
   \nonumber\\
   &  \qquad\quad {}
   - (1 - 2 \log\beta) \, V^{\nf + 1 [2,0]}_{a_1 a_2, a_0}
   - (1 + 2 \log^2\beta) \, V^{\nf + 1 [2,1]}_{a_1 a_2, a_0}
   + v^{I}_{a_1 a_2, a_0}
   \,,
\end{align}
where we defined
\begin{align}
   \label{eq:VI-yint}
      \int_{y_\beta}^{y_\alpha} \!\! \text{d}^2 y \,
      \frac{1}{\pi y^2}\, V^{I}_{a_1 a_2, a_0}(z_1, z_2, y \ms \mQ)
   &= v^{I}_{a_1 a_2, a_0}(z_1, z_2)
   \,.
\end{align}
Since $V^{I}$ tends to zero for both $y\to 0$ and $y\to \infty$, the $y$ integral in \eqref{eq:VI-yint} is finite for $\alpha \to 0$ and $\beta \to \infty$.  \rev{The r.h.s.\ of \eqn{\eqref{eq:VI-yint}} is evaluated in this limit, as are the integrals in \tab{\ref{tab:sumrule-integrals}}.}
%
%
\paragraph{Momentum sum rule.}
For $\mu = \mQ$, the momentum sum rule \eqref{eq:momsum-8} simplifies to
\begin{align}
   \label{eq:momsum-9}
   &  \sum_{a_2} \int\limits_2 X_2
      \int_{y_\beta}^{y_\alpha} \!\! \text{d}^2 y
      \frac{1}{\pi y^2} \, V^{Q (2)}_{a_1 a_2, a_0}(\mu = \mQ)
   = (1 - X) \ms A_{a_1 a_0}^{Q [2,0]}
   \nonumber \\
   & \qquad {}
       + \sum_{a_2} \int\limits_2 X_2 \,
         \Bigl[\ms
            U^{\nf (2)}_{a_1 a_2, a_0}(\alpha^{-1})
            - U^{\nf + 1 (2)}_{a_1 a_2, a_0}(\beta^{-1})
         \Bigr]
   \,.
\end{align}
Here we have replaced the two-loop PDF matching kernel $A^{Q (2)}$ with its non-logarithmic part $A^{Q [2,0]}$, given that the other terms in \eqref{eq:A-general-logs} vanish for $\mu = \mQ$.  The one-loop matching kernels $A^{(1)}_{\smash{\alpha}}$ and $A^{Q (1)}$ are zero at that point.

Inserting \eqref{eq:Vm-yint} for the $y$ integral on the l.h.s.\ of \eqn{\eqref{eq:momsum-9}} and the decomposition \eqref{eq:U-logarithms} for the two-loop $U$ kernels on its r.h.s., we obtain
\begin{align}
   \label{eq:momsum-10}
   &  \sum_{a_2} \int\limits_2 X_2 \,
      \Bigl[
         (1 - 2 \log\alpha) \, V^{\nf [2,0]}_{a_1 a_2, a_0} +
         (1 + 2 \log^2\alpha) \, V^{\nf [2,1]}_{a_1 a_2, a_0}
   \nonumber\\
   &  \hphantom{\sum_{a_2} \int X_2 \Bigl[} {}
      - (1 - 2 \log\beta) V^{\nf + 1 [2,0]}_{a_1 a_2, a_0}
      - (1 + 2 \log^2\beta) V^{\nf + 1 [2,1]}_{a_1 a_2, a_0}
      + v^{I}_{a_1 a_2, a_0}
      \Bigr]
   \nonumber\\[0.5em]
   & \quad
      = (1 - X) \ms A_{a_1 a_0}^{Q [2,0]}
      + \sum_{a_2} \int\limits_2 X_2
      \left(
         U^{\nf [2,0]}_{a_1 a_2, a_0} -
         2 \log\alpha \; U^{\nf [2,1]}_{a_1 a_2, a_0} +
         4 \log^2\alpha \; U^{\nf [2,2]}_{a_1 a_2, a_0}
      \right)
   \nonumber\\
   & \qquad {}
      - \sum_{a_2} \int\limits_2 X_2
      \left(
         U^{\nf + 1 [2,0]}_{a_1 a_2, a_0} -
         2 \log\beta \; U^{\nf + 1 [2,1]}_{a_1 a_2, a_0} +
         4 \log^2\beta \; U^{\nf + 1 [2,2]}_{a_1 a_2, a_0}
      \right)
   \,.
\end{align}
Using the relation \eqref{eq:relation-U-V} between $U^{[k, \ell\ms]}$ and $V^{[k, \ell - 1]}$, we find that all terms depending on $\alpha$ or $\beta$ cancel and obtain the momentum sum rule
\begin{align}
   \label{eq:momsum-11}
      \sum_{a_2} \int\limits_2 X_2 \, v^{I}_{a_1 a_2, a_0}
   &= (1 - X) \ms A_{a_1 a_0}^{Q [2,0]}
      + \sum_{a_2} \int\limits_2 X_2
      \left(
         U^{\nf [2,0]}_{a_1 a_2, a_0} - V^{\nf [2,0]}_{a_1 a_2, a_0} -
         V^{\nf [2,1]}_{a_1 a_2, a_0}
      \right)
   \nonumber\\
   & \quad {}
      - \sum_{a_2} \int\limits_2 X_2
      \left(
         U^{\nf + 1 [2,0]}_{a_1 a_2, a_0} - V^{\nf + 1 [2,0]}_{a_1 a_2, a_0} -
         V^{\nf + 1 [2,1]}_{a_1 a_2, a_0}
      \right)
\end{align}
for the interpolating function $v^I$.  Notice that this relation only depends on one momentum fraction and does not involve any scale or quark mass.
%
%
\paragraph{Number sum rule.}
For $\mu = \mQ$, the number sum rule \eqref{eq:numsum-8} has the form
\begin{align}
   \label{eq:numsum-9}
   &  \int\limits_2 \int_{y_{\beta}}^{y_{\alpha}} \!\!
      \text{d}^2 y
      \frac{1}{\pi y^2} \, V^{Q, \nf (2)}_{a_1 {a_2}_v, a_0}(\mu = \mQ)
   = \bigl( \delta_{a_1 \abar_2} - \delta_{a_1 a_2}
          - \delta_{a_2 \abar_0} + \delta_{a_2 a_0} \bigr) \,
      A^{Q [2,0]}_{a_1 a_0}
   \nonumber\\
   &  \qquad
   + \int\limits_2
      \Bigl[\ms
         U^{\nf (2)}_{a_1 {a_2}_v, a_0}(\alpha^{-1})
         - U^{\nf + 1 (2)}_{a_1 {a_2}_v, a_0}(\beta^{-1})
      \ms\Bigr] \,.
\end{align}
Following the same steps that lead from \eqref{eq:momsum-9} to \eqref{eq:momsum-10}, we obtain
\begin{align}
   \label{eq:numsum-11}
      \int\limits_2 v^{I}_{a_1 {a_2}_v, a_0}
   &= \bigl( \delta_{a_1 \abar_2} - \delta_{a_1 a_2}
          - \delta_{a_2 \abar_0} + \delta_{a_2 a_0} \bigr) \,
      A_{a_1 a_0}^{Q [2,0]}
      + \int\limits_2
      \left(
         U^{\nf [2,0]}_{a_1 {a_2}_v, a_0}
         - V^{\nf [2,0]}_{a_1 {a_2}_v, a_0}
         - V^{\nf [2,1]}_{a_1 {a_2}_v, a_0}
      \right)
   \nonumber\\
   &  \quad {}
      - \int\limits_2
      \left(
         U^{\nf + 1 [2,0]}_{a_1 {a_2}_v, a_0}
         - V^{\nf + 1 [2,0]}_{a_1 {a_2}_v, a_0}
         - V^{\nf + 1 [2,1]}_{a_1 {a_2}_v, a_0}
      \right)
   \,.
\end{align}
%
%
\subsection{Explicit determination of the NLO kernel for \texorpdfstring{$g \to q \qbar$}{g -> q qbar}}
\label{sec:Vqqbarg-massive-NLO}
\begin{figure}[ht]
   \centering
   \includegraphics[width=0.4\linewidth]{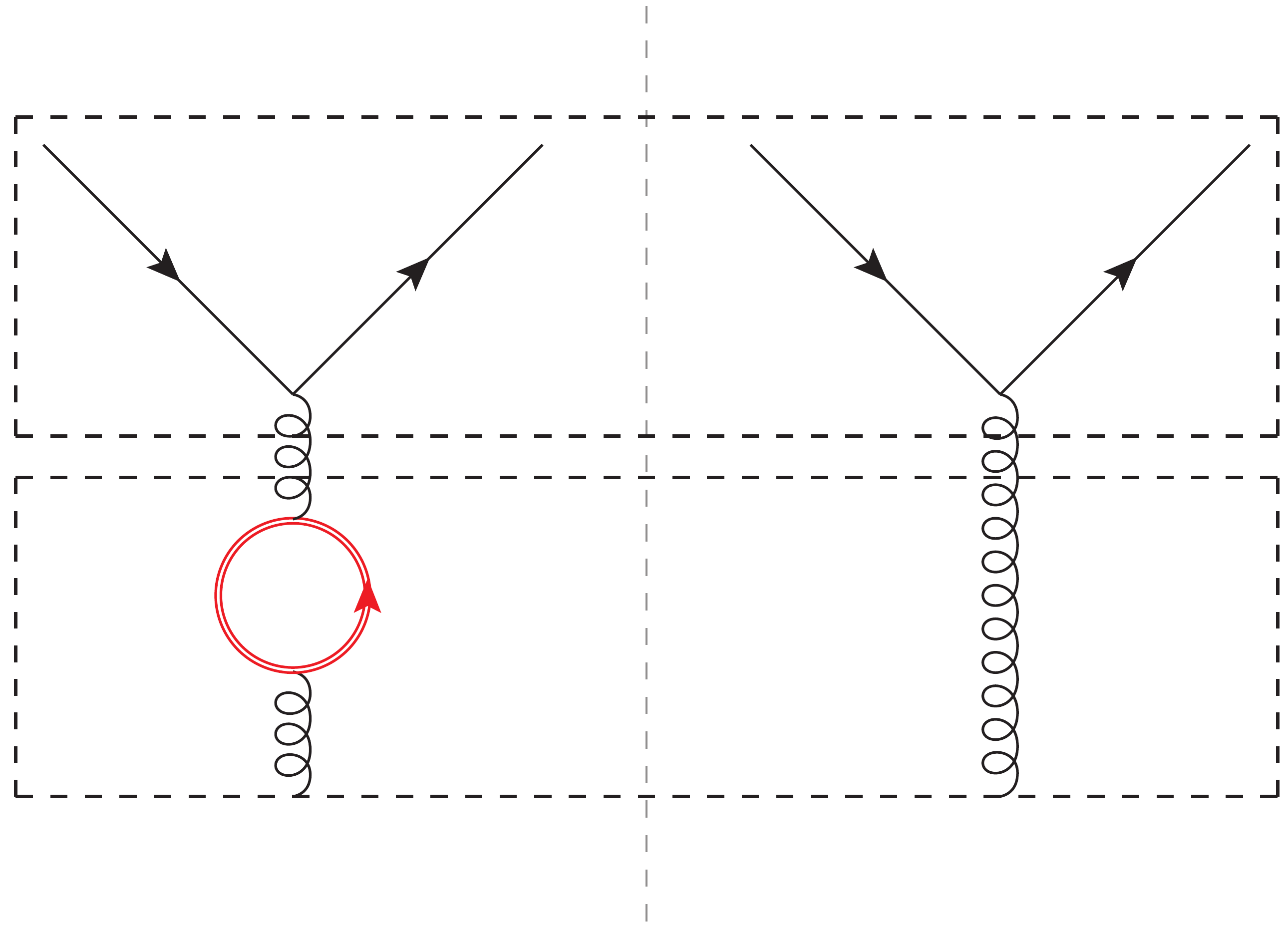}
   \caption{\label{fig:Vmqqbarg2-heavy}Heavy-quark contribution to $V^{Q (2)}_{q \qbar, g}$. The heavy quark is indicated by a red double line.}
\end{figure}

The constraints discussed in the previous subsections can be used to fully determine the kernel $V_{q \qbar, g}^{Q (2)}$ for a gluon splitting into a light $q \qbar$ pair. To this end it is crucial to realise that this kernel can be written as
\begin{align}
   \label{eq:Vmqqbarg2-1}
      V^{Q (2)}_{q \qbar, g}
   &= V^{(2)}_{q \qbar, g} +
      v^{Q}_{q \qbar, g} \,,
\end{align}
where $V^{(2)}$ is the massless kernel.  The term $v^{Q}$ contains the contributions from the graph in \fig{\ref{fig:Vmqqbarg2-heavy}} and its complex conjugate, together with the $\msbar$ counterterm of the heavy-quark loop.
Since this is a virtual graph with no lines crossing the final-state cut, $v^{Q}$ includes the momentum conservation constraint $\delta(1 - z_1 - z_2)$, just like the LO splitting kernels \eqref{eq:DPD-split-LO}.  As indicated in the figure, $v^{Q}$ factorises into two parts,
\begin{align}
  \label{eq:vmqqbarg-factorized}
      v^{Q}_{q \qbar, g}(z_1, z_2, y, \mQ; \mu)
   &= V^{(1)}_{q \qbar, g}(z_1, z_2, y \ms \mQ) \;
      g(\mQ; \mu) \,,
\end{align}
where the upper part is simply given by the LO $g \to q \qbar$ splitting kernel.
The function $g$ that represents the lower part cannot depend on $y$, since this variable is defined in terms of the lines in the upper part of the graph (see \sect{4.1} of \cite{Diehl:2019rdh} for details).  Likewise, $g$ cannot depend on the momentum fractions $z_1$ and $z_2$ of the lines in the upper part.
It is straightforward to compute it from the renormalised one-loop graph.  Here, we show instead how its expression can be obtained from the results we already derived so far.

The $\mu$ dependence of $g$ is readily obtained from \eqref{eq:Vmqqbarg2-1} and \eqref{eq:vmqqbarg-factorized}:
\begin{align}
   \label{eq:g-RGE-1}
      \frac{\text{d}}{\text{d}\log\mu^2}
      \left(
         V^{(1)}_{q \qbar, g} \; g(\mu)
      \right)
   &= \frac{\text{d}}{\text{d}\log\mu^2}
      \left(
         V^{Q (2)}_{q \qbar, g}(\mu) - V^{(2)}_{q \qbar, g}(\mu)
      \right)
    = \frac{\Delta \beta_0}{2} V^{(1)}_{q \qbar, g} \,,
\end{align}
where we indicated only the $\mu$ dependence of the functions.  In the last step, we used \eqs{\eqref{eq:VnF-logs}}, \eqref{eq:RGE-4}, and \eqref{eq:RGE-NLO-qqbarg}.
The solution of \eqref{eq:g-RGE-1} is
\begin{align}
   \label{eq:g-RGE-2}
      g(\mQ; \mu)
   &= g_0 + \frac{\Delta \beta_0}{2} \log \frac{\mu^2}{\mQ^2}
      \,.
\end{align}
The integration constant $g_0$ can be fixed using the number sum rule \eqref{eq:numsum-8} for $V^{Q (2)}_{q q_v, g \vphantom{\qbar}} = {}- V^{Q (2)}_{q \qbar, g}$.  Evaluating this sum rule at $\mu = \mQ$, we obtain
\begin{align}
   \label{eq:numsum-qqg}
   \int\limits_2 \int_{y_{\beta}}^{y_{\alpha}} \!\!
      \text{d}^2 y \frac{1}{\pi y^2} \,
      \biggl[
         V^{[2,0]}_{q \qbar, g}
         + \log \frac{\mQ^2}{\mu_y^2} \; V^{[2,1]}_{q \qbar, g}
         + V^{(1)}_{q \qbar, g} \; g_0
      \ms\biggr]
   &= \int\limits_2
      \Bigl[\ms
         U^{(2)}_{q \qbar, g}(\alpha^{-1})
         - U^{(2)}_{q \qbar, g}(\beta^{-1})
      \ms\Bigr]
\end{align}
where we used that $A^{Q (2)}_{q g} = 0$ and that $U^{(2)}_{q \qbar, g}$ is independent of $\nf$.  The evaluation of both sides proceeds in full analogy to the steps leading from \eqref{eq:momsum-9} to \eqn{\eqref{eq:momsum-10}}.  Using the relation \eqref{eq:relation-U-V} between $U$ and $V$ kernels, we find that the sum rule integral over $g_0$ must be zero.  Since $g_0$ is a constant, this implies
\begin{align}
   \label{eq:g-nonRGE}
      g_0
   &= 0\,.
\end{align}
The full expression for $V^{Q (2)}_{q \qbar, g}$ thus reads
\begin{align}
   \label{eq:Vmqqbarg2-2}
   V^{Q (2)}_{q \qbar, g}
   &= V^{(2)}_{q \qbar, g}
      + \frac{\Delta \beta_0}{2} \log \frac{\mu^2}{\mQ^2} \, V^{(1)}_{q \qbar, g}
    = V^{(2)}_{q \qbar, g}
      + V^{(1)}_{q \qbar, g} \conv{12} A^{Q (1)}_{g g}
   \,.
\end{align}
Comparing this result with the general parametrisation \eqref{eq:Vm-qqbarg-explicit}, we find that the interpolating function $V^{I}_{q \qbar, g}$ is zero, as we anticipated in \eqref{eq:VI-qqbarg-zero}.

Whilst the sum rules are only valid for unpolarised DPDs, a representation analogous to \eqref{eq:Vmqqbarg2-1} and \eqref{eq:vmqqbarg-factorized} holds also for polarised quarks.  Since it involves the same function $g$ for the heavy-quark loop on a gluon line, the result \eqref{eq:Vmqqbarg2-2} generalises to longitudinal or transverse polarisation of the quark and antiquark, with the appropriate $V$ kernels for polarised $g\to q \qbar$ splitting.
%
%
%
\section{A model for the unpolarised massive splitting kernels at NLO}
\label{sec:NLO-kernels-model}
In this section, we formulate a model ansatz for the interpolation part $V^{I}$ of the massive two-loop kernels, which may be used as long as these kernels have not been computed.  The interpolating functions must satisfy the limiting behaviour \eqref{eq:VI-limits} at small and large $y$, which does not provide any constraints on their form at intermediate $y \sim 1/\mQ$.  For unpolarised partons, such constraints are provided by the sum rules \eqref{eq:momsum-11} and \eqref{eq:numsum-11} derived in the previous section, which involve an integral over $y$.  We therefore limit our model ansatz to the unpolarised case.

To begin with, we set
\begin{align}
   \label{eq:VI-param}
      V^{I}_{a_1 a_2, a_0}(z_1, z_2, y, \mQ)
   &= k_{0 0}(y \ms \mQ) \, v^{I}_{a_1 a_2, a_0}(z_1, z_2) \,,
\end{align}
which goes to zero for $y\to 0$ and $y\to \infty$, and which is consistent with the normalisation condition \eqref{eq:VI-yint} according to the rightmost entry in table~\ref{tab:sumrule-integrals}.
The form \eqref{eq:VI-param} is inspired by the result for \eqref{eq:VQQbarg-LO} of the LO kernels, which contain $k_{0 0}(y \ms \mQ)$ multiplied by a function of $z_1$ and~$z_2$.  To be clear, we do \emph{not} expect that the NLO splitting graphs with massive lines yield such a simple factorised form of the $y$ dependence, so that our ansatz is only guided by simplicity at this stage.
%
%
\subsection{Constraints on the interpolating functions}
\label{sec:NLO-kernels-model-determination}
The factorised ansatz \eqref{eq:VI-param} allows us to directly use the sum rule constraints \eqref{eq:momsum-11} and \eqref{eq:numsum-11} on the two-parameter functions $v^{I}_{a_1 a_2, a_0}(z_1, z_2)$.  To determine these is not a trivial task, since in general the kernel for given parton labels $(a_1 a_2, a_0)$ appears in different sum rules.  In the following we present a sequential construction that ensures that the sum rule constraints are satisfied for all parton channels.
%
%
\subsubsection*{First group: \texorpdfstring{$V^{Q (2)}_{Q \smash{\Qbar},q} + V^{Q (2)}_{Q q,q}$}{} and \texorpdfstring{$V^{Q (2)}_{q Q,q} + V^{Q (2)}_{\vphantom{Q} q g,q}$}{}}
\label{sec:VmQbq2}
The kernels for the channels $q \to Q \Qbar$, $q \to Q q$, and $q \to q g$ have to be treated simultaneously, because they enter together in two distinct momentum sum rules. In addition to these momentum sum rules, one finds three number sum rules for
\begin{align}
   V^{Q (2)}_{Q q_v,q} &= V^{Q (2)}_{Q q,q} \,,
   &
   V^{Q (2)}_{Q Q_v,q} &= - V^{Q (2)}_{Q \Qbar,q} \,,
   &
   V^{Q (2)}_{g q_v,q} &= V^{Q (2)}_{g q,q} \,.
\end{align}
\paragraph{Sum rule constraints.} The momentum sum rules for the interpolating functions read
\begin{align}
   \label{eq:momsum-Qbq2}
      \int\limits_2 X_2 \,
      \Bigl( v^{I}_{Q \smash{\Qbar}, q} + v^{I}_{Q q, q} \Bigr)
   &= (1 - X) \ms A_{Q q}^{Q [2,0]} - \int\limits_2 X_2 \,
      \Bigl(
         U^{[2,0]}_{q' \qbar'\!, q} -
         V^{[2,0]}_{q' \qbar'\!, q} -
         V^{[2,1]}_{q' \qbar'\!, q}
   \nonumber\\[-0.5em]
   &  \hphantom{{} = (1 - X) A_{Q q}^{Q [2,0]} - \int\limits_2 X_2 \Bigl(} +
         U^{[2,0]}_{q' q, q} -
         V^{[2,0]}_{q' q, q} -
         V^{[2,1]}_{q' q, q}
   \Bigr) \,,
   \\[0.5em]
   \label{eq:momsum-qbq2}
      \int\limits_2 X_2 \,
      \Bigl( 2 v^{I}_{q Q, q} + v^{I}_{q g, q} \Bigr)
   &= (1 - X) \ms A_{q q}^{Q [2,0]}
      - 2 \int\limits_2 X_2 \,
      \Bigl(
         U^{[2,0]}_{q q', q} -
         V^{[2,0]}_{q q', q} -
         V^{[2,1]}_{q q', q}
      \Bigr)
   \nonumber\\[-0.5em]
   &  \hphantom{{}= (1 - X) \ms A_{q q}^{Q [2,0]}}
      {}- \Delta\beta_0 \int\limits_2 X_2 \,
      \Bigl(
         U^{\beta \ms [2,0]}_{q g, q} -
         V^{\beta [2,0]}_{q g, q} -
         V^{(1)}_{q g, q}
      \Bigr)
   \,,
\end{align}
where we used \eqref{eq:V-kernel-nf-dep}, \eqref{eq:V-beta-21} and \eqref{eq:U-kernel-nf-dep} to simplify the difference of $V$ or $U$ kernels for $\nf + 1$ and $\nf$.
In \eqref{eq:momsum-qbq2} we also used $v^{I}_{q Q, q} + v^{I}_{q \smash{\Qbar}\vphantom{Q}, q} = 2 v^{I}_{q Q, q}$ and corresponding relations for the $V$ and $U$ kernels on the r.h.s.
The relevant number sum rules are given by
\begin{align}
   \label{eq:numsum-QQbarq2}
      \int\limits_2 v^{I}_{Q\smash{\Qbar}, q}
   &= A_{Q q}^{Q [2,0]} - \int\limits_2
      \Bigl(
         U^{[2,0]}_{q' \qbar'\!, q} -
         V^{[2,0]}_{q' \qbar'\!, q} -
         V^{[2,1]}_{q' \qbar'\!, q}
      \Bigr)
   \,,
   \\[0.5em]
   \label{eq:numsum-Qqq2}
      \int\limits_2 v^{I}_{Q q, q}
   &= A_{Q q}^{Q [2,0]} - \int\limits_2
      \Bigl(
         U^{[2,0]}_{q' q, q} -
         V^{[2,0]}_{q' q, q} -
         V^{[2,1]}_{q' q, q}
      \Bigr)
   \,,
   \\[0.5em]
   \label{eq:numsum-gqq2}
      \int\limits_2 v^{I}_{g q, q}
   &= A_{g q}^{Q [2,0]}
      - \Delta\beta_0 \int\limits_2
      \Bigl(
         U^{\beta \ms [2,0]}_{g q, q} -
         V^{\beta \ms [2,0]}_{g q, q} -
         V^{(1)}_{g q, q}
      \Bigr)
   \,.
\end{align}
Since the only heavy-quark contribution to $\smash{V^{Q(2)}_{g q, q}}$ is due to the virtual graph in \fig{\ref{subfig:qqg}} and its complex conjugate, the function $v^{I}_{g q, q}$ includes a factor $\delta(1 - z_1 - z_2)$.  It is therefore uniquely fixed by~\eqref{eq:numsum-gqq2} and reads
\begin{align}
   \label{eq:Vmgqq2}
      v^{I}_{g q, q}
   &=
      \delta(1 - z_1 - z_2) \, A_{g q}^{Q [2,0]}(z_1)
      - \Delta \beta_0 \,
      \Bigl(
         U^{\beta \ms [2,0]}_{g q, q} -
         V^{\beta \ms [2,0]}_{g q, q} -
         V^{(1)}_{g q, q}
      \Bigr)
   \,,
\end{align}
where for definiteness we have specified the argument of $A^{Q [2,0]}_{g q}$.
Given that $v^{I}_{g q, q}(z_1, z_2) =v^{I}_{q g, q}(z_2, z_1)$, the momentum sum rule in \eqref{eq:momsum-qbq2} reduces to
\begin{align}
   \label{eq:momsum-qbq2-2}
      \int\limits_2 \! X_2 \, v^{I}_{q Q, q}
   &= (1 - X)
      \left(
         A_{q q}^{Q [2,0]}(z_1) -
         A_{g q}^{Q [2, 0]}(1 - z_1)
      \right)
      - \int\limits_2 X_2
      \Bigl(
         U^{[2,0]}_{q q', q} -
         V^{[2,0]}_{q q', q} -
         V^{[2,1]}_{q q', q}
      \Bigr)
   \,.
\end{align}

The next step is to fix ${v^{I}_{Q \smash{\Qbar}, q}}$ and ${v^{I}_{Q q, q}}$.  To this end, we make an ansatz
\begin{align}
   \label{eq:ansatz-qQQbar}
      v^{I}_{Q\smash{\Qbar}, q}(z_1, z_2)
   &= C_F \ms T_F \sum_i c_i \, f_i(z_1, z_2) \,,
   \\
   \label{eq:ansatz-qQq}
      v^{I}_{Q q, q}(z_1, z_2)
   &= C_F \ms T_F \sum_i d_i \, g_i(z_1, z_2) \,,
   \\
   \label{eq:ansatz-qqQ}
      v^{I}_{q Q, q}(z_1, z_2)
   &= C_F \ms T_F \sum_i d_i \, g_i(z_2, z_1) \,,
\end{align}
where $c_i, d_i$ are numerical coefficients and $f_i, g_i$ are basis functions.  We have made the colour factor $C_F \ms T_F$ explicit, which is readily obtained from the Feynman graphs in \figs{\ref{subfig:qQQbar}} and \ref{subfig:qqQ}.  The function $v^{I}_{Q\smash{\Qbar}, q}$ must be symmetric under $z_1 \leftrightarrow z_2$, so that we require the same symmetry for the individual basis functions $f_i$.  Of course, \eqref{eq:ansatz-qqQ} follows from \eqref{eq:ansatz-qQq}.

The set of basis functions $f_i$ and $g_i$ must be large enough to admit a solution of the integral equations \eqref{eq:momsum-Qbq2}, \eqref{eq:numsum-QQbarq2}, \eqref{eq:numsum-Qqq2}, and \eqref{eq:momsum-qbq2-2}.  As a first attempt, one may take the functions that appear in the corresponding massless kernels, i.e.\ in $V^{[2,0]}_{q' \qbar'\!, q}$ and $V^{[2,1]}_{q' \qbar'\!, q}$ for $f_i$, and in $V^{[2,0]}_{q' q,q}$ and $V^{[2,1]}_{q' q,q}$ for $g_i$, given that the massive two-loop graphs for these channels have the same topology as their massless counterparts.  It turns out that this set is insufficient.  We do, however, obtain solution to all constraints when including the functions that appear in $U^{[2,0]}_{q' \qbar'\!, q}$ in the set of $f_i$, and those appearing in $U^{[2,0]}_{q' q,q}$ in the set of $g_i$.  This choice is motivated by the observation that these kernels enter the integral equations in the same manner as the functions $v^{I}$.  The resulting basis functions are specified in \app{\ref{app:basis-functions}}.

\paragraph{Kinematic constraints.}  It is plausible to expect that the DPDs computed with massive splitting graphs are not more singular than their massless counterparts in certain kinematical limits.  A detailed analysis for the massless case has been given in \sects{4.2} to 4.5 of \cite{Diehl:2021wpp} for the limits $x_1 + x_2 \to 1$, $x_1 + x_2 \ll 1$, $x_1 \ll 1$, and $x_2 \ll 1$.

In the three latter cases, it turns out that individual terms in the massless kernels can be more singular than the sum over all terms.  In our ansatz for the massive kernels, we should therefore enforce the same type of cancellation between super-leading terms as happens in the massless case.

To discuss the singular behaviour, it is useful to change variables
\begin{align}
   \label{eq:u-z-def}
   z_1 &= z \ms u \,,
   &
   z_2 &= z \ms \bar{u} = z \ms (1 - u) \,,
\intertext{in the splitting kernels.  With the form \eqref{eq:Mellin-12} of the convolution $\otimes_{12}$, one readily finds}
   u &= x_1 / (x_1 + x_2) \,,
   &
   \bar{u} &= x_2 / (x_1 + x_2)
\end{align}
for the momentum fractions in the splitting DPD.

In the limit of small $x_1 + x_2$, one finds that the convolution of a $1 \to 2$ kernel with a PDF is enhanced if the kernels grow at least like $\sim z^{-2}$ for small $z$.  Some of the individual basis functions $f_i$ and $g_i$ are in fact more singular than the complete massless kernels, and one finds super-leading terms
\begin{align}
   \label{eq:kinematic-qQQbar}
      \sim \frac{\log^2(z)}{z^2} \,,
   && \sim \frac{\log(z)}{z^2}
   && \text{ in some } f_i
   \,.
\end{align}
The massless kernel for $q\to q' q$ is finite for $z\to 0$, whilst individual terms in the basis functions behave like
\begin{align}
   \label{eq:kinematic-qQq}
      \sim \log^2(z) \,,
   && \sim \log(z)
   && \text{ in some } g_i
   \,.
\end{align}
In this case, we impose the constraint that the massive splitting kernel should not be more singular for $z\to 0$ than its massless counterpart, even though this does not enhance the resulting DPD at small $x_1 + x_2$.  Requiring that terms of the form shown in \eqref{eq:kinematic-qQQbar} and \eqref{eq:kinematic-qQq} cancel in \eqref{eq:ansatz-qQQbar}, \eqref{eq:ansatz-qQq}, and \eqref{eq:ansatz-qqQ} imposes constraints on the coefficients $c_i$ and $d_i$.

The limits $x_1 \ll 1$ and $x_2 \ll 1$ (with generic values of the other momentum fraction) correspond to $u \ll 1$ and $\bar{u} \ll 1$, respectively.  As explained in \sect{4.4} of \cite{Diehl:2021wpp}, a singular behaviour of the splitting DPD for $u \ll 1$ arises from terms in the kernels that grow at least like $1/u$, and from terms with a power of $1 - z_2 = 1 - (1 - u) z$ in the denominator.  Inspection of our basis functions reveals that some of them lead to a super-leading behaviour
\begin{align}
   \label{eq:kinematic-Qqq}
      \sim \frac{\log u}{u} \,,
   && \sim\frac{\log^2 u}{u}
\end{align}
for the convolution of some $g_i$ with a PDF in the limit $u \ll 1$.  Requiring that these terms cancel places additional constraints on the coefficients~$d_i$.

The analysis in \cite{Diehl:2021wpp} also includes the triple Regge limits $x_1 \ll x_1 + x_2 \ll 1$ and $x_2 \ll x_1 + x_2 \ll 1$, which respectively correspond to taking $u \ll 1$ or $\bar{u} \ll 1$ in the small-$z$ limit of the kernels. It turns out that these limits yield no additional constraints for the kernels at hand.

The kinematic requirements just specified, together with the integral equations \eqref{eq:momsum-Qbq2}, \eqref{eq:numsum-QQbarq2}, \eqref{eq:numsum-Qqq2}, and \eqref{eq:momsum-qbq2-2} result in a system of linear equations for the coefficients $c_i$ and $d_i$.  With our choice of basis functions, this system is under-constrained and admits a plethora of solutions.  In \sect{\ref{sec:unique-solution}} an algorithm for picking a unique solution will be presented.
%
%
\subsubsection*{Second group: \texorpdfstring{$V^{Q (2)}_{Q \smash{\Qbar},g} + V^{Q (2)}_{Q g,g}$}{} and \texorpdfstring{$V^{Q (2)}_{g Q,g} + V^{Q, \nf (2)}_{g g, g\vphantom{Q}}$}{}}
\label{sec:VmQbg2}
For the channels $g \to Q \Qbar$, $g \to Q g$, and $g \to g g$, one finds again several sum rules that have to be fulfilled simultaneously, namely two momentum sum rules and a number sum rule for
\begin{align}
   V^{Q (2)}_{Q Q_v,g} = {}- V^{Q (2)}_{Q \smash{\Qbar},g}
   \,.
\end{align}

\paragraph{Sum rule constraints.} The momentum sum rules for the interpolating functions read
\begin{align}
   \label{eq:momsum-Qbg2}
      \int\limits_2 X_2 \,
      \Bigl(
         v^{I}_{Q \smash{\Qbar}, g} + v^{I}_{Q g, g}
      \Bigr)
   &= (1 - X) \ms A_{Q g}^{Q [2,0]} - \int\limits_2 X_2 \,
      \Bigl(
         U^{[2,0]}_{q' \qbar', g} -
         V^{[2,0]}_{q' \qbar', g} -
         V^{[2,1]}_{q' \qbar', g}
   \nonumber\\[-0.5em]
   &  \hphantom{{} = (1 - X) A_{Q g}^{Q [2,0]} - \int\limits_2 X_2 \Bigl(} +
         U^{[2,0]}_{q g, g} - V^{[2,0]}_{q g, g} - V^{[2,1]}_{q g, g}
      \Bigr) \,,
   \\[0.5em]
   \label{eq:momsum-gbg2}
      \int\limits_2 X_2 \,
      \Bigl(
         2 v^{I}_{g Q, g} + v^{I}_{g g, g}
      \Bigr)
   &= (1 - X) \ms A_{g g}^{Q [2,0]}
      - 2 \int\limits_2 X_2 \,
      \Bigl(
         U^{[2,0]}_{g q, g} - V^{[2,0]}_{g q, g} - V^{[2,1]}_{g q, g}
      \Bigr)
   \nonumber\\[-0.5em]
   &  \hphantom{{} = (1 - X) \ms A_{g g}^{Q [2,0]}}
      - \Delta\beta_0 \int\limits_2 X_2 \,
      \Bigl(
         U^{\beta \ms [2,0]}_{g g, g} -
         V^{\beta \ms [2,0]}_{g g, g} -
         V^{(1)}_{g g, g}
      \Bigr) \,,
\end{align}
whilst the number sum rule for $v^{I}_{Q \smash{\Qbar}, g}$ is given by
\begin{align}
   \label{numsum-QQbarg2}
      \int\limits_2 v^{I}_{Q\smash{\Qbar},g}
   &= A_{Q g}^{Q [2,0]} - \int\limits_2
      \Bigl(
         U^{[2,0]}_{q \qbar, g} -
         V^{[2,0]}_{q \qbar, g} -
         V^{[2,1]}_{q \qbar, g}
      \Bigr) \,.
\end{align}
Our strategy for this case is the same as for the previous one, and we make the  ansatz
\begin{align}
   \label{eq:ansatz-gQQbar}
      v^{I}_{Q\smash{\Qbar}, g}(z_1, z_2)
   &= C_A \ms T_F \sum_i c_i^A \ms f_i^A(z_1, z_2) +
      C_F \ms T_F \sum_i c_i^F \ms f_i^F(z_1, z_2) \,,
   \\
   \label{eq:ansatz-gQg}
      v^{I}_{Q g, g}(z_1, z_2)
   &= C_A \ms T_F \sum_i d_i^A \ms g_i^A(z_1, z_2) +
      C_F \ms T_F \sum_i d_i^F \ms g_i^F(z_1, z_2) \,,
   \\
   \label{eq:ansatz-ggQ}
      v^{I}_{g Q, g}(z_1, z_2)
   &= C_A \ms T_F \sum_i d_i^A \ms g_i^A(z_2, z_1) +
      C_F \ms T_F \sum_i d_i^F \ms g_i^F(z_2, z_1) \,,
   \\
   \label{eq:ansatz-ggg}
      v^{I}_{g g, g}(z_1, z_2)
   &= C_A \ms T_F \sum_i e_i \ms h_i(\rev{z_1, z_2})
   \,,
\end{align}
where the different colour factors are again deduced from the relevant Feynman graphs.
As before, we take as basis functions the terms that appear in the massless kernels $V^{[2,0]}$, $V^{[2,1]}$, and $U^{[2,0]}$ for the relevant parton channel.  For $f_i^F$, we must include additional functions pro\-portional to $(1 - z_1 - z_2)^{-1}$ in order to be able to fulfil all constraints.  A simplification occurs for $v^{I}_{g g, g}$, which includes a factor $\delta(1 - z_1 - z_2)$ according to \eqref{eq:VI-ggg-delta}.  Only basis functions with this structure are therefore needed in \eqref{eq:ansatz-ggg}.
Details on the resulting set of basis functions are given in \app{\ref{app:basis-functions}}.

\paragraph{Kinematic constraints.}  As in the previous case, we obtain additional constraints by requiring that the singular behaviour of massive kernels and the resulting splitting DPDs should not be stronger than for their massless counterparts.

For the limit of small $x_1 + x_2$, we find super-leading terms in the small-$z$ behaviour of the basis functions, namely
\begin{align}
   \label{eq:kinematic-gQQbarA}
      \sim \frac{\log^2(z)}{z^2} \,,
   && \sim \frac{\log(z)}{z^2}
   &&& \text{ for some } f_i^A
\intertext{and}
   \label{eq:kinematic-gQg-1}
      \sim \frac{\log^2(z)}{z} \,,
   && \sim \frac{\log(z)}{z}
   &&& \text{ for some } g_i^A \text{ and } g_i^F.
\end{align}
We require that these terms cancel in $v^{I}_{Q\smash{\Qbar}, g}$ and $v^{I}_{Q g, g}$ (and thus also in $v^{I}_{g Q, g}$).

For the function $v^{I}_{Q g, g}$, we find further constraints in limits of small $x_1$ or small $x_2$.  The convolution of some basis functions $g_i^A$ and $g_i^F$ with a PDF yields super-leading terms
\begin{align}
   \label{eq:kinematic-gQg-3}
   &  \sim \frac{\log^2 u}{u} \,,
   && \sim \frac{\log u}{u}
   && \text{ for } u\to 0 \,,
   \nonumber\\
   &  \sim \frac{\log^2(1-u)}{1-u} \,,
   && \sim \frac{\log(1-u)}{1-u}
   && \text{ for } u\to 1 \,,
\end{align}
and we require that these cancel in the convolution of $v^{I}_{Q g, g}$ with a PDF.

For the same kernel, we also obtain constraints in the triple Regge limit $x_2 \ll x_1+x_2 \ll 1$, given that some of basis functions show a super-leading behaviour
\begin{align}
   \label{eq:kinematic-gQg-2}
      \sim \frac{\log^2(1 - u)}{z (1 - u)} \,,
   && \sim \frac{\log(1 - u)}{z (1 - u)} \,,
   && \sim \frac{1}{z (1 - u)} \,,
\end{align}
for $z\to 0$ and $u \to 1$.  This puts additional constraints on the coefficients $d_i^A$ and $d_i^F$.

The combination of all sum rules and kinematic constraints leads to a system of linear equations for the coefficients $c_i^A$, $c_i^F$, $d_i^A$, $d_i^F$, and $e_i$ in \eqs{\eqref{eq:ansatz-gQQbar}} to \eqref{eq:ansatz-ggg} that is under-determined and admits a large number of solutions.
%
%
\subsection{A particular solution for the kernels}
\label{sec:unique-solution}
As just stated, the constraints from the sum rules and kinematic limits for the interpolating functions $v^{I}$ do not uniquely fix the coefficients in our expansion on basis functions.  Since we are building a model ansatz, we are free to impose additional requirements.  It appears natural to require that the different expansion coefficients do not become too large overall, given that this is true if one expands the massless kernels on the same basis functions.

To turn this into a criterion that leads to a unique solution, let us first build some intuition in a simple toy example.  We consider a system with three coefficients $c_1$, $c_2$, and $c_3$ and one linear constraint
\begin{align}
    c_3
  &= A c_1 + B c_2 + C \,,
\end{align}
where $A$, $B$, and $C$ are fixed.  In geometric terms, the solutions of this under-determined system form a plane $(c_1, c_2, A c_1 + B c_2 + C)$ in the three-dimensional $(c_1, c_2, c_3)$ space.  $c_1$ and $c_2$ can then be fixed by choosing the point on this plane that has a minimal distance from the origin $(0, 0, 0)$, which in a global sense minimises the size of the coefficients. This point is obtained by minimising the length $L$ of the vector $(c_1, c_2, A c_1 + B c_2 + C)$, which is given by
\begin{align}
      L^2
   &= c_1^2 + c_2^2 + (A c_1 + B c_2 + C)^2 \,.
\end{align}
The minimum of $L$ is obtained by requiring
\begin{align}
      \frac{\text{d} L^2}{\text{d} c_1}
   &= 0\,,
   && \frac{\text{d} L^2}{\text{d} c_2}
    = 0\,,
\end{align}
which yields two more linear constraints for the coefficients $c_1$ and $c_2$.

This approach is readily generalised to the case where one has $n$ coefficients $c_1, \ldots , c_n$ and $m < n$ linear constraints. Without loss of generality, one can use the constraints to write the first $m$ coefficients $c_1, \ldots , c_m$ as linear combinations of the remaining $n - m$ coefficients $c_{m + 1}, \ldots , c_n$, i.e.
\begin{align}
      c_i
   &= f_i(c_{m + 1}, \ldots , c_n)
   && \text{for } 1 \le i \le m\,.
\end{align}
The length of the vector $(c_1, \ldots , c_n)$ is then given by
\begin{align}
      L^2
   &= \sum_{i = 1}^n c_i^2 \,,
\end{align}
and the linear constraints that fix the remaining coefficients $c_{m + 1}, \ldots , c_n$ are
\begin{align}
      \frac{\text{d} L^2}{\text{d} c_i}
   &= 0\,,
   && \text{for } m + 1 \le i \le n \,.
\end{align}

\begin{figure}
   \begin{center}
      \subfigure[\label{subfig:coefficients-q-c} $c_i$]{
         \includegraphics[width=0.475\linewidth]{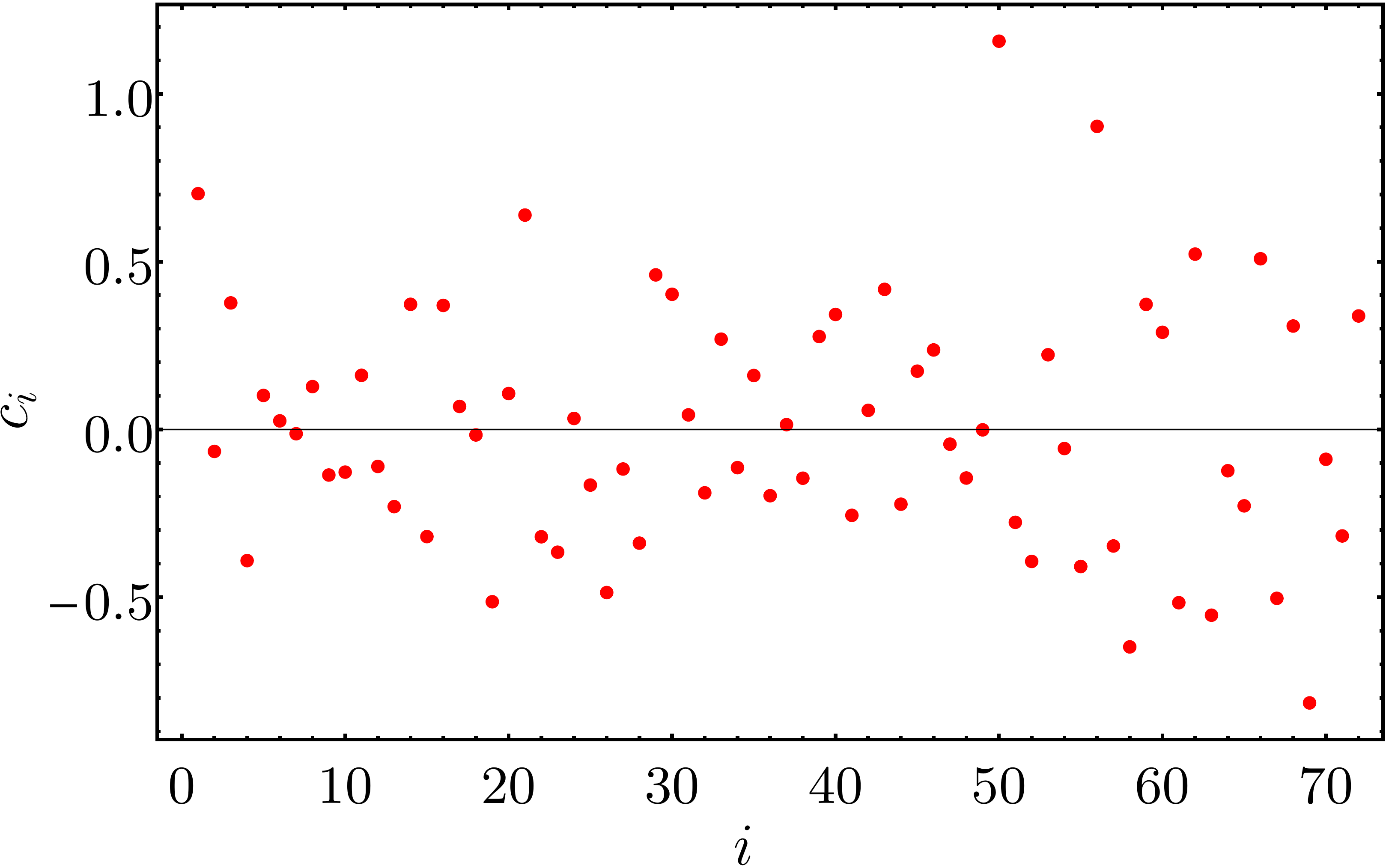}
      }
      \subfigure[\label{subfig:coefficients-q-d} $d_i$]{
         \includegraphics[width=0.475\linewidth]{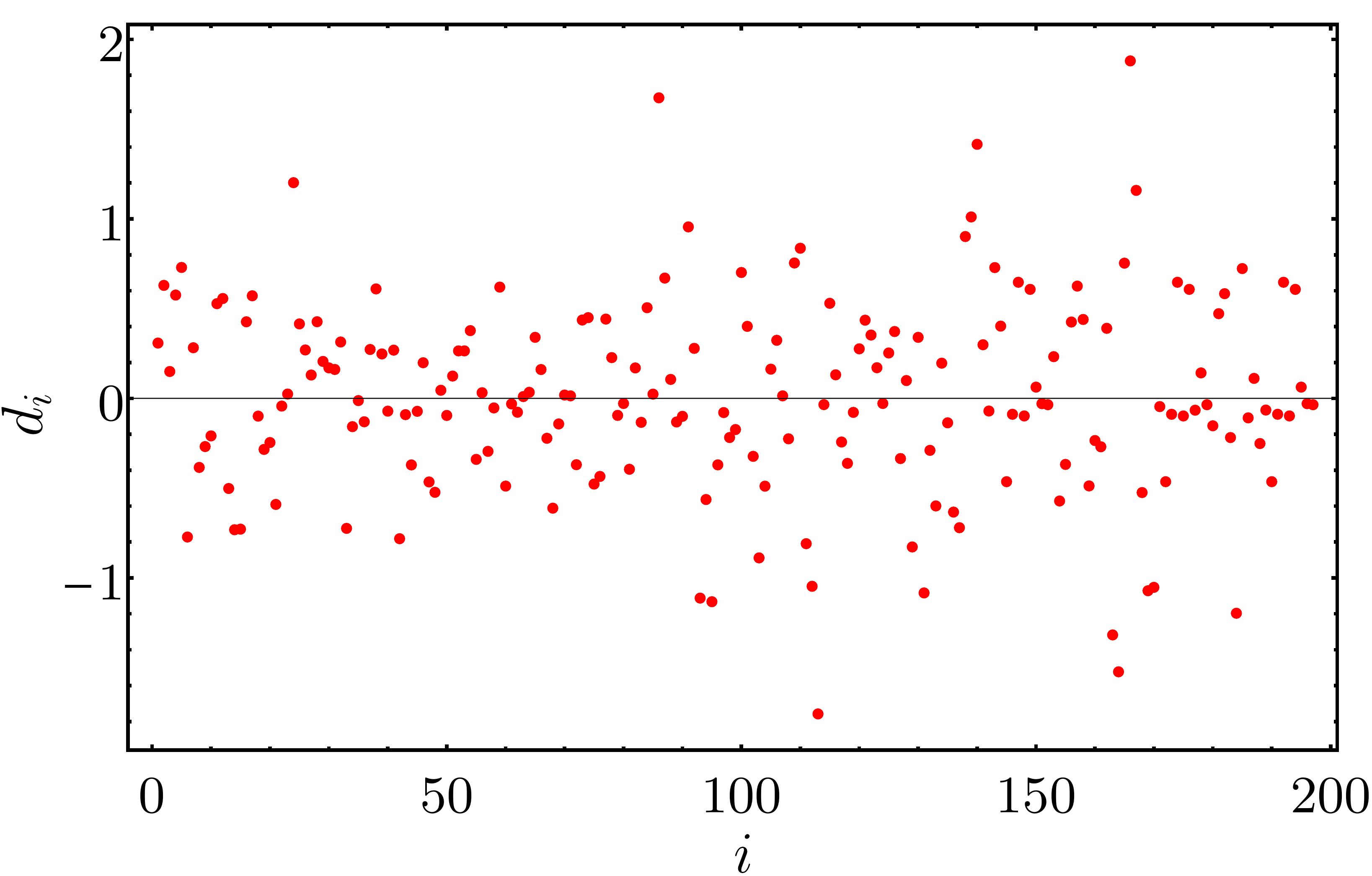}
      }
      \caption{\label{fig:coefficients-q}The coefficients $c_i$ and $d_i$ in the expansions \eqref{eq:ansatz-qQQbar}, \eqref{eq:ansatz-qQq}, and \eqref{eq:ansatz-qqQ}, obtained with the criterion of minimal length explained in the text.}
   \end{center}
\end{figure}

\begin{figure}
   \begin{center}
      \subfigure[\label{subfig:coefficients-g-a} $c_i^A$]{
         \includegraphics[width=0.475\linewidth]{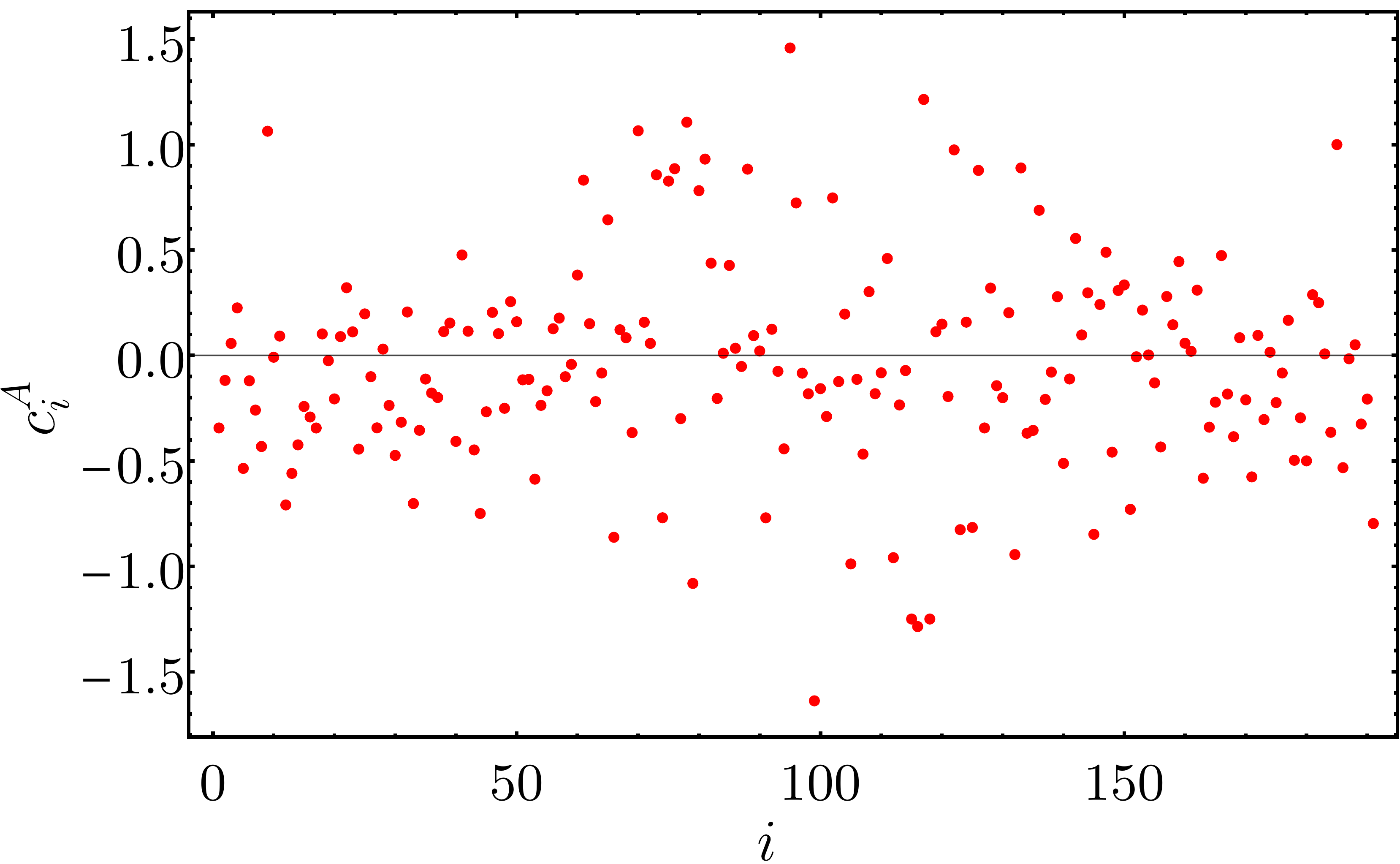}
      }
      \subfigure[\label{subfig:coefficients-g-b} $c_i^F$]{
         \includegraphics[width=0.475\linewidth]{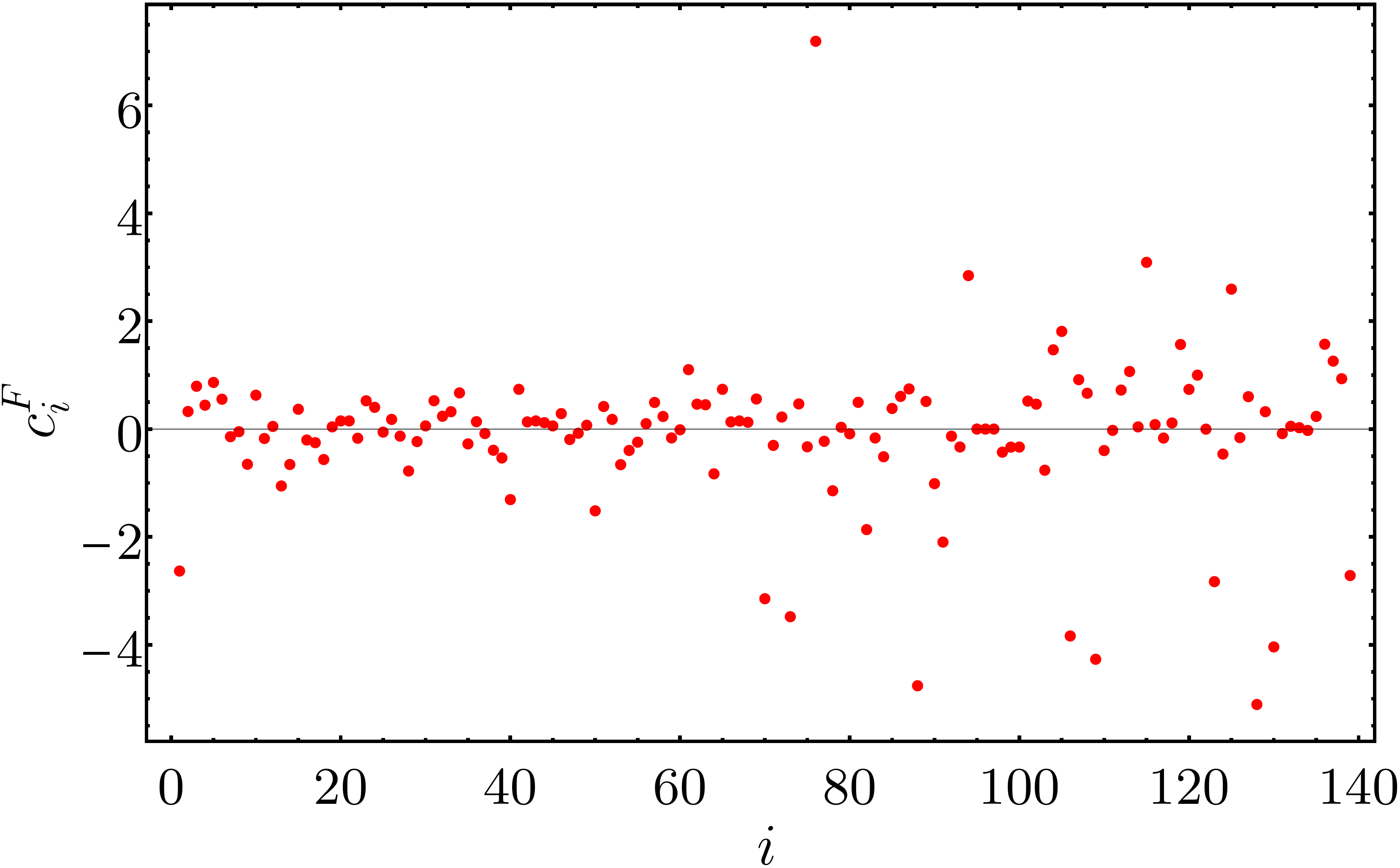}
      }
      \\
      \subfigure[\label{subfig:coefficients-g-c} $d_i^A$]{
         \includegraphics[width=0.475\linewidth]{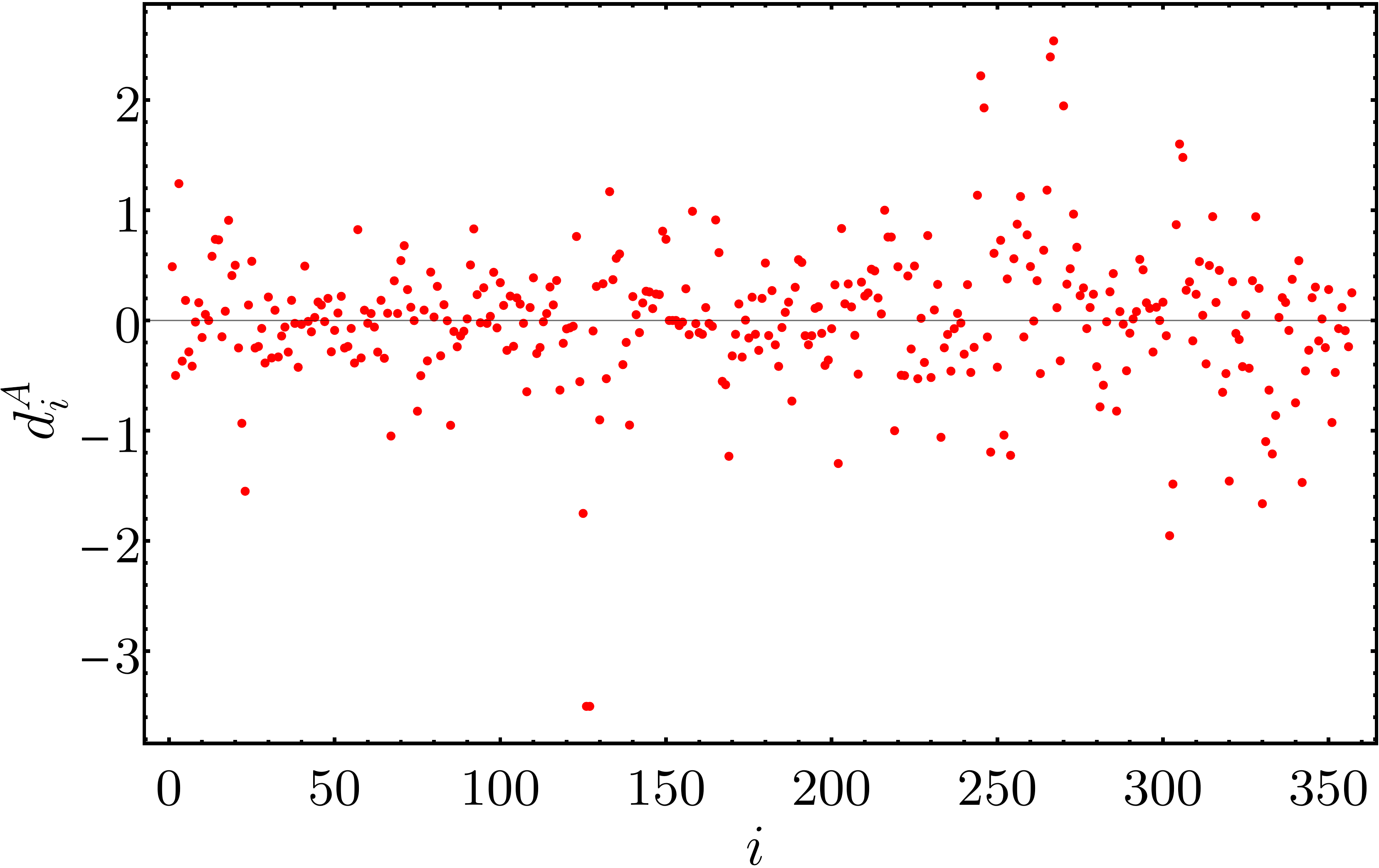}
      }
      \subfigure[\label{subfig:coefficients-g-d} $d_i^F$]{
         \includegraphics[width=0.475\linewidth]{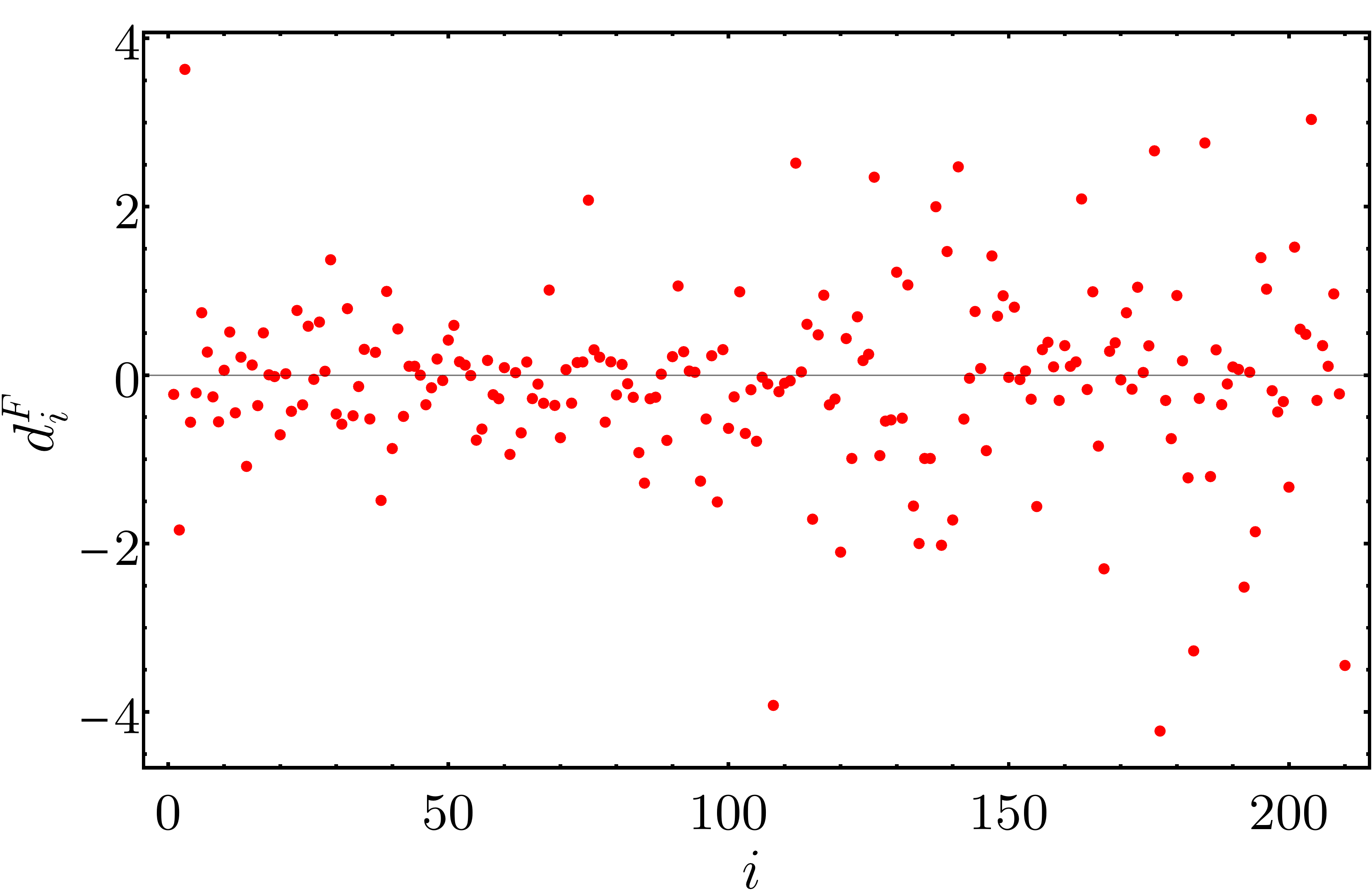}
      }
      \caption{\label{fig:coefficients-g}As the previous figure, but for the coefficients $c_i^A$, $c_i^F$, $d_i^A$, and $d_i^F$ in the expansions \eqref{eq:ansatz-gQQbar}, \eqref{eq:ansatz-gQg}, and \eqref{eq:ansatz-ggQ}.  Not shown are the three coefficients $e_i$ in \eqref{eq:ansatz-ggg}, which have the values $e_{1} \approx -1.1$, $e_{2} \approx 2.3$, $e_{3} \approx -0.45$.}
   \end{center}
\end{figure}

The expansion coefficients for the interpolating functions $v^{I}$ that we obtain  with this procedure are given in the ancillary files associated with this paper on the \href{https://arxiv.org}{arXiv}.  We visualise them in \figs{\ref{fig:coefficients-q} and \ref{fig:coefficients-g}} and find that they are all of order unity, as desired.
%
%
\subsection{Numerical illustration}
\label{sec:nlo_numerics}
With a model for the massive NLO kernels at hand, it is natural to ask to which extent the shortcomings of the LO description discussed in \sect{\ref{sec:LO-numerics}} are mitigated when splitting is computed at NLO.

Investigating the interplay between NLO splitting and DGLAP evolution requires the numerical evaluation of the convolution integral \eqref{eq:Mellin-12} on interpolation grids in $x_1, x_2$, and~$y$ for all flavor combinations in $V_{a_1 a_2, a_0}$.  An efficient and accurate implementation of this is nontrivial and beyond the scope of the present work.
Instead, we will study DPDs at the scale where the splitting formula is evaluated, i.e.\ at $\mu = \mu_{y^*}$ with $\mu_{y^*}$ specified in \eqref{eq:mu-y-star}, \eqref{eq:y-star} and the surrounding text.  For this it is sufficient to evaluate the convolutions \eqref{eq:Mellin-12} at selected values of $x_1$ and $x_2$ and for selected flavors, which can be done with standard quadrature routines.

This allows us to address the discontinuity at $\mu_y = \beta \ms m_b$ which we observed for $F_{g b}$ at LO in \figs{\ref{subfig:Fgb-massive-2}} and \ref{subfig:Fgb-massive-4}.  This discontinuity increases with $\beta$ and led us to use the rather small value $\beta = 2$.  We recall that in the massive scheme, five-flavor DPDs are obtained from the convolution of the massive kernels $V^{b}$ with four-flavor PDFs if $\beta m_c < \mu_y < \beta m_b$, and from the convolution of massless kernels $V^{5}$ with five-flavor PDFs if $\mu_y > \beta m_b$ (see \fig{\protect\ref{fig:scheme-two-massive}}).  With LO kernels, the description for $\mu_y < \beta m_b$ is then missing the contribution from the graph in \fig{\ref{subfig:Fgb-massless-scheme}}.  This deficiency is cured at NLO, where that graph is taken into account via the splitting kernel for $g \to g b$.

\begin{figure}
   \begin{center}
      \subfigure[\label{subfig:F_gb_low_muy}
         $y^2 F_{b g}$ for $\mu_y \le m_b$]{
         \includegraphics[width=0.48\linewidth, trim=0 0 0 50, clip]{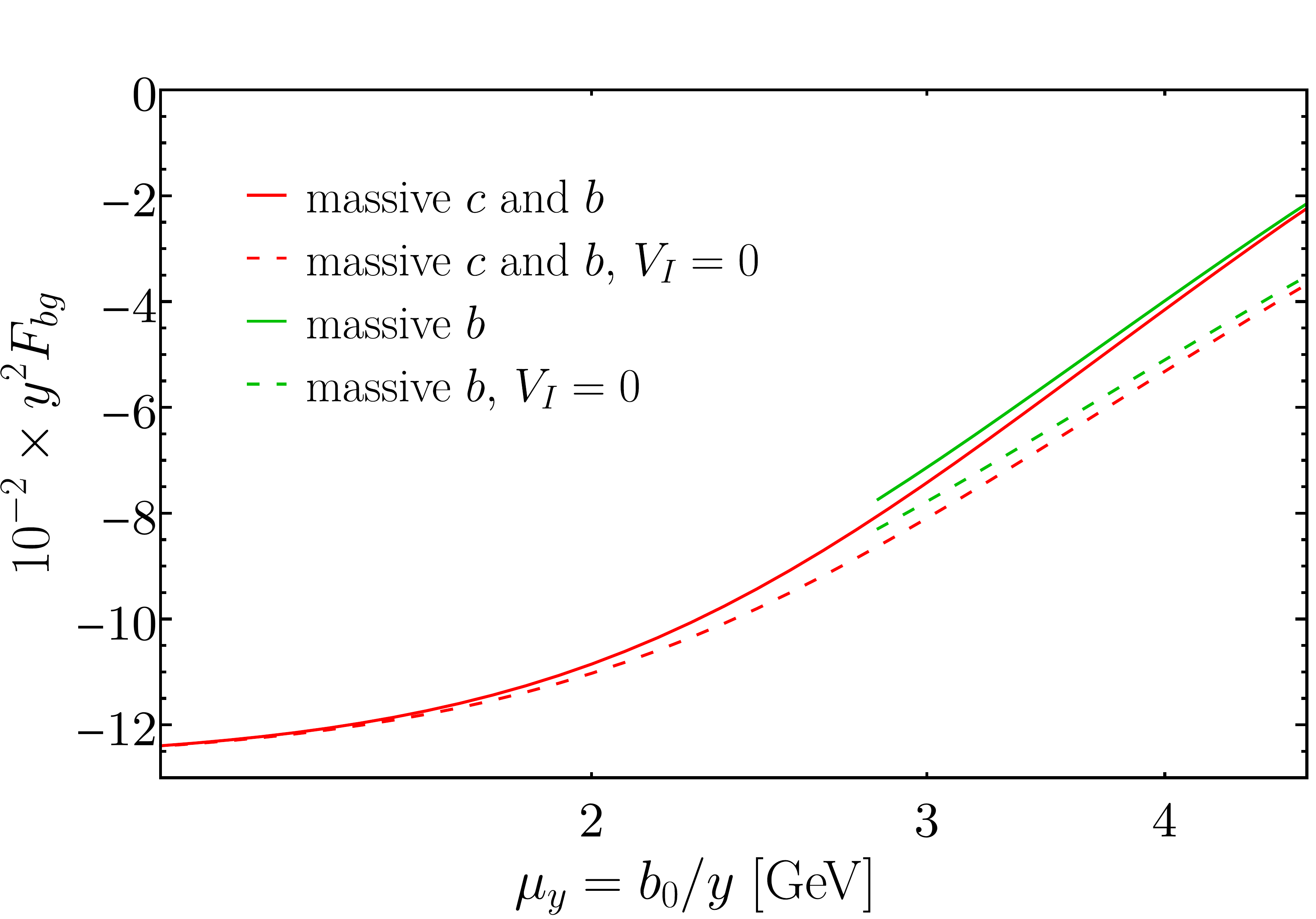}
      }
      \hfill
      \subfigure[\label{subfig:F_gb_high_muy}
         $y^2 F_{b g}$ for $\mu_y \ge m_b$]{
         \includegraphics[width=0.475\linewidth, trim=0 0 0 50, clip]{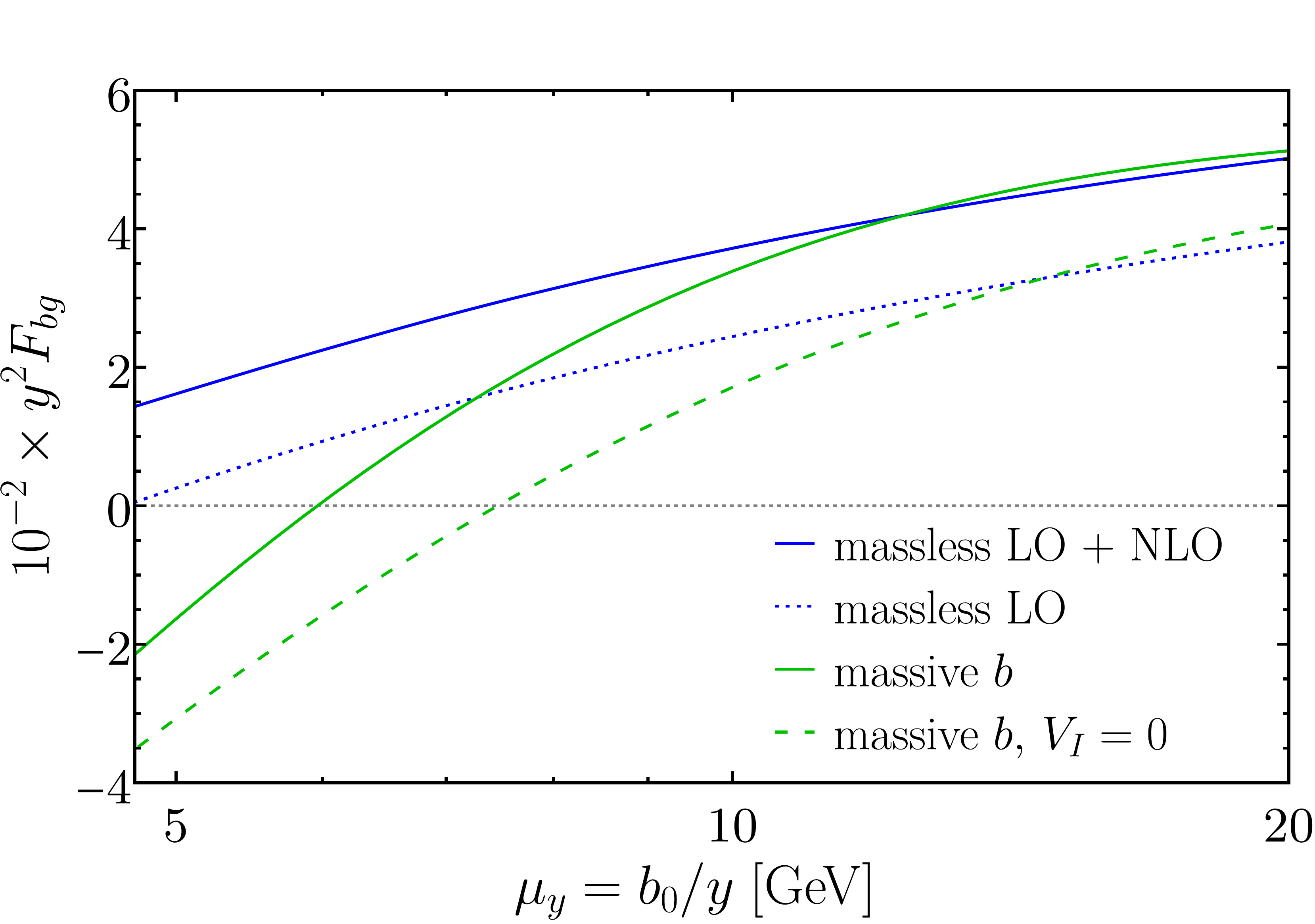}
      }
      \caption{\label{fig:F_gb} The $g b$ distribution for $\nf = 5$ at the scale $\mu = \mu_{y^*}$ where the splitting formula \protect\eqref{eq:small-y-DPD} is evaluated.  The momentum fractions $x_1$ and $x_2$ correspond to the dijet production setting~\eqref{eq:dijet-setting}.  The splitting is computed from $V^{5} \otimes f^{5}$ (massless, blue curves), from $V^{b} \otimes f^{4}$ (massive $b$, green curves), or from $V^{c b} \otimes f^{3}$ (massive $c$ and $b$, red curves).  With massive splitting kernels, the $g b$ distribution starts at NLO, and the interpolating part $V^I$ of the kernels is modelled as described in the present section.  We have multiplied the DPD with $y^2$ in order to remove the leading power behaviour $F_{g b} \propto 1/y^2$ at large $\mu_y$.}
   \end{center}
\end{figure}

In \fig{\ref{fig:F_gb}} we show $F_{g b}$ for $\nf = 5$ active flavors at the scale $\mu = \mu_{y^*}$, computed with different treatments of the charm and bottom masses.  We note that for $\mu_{y} \ge m_b$ one has $\mu_{y^*} \approx \mu_{y}$.  For the PDFs in the splitting formula we take the NLO distributions of the MSHT20 parametrisation, with $\nf = 3, 4$, or $5$ active flavours as appropriate.\footnote{%
The respective LHAPDF set names are \texttt{MSHT20nlo\_nf3}, \texttt{MSHT20nlo\_nf4}, and \texttt{MSHT20nlo\_as180}.}
For the strong coupling we use the value corresponding to these distributions, namely $\alpha_s(M_Z) = 0.118$ for $\nf = 5$.  The quark masses are as given in \eqref{eq:MSHT-masses}.

Comparing the solid and dotted blue curves in \fig{\ref{subfig:F_gb_high_muy}}, we see that even with massless kernels the contribution from NLO splitting to $F_{g b}$ is important.  This is because the LO term goes with $f_b$ evaluated at the scale $\mu_{y^*}$, which is much smaller than $f_g$ and even vanishes when $\mu_{y^*} = m_b$.  With massive splitting kernels, the LO term is zero and the NLO result originates entirely from $f_g$.

Most importantly, we see in \fig{\ref{subfig:F_gb_high_muy}} that the NLO results computed with $V^{b}$ or $V^{5}$ are close to each other in the entire region $2 m_b \le \mu_y \le 4 m_b$.  Switching from one to the other description in this region (i.e.\ taking $2 \le \beta \le 4$) hence results in only a small discontinuity of $F_{g b}(x_1, x_2, y; \mu_{y^*})$ in $y$.  This discontinuity should become even smaller when the DPDs are evolved to higher scales, e.g.\ to $\mu = 25 \gev$ as in \fig{\ref{fig:F-alpha-beta-comp}}.  This is because under evolution $F_{g b}$ receives a substantial contribution from the convolution of $F_{g g}$ with the $g \to b$ splitting kernel.  We verified that in the kinematics of \fig{\ref{subfig:F_gb_high_muy}} $F_{g g}$ barely changes when switching from massive to massless kernels.  The upshot of our discussion is that with NLO kernels one can indeed take larger values of $\beta$ and thus extend the region in which mass effects are taken into account.

To exhibit the importance of the interpolating part $V^{I}$ of the NLO kernel in our model, we also show in \fig{\ref{fig:F_gb}} the results obtained when setting this part to zero.  Comparison between the solid and dashed curves reveals that for decreasing $\mu_y$, i.e.\ when the $b$ quark decouples, the impact of $V^{I}$ is very mild.  By contrast, the interpolating kernel remains more important for increasing $\mu_y \ms$: even at $\mu_y = 20 \gev$ it accounts for about 20\% of the full NLO result.  This is not surprising, because we modelled the $y$ dependence as $V^{I}_{\ms g \smash{b, b}} \propto k_{00}^{}(y \ms m_b)$, which decreases like $y \exp(- 2 y\ms m_b)$ for $y\to \infty$ but only like $y^2 \log^2(y\ms m_b)$ for $y\to 0$.  We recall that a term with $k_{00}^{}(y \ms m_b)$ actually appears in the massive LO kernel \eqref{eq:VQQbarg-LO} for $g\to b \bbar$ splitting.

Taking the difference between our model for $V^{I}$ and $V^{I} = 0$ as an estimate of the model uncertainty, we conclude that taking NLO effects into account should be worthwhile even if the interpolating part of the massive kernels has not yet been calculated.

%
\section{Conclusions}
\label{sec:conclusions}

We have investigated quark mass effects for the $1\to 2$ splitting process in DPDs, which dominates the distributions in the limit of small interparton distances $y$.  Such mass effects should be taken into account if $y$ is comparable to the inverse $1/ \mQ$ of a heavy-quark mass.  We have set up schemes to compute the splitting part of DPDs in the full range of perturbatively small distances $y$.  In the massive scheme, $1\to 2$ splitting kernels including mass effects are used in a $y$ interval around $1 /\mQ$.  For smaller $y$, the quark $Q$ is treated as massless in the splitting process, whereas for large $y$, the DPDs including the heavy flavour are obtained by standard flavour matching, which involves the same matching kernels as flavour matching of PDFs.  A graphical representation of these three regimes is given in \fig{\ref{fig:splitting-regions}}.  The massive $1\to 2$ splitting kernels at LO accuracy can be found in \eqs{\eqref{eq:VQQbarg-LO}} and \eqref{eq:VQ-light-LO}.  A simpler, purely massless scheme, is formulated in which only massless $1\to 2$ splitting kernels are used.  The different schemes are depicted in \figs{\ref{fig:scheme-massless}} and \ref{fig:scheme-one-massive} for a single heavy flavour, and in \fig{\ref{fig:scheme-two-massive}} for the combination of charm and bottom quarks.

We numerically studied the different schemes at LO accuracy, focusing on the $y$ dependence of the DPDs and on the double parton luminosities that appear in physical cross sections, both evolved to scales much larger than the masses of the active quarks.  This is done using the \textsc{ChiliPDF} library, which we plan to make public in the future.  We find a moderate dependence of parton luminosities on the parameters $\alpha$ and $\beta$ that specify at which~$y$ the calculation switches between the three regimes of the massive scheme described above.  Luminosities calculated in the massless scheme roughly approximate the massive results if the scheme parameter $\gamma$ takes values between $1/2$ and $1$, with deviations that can strongly depend on the parton momentum fractions.  We furthermore see a substantial dependence of our results on the scale at which the DPDs are computed using the LO splitting formula, which hints at large corrections from higher-order corrections.

Whilst the calculation of massive splitting kernels at NLO is beyond the scope of this work, we derived several properties of these kernels, including their limiting values for small or large $y$, and their dependence on $\nf$ and the renormalisation scale $\mu$.  These properties are encoded in explicit parametrisations, given in \eqref{eq:Vm-qqbarg-explicit} to \eqref{eq:VI-qqbarg-zero} for one heavy flavour and in \eqref{eq:Vcb-qqbarg-explicit} to \eqref{eq:Vcb-bbbarg-explicit} for charm and bottom.  These results hold for all parton polarisations.  The non-trivial part of the NLO kernels is described by interpolating kernels $V^{I}_{a_1 a_2, a_0}$ that depend on two momentum fractions and on $y \ms \mQ$.

For unpolarised partons, DPDs obey momentum and number sum rules.  Using that these must hold for both $\nf$ and $\nf + 1$ active flavours, we derived corresponding sum rules for the massive splitting kernels.  Their all-order form is given in \eqs{\eqref{eq:momsum-7}} and \eqref{eq:numsum-7}, and their NLO part in \eqs{\eqref{eq:momsum-8}}, \eqref{eq:numsum-8}, \eqref{eq:momsum-11}, and \eqref{eq:numsum-11}.  Using these sum rule constraints, we have constructed a model ansatz for the unpolarised massive splitting kernels at two loops.  \rev{Within this model, we find that the inclusion of NLO effects reduces unphysical discontinuities when switching from massive to massless $b$ quarks in the calculation of $F_{b g}$.}

It would of course be preferable to have an explicit calculation of these kernels. \rev{This must, however, be left to future work.}

\appendix
%
\graphicspath{{Figures/Appendix/}}
%
%
\section{Numerical studies for \texorpdfstring{$W$}{W} pair production}
\label{app:add-numerics}
In \sects{\ref{sec:LO-DPDs}} and \ref{sec:LO-DPD-lumis} we presented numerical results for the two kinematic settings in \eqref{eq:dijet-setting} and \eqref{eq:ttbar-setting}.  Here we show and discuss some analogous plots for the setting in \eqref{eq:WW-setting}, which corresponds to the production of two $W$ bosons at the LHC.  This is because of the phenomenological importance of $W$-pair production, and it shows to which extent mass effects for charm and bottom persist in DPDs at the scale $\mu = m_W$ and in the associated parton luminosities.  Compared with the dijet production setting \eqref{eq:dijet-setting}, the DPDs in the $W$ production setting are $(i)$ evolved to a higher scale and $(ii)$ evaluated at parton momentum fractions that are larger by a factor $m_W / (25 \gev) \approx 3.1$.

\begin{figure}[t!]
   \begin{center}
   \subfigure[\label{subfig:Fccbar-massive-24-WW}${c \cbar}$]{
         \includegraphics[width=0.475\linewidth, trim=0 0 0 50, clip]{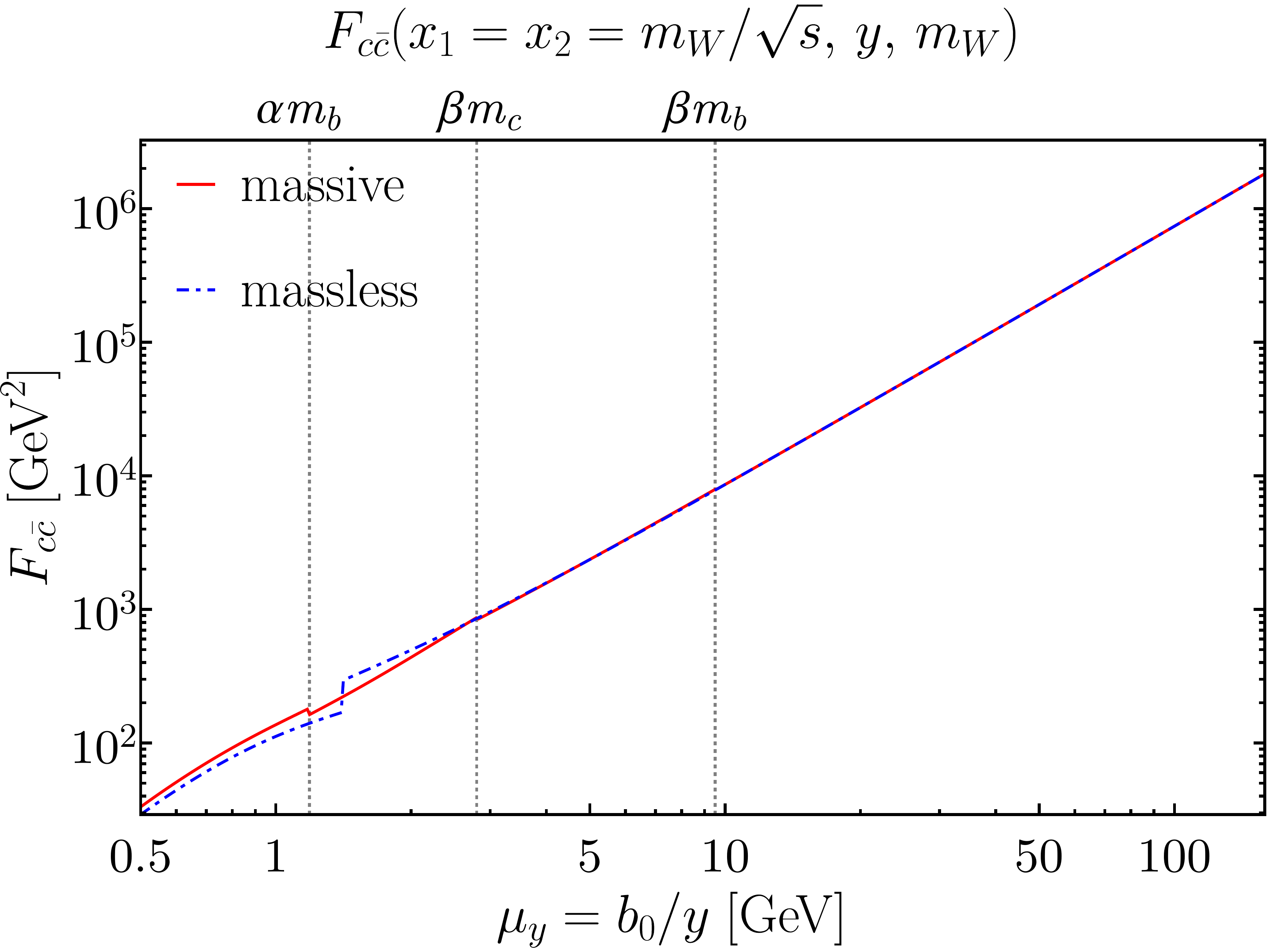}
      }
      \subfigure[\label{subfig:Fcb-massive-24-WW}${c b}$]{
         \includegraphics[width=0.475\linewidth, trim=0 0 0 50, clip]{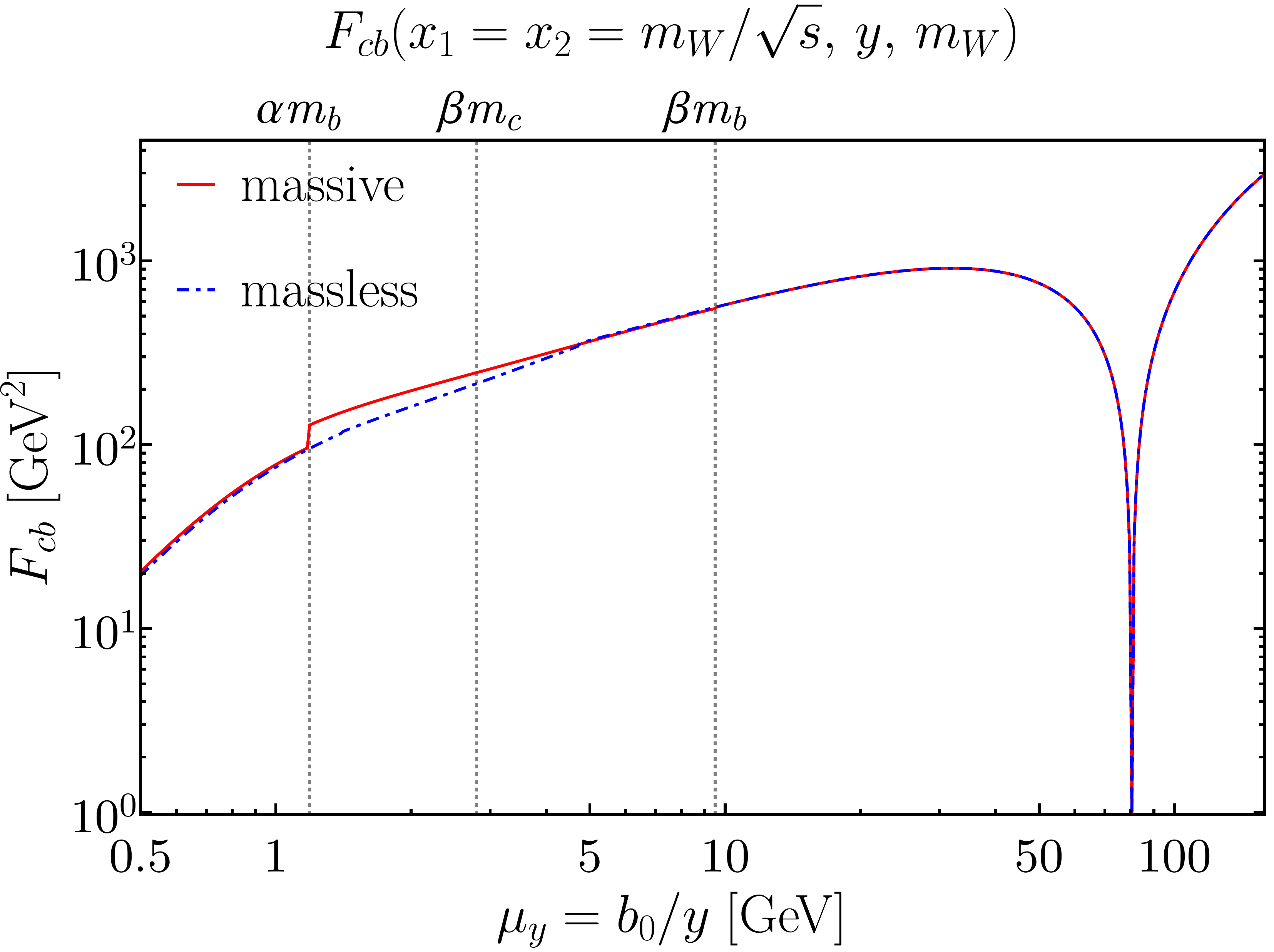}
      }
   \end{center}%
   \caption{\label{fig:F-massive-24-WW} $\nf = 5$ splitting DPDs at $\mu = m_W$ for the $W$ production setting \protect\eqref{eq:WW-setting}.  Solid lines are for the massive scheme of section \ref{sec:two-massive} with our preferred values $\alpha = 1/4$ and $\beta = 2$, and dashed lines are for the massless scheme with $\gamma = 1$.  The momentum fractions are $x_1 = x_2 \approx 5.7 \times 10^{-3}$ according to \protect\eqref{eq:mom-fracs-central-rap}.  Corresponding plots for the dijet production setting are shown in panels (a) and (c) of \fig{\protect\ref{fig:F-massive-24}}.}
\end{figure}

As one may expect, discontinuities in the $y$ dependence of the splitting DPDs are generally weaker at $\mu = m_W$ than at $\mu = 25 \gev$.  This is  clearly seen for the $c b$ distribution by comparing \figs{\ref{subfig:Fcb-massive-24-WW}} and \ref{subfig:Fcb-massive-24}.  Nonetheless, \fig{\ref{fig:F-massive-24-WW}} shows that discontinuities for $\mu_y$ just above $1 \gev$ persist even after evolution to the electroweak scale.

For double parton luminosities, we consider the two parton combinations $c \cbar  \bbar b$ and $c \bbar \sbar c$, which respectively contribute to opposite-sign and like-sign $W$-pair production.  The following plots for $c \cbar  \bbar b$ can be compared with the ones for $c \cbar b \bbar$ in the dijet production setting, because with splitting, evolution and flavour matching evaluated at LO, DPDs for $\bbar b$ and $b \bbar$ are identical.

\begin{figure}[p!]
   \begin{center}
      \subfigure[\label{subfig:Lccbarbbarb-WW}${c \cbar \bbar b}$]{
         \includegraphics[width=0.475\linewidth, trim=0 0 0 50, clip]{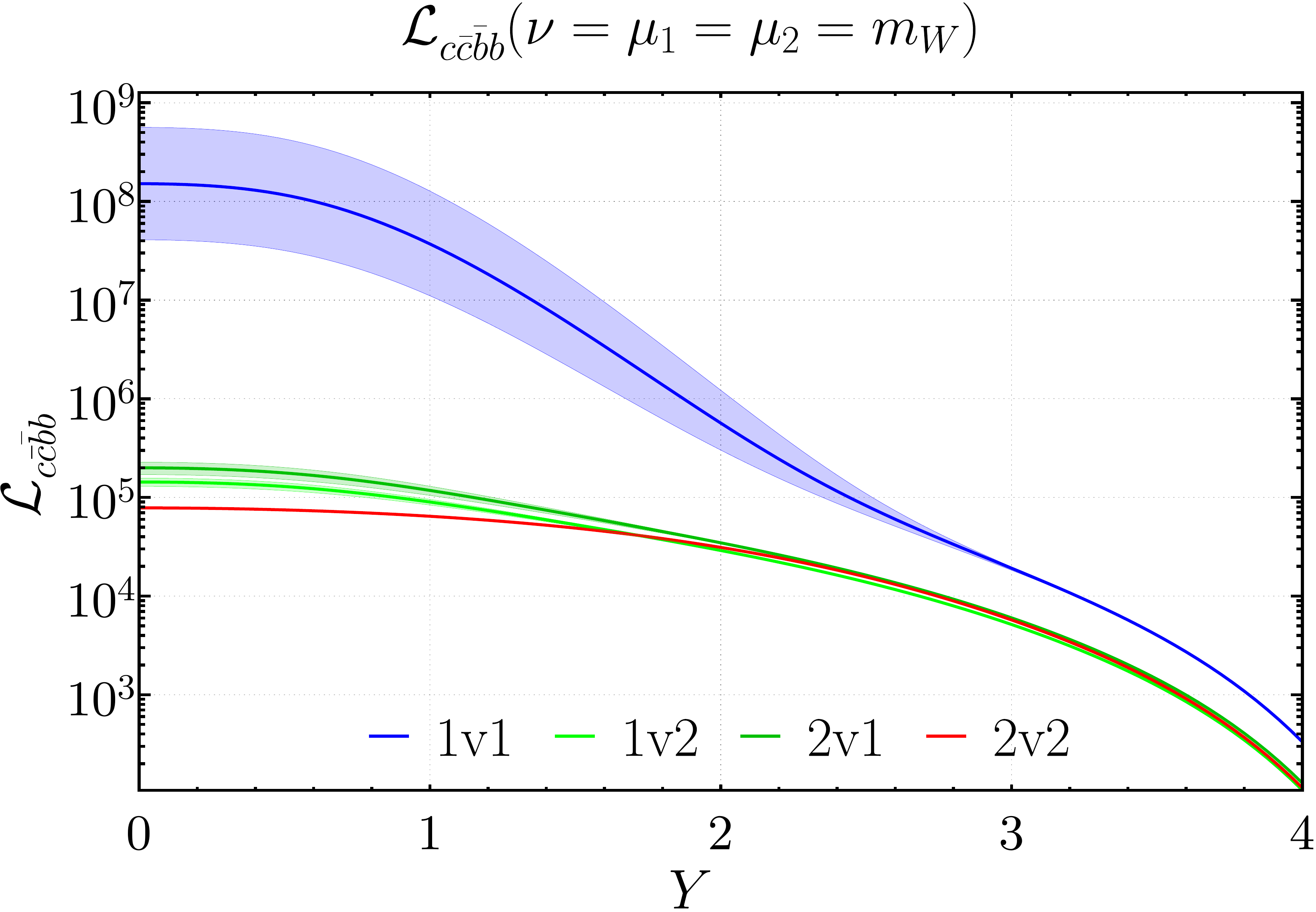}
      }
      \subfigure[\label{subfig:Lcbbarsbarc-WW}${c \bbar \sbar c}$]{
         \includegraphics[width=0.475\linewidth, trim=0 0 0 50, clip]{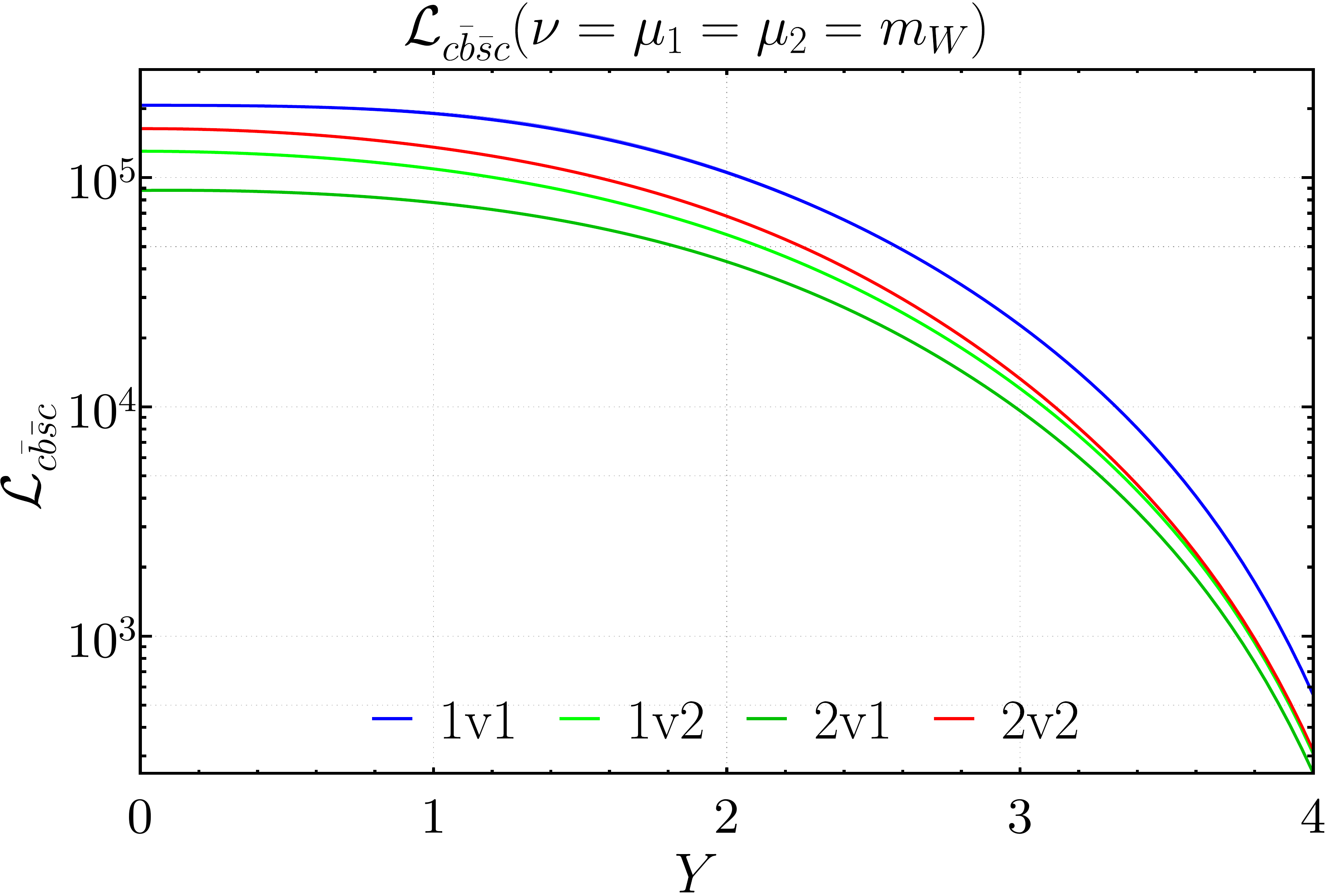}
      }
   \end{center}
   \caption{\label{fig:lumis-WW-contribs} Double parton luminosities at $\mu = m_W$ for the $W$ production setting, computed in the massive scheme of section \ref{sec:two-massive} with $\alpha = 1/4$ and $\beta = 2$.  Central values correspond to a cutoff parameter $\nu = \mu$, and bands to the variation of $\nu$ between $\mu/2$ and $2 \mu$.}
   \begin{center}
      \subfigure[\label{subfig:Lbbbarcbarc-WW-1v1}${c \cbar \bbar b}, \, {\text{1v1}}$]{
         \includegraphics[width=0.475\linewidth, trim=0 0 0 55, clip]{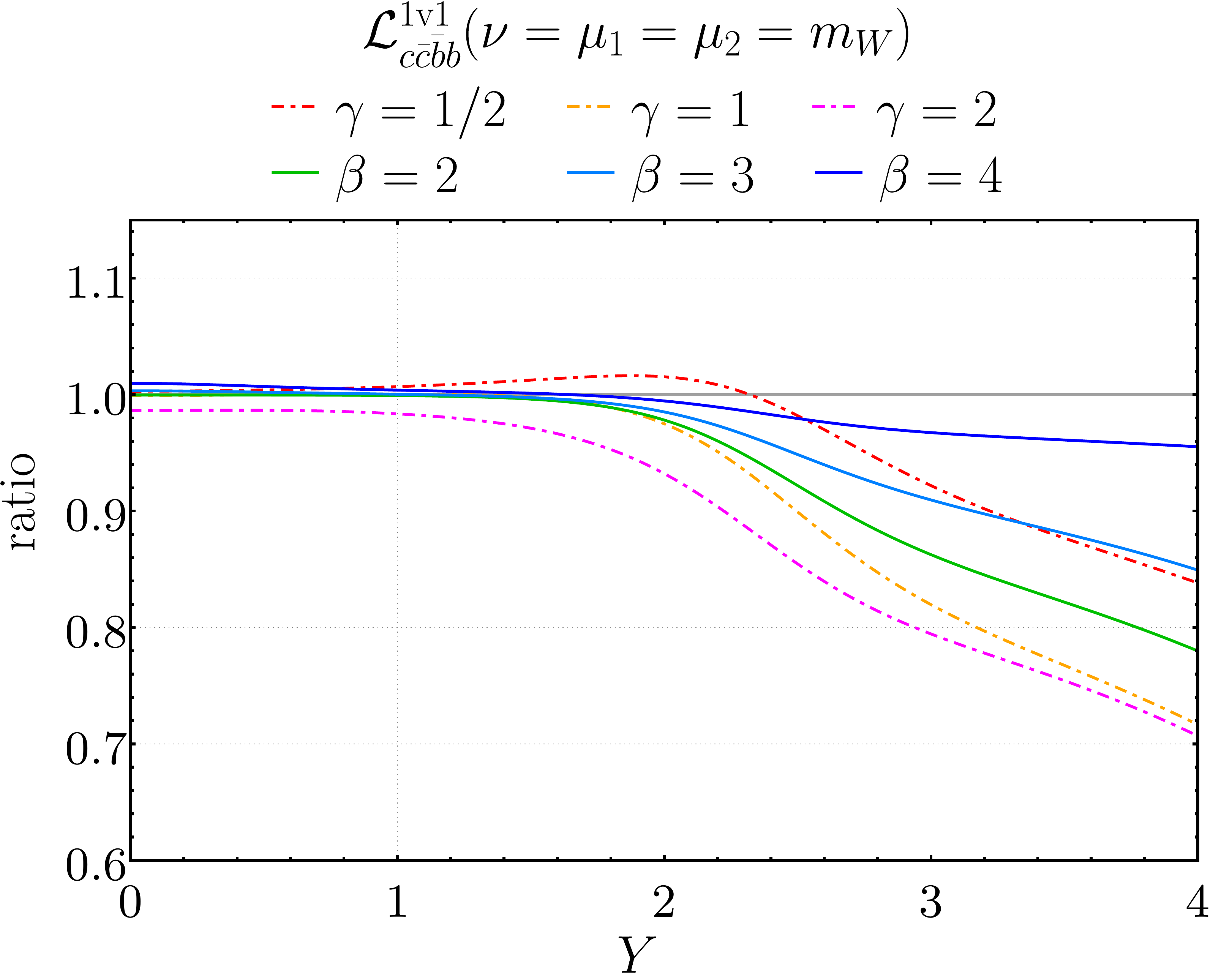}
      }
      \subfigure[\label{subfig:Lbbbarcbarc-WW-1v2}${c \cbar \bbar b}, \, {\text{1v2}}$]{
         \includegraphics[width=0.475\linewidth, trim=0 0 0 55, clip]{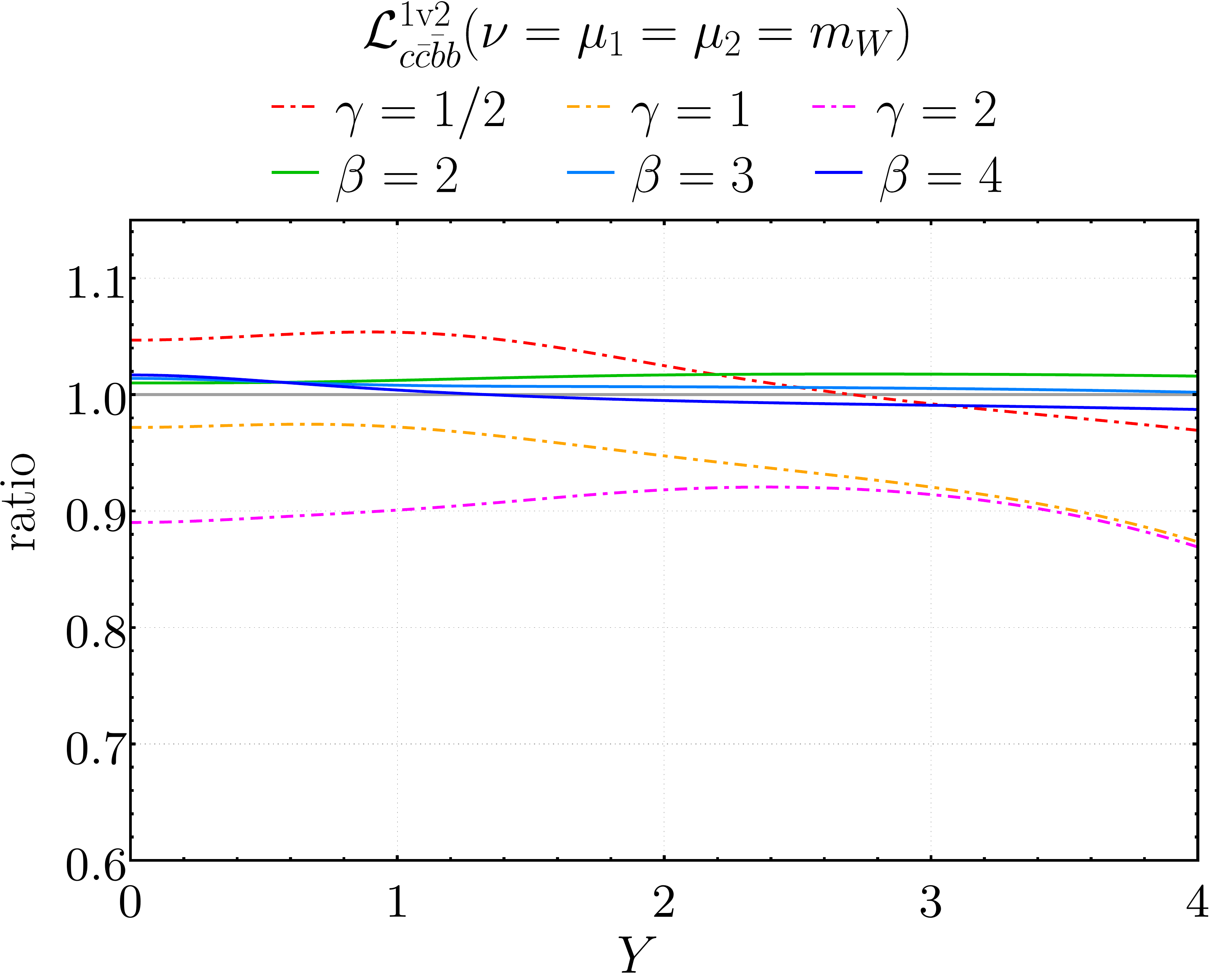}
      }
      \subfigure[\label{subfig:Lbbbarcbarc-WW-2v1}${c \cbar \bbar b}, \, {\text{2v1}}$]{
         \includegraphics[width=0.475\linewidth, trim=0 0 0 55, clip]{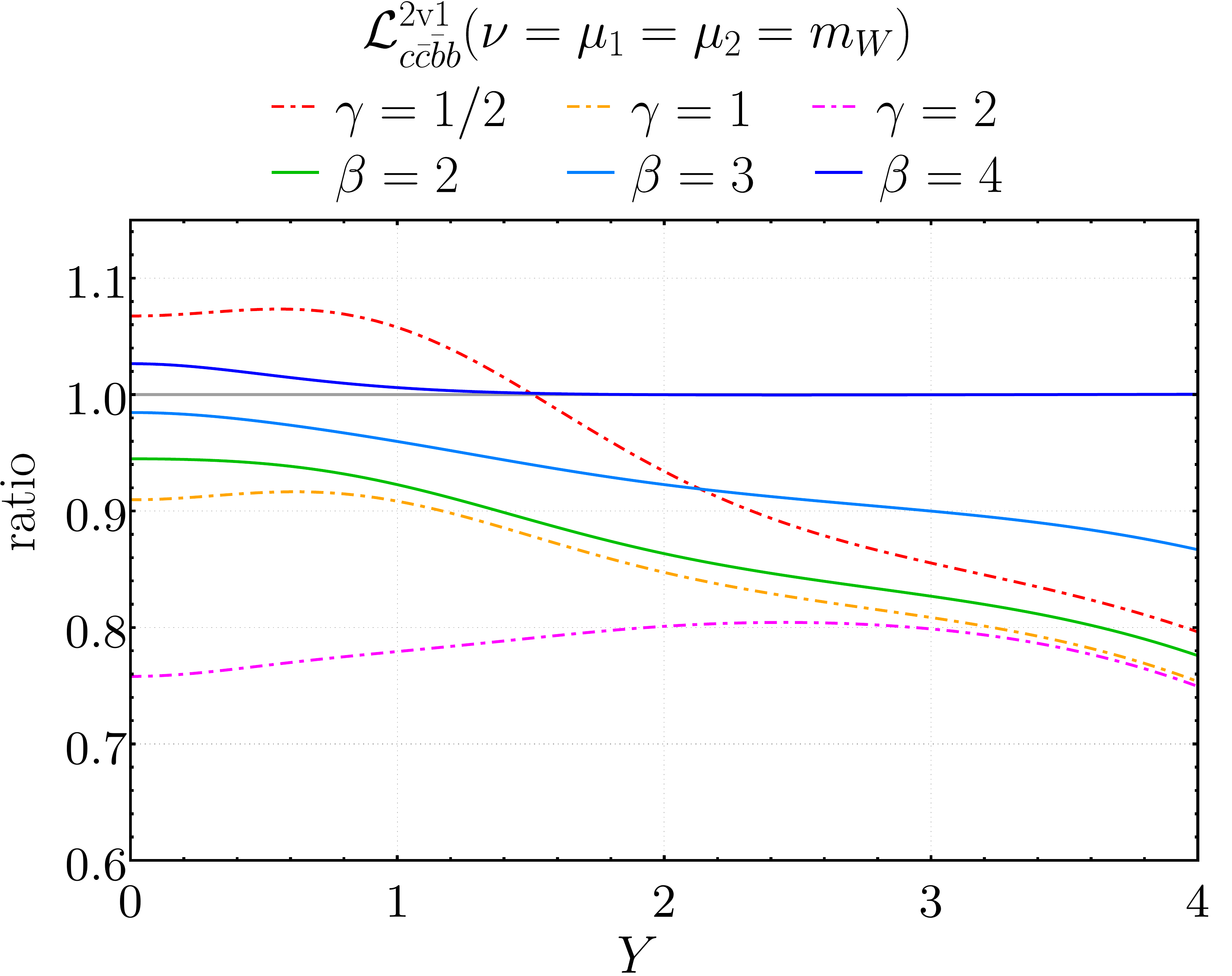}
      }
   \end{center}
   \caption{\label{fig:Lbbbarcbarc-WW} Double parton luminosity ratios \protect\eqref{eq:lumi-ratios} for $c \cbar \bbar b$ in the $W$ production setting with $\nu = \mu = 25 \gev$.  Solid (dashed) lines correspond to the numerator computed in the (massless) scheme.  The corresponding plot for $c \cbar b \bbar$ in the dijet production setting is shown in \fig{\protect\ref{fig:Lccbarbbbar-jets}}.}
\end{figure}

It is therefore not surprising that the qualitative behaviour of the $c \cbar \bbar b$ luminosity in \fig{\ref{subfig:Lccbarbbarb-WW}} is the same as for $c \cbar b \bbar$ in \fig{\ref{subfig:Lccbarbbbar-jets}}.  In particular, the 1v1 contribution is dominant and has a large $\nu$ dependence at low $Y$.  In the $c \bbar \sbar c$ channel, the $\nu$ dependence is very small everywhere, as expected for a parton combination that cannot be directly produced by $1\to 2$ splitting.

The scheme dependence of the ${c \cbar \bbar b}$ luminosity is shown in \fig{\ref{fig:Lbbbarcbarc-WW}}.  Compared with the dijet production setting in \fig{\ref{fig:Lccbarbbbar-jets}}, the qualitative behaviour of the ratios is very similar, whilst their deviation from unity is somewhat reduced in most cases.  In both settings, the smallest deviations in the massless scheme are obtained for $\gamma = 1/2$.

\begin{figure}[p!]
   \begin{center}
      \subfigure[\label{subfig:Lbbbarcbarc-WW-scale}${c \cbar \bbar b}$]{
         \includegraphics[width=0.475\linewidth, trim=0 0 0 50, clip]{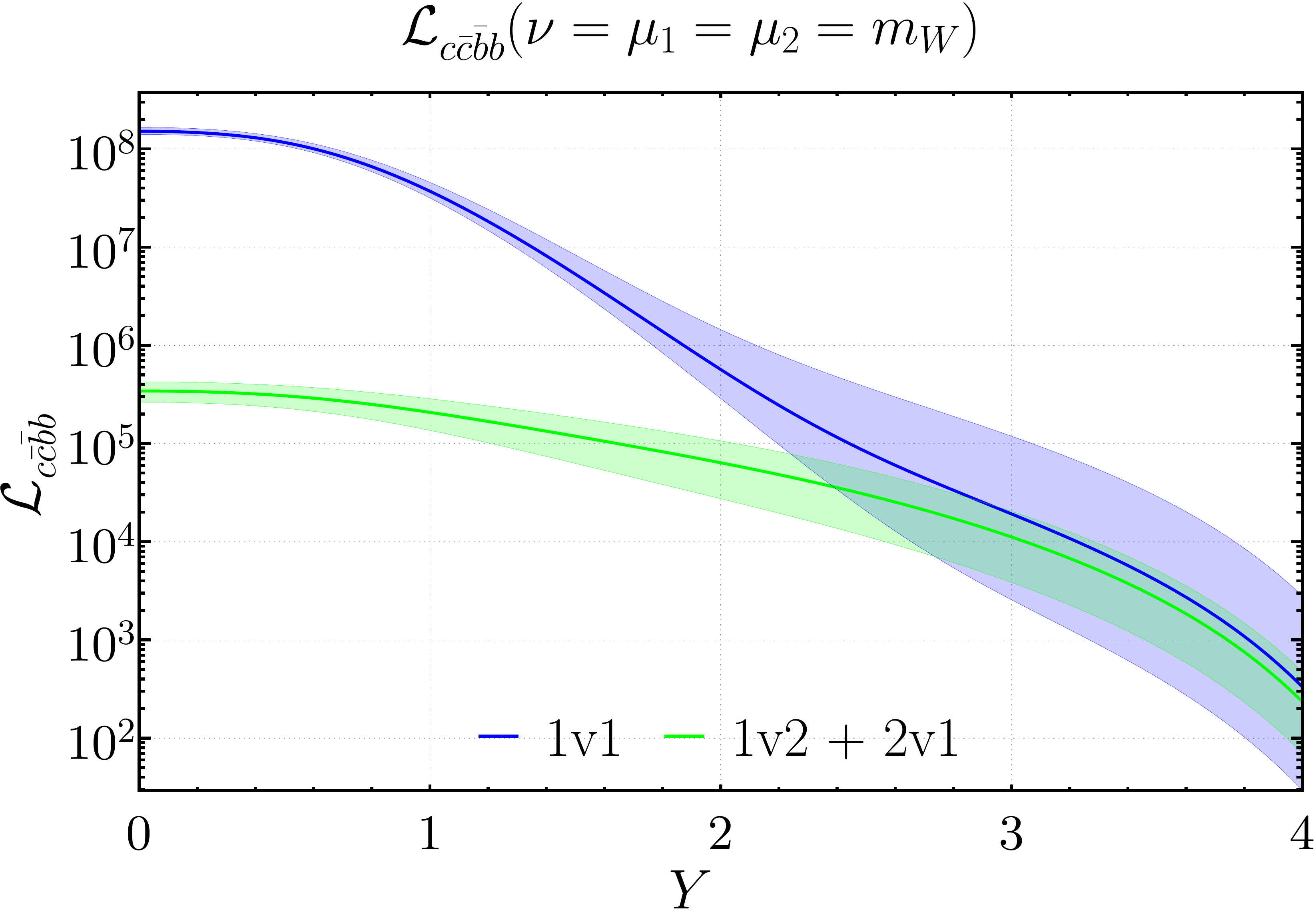}
      }
      \subfigure[\label{subfig:Lcbbarsbarc-WW-scale}${c \bbar \sbar c}$]{
         \includegraphics[width=0.475\linewidth, trim=0 0 0 50, clip]{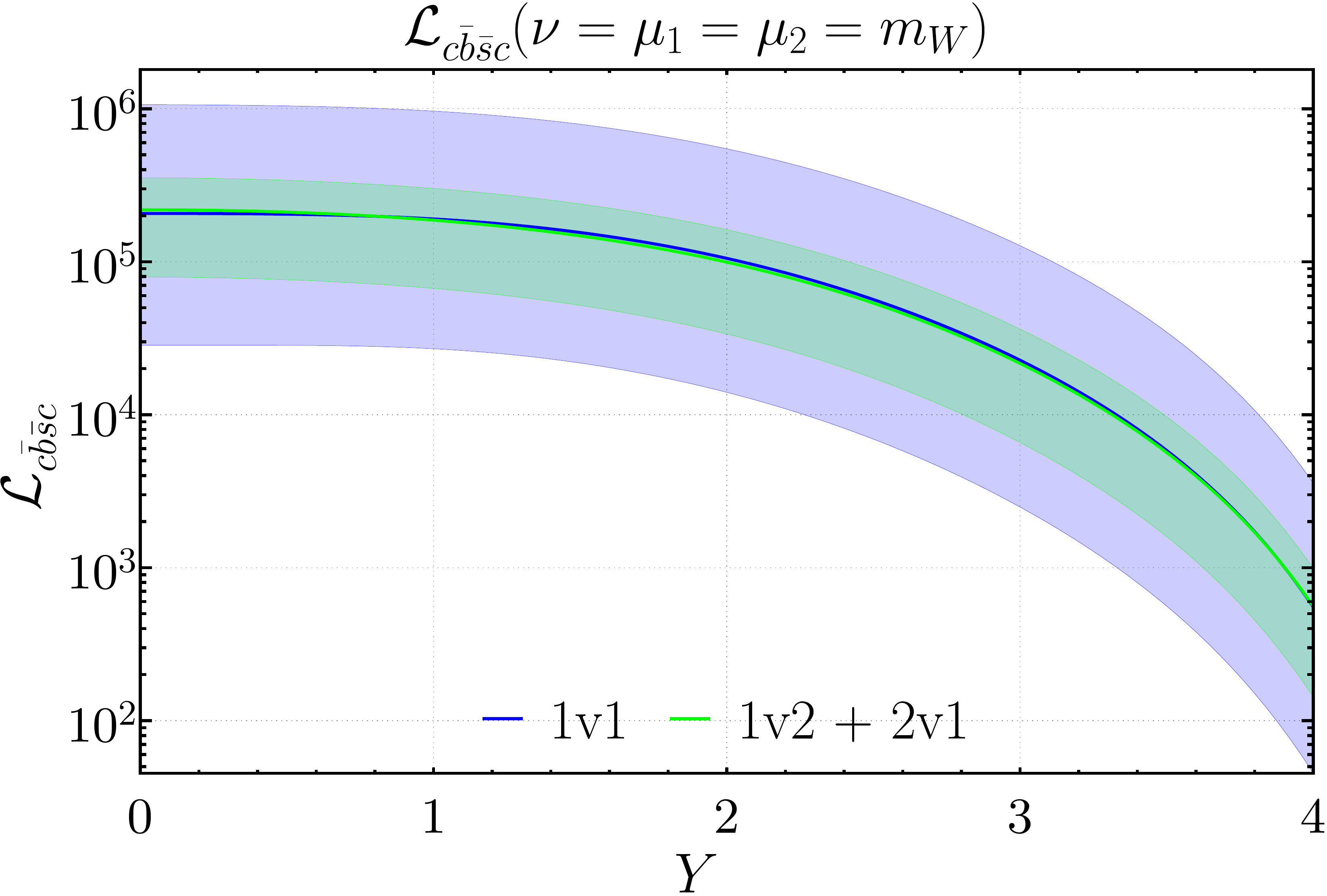}
      }
   \end{center}
   \caption{\label{fig:lumis-WW-scale} Splitting scale variation of the double parton luminosities shown in \fig{\protect\ref{fig:lumis-WW-contribs}} for the $W$ production setting.  Central curves are for $\mu_{\text{split}} = \mu_{y^*}$, and bands correspond to the variation specified in \protect\eqref{eq:mu-split-variation}.}
   \begin{center}
      \subfigure[\label{subfig:Lbbbarsbars-WW-match-LO}${c \cbar \bbar b}$ with LO flavour matching]{
         \includegraphics[width=0.475\linewidth, trim=0 0 0 50, clip]{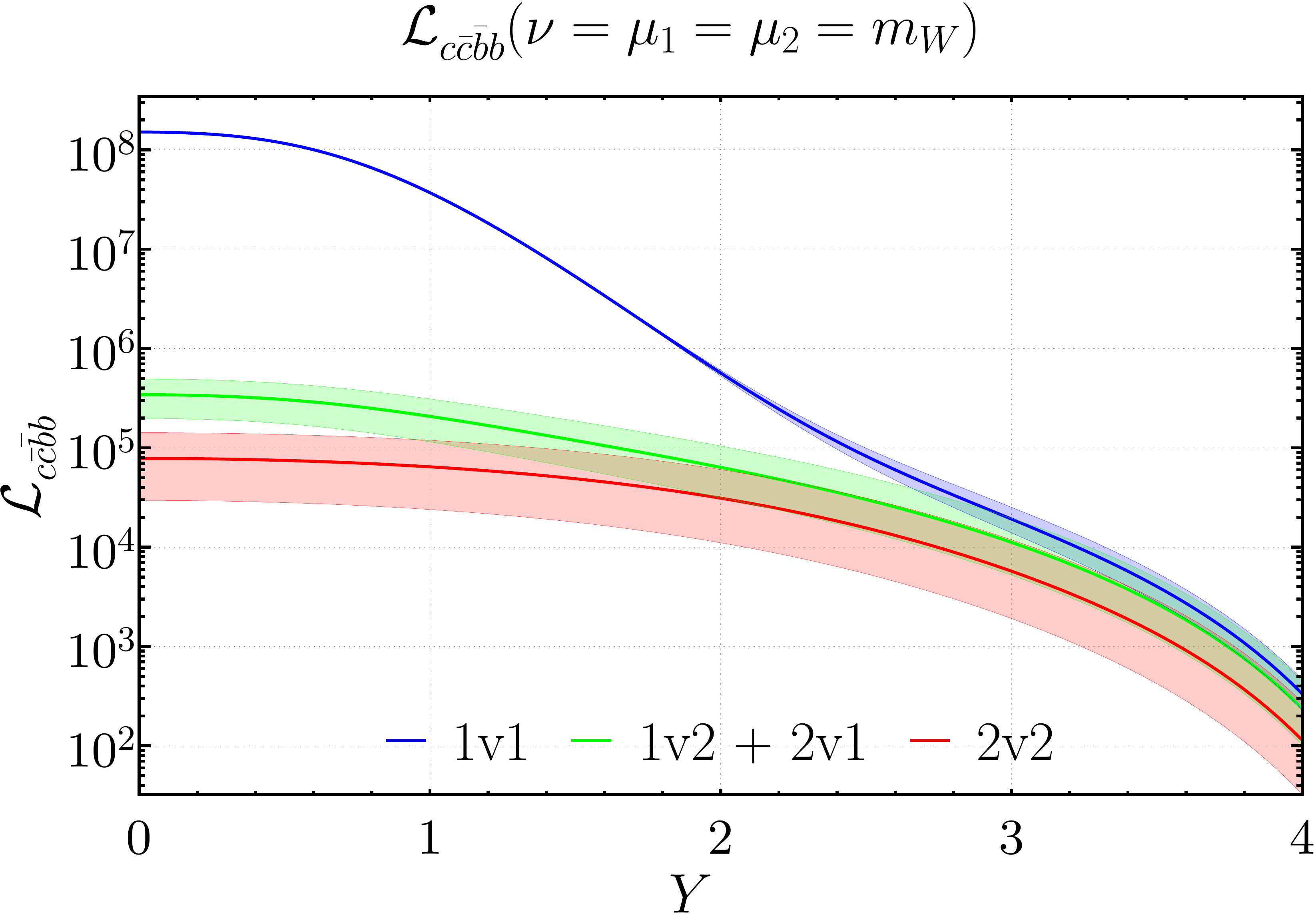}
      }
      \subfigure[\label{subfig:Lbbbarsbars-WW-match-NLO}${c \cbar \bbar b}$ with NLO flavour matching]{
         \includegraphics[width=0.475\linewidth, trim=0 0 0 50, clip]{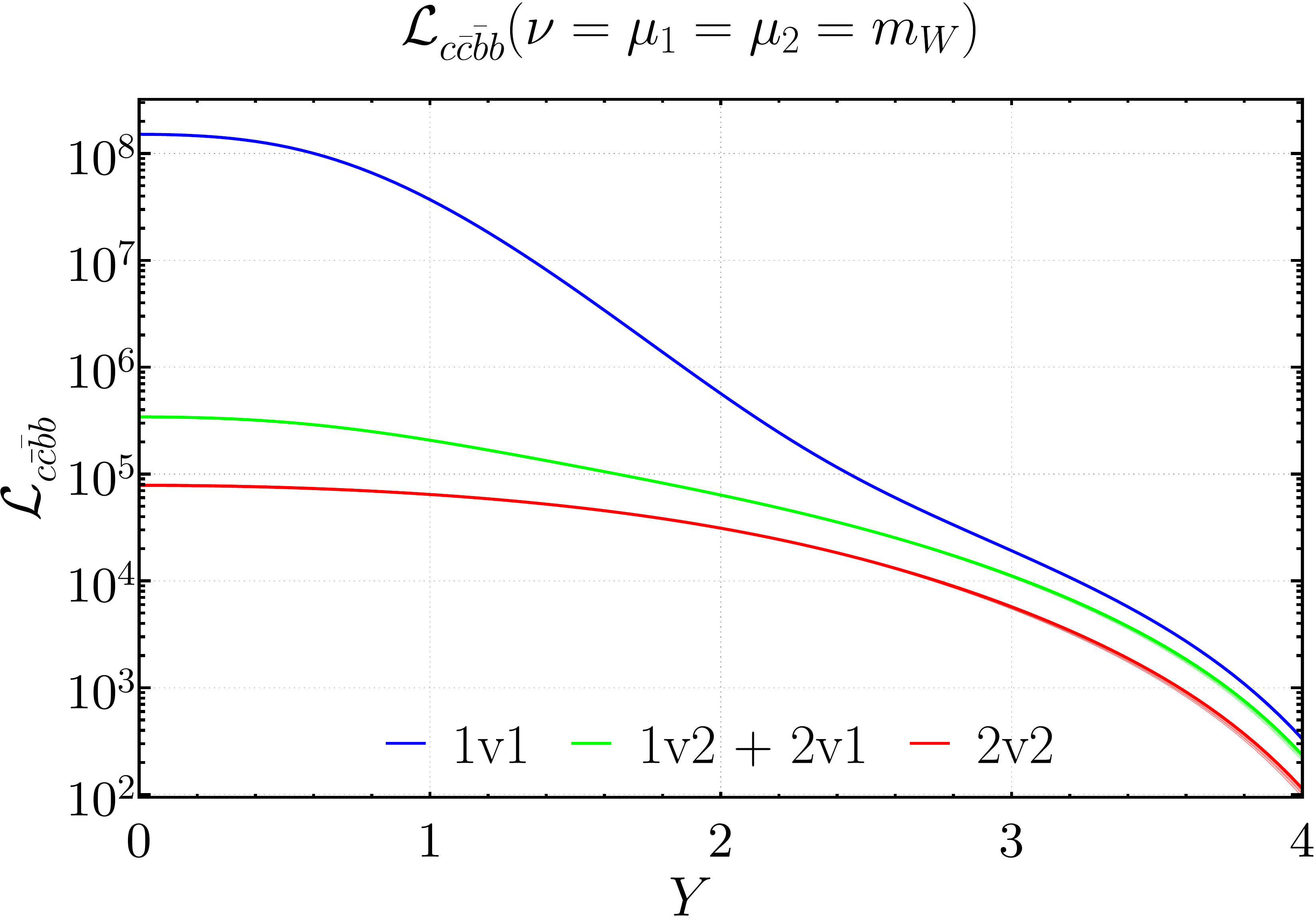}
      }
      \\
      \subfigure[\label{subfig:Lcbsbardbar-WW-match-LO}${c \bbar \sbar c}$ with LO flavour matching]{
         \includegraphics[width=0.475\linewidth, trim=0 0 0 47, clip]{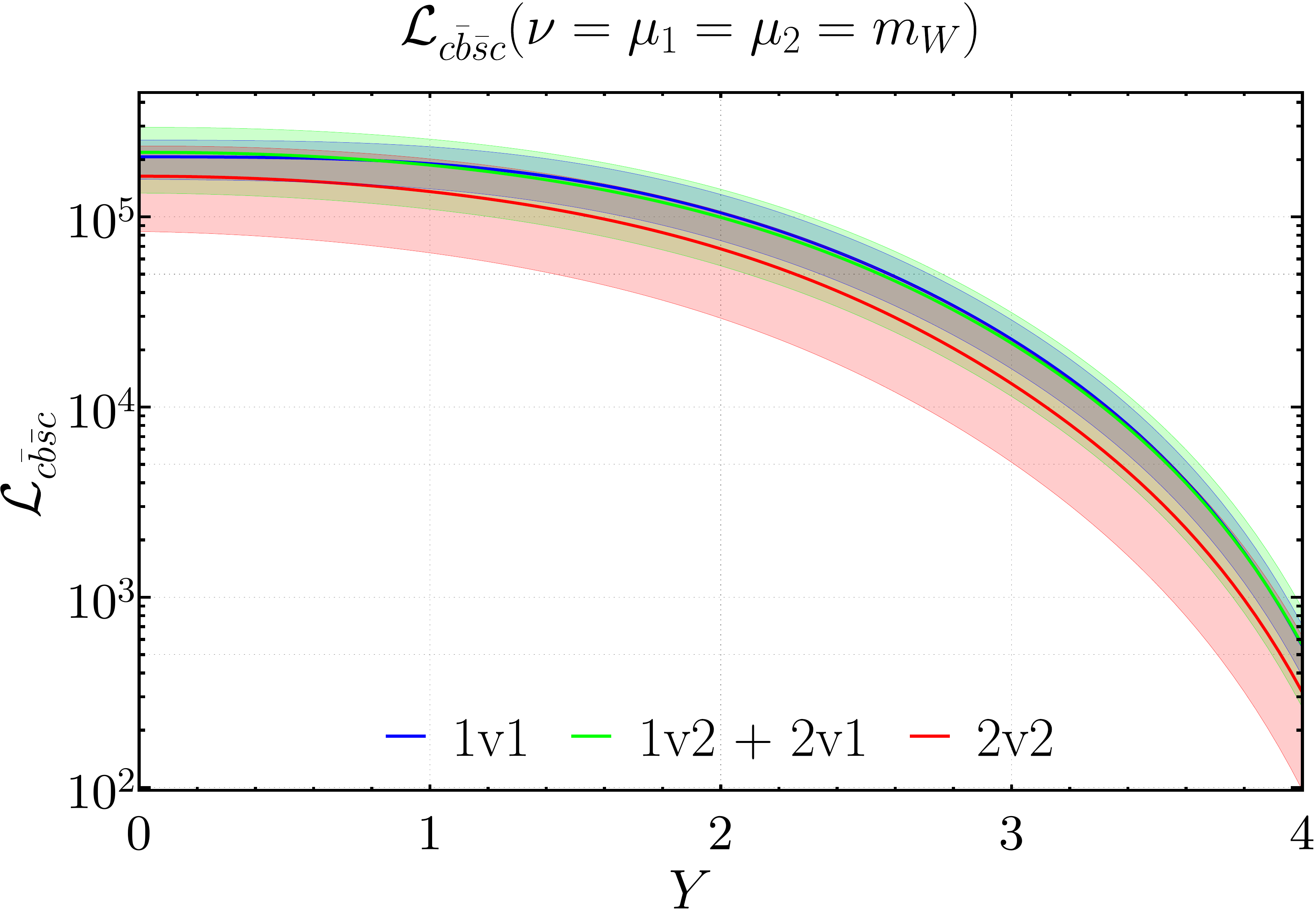}
      }
      \subfigure[\label{subfig:Lcbsbardbar-WW-match-NLO}${c \bbar \sbar c}$ with NLO flavour matching]{
         \includegraphics[width=0.475\linewidth, trim=0 0 0 47, clip]{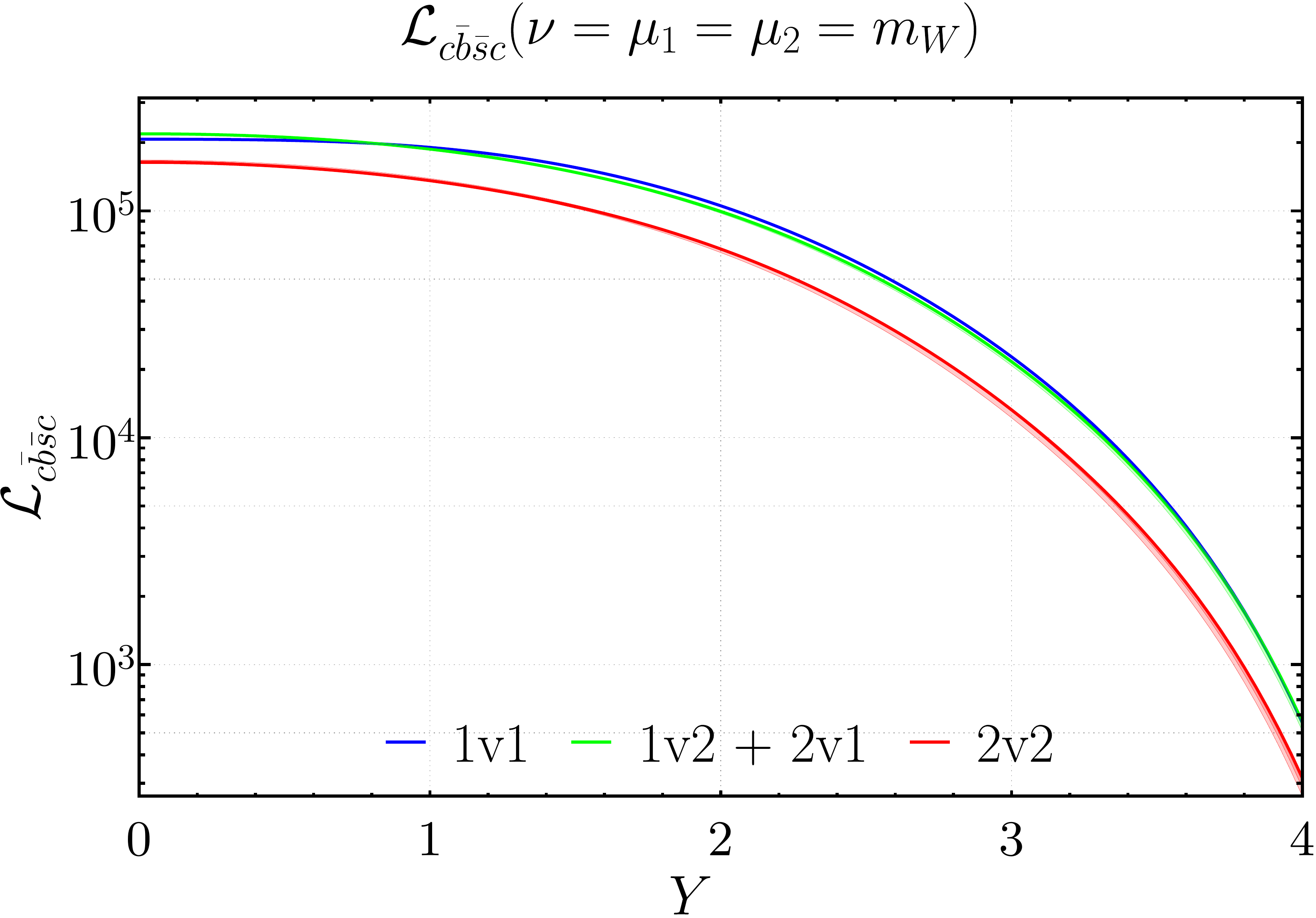}
      }
   \end{center}
   \caption{\label{fig:lumis-WW-match} Dependence on the flavour matching scale of the double parton luminosities shown in \fig{\protect\ref{fig:lumis-WW-contribs}} for the $W$ production setting.  Central curves are for $\mu_{Q} = \mQ$, and bands correspond to the variation specified in \protect\eqref{eq:mu-Q-variation}. }
\end{figure}

The dependence of the luminosities on the splitting scale (\fig{\ref{fig:lumis-WW-scale}}) follows the same pattern we found in the two other settings.  For ${c \cbar \bbar b}$, which can be produced directly by gluon splitting in both DPDs, the scale variation is weak at low $Y$ and grows towards large $Y$.  By contrast, a strong scale variation with little $Y$ dependence is seen for ${c \bbar \sbar c}$, just like in panels (b) to (d) of \fig{\ref{fig:lumis-jets-scale}}.  The band for the sum of 1v2 and 2v1 contributions is quite asymmetric in \fig{\ref{subfig:Lcbbarsbarc-WW-scale}}, which indicates important contributions from large $y$ also in this setting.

For the matching scale dependence, we observe large effects in \fig{\ref{fig:lumis-WW-match}} when flavour matching is performed at LO, even though the matching scales for charm and bottom are much smaller than the scale $\mu = m_W$ at which the DPDs are evaluated.  As in the kinematic settings for dijet or $t \tbar$ production, the scale variation becomes small when the NLO matching coefficients are included in the calculation.

\begin{figure}[t!]
   \begin{center}
      \subfigure[\label{subfig:Fbbbar-scale-Y0}$F_{\ms \bbar b}$ for $Y = 0$]{
         \includegraphics[width=0.475\linewidth, trim=0 0 0 50, clip]{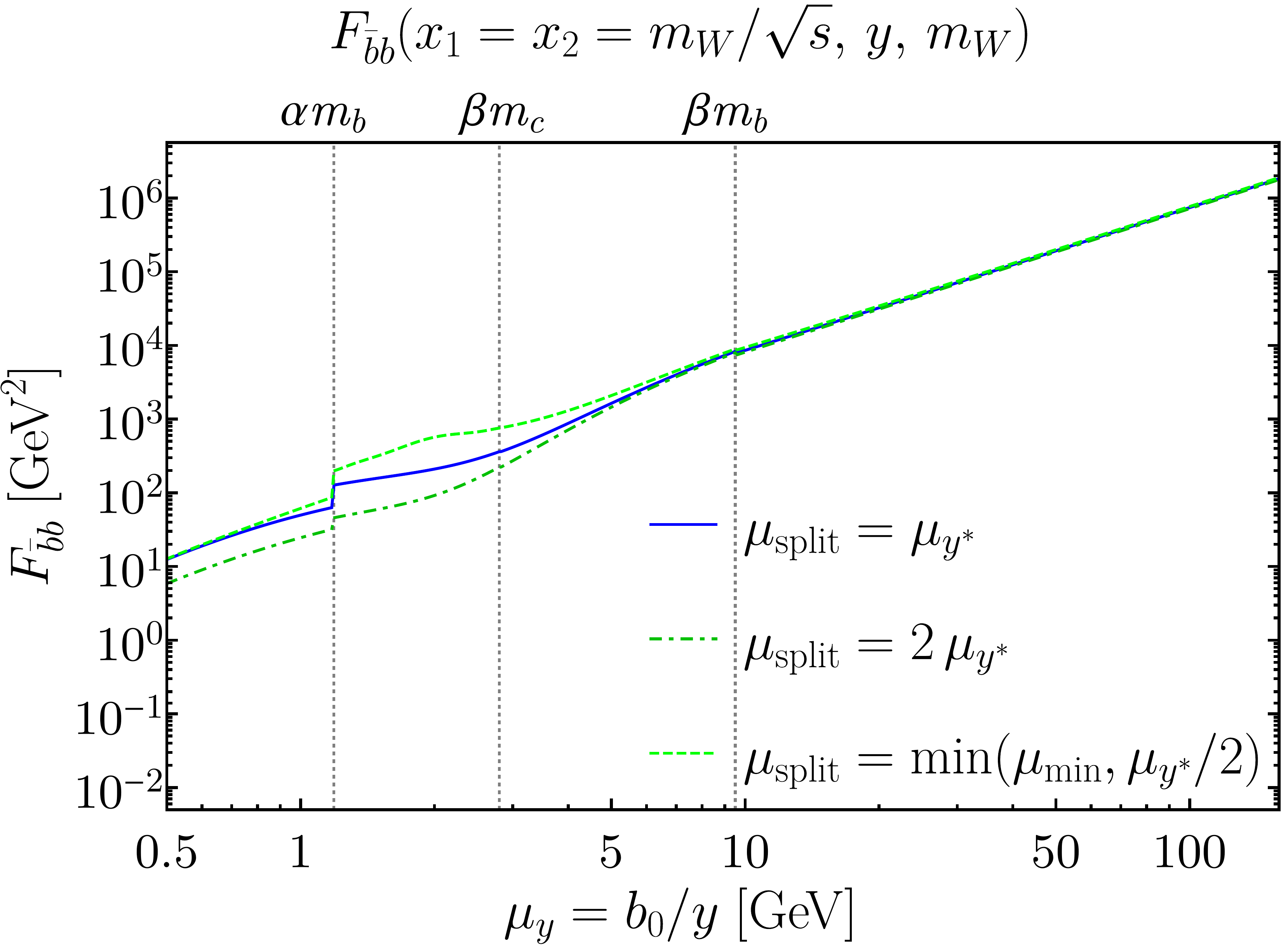}
      }
      \subfigure[\label{subfig:Fbbbar-scale-Y0-g-qqbar}$F_{\ms \bbar b}$ for $Y = 0$, only $V_{\, \smash{\bbar} b, g}$]{
         \includegraphics[width=0.475\linewidth, trim=0 0 0 50, clip]{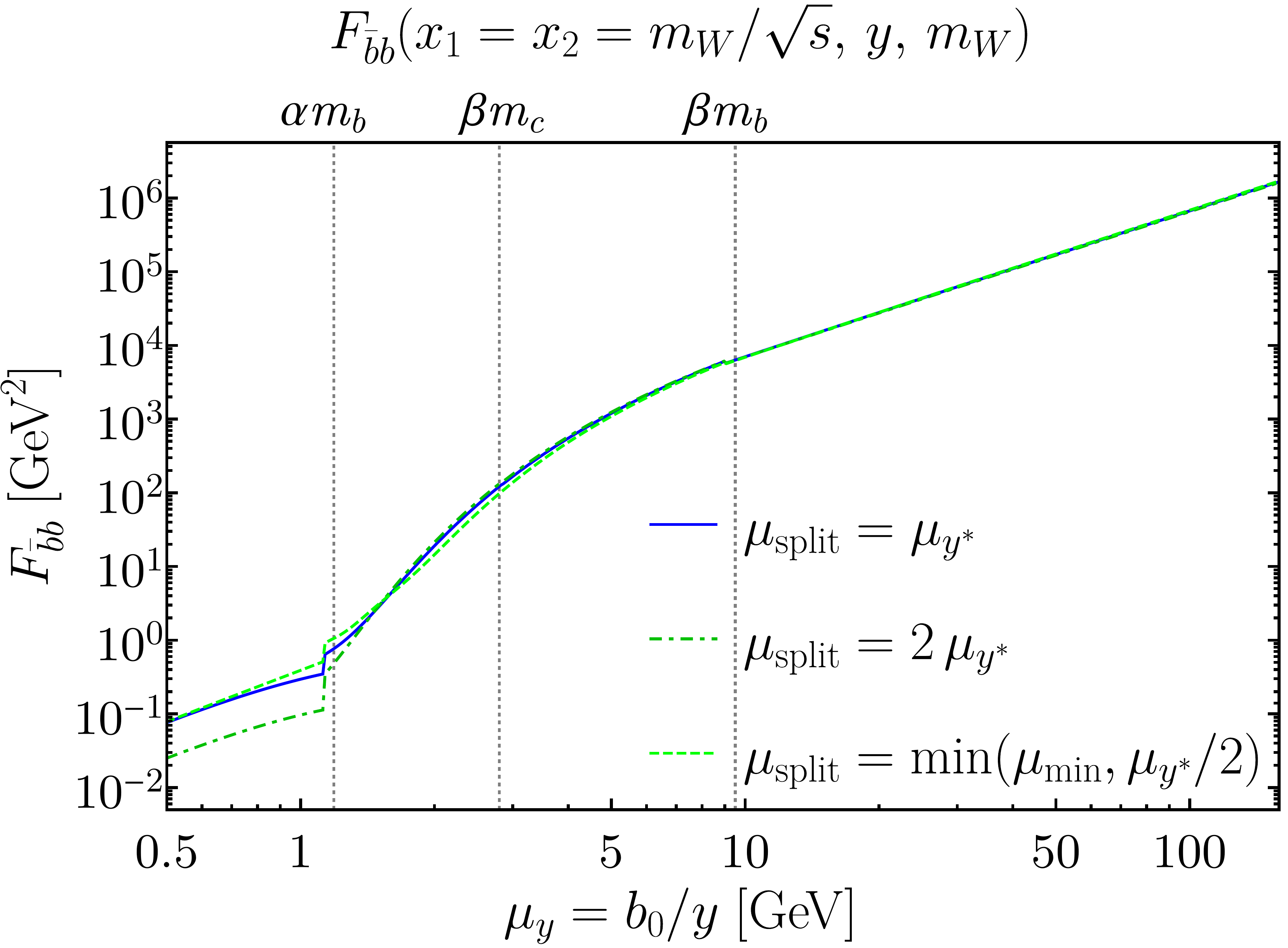}
      }
      \\
      \subfigure[\label{subfig:Fbbbar-scale-Y3}$F_{\ms \bbar b}$ for $Y = 3$]{
         \includegraphics[width=0.475\linewidth, trim=0 0 0 50, clip]{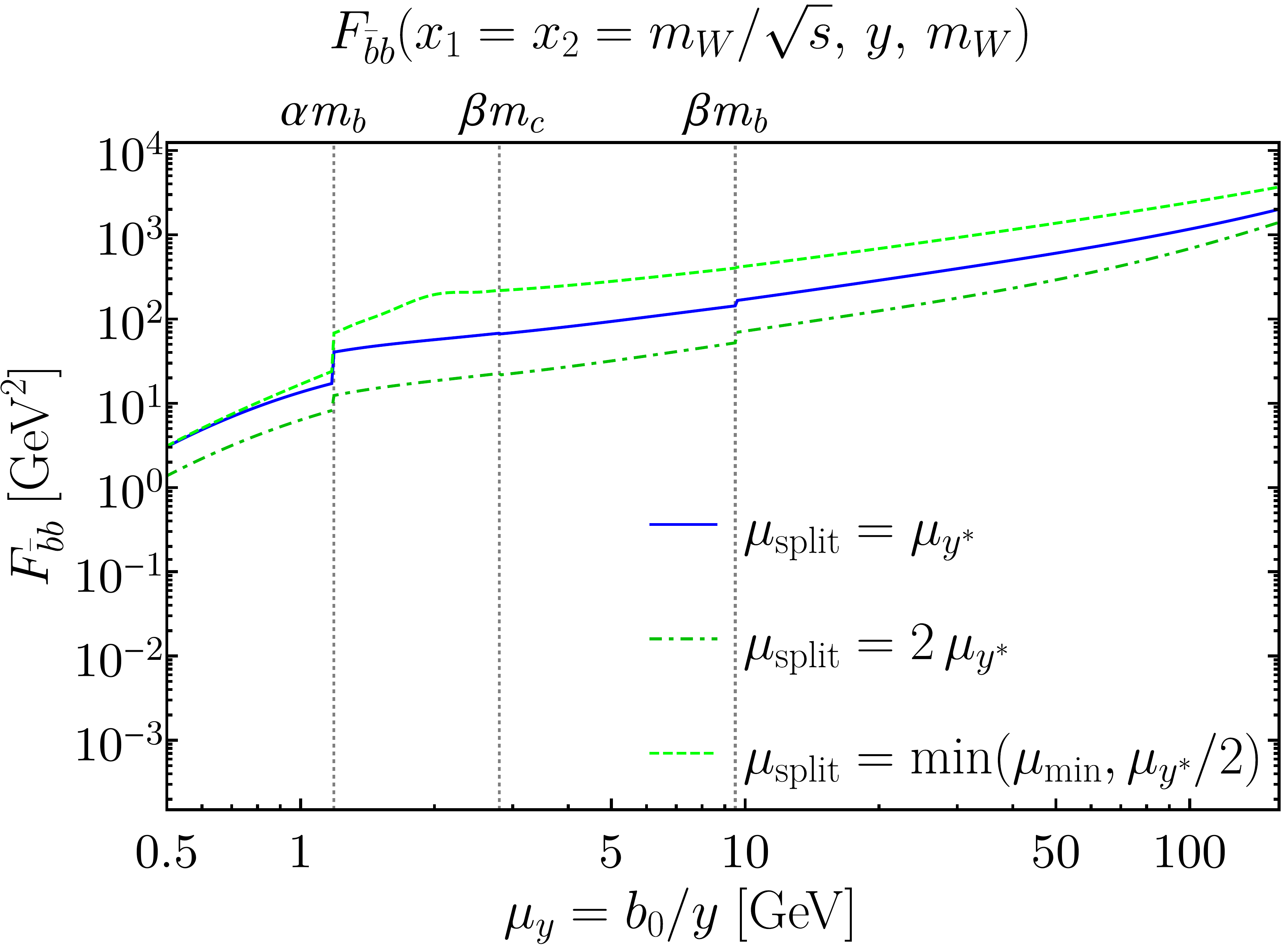}
      }
      \subfigure[\label{subfig:Fbbbar-scale-Y3-g-qqbar}$F_{\ms \bbar b}$ for $Y = 3$, only $V_{\, \smash{\bbar} b, g}$]{
         \includegraphics[width=0.475\linewidth, trim=0 0 0 50, clip]{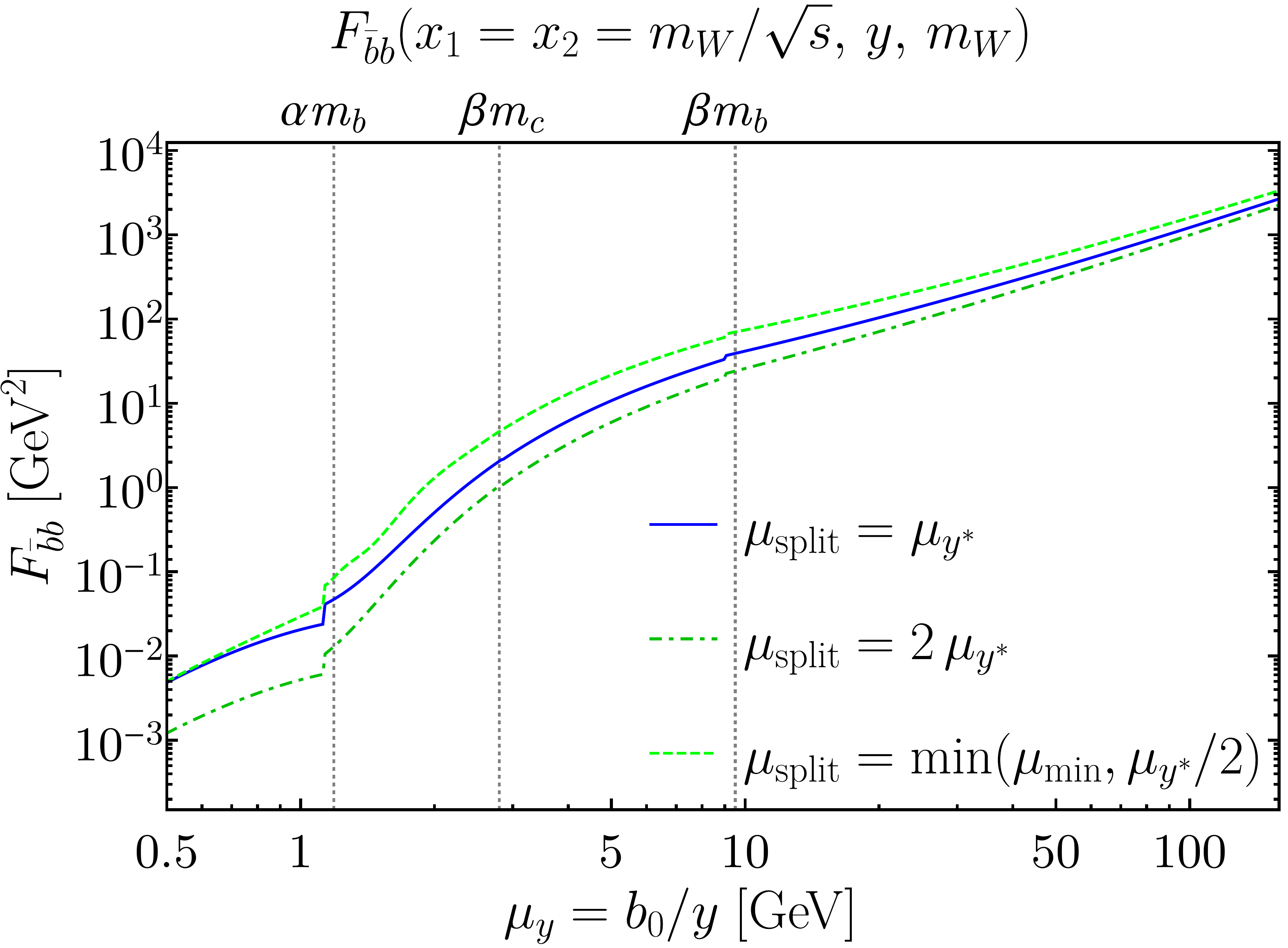}
      }
   \end{center}
   \caption{\label{fig:Fbbbar-scale} Dependence on $\mu_{\text{split}}$ of the $\bbar b$ splitting DPD in the $W$ production setting, computed with $\alpha = 1/4$ and $\beta = 2$ in the massive scheme.  The parton momentum fractions correspond to the $c \cbar \bbar b$ luminosity at $Y=0$ (top row) or at $Y=3$ (bottom row).  According to \eqref{eq:mom-fracs-opposite-rap} this means $x_{1} = x_{2} \approx 5.7 \times 10^{-3}$ for $Y=0$ and $x_{1} \approx 2.9 \times 10^{-4}$, $x_{2} \approx 0.12$ for $Y=3$.
   In the left column the complete result is shown, whereas in the right column only the contribution from $g \to \bbar b$ is taken into account in the DPD splitting formula.}
\end{figure}

Let us finally investigate the reason for the peculiar splitting scale dependence in channels with a heavy $Q \Qbar$ pair in both DPDs, seen in  \figs{\ref{subfig:Lccbarbbbar-jets-scale}}, \ref{subfig:Lttbartbart-ttbar-scale}, and \ref{subfig:Lbbbarcbarc-WW-scale}.  In the left panels of \fig{\ref{fig:Fbbbar-scale}}, we show the ${\bbar b}$ splitting DPD with momentum fractions that correspond to the $c \cbar \bbar b$ luminosity at $Y=0$ or $Y=3$ in the $W$ production setting.  For $Y=0$ there is some dependence on the splitting scale at low $\mu_y$, but in this region the DPD is very small and has little influence on $\mathcal{L}_{c \cbar \smash{\bbar} b}$.  For $Y=3$, there is a strong splitting scale dependence over the full range of $\mu_y$, which results in a strong scale dependence of $\mathcal{L}_{c \cbar \smash{\bbar} b}$.

To elucidate the origin of this behaviour, we show in the right panels of \fig{\ref{fig:Fbbbar-scale}} what happens if we set all DPD splitting kernels other than $V_{\, \smash{\bbar} b, g}$ to zero.  For $Y=0$ this has almost no impact on the DPD if $\mu_y$ is above $10 \gev$, whereas for $Y=3$ it significantly decreases the DPD unless $\mu_y$ is close to the final scale $\mu$.  This means that for $x_{1} = x_{2}$ (corresponding to $Y=0$) the $\bbar b$ distribution at large $\mu_y$ arises almost entirely from direct $g\to \bbar b$ splitting.  For $x_{1} \ll x_{2}$ (corresponding to $Y=3$), a significant part of $F_{\bbar b}$ originates from other splitting channels such as $F_{g b}$ and $F_{g g}$ by DGLAP evolution.  This is not surprising because the $1\to 2$ splitting kernels for these channels are enhanced for asymmetric momentum fractions with a soft emitted gluon.  This kinematic enhancement partially compensates the factor of $a_s$ that comes with each DGLAP evolution step.

With DGLAP evolution after the $1\to 2$ splitting process playing an important role for large $Y$, changing $\mu_{\text{split}}$ has a big impact because it significantly changes the interval $[\mu_{\text{split}}, \mu]$ in which the DPD can evolve.  For $Y=0$, DPD evolution after the $1\to 2$ splitting plays a minor role at high $\mu_y$, and the plot in \fig{\ref{subfig:Fbbbar-scale-Y0}} shows that the resulting splitting scale dependence is very weak.

The preceding analysis is corroborated by the $y$ dependence of the DPD at large $\mu_y$.  One can read off in \fig{\ref{fig:Fbbbar-scale}} that for $Y=0$ it is very close to the $1/y^2$ behaviour of the massless DPD splitting formula, whereas it is much flatter (close to $1/y$) for $Y=3$.  The significant flattening of the $y$ dependence of splitting DPDs by DGLAP evolution was already observed in \cite{Diehl:2017kgu}.
%
%
%
\section{Basis functions for the model in \sect{\protect\ref{sec:NLO-kernels-model}}}
\label{app:basis-functions}
In this appendix, we specify the basis functions that appear in the decompositions \eqref{eq:ansatz-qQQbar} to \eqref{eq:ansatz-qqQ} and \eqref{eq:ansatz-gQQbar} to \eqref{eq:ansatz-ggg}.  Given their large number, we discuss their general structure and common building blocks here.  An explicit list of all basis functions can be found in the ancillary files associated with this paper on the \href{https://arxiv.org}{arXiv}.

Note that the kernels for NLO channels are functions, whereas the kernels for the LO channels $g \to Q \Qbar$ and $g \to g g$ include plus- or $\delta$-distributions.  This reflects the structure of the corresponding massless kernels.

%
\subsection*{Basis functions for \texorpdfstring{$v^{I}_{Q \Qbar, q}$}{}}
\label{app:Basis-vRQQbarq}
The overall structure of the $72$ basis functions for this kernel is rather simple. One finds rational functions of the form
\begin{align}
      \frac{(z_1^2 + z_2^2)^n}{(z_1 + z_2)^m}
\end{align}
with $n = 0, 1$ and $m = 0, \ldots, 4$, which are multiplied by logarithms and products of logarithms. The arguments of these logarithms are
\begin{align}
      z_1 \,, \quad z_2 \,, \quad z_1 + z_2 \,, \quad 1 - z_1 - z_2 \,.
\end{align}
Since the overall $q \to Q \Qbar$ kernel is symmetric under $z_1 \leftrightarrow z_2$, each basis function is symmetrised if necessary.
%
%
\subsection*{Basis functions for \texorpdfstring{$v^{I}_{Q q, q}$}{}}
\label{app:Basis-vRQqq}
The structure of the $197$ basis functions is again rather simple.  The involve rational prefactors
\begin{align}
      \frac{z_1^n}{(1 - z_2)^m}
\end{align}
with $n = 0, 2$ and $m = 0, \ldots, 4$, which are multiplied by logarithms, products of logarithms, or dilogarithms. The logarithms encountered in this channel have the arguments
\begin{align}
      z_1 \,, \quad z_2 \,, \quad 1 - z_2 \,, \quad z_1 + z_2 \,, \quad
      1 - z_1 - z_2 \,,
\end{align}
whilst the arguments of the dilogarithms read
\begin{align}
      z_2 \,, \quad -\frac{z_1}{z_2} \,, \quad \frac{z_1}{1 - z_2} \,, \quad
      z_1 + z_2 \,.
\end{align}
%
%
\subsection*{Basis functions for \texorpdfstring{$v^{I}_{Q \Qbar, g}$}{}}
\label{app:Basis-vRQQbarg}
For this kernel, one has two sets of basis functions for the colour factors $C_A$ and $C_F$.  Each set includes members with plus- and $\delta$-distributions.
%
%
\paragraph{Colour factor \texorpdfstring{$C_A$}{}.}
\label{app:Basis-vRQQbarg-CA}
The regular basis functions can be decomposed into rational prefactors
\begin{align}
      \frac{z_1^m}{(z_1 + z_2)^n} \,, \quad \frac{z_2^m}{(z_1 + z_2)^n}
\end{align}
with $m = 0, 1, 2$ and $n = 0, \ldots, 4$, multiplied by logarithms, products of logarithms, or dilogarithms. The arguments of the logarithms read
\begin{align}
   \label{eq:log-args-vRQQbarg}
      z_1 \,, \quad z_2 \,, \quad 1 - z_1 \,, \quad 1 - z_2 \,, \quad z_1 + z_2
      \,, \quad 1 - z_1 - z_2 \,,
\end{align}
and those of the dilogarithms are given by
\begin{align}
   \label{eq:dilog-args-vRQQbarg}
      z_1 \,, \quad z_2 \,, \quad -\frac{z_1}{z_2} \,, \quad -\frac{z_2}{z_1}
      \,, \quad \frac{z_1}{1 - z_2} \,, \quad \frac{z_2}{1 - z_1} \,, \quad
      z_1 + z_2 \,.
\end{align}
In addition, one finds terms with a denominator $(1 - z_1 - z_2)$. These terms have combinations of logarithms and dilogarithms as numerators, which vanish sufficiently fast for $z_1 + z_2 \to 1$ such that the singularity of the denominator is cancelled of becomes integrable.

The basis functions that contain a $\delta$-distribution can be written as a prefactor
\begin{align}
   \delta(1 - z_1 - z_2) \,
      \bigl[ z_1 (1 - z_1) \bigr]^n
\end{align}
with $n = 0, 1$, multiplied by products of up to three logarithms, dilogarithms, products of dilogarithms and logarithms, and trilogarithms. The arguments of the logarithms, dilogarithms, and trilogarithms read
\begin{align}
      z_1 \,, \quad 1 - z_1 \,.
\end{align}
Here it has been used that one can replace $z_2$ by $1 -z_1$ due to the factor $\delta(1 - z_1 - z_2)$.

Finally one finds plus-distribution terms with the simple structure
\begin{align}
      \frac{z_1^n + z_2^n}{ [1 - z_1 - z_2]_+}
\end{align}
with $n = 0, 1, 2$.

Since the basis functions should be symmetric under $z_1 \leftrightarrow z_2$ (or under $z_1 \leftrightarrow 1 - z_1$ for the $\delta$-distribution terms), it is understood that an appropriate symmetrisation is performed where necessary.
%
%
\paragraph{Colour factor \texorpdfstring{$C_F$}{}.}
\label{app:Basis-vRQQbarg-CF}
In this colour channel, the regular basis functions can again be decomposed into a rational prefactor
\begin{align}
      \frac{z_1^m}{(1 - z_2)^n} \,, \quad \frac{z_2^m}{(1 - z_1)^n}
\end{align}
with $m = 0, 1$ and $n = 0, 1, 2$, multiplied by logarithms, products of logarithms, and dilogarithms with the same arguments as given in \eqref{eq:log-args-vRQQbarg} and \eqref{eq:dilog-args-vRQQbarg} for the colour factor $C_A$. In addition to these terms, one finds again terms with a denominator $1 - z_1 - z_2$ and combinations of logarithms and dilogarithms in the numerator that lead to a finite value or an integrable singularity of the function in the $z_1 + z_2 \to 1$ limit.

The structure of the $\delta$- and plus-distribution terms is identical to the ones for the colour factor $C_A$.  Again, an appropriate symmetrisation of the basis functions is understood.
%
%
\subsection*{Basis functions for \texorpdfstring{$v^{I}_{Q g, g}$}{}}
\label{app:Basis-vRQgg}
Also for this kernel, one has two sets of basis functions for the colour factors
$C_A$ and $C_F$.
%
%
\paragraph{Colour factor \texorpdfstring{$C_A$}{}.}
\label{app:Basis-vRQgg-CA}
The basis functions multiplying $C_A$ have the same structure encountered as the channels discussed so far, with rational prefactors multiplied by logarithms, products of logarithms, and dilogarithms. The rational prefactors are
\begin{align}
      \frac{z_1^m}{(1 - z_2)^n} \,, \quad \frac{z_1^m}{z_2} \,, \quad z_2
\end{align}
with $m = 0, 1, 2$ and $n = 0, 1, \ldots, 4$.  The arguments of the logarithms read
\begin{align}
      z_1 \,, \quad z_2 \,, \quad 1 - z_1 \,, \quad 1 - z_2 \,, \quad z_1 + z_2
      \,, \quad 1 - z_1 - z_2 \,,
\end{align}
whilst the arguments of the dilogarithms are given by
\begin{align}
      z_1 \,, \quad z_2 \,, \quad -\frac{z_2}{z_1}
      \,, \quad \frac{z_1}{1 - z_2} \,, \quad \frac{z_2}{1 - z_1} \,, \quad
      z_1 + z_2 \,.
\end{align}
%
%
\paragraph{Colour factor \texorpdfstring{$C_F$}{}.}
\label{app:Basis-vRQgg-CF}
The basis functions multiplying $C_F$ can again be written as discussed above, with rational prefactors
\begin{align}
      z_1 \,, \quad z_2 \,, \quad \frac{z_2^m}{(1 - z_1)^n} \,, \quad
      \frac{z_1^m}{(z_1 + z_2)^n} \,,
\end{align}
where $m = 0, 1, 2$ and $n = 1, 2$.  These prefactors are multiplied by logarithms and products of logarithms with arguments
\begin{align}
      z_1 \,, \quad z_2 \,, \quad 1 - z_1 \,, \quad z_1 + z_2
      \,, \quad 1 - z_1 - z_2 \,,
\end{align}
and by dilogarithms with arguments
\begin{align}
      z_1 \,,\quad -\frac{z_1}{z_2} \,, \quad \frac{z_2}{1 - z_1} \,, \quad
      z_1 + z_2 \,.
\end{align}
%
%
\subsection*{Basis functions for \texorpdfstring{$v^{I}_{g g, g}$}{}}
\label{app:Basis-vRggg-CA}
Due to the kinematic constraint in \eqref{eq:VI-ggg-delta}, the set of basis functions is very small in this case:
\begin{align}
      h_1(z_1, z_2)
   &= \delta(1 - z_1 - z_2) \,,
   \nonumber\\[0.2em]
      h_2(z_1, z_2)
   &= \delta(1 - z_1 - z_2)
      \left(
         \frac{1}{1 - z_1} + \frac{1}{z_1}
      \right) \,,
   \nonumber\\[0.2em]
      h_3(z_1, z_2)
   &= \delta(1 - z_1 - z_2) \,
      \bigl[
         z_1 \log (1 - z_1) + (1 - z_1) \log z_1
      \bigr]\,.
\end{align}


\section*{Acknowledgements}

This work is in part supported by the Deutsche Forschungsgemeinschaft (DFG, German Research Foundation) -- grant number 409651613 (Research Unit FOR 2926) and grant number 491245950.  The work of RN is supported by the ERC Starting Grant REINVENT-714788.
The Feynman graphs in this manuscript were produced with JaxoDraw \cite{Binosi:2003yf, Binosi:2008ig}.  The numerical studies have been performed using the \textsc{ChiliPDF} library \cite{Diehl:2021gvs, Diehl:2023cth}, which is under development.  We gratefully acknowledge the contributions of our collaborators Florian Fabry, Oskar Grocholski, Mees van Kampen, and Frank Tackmann to this project.


\phantomsection
\addcontentsline{toc}{section}{References}

\bibliographystyle{JHEP}
\bibliography{massive-split.bib}

\providecommand{\href}[2]{#2}\begingroup\raggedright\begin{thebibliography}{10}

\bibitem{Gaunt:2009re}
J.R.~Gaunt and W.J.~Stirling, \emph{{Double Parton Distributions Incorporating
  Perturbative QCD Evolution and Momentum and Quark Number Sum Rules}},
  \href{https://doi.org/10.1007/JHEP03(2010)005}{\emph{JHEP} {\bfseries 03}
  (2010) 005} [\href{https://arxiv.org/abs/0910.4347}{{\ttfamily 0910.4347}}].

\bibitem{Blok:2010ge}
B.~Blok, Y.~Dokshitzer, L.~Frankfurt and M.~Strikman, \emph{{The Four jet
  production at LHC and Tevatron in QCD}},
  \href{https://doi.org/10.1103/PhysRevD.83.071501}{\emph{Phys. Rev. D}
  {\bfseries 83} (2011) 071501}
  [\href{https://arxiv.org/abs/1009.2714}{{\ttfamily 1009.2714}}].

\bibitem{Gaunt:2011xd}
J.R.~Gaunt and W.~Stirling, \emph{{Double Parton Scattering Singularity in
  One-Loop Integrals}},
  \href{https://doi.org/10.1007/JHEP06(2011)048}{\emph{JHEP} {\bfseries 06}
  (2011) 048} [\href{https://arxiv.org/abs/1103.1888}{{\ttfamily 1103.1888}}].

\bibitem{Ryskin:2011kk}
M.G.~Ryskin and A.M.~Snigirev, \emph{{A Fresh look at double parton
  scattering}}, \href{https://doi.org/10.1103/PhysRevD.83.114047}{\emph{Phys.
  Rev.} {\bfseries D83} (2011) 114047}
  [\href{https://arxiv.org/abs/1103.3495}{{\ttfamily 1103.3495}}].

\bibitem{Blok:2011bu}
B.~Blok, Y.~Dokshitser, L.~Frankfurt and M.~Strikman, \emph{{pQCD physics of
  multiparton interactions}},
  \href{https://doi.org/10.1140/epjc/s10052-012-1963-8}{\emph{Eur. Phys. J.}
  {\bfseries C72} (2012) 1963}
  [\href{https://arxiv.org/abs/1106.5533}{{\ttfamily 1106.5533}}].

\bibitem{Diehl:2011yj}
M.~Diehl, D.~Ostermeier and A.~Sch{\"a}fer, \emph{{Elements of a theory for
  multiparton interactions in QCD}},
  \href{https://doi.org/10.1007/JHEP03(2012)089}{\emph{JHEP} {\bfseries 03}
  (2012) 089} [\href{https://arxiv.org/abs/1111.0910}{{\ttfamily 1111.0910}}].

\bibitem{Manohar:2012jr}
A.V.~Manohar and W.J.~Waalewijn, \emph{{A QCD Analysis of Double Parton
  Scattering: Color Correlations, Interference Effects and Evolution}},
  \href{https://doi.org/10.1103/PhysRevD.85.114009}{\emph{Phys. Rev.}
  {\bfseries D85} (2012) 114009}
  [\href{https://arxiv.org/abs/1202.3794}{{\ttfamily 1202.3794}}].

\bibitem{Manohar:2012pe}
A.V.~Manohar and W.J.~Waalewijn, \emph{{What is Double Parton Scattering?}},
  \href{https://doi.org/10.1016/j.physletb.2012.05.044}{\emph{Phys. Lett. B}
  {\bfseries 713} (2012) 196}
  [\href{https://arxiv.org/abs/1202.5034}{{\ttfamily 1202.5034}}].

\bibitem{Ryskin:2012qx}
M.G.~Ryskin and A.M.~Snigirev, \emph{{Double parton scattering in double
  logarithm approximation of perturbative QCD}},
  \href{https://doi.org/10.1103/PhysRevD.86.014018}{\emph{Phys. Rev.}
  {\bfseries D86} (2012) 014018}
  [\href{https://arxiv.org/abs/1203.2330}{{\ttfamily 1203.2330}}].

\bibitem{Gaunt:2012dd}
J.R.~Gaunt, \emph{{Single Perturbative Splitting Diagrams in Double Parton
  Scattering}}, \href{https://doi.org/10.1007/JHEP01(2013)042}{\emph{JHEP}
  {\bfseries 01} (2013) 042} [\href{https://arxiv.org/abs/1207.0480}{{\ttfamily
  1207.0480}}].

\bibitem{Blok:2013bpa}
B.~Blok, Y.~Dokshitzer, L.~Frankfurt and M.~Strikman, \emph{{Perturbative QCD
  correlations in multi-parton collisions}},
  \href{https://doi.org/10.1140/epjc/s10052-014-2926-z}{\emph{Eur. Phys. J. C}
  {\bfseries 74} (2014) 2926}
  [\href{https://arxiv.org/abs/1306.3763}{{\ttfamily 1306.3763}}].

\bibitem{Diehl:2015bca}
M.~Diehl, J.R.~Gaunt, D.~Ostermeier, P.~Pl\"o\ss{}l and A.~Sch\"afer,
  \emph{{Cancellation of Glauber gluon exchange in the double Drell-Yan
  process}}, \href{https://doi.org/10.1007/JHEP01(2016)076}{\emph{JHEP}
  {\bfseries 01} (2016) 076}
  [\href{https://arxiv.org/abs/1510.08696}{{\ttfamily 1510.08696}}].

\bibitem{Diehl:2017kgu}
M.~Diehl, J.R.~Gaunt and K.~Sch\"onwald, \emph{{Double hard scattering without
  double counting}}, \href{https://doi.org/10.1007/JHEP06(2017)083}{\emph{JHEP}
  {\bfseries 06} (2017) 083}
  [\href{https://arxiv.org/abs/1702.06486}{{\ttfamily 1702.06486}}].

\bibitem{Diehl:2018wfy}
M.~Diehl and R.~Nagar, \emph{{Factorisation of soft gluons in multiparton
  scattering}}, \href{https://doi.org/10.1007/JHEP04(2019)124}{\emph{JHEP}
  {\bfseries 04} (2019) 124}
  [\href{https://arxiv.org/abs/1812.09509}{{\ttfamily 1812.09509}}].

\bibitem{Cabouat:2019gtm}
B.~Cabouat, J.R.~Gaunt and K.~Ostrolenk, \emph{{A Monte-Carlo Simulation of
  Double Parton Scattering}},
  \href{https://doi.org/10.1007/JHEP11(2019)061}{\emph{JHEP} {\bfseries 11}
  (2019) 061} [\href{https://arxiv.org/abs/1906.04669}{{\ttfamily
  1906.04669}}].

\bibitem{Cabouat:2020ssr}
B.~Cabouat and J.R.~Gaunt, \emph{{Combining single and double parton
  scatterings in a parton shower}},
  \href{https://doi.org/10.1007/JHEP10(2020)012}{\emph{JHEP} {\bfseries 10}
  (2020) 012} [\href{https://arxiv.org/abs/2008.01442}{{\ttfamily
  2008.01442}}].

\bibitem{Abe:1997xk}
{\scshape CDF} collaboration, \emph{{Double parton scattering in $\bar{p}p$
  collisions at $\sqrt{s} = 1.8 $TeV}},
  \href{https://doi.org/10.1103/PhysRevD.56.3811}{\emph{Phys. Rev. D}
  {\bfseries 56} (1997) 3811}.

\bibitem{Abazov:2015nnn}
{\scshape D0} collaboration, \emph{{Study of double parton interactions in
  diphoton + dijet events in $p\bar{p}$ collisions at $\sqrt{s} = 1.96$ TeV}},
  \href{https://doi.org/10.1103/PhysRevD.93.052008}{\emph{Phys. Rev. D}
  {\bfseries 93} (2016) 052008}
  [\href{https://arxiv.org/abs/1512.05291}{{\ttfamily 1512.05291}}].

\bibitem{Aaij:2016bqq}
{\scshape LHCb} collaboration, \emph{{Measurement of the J/$\psi$ pair
  production cross-section in pp collisions at $ \sqrt{s}=13 $ TeV}},
  \href{https://doi.org/10.1007/JHEP06(2017)047}{\emph{JHEP} {\bfseries 06}
  (2017) 047} [\href{https://arxiv.org/abs/1612.07451}{{\ttfamily
  1612.07451}}].

\bibitem{Aaboud:2018tiq}
{\scshape ATLAS} collaboration, \emph{{Study of the hard double-parton
  scattering contribution to inclusive four-lepton production in $pp$
  collisions at $\sqrt s=$ 8 TeV with the ATLAS detector}},
  \href{https://doi.org/10.1016/j.physletb.2019.01.062}{\emph{Phys. Lett.}
  {\bfseries B790} (2019) 595}
  [\href{https://arxiv.org/abs/1811.11094}{{\ttfamily 1811.11094}}].

\bibitem{CMS:2022pio}
{\scshape CMS} collaboration, \emph{{Observation of same-sign WW production
  from double parton scattering in proton-proton collisions at $\sqrt{s}$ = 13
  TeV}},  \href{https://arxiv.org/abs/2206.02681}{{\ttfamily 2206.02681}}.

\bibitem{Fedkevych:2020cmd}
O.~Fedkevych and A.~Kulesza, \emph{{Double parton scattering in four-jet
  production in proton-proton collisions at the LHC}},
  \href{https://doi.org/10.1103/PhysRevD.104.054021}{\emph{Phys. Rev. D}
  {\bfseries 104} (2021) 054021}
  [\href{https://arxiv.org/abs/2008.08347}{{\ttfamily 2008.08347}}].

\bibitem{Bartalini:2017jkk}
P.~Bartalini and J.R.~Gaunt, eds., \emph{{Multiple Parton Interactions at the
  LHC}}, vol.~29, WSP (2019),
  \href{https://doi.org/10.1142/10646}{10.1142/10646}.

\bibitem{Chang:2012nw}
H.-M.~Chang, A.V.~Manohar and W.J.~Waalewijn, \emph{{Double Parton Correlations
  in the Bag Model}},
  \href{https://doi.org/10.1103/PhysRevD.87.034009}{\emph{Phys. Rev. D}
  {\bfseries 87} (2013) 034009}
  [\href{https://arxiv.org/abs/1211.3132}{{\ttfamily 1211.3132}}].

\bibitem{Rinaldi:2013vpa}
M.~Rinaldi, S.~Scopetta and V.~Vento, \emph{{Double parton correlations in
  constituent quark models}},
  \href{https://doi.org/10.1103/PhysRevD.87.114021}{\emph{Phys. Rev. D}
  {\bfseries 87} (2013) 114021}
  [\href{https://arxiv.org/abs/1302.6462}{{\ttfamily 1302.6462}}].

\bibitem{Broniowski:2013xba}
W.~Broniowski and E.~Ruiz~Arriola, \emph{{Valence double parton distributions
  of the nucleon in a simple model}},
  \href{https://doi.org/10.1007/s00601-014-0840-4}{\emph{Few Body Syst.}
  {\bfseries 55} (2014) 381} [\href{https://arxiv.org/abs/1310.8419}{{\ttfamily
  1310.8419}}].

\bibitem{Rinaldi:2014ddl}
M.~Rinaldi, S.~Scopetta, M.~Traini and V.~Vento, \emph{{Double parton
  correlations and constituent quark models: a Light Front approach to the
  valence sector}}, \href{https://doi.org/10.1007/JHEP12(2014)028}{\emph{JHEP}
  {\bfseries 12} (2014) 028} [\href{https://arxiv.org/abs/1409.1500}{{\ttfamily
  1409.1500}}].

\bibitem{Broniowski:2016trx}
W.~Broniowski, E.~Ruiz~Arriola and K.~Golec-Biernat, \emph{{Generalized Valon
  Model for Double Parton Distributions}},
  \href{https://doi.org/10.1007/s00601-016-1087-z}{\emph{Few Body Syst.}
  {\bfseries 57} (2016) 405}
  [\href{https://arxiv.org/abs/1602.00254}{{\ttfamily 1602.00254}}].

\bibitem{Kasemets:2016nio}
T.~Kasemets and A.~Mukherjee, \emph{{Quark-gluon double parton distributions in
  the light-front dressed quark model}},
  \href{https://doi.org/10.1103/PhysRevD.94.074029}{\emph{Phys. Rev. D}
  {\bfseries 94} (2016) 074029}
  [\href{https://arxiv.org/abs/1606.05686}{{\ttfamily 1606.05686}}].

\bibitem{Rinaldi:2016jvu}
M.~Rinaldi, S.~Scopetta, M.C.~Traini and V.~Vento, \emph{{Correlations in
  Double Parton Distributions: Perturbative and Non-Perturbative effects}},
  \href{https://doi.org/10.1007/JHEP10(2016)063}{\emph{JHEP} {\bfseries 10}
  (2016) 063} [\href{https://arxiv.org/abs/1608.02521}{{\ttfamily
  1608.02521}}].

\bibitem{Rinaldi:2016mlk}
M.~Rinaldi and F.A.~Ceccopieri, \emph{{Relativistic effects in model
  calculations of double parton distribution function}},
  \href{https://doi.org/10.1103/PhysRevD.95.034040}{\emph{Phys. Rev. D}
  {\bfseries 95} (2017) 034040}
  [\href{https://arxiv.org/abs/1611.04793}{{\ttfamily 1611.04793}}].

\bibitem{Rinaldi:2018zng}
M.~Rinaldi, S.~Scopetta, M.~Traini and V.~Vento, \emph{{A model calculation of
  double parton distribution functions of the pion}},
  \href{https://doi.org/10.1140/epjc/s10052-018-6256-4}{\emph{Eur. Phys. J. C}
  {\bfseries 78} (2018) 781}
  [\href{https://arxiv.org/abs/1806.10112}{{\ttfamily 1806.10112}}].

\bibitem{Courtoy:2019cxq}
A.~Courtoy, S.~Noguera and S.~Scopetta, \emph{{Double parton distributions in
  the pion in the Nambu--Jona-Lasinio model}},
  \href{https://doi.org/10.1007/JHEP12(2019)045}{\emph{JHEP} {\bfseries 12}
  (2019) 045} [\href{https://arxiv.org/abs/1909.09530}{{\ttfamily
  1909.09530}}].

\bibitem{Broniowski:2019rmu}
W.~Broniowski and E.~Ruiz~Arriola, \emph{{Double parton distribution of valence
  quarks in the pion in chiral quark models}},
  \href{https://doi.org/10.1103/PhysRevD.101.014019}{\emph{Phys. Rev. D}
  {\bfseries 101} (2020) 014019}
  [\href{https://arxiv.org/abs/1910.03707}{{\ttfamily 1910.03707}}].

\bibitem{Bali:2020mij}
G.S.~Bali, L.~Castagnini, M.~Diehl, J.R.~Gaunt, B.~Gl\"a\ss{}le, A.~Sch\"afer
  et~al., \emph{{Double parton distributions in the pion from lattice QCD}},
  \href{https://doi.org/10.1007/JHEP02(2021)067}{\emph{JHEP} {\bfseries 02}
  (2021) 067} [\href{https://arxiv.org/abs/2006.14826}{{\ttfamily
  2006.14826}}].

\bibitem{Bali:2021gel}
G.S.~Bali, M.~Diehl, B.~Gl\"a\ss{}le, A.~Sch\"afer and C.~Zimmermann,
  \emph{{Double parton distributions in the nucleon from lattice QCD}},
  \href{https://doi.org/10.1007/JHEP09(2021)106}{\emph{JHEP} {\bfseries 09}
  (2021) 106} [\href{https://arxiv.org/abs/2106.03451}{{\ttfamily
  2106.03451}}].

\bibitem{Golec-Biernat:2014bva}
K.~Golec-Biernat and E.~Lewandowska, \emph{{How to impose initial conditions
  for QCD evolution of double parton distributions?}},
  \href{https://doi.org/10.1103/PhysRevD.90.014032}{\emph{Phys. Rev. D}
  {\bfseries 90} (2014) 014032}
  [\href{https://arxiv.org/abs/1402.4079}{{\ttfamily 1402.4079}}].

\bibitem{Golec-Biernat:2015aza}
K.~Golec-Biernat, E.~Lewandowska, M.~Serino, Z.~Snyder and A.M.~Stasto,
  \emph{{Constraining the double gluon distribution by the single gluon
  distribution}},
  \href{https://doi.org/10.1016/j.physletb.2015.09.067}{\emph{Phys. Lett. B}
  {\bfseries 750} (2015) 559}
  [\href{https://arxiv.org/abs/1507.08583}{{\ttfamily 1507.08583}}].

\bibitem{Diehl:2020xyg}
M.~Diehl, J.~Gaunt, D.~Lang, P.~Pl\"o\ss{}l and A.~Sch\"afer, \emph{{Sum rule
  improved double parton distributions in position space}},
  \href{https://doi.org/10.1140/epjc/s10052-020-8038-z}{\emph{Eur. Phys. J. C}
  {\bfseries 80} (2020) 468}
  [\href{https://arxiv.org/abs/2001.10428}{{\ttfamily 2001.10428}}].

\bibitem{Fedkevych:2022myf}
O.~Fedkevych and J.R.~Gaunt, \emph{{On sum rules for double and triple parton
  distribution functions and Pythia\textquoteright{}s model of multiple parton
  interactions}}, \href{https://doi.org/10.1007/JHEP02(2023)090}{\emph{JHEP}
  {\bfseries 02} (2023) 090}
  [\href{https://arxiv.org/abs/2208.08197}{{\ttfamily 2208.08197}}].

\bibitem{Golec-Biernat:2022wkx}
K.~Golec-Biernat and A.M.~Sta\'sto, \emph{{Momentum sum rule and factorization
  of double parton distributions}},
  \href{https://doi.org/10.1103/PhysRevD.107.054020}{\emph{Phys. Rev. D}
  {\bfseries 107} (2023) 054020}
  [\href{https://arxiv.org/abs/2212.02289}{{\ttfamily 2212.02289}}].

\bibitem{Snigirev:2014eua}
A.~Snigirev, N.~Snigireva and G.~Zinovjev, \emph{{Perturbative and
  nonperturbative correlations in double parton distributions}},
  \href{https://doi.org/10.1103/PhysRevD.90.014015}{\emph{Phys. Rev. D}
  {\bfseries 90} (2014) 014015}
  [\href{https://arxiv.org/abs/1403.6947}{{\ttfamily 1403.6947}}].

\bibitem{Blok:2015rka}
B.~Blok and P.~Gunnellini, \emph{{Dynamical approach to MPI four-jet production
  in Pythia}}, \href{https://doi.org/10.1140/epjc/s10052-015-3520-8}{\emph{Eur.
  Phys. J. C} {\bfseries 75} (2015) 282}
  [\href{https://arxiv.org/abs/1503.08246}{{\ttfamily 1503.08246}}].

\bibitem{Blok:2015afa}
B.~Blok and P.~Gunnellini, \emph{{Dynamical approach to MPI in W+dijet and
  Z+dijet production within the PYTHIA event generator}},
  \href{https://doi.org/10.1140/epjc/s10052-016-4035-7}{\emph{Eur. Phys. J. C}
  {\bfseries 76} (2016) 202}
  [\href{https://arxiv.org/abs/1510.07436}{{\ttfamily 1510.07436}}].

\bibitem{Golec-Biernat:2014nsa}
K.~Golec-Biernat and E.~Lewandowska, \emph{{Electroweak boson production in
  double parton scattering}},
  \href{https://doi.org/10.1103/PhysRevD.90.094032}{\emph{Phys. Rev. D}
  {\bfseries 90} (2014) 094032}
  [\href{https://arxiv.org/abs/1407.4038}{{\ttfamily 1407.4038}}].

\bibitem{Gaunt:2014rua}
J.R.~Gaunt, R.~Maciula and A.~Szczurek, \emph{{Conventional versus
  single-ladder-splitting contributions to double parton scattering production
  of two quarkonia, two Higgs bosons and $c \bar c c \bar c$}},
  \href{https://doi.org/10.1103/PhysRevD.90.054017}{\emph{Phys. Rev. D}
  {\bfseries 90} (2014) 054017}
  [\href{https://arxiv.org/abs/1407.5821}{{\ttfamily 1407.5821}}].

\bibitem{Pietrulewicz:2017gxc}
P.~Pietrulewicz, D.~Samitz, A.~Spiering and F.J.~Tackmann, \emph{{Factorization
  and Resummation for Massive Quark Effects in Exclusive Drell-Yan}},
  \href{https://doi.org/10.1007/JHEP08(2017)114}{\emph{JHEP} {\bfseries 08}
  (2017) 114} [\href{https://arxiv.org/abs/1703.09702}{{\ttfamily
  1703.09702}}].

\bibitem{Chetyrkin:1997sg}
K.G.~Chetyrkin, B.A.~Kniehl and M.~Steinhauser, \emph{{Strong coupling constant
  with flavor thresholds at four loops in the $\msbar$ scheme}},
  \href{https://doi.org/10.1103/PhysRevLett.79.2184}{\emph{Phys. Rev. Lett.}
  {\bfseries 79} (1997) 2184}
  [\href{https://arxiv.org/abs/hep-ph/9706430}{{\ttfamily hep-ph/9706430}}].

\bibitem{Buza:1996wv}
M.~Buza, Y.~Matiounine, J.~Smith and W.L.~van Neerven, \emph{{Charm
  electroproduction viewed in the variable flavor number scheme versus fixed
  order perturbation theory}},
  \href{https://doi.org/10.1007/BF01245820}{\emph{Eur. Phys. J. C} {\bfseries
  1} (1998) 301} [\href{https://arxiv.org/abs/hep-ph/9612398}{{\ttfamily
  hep-ph/9612398}}].

\bibitem{Ablinger:2014lka}
J.~Ablinger, J.~Bl\"umlein, A.~De~Freitas, A.~Hasselhuhn, A.~von Manteuffel,
  M.~Round et~al., \emph{{The Transition Matrix Element $A_{gq}(N)$ of the
  Variable Flavor Number Scheme at $O(\alpha_s^3)$}},
  \href{https://doi.org/10.1016/j.nuclphysb.2014.02.007}{\emph{Nucl. Phys. B}
  {\bfseries 882} (2014) 263}
  [\href{https://arxiv.org/abs/1402.0359}{{\ttfamily 1402.0359}}].

\bibitem{Behring:2014eya}
A.~Behring, I.~Bierenbaum, J.~Bl\"umlein, A.~De~Freitas, S.~Klein and
  F.~Wi\ss{}brock, \emph{{The logarithmic contributions to the $O(\alpha^3_s)$
  asymptotic massive Wilson coefficients and operator matrix elements in deeply
  inelastic scattering}},
  \href{https://doi.org/10.1140/epjc/s10052-014-3033-x}{\emph{Eur. Phys. J. C}
  {\bfseries 74} (2014) 3033}
  [\href{https://arxiv.org/abs/1403.6356}{{\ttfamily 1403.6356}}].

\bibitem{Ablinger:2014vwa}
J.~Ablinger, A.~Behring, J.~Bl\"umlein, A.~De~Freitas, A.~Hasselhuhn, A.~von
  Manteuffel et~al., \emph{{The 3-Loop Non-Singlet Heavy Flavor Contributions
  and Anomalous Dimensions for the Structure Function $F_2(x,Q^2)$ and
  Transversity}},
  \href{https://doi.org/10.1016/j.nuclphysb.2014.07.010}{\emph{Nucl. Phys. B}
  {\bfseries 886} (2014) 733}
  [\href{https://arxiv.org/abs/1406.4654}{{\ttfamily 1406.4654}}].

\bibitem{Diehl:2019rdh}
M.~Diehl, J.R.~Gaunt, P.~Pl\"o\ss{}l and A.~Sch\"afer, \emph{{Two-loop
  splitting in double parton distributions}},
  \href{https://doi.org/10.21468/SciPostPhys.7.2.017}{\emph{SciPost Phys.}
  {\bfseries 7} (2019) 017} [\href{https://arxiv.org/abs/1902.08019}{{\ttfamily
  1902.08019}}].

\bibitem{Diehl:2021wpp}
M.~Diehl, J.R.~Gaunt and P.~Pl\"o\ss{}l, \emph{{Two-loop splitting in double
  parton distributions: the colour non-singlet case}},
  \href{https://doi.org/10.1007/JHEP08(2021)040}{\emph{JHEP} {\bfseries 08}
  (2021) 040} [\href{https://arxiv.org/abs/2105.08425}{{\ttfamily
  2105.08425}}].

\bibitem{Bacchetta:2015ora}
A.~Bacchetta, M.G.~Echevarria, P.J.G.~Mulders, M.~Radici and A.~Signori,
  \emph{{Effects of TMD evolution and partonic flavor on $e^{+} e^{-}$
  annihilation into hadrons}},
  \href{https://doi.org/10.1007/JHEP11(2015)076}{\emph{JHEP} {\bfseries 11}
  (2015) 076} [\href{https://arxiv.org/abs/1508.00402}{{\ttfamily
  1508.00402}}].

\bibitem{Diehl:2021gvs}
M.~Diehl, R.~Nagar and F.J.~Tackmann, \emph{{ChiliPDF: Chebyshev interpolation
  for parton distributions}},
  \href{https://doi.org/10.1140/epjc/s10052-022-10223-1}{\emph{Eur. Phys. J. C}
  {\bfseries 82} (2022) 257}
  [\href{https://arxiv.org/abs/2112.09703}{{\ttfamily 2112.09703}}].

\bibitem{Diehl:2023cth}
M.~Diehl, R.~Nagar, P.~Pl{\"o}{\ss}l and F.J.~Tackmann, \emph{{Evolution and
  interpolation of double parton distributions using Chebyshev grids}},
  \href{https://doi.org/10.1140/epjc/s10052-023-11692-8}{\emph{Eur. Phys. J. C}
  {\bfseries 83} (2023) 536}
  [\href{https://arxiv.org/abs/2305.04845}{{\ttfamily 2305.04845}}].

\bibitem{Kulesza:1999zh}
A.~Kulesza and W.~Stirling, \emph{{Like sign $W$ boson production at the LHC as
  a probe of double parton scattering}},
  \href{https://doi.org/10.1016/S0370-2693(99)01512-9}{\emph{Phys. Lett. B}
  {\bfseries 475} (2000) 168}
  [\href{https://arxiv.org/abs/hep-ph/9912232}{{\ttfamily hep-ph/9912232}}].

\bibitem{Gaunt:2010pi}
J.R.~Gaunt, C.-H.~Kom, A.~Kulesza and W.~Stirling, \emph{{Same-sign W pair
  production as a probe of double parton scattering at the LHC}},
  \href{https://doi.org/10.1140/epjc/s10052-010-1362-y}{\emph{Eur. Phys. J. C}
  {\bfseries 69} (2010) 53} [\href{https://arxiv.org/abs/1003.3953}{{\ttfamily
  1003.3953}}].

\bibitem{Ceccopieri:2017oqe}
F.A.~Ceccopieri, M.~Rinaldi and S.~Scopetta, \emph{{Parton correlations in
  same-sign $W$ pair production via double parton scattering at the LHC}},
  \href{https://doi.org/10.1103/PhysRevD.95.114030}{\emph{Phys. Rev. D}
  {\bfseries 95} (2017) 114030}
  [\href{https://arxiv.org/abs/1702.05363}{{\ttfamily 1702.05363}}].

\bibitem{Cotogno:2018mfv}
S.~Cotogno, T.~Kasemets and M.~Myska, \emph{{Spin on same-sign $W$-boson pair
  production}}, \href{https://doi.org/10.1103/PhysRevD.100.011503}{\emph{Phys.\
  Rev.\ D} {\bfseries 100} (2019) 011503}
  [\href{https://arxiv.org/abs/1809.09024}{{\ttfamily 1809.09024}}].

\bibitem{Cotogno:2020iio}
S.~Cotogno, T.~Kasemets and M.~Myska, \emph{{Confronting same-sign W-boson
  production with parton correlations}},
  \href{https://doi.org/10.1007/JHEP10(2020)214}{\emph{JHEP} {\bfseries 10}
  (2020) 214} [\href{https://arxiv.org/abs/2003.03347}{{\ttfamily
  2003.03347}}].

\bibitem{Sirunyan:2019zox}
{\scshape CMS} collaboration, \emph{{Evidence for $\text {W}\text {W}$
  production from double-parton interactions in proton--proton collisions at
  $\sqrt{s} = 13 \,\text {TeV} $}},
  \href{https://doi.org/10.1140/epjc/s10052-019-7541-6}{\emph{Eur. Phys. J. C}
  {\bfseries 80} (2020) 41} [\href{https://arxiv.org/abs/1909.06265}{{\ttfamily
  1909.06265}}].

\bibitem{Bailey:2020ooq}
S.~Bailey, T.~Cridge, L.A.~Harland-Lang, A.D.~Martin and R.S.~Thorne,
  \emph{{Parton distributions from LHC, HERA, Tevatron and fixed target data:
  MSHT20 PDFs}},
  \href{https://doi.org/10.1140/epjc/s10052-021-09057-0}{\emph{Eur. Phys. J. C}
  {\bfseries 81} (2021) 341}
  [\href{https://arxiv.org/abs/2012.04684}{{\ttfamily 2012.04684}}].

\bibitem{H1:2015ubc}
{\scshape H1, ZEUS} collaboration, \emph{{Combination of measurements of
  inclusive deep inelastic ${e^{\pm }p}$ scattering cross sections and QCD
  analysis of HERA data}},
  \href{https://doi.org/10.1140/epjc/s10052-015-3710-4}{\emph{Eur. Phys. J. C}
  {\bfseries 75} (2015) 580}
  [\href{https://arxiv.org/abs/1506.06042}{{\ttfamily 1506.06042}}].

\bibitem{NNPDF:2021njg}
{\scshape NNPDF} collaboration, \emph{{The path to proton structure at 1\%
  accuracy}}, \href{https://doi.org/10.1140/epjc/s10052-022-10328-7}{\emph{Eur.
  Phys. J. C} {\bfseries 82} (2022) 428}
  [\href{https://arxiv.org/abs/2109.02653}{{\ttfamily 2109.02653}}].

\bibitem{Buckley:2014ana}
A.~Buckley, J.~Ferrando, S.~Lloyd, K.~Nordstr\"om, B.~Page, M.~R\"ufenacht
  et~al., \emph{{LHAPDF6: parton density access in the LHC precision era}},
  \href{https://doi.org/10.1140/epjc/s10052-015-3318-8}{\emph{Eur. Phys. J. C}
  {\bfseries 75} (2015) 132} [\href{https://arxiv.org/abs/1412.7420}{{\ttfamily
  1412.7420}}].

\bibitem{Diehl:2018kgr}
M.~Diehl, P.~Pl\"o\ss{}l and A.~Sch\"afer, \emph{{Proof of sum rules for double
  parton distributions in QCD}},
  \href{https://doi.org/10.1140/epjc/s10052-019-6777-5}{\emph{Eur. Phys. J. C}
  {\bfseries 79} (2019) 253}
  [\href{https://arxiv.org/abs/1811.00289}{{\ttfamily 1811.00289}}].

\bibitem{Ablinger:2017xml}
J.~Ablinger, J.~Bl\"umlein, A.~De~Freitas, C.~Schneider and K.~Sch\"onwald,
  \emph{{The two-mass contribution to the three-loop pure singlet operator
  matrix element}},
  \href{https://doi.org/10.1016/j.nuclphysb.2017.12.018}{\emph{Nucl. Phys. B}
  {\bfseries 927} (2018) 339}
  [\href{https://arxiv.org/abs/1711.06717}{{\ttfamily 1711.06717}}].

\bibitem{Blumlein:2018jfm}
J.~Bl\"umlein, A.~De~Freitas, C.~Schneider and K.~Sch\"onwald, \emph{{The
  Variable Flavor Number Scheme at Next-to-Leading Order}},
  \href{https://doi.org/10.1016/j.physletb.2018.05.054}{\emph{Phys. Lett. B}
  {\bfseries 782} (2018) 362}
  [\href{https://arxiv.org/abs/1804.03129}{{\ttfamily 1804.03129}}].

\bibitem{Binosi:2003yf}
D.~Binosi and L.~Theussl, \emph{{JaxoDraw: A Graphical user interface for
  drawing Feynman diagrams}},
  \href{https://doi.org/10.1016/j.cpc.2004.05.001}{\emph{Comput. Phys. Commun.}
  {\bfseries 161} (2004) 76}
  [\href{https://arxiv.org/abs/hep-ph/0309015}{{\ttfamily hep-ph/0309015}}].

\bibitem{Binosi:2008ig}
D.~Binosi, J.~Collins, C.~Kaufhold and L.~Theussl, \emph{{JaxoDraw: A Graphical
  user interface for drawing Feynman diagrams. Version 2.0 release notes}},
  \href{https://doi.org/10.1016/j.cpc.2009.02.020}{\emph{Comput. Phys. Commun.}
  {\bfseries 180} (2009) 1709}
  [\href{https://arxiv.org/abs/0811.4113}{{\ttfamily 0811.4113}}].

\end{thebibliography}\endgroup

\end{document}